\documentclass[useAMS,usenatbib]{mn2e}
\usepackage[dvips]{graphicx}

\newcommand{\bz}{$\langle B_z \rangle$}
\newcommand{\nz}{$\langle N_z \rangle$}

\newcommand{\kms}{km\,s$^{-1}$}

\newcommand{\vsini}{$v \sin i$}
\newcommand{\teff}{$T_{\rm eff}$}

\title[Magnetometry and Rotation of Early B-type Stars]{The Magnetic Early B-type Stars I: Magnetometry and Rotation}
\author[M.\ Shultz et al.]{M.\ E.\ Shultz$^{1,2,3,4}$\thanks{E-mail:
matthew.shultz@physics.uu.se},
G.\ A.\ Wade$^2$,
Th.\ Rivinius$^3$,
C.\ Neiner$^5$,
E.\ Alecian$^{5}$,
\newauthor{D.\ Bohlender$^6$, D.\ Monin$^6$, J.\ Sikora$^{1,2}$}
\newauthor{and the MiMeS and BinaMIcS Collaborations}
\footnotemark[1]\thanks{Based on observations obtained at the Canada-France-Hawaii Telescope (CFHT) which is operated by the National Research Council of Canada, the Institut National des Sciences de l'Univers of the Centre National de la Recherche Scientifique of France, and the University of Hawaii; and at the La Silla Observatory, ESO Chile with the MPA 2.2 m telescope.}\\
$^1$Department of Physics, Engineering Physics \& Astronomy, Queen's University, Kingston, ON Canada, K7L 3N6 \\
$^2$Department of Physics, Royal Military College of Canada, Kingston, Ontario K7K 7B4, Canada\\
$^3$ESO - European Organisation for Astronomical Research in the Southern Hemisphere, Casilla 19001, Santiago 19, Chile\\
$^4$Department of Physics and Astronomy, Uppsala University, Box 516, Uppsala 75120 \\
$^5$LESIA, Observatoire de Paris, PSL Research University, CNRS, Sorbonne Universités, UPMC Univ. Paris 06, Univ. Paris Diderot,\\
Sorbonne Paris Cité, 5 place Jules Janssen, 92195 Meudon, France\\
$^6$National Research Council of Canada, Herzberg Institute of Astronomy and Astrophysics, 5071 West Saanich Road, Victoria, BC V9E 2E7,\\
}
\begin{document}

\date{}

\pagerange{\pageref{firstpage}--\pageref{lastpage}} \pubyear{2002}

\maketitle

\label{firstpage}

\begin{abstract}
The rotational and magnetic properties of many magnetic hot stars are poorly characterized, therefore the MiMeS and BinaMIcS collaborations have collected extensive high-dispersion spectropolarimetric datasets of these targets. We present longitudinal magnetic field measurements \bz~for 52 early B-type stars (B5 to B0), with which we attempt to determine their rotational periods $P_{\rm rot}$. Supplemented with high-resolution spectroscopy, low-resolution DAO circular spectropolarimetry, and archival Hipparcos photometry, we determined $P_{\rm rot}$ for 10 stars, leaving only 5 stars for which $P_{\rm rot}$ could not be determined. Rotational ephemerides for 14 stars were refined via comparison of new to historical magnetic measurements. The distribution of $P_{\rm rot}$ is very similar to that observed for the cooler Ap/Bp stars. We also measured \vsini~and $v_{\rm mac}$ for all stars. Comparison to non-magnetic stars shows that \vsini~is much lower for magnetic stars, an expected consequence of magnetic braking. We also find evidence that $v_{\rm mac}$ is lower for magnetic stars. LSD profiles extracted using single-element masks revealed widespread, systematic discrepancies in \bz~between different elements: this effect is apparent only for chemically peculiar stars, suggesting it is a consequence of chemical spots. Sinusoidal fits to H line \bz~measurements (which should be minimally affected by chemical spots), yielded evidence of surface magnetic fields more complex than simple dipoles in 6 stars for which this has not previously been reported; however, in all 6 cases the second- and third-order amplitudes are small relative to the first-order (dipolar) amplitudes. 
\end{abstract}


\section{Introduction}

Magnetic fields are detected in approximately 10\% of stars with spectral types earlier than about A5, most prominently the chemically peculiar Ap/Bp stars and the Of?p stars \citep[e.g.][]{grun2012c,2016MNRAS.456....2W,2017MNRAS.465.2432G}. In contrast to the magnetic fields of cool stars, which tend to display complex surface topologies as well as magnetic activity cycles associated with dynamos, the magnetic fields of early-type stars tend to be strong (several kG), stable over timespans of up to decades, and topologically simple i.e.\ primarily dipolar \citep{dl2009}. Their striking observational properties, in combination with recent advances in the theoretical understanding of the behaviour of magnetic fields in stellar radiative zones \citep[e.g.][]{2004Natur.431..819B,2008MNRAS.386.1947B,2009MNRAS.397..763B,2005AA...440..653M,2010ApJ...724L..34D}, strongly support the hypothesis that massive star magnetic fields are fossil fields, i.e.\ relics of an earlier stage in stellar evolution \citep[e.g.][]{2015IAUS..305...61N,2017arXiv170510650A}. 

Until recently fossil fields were only known to exist in Ap stars and in a few He-weak and He-strong Bp stars: the first magnetic field detection in a chemically normal B-type star was for $\beta$ Cep \citep{henrichs2000}, while knowledge of Ap star magnetism dates to the beginning of stellar magnetometry \citep{babcock1947}, and the magnetic nature of He-weak and He-strong Bp stars has been known since the 1970s \citep{1976ApJ...203..171W,lb1978,1979ApJ...228..809B}. The advent of high-dispersion spectropolarimeters optimized for stellar magnetometry and mounted on 2- and 4-m class telescopes enabled the detection of fossil magnetic fields in a larger number of targets across a broader range of spectral types. The Magnetism in Massive Stars (MiMeS) survey systematically observed stars earlier than about spectral type B5 \citep{2016MNRAS.456....2W}. In addition to demonstrating that the Of?p stars are invariably magnetic \citep[e.g.][]{2010MNRAS.407.1423M,wade2011,wade2012a,2015MNRAS.447.2551W,2017MNRAS.465.2432G}, the MiMeS survey also detected magnetic fields in a large number of B-type stars. The MiMeS large programs were followed up by the Binarity and Magnetic Interactions in various classes of Stars (BinaMIcS) and the B-fields in OB stars (BOB) large programs, which respectively searched for magnetic fields in close binary systems containing at least one hot star, and for relatively weak magnetic fields \citep{2015IAUS..307..330A,2015A&A...582A..45F,2017A&A...599A..66S}. 

Many hot, magnetic stars display variable emission in H Balmer, Paschen, and Brackett lines \citep[e.g.][]{1974ApJ...191L..95W,grun2012,2015A&A...578A.112O}, UV resonance lines \citep{barker1982,smithgroote2001}, hard and overluminous X-rays \citep{1997ApJ...478L..87G,oskinova2011,naze2014,2017MNRAS.467.2820L}, and radio emission \citep{1987ApJ...322..902D,1992ApJ...393..341L,2015MNRAS.452.1245C,2017MNRAS.467.2820L}. These emissions originate in their circumstellar magnetospheres, which are classified as either dynamical or centrifugal magnetospheres depending on whether rotational support plays a negligible or dominant role in shaping the corotating, magnetically confined stellar wind \citep{petit2013}. In dynamical magnetospheres plasma returns to the photosphere on dynamical timescales due to gravitational infall, thus they show emission only when the mass-loss rate is high enough to replenish the magnetosphere at the same rate that it is emptied due to gravity. Therefore in general the only stars with dynamical magnetospheres in emission are the magnetic O-type stars \citep{petit2013}. In centrifugal magnetospheres, rotational support of the corotating plasma enables accumulation of plasma over longer timescales; thus the magnetic B-type stars, which have weaker stellar winds, generally show emission only when they have centrifugal magnetospheres \citep{petit2013}. While this classification scheme has been successful in explaining the qualitative differences in the emission properties of magnetic OB stars \citep{petit2013}, our understanding of these systems is hampered by substantial uncertainty regarding the rotational and magnetic properties of magnetic early-type stars, as a large number of these stars are fairly recent magnetic detections. Precise characterization of rotational and magnetic properties is particularly important for the magnetic B-type stars, which due to their relatively weak winds should only show magnetospheric emission when rapid rotation enables the formation of a centrifugal magnetosphere \citep{town2005c,ud2006,petit2013}. 

In this paper (hereafter Paper I), we present an extensive new database of high spectral resolution circular spectropolarimetry for the magnetic early B-type stars, with which (supplemented with low-resolution spectropolarimetry, high-resolution spectroscopy, and archival photometry) we attempt to determine their rotational periods, measure line broadening parameters, and characterize the properties of their longitudinal magnetic field curves. These measurements will form the empirical underpinning of the oblique rotator models and magnetospheric parameters, which will be presented in Paper II. \S~2 gives an overview of the observing programs and observations. \S~3 outlines the magnetic measurements, conducted using multi- and single-element least-squares deconvolution profiles, and with H lines. In \S~4 line broadening parameters are measured and rotational periods determined. \S~5 examines the longitudinal magnetic field curves with the goal of identifying stars with significant contributions from higher-order multipoles to their surface magnetic field topologies, and discusses the distributions of rotational properties in comparison to non-magnetic stars of similar spectral type, and to the magnetic stars of other spectral types. Conclusions and next steps are outlined in \S~6. Periodograms and longitudinal magnetic field curves of individual stars are given in the Appendix.


\section{Observations}

\begin{table*}
\caption[Summary of the sample stars and available observations.]{ Summary of the sample stars and available spectropolarimetric, spectroscopic, and photometric observations. Stars are listed in order of HD number, with the exception of ALS 3694 which is not in the Henry Draper catalogue. The same observations were used for HD 136504A and HD136504B, therefore these are not counted twice in the table. Remarks indicate the type of chemical peculiarity (He-strong or He-weak), the star's binary status (SB1/2/3), and/or pulsational variability ($\beta$ Cep or SPB). The next four columns give the number of high-dispersion spectropolarimetric observations obtained with ESPaDOnS (E), Narval (N), HARPSpol (H), and the total number (T) of magnetic measurements for each star, once low-signal to noise (S/N) observations have been removed and spectral binning has been performed. The next column gives the number of DAO (D) observations obtained with the low-dispersion dimaPol spectropolarimeter. The number of spectroscopic observations obtained with FEROS (F) are given in the next column. The final column gives the number of archival Hipparcos (Hip) photometric measurements.}
\centering
\begin{tabular}{l l l l | r r r r r | r | r | r}  
\label{obstab}
HD No. & Star & Spec. Type & Remarks & E & N & H & T & $\langle$S/N$\rangle$ & D & F & Hip \\
\hline
  3360 & $\zeta$\,Cas       & B2\,IV              & SPB &  -- & 56 &  -- & 56 & 1771 &  1 &  -- & 194 \\
 23478 &   & B3\,IV & He-s & 14 &  -- &  -- & 10 &  555 &  11 &  -- &  74 \\
 25558 & 40\,Tau            & B3\,V               & SB2, SPB & 12 & 19 &  -- & 31 &  994 &  -- &  -- &  90 \\
 35298 &                    & B3\,Vw              & He-w &  10 &  2 &  -- &  12 &  541 & 16 &  -- & 119 \\
 35502 &                    & B5\,V               & SB3, He-w &  4 & 20 &  -- & 24 &  486 &  24 & 33 &  98 \\
 35912 & HR 1820            & B2/3\,V             &           &  12 & -- & -- & 6 & 317 & 3 & -- & 130 \\
 36485 & $\delta$\,Ori\,C   & B3\,Vp              & SB2, He-s & 13 &  -- &  -- & 11 &  407 &  10 & 10 &  -- \\ 
 36526 & V1099\,Ori         & B8\,Vp              & He-w & 11 &  -- &  -- & 10 &  318 &  14 &  -- &  -- \\ 
 36982 & LP\,Ori            & B1.5\,Vp            & HeBe, He-s & 18 &  5 &  -- & 14 &  471 &  -- &  -- &  -- \\ 
 37017 & V\,1046\,Ori       & B1.5-2.5\,IV-Vp     & SB2, He-s & 10 &  -- &  -- &  8 &  854 &  -- & 33 &  95 \\
 37058 & V\,359\,Ori        & B3\,VpC             & He-w & 15 &  -- &  -- & 10 &  597 &  3 &  -- &  -- \\ 
 37061 & NU\,Ori            & B0.5\,V             & SB2 & 24 &  -- &  -- & 11 & 1132 & 1 &  -- &  97 \\
 37479 & $\sigma$\,Ori\,E   & B2\,Vp              & He-s &  2 & 16 &  -- & 18 &  670 &  -- & 36 &  -- \\ 
 37776 & V\,901\,Ori        & B2\,Vp              & He-s & 13 & 13 &  -- & 26 &  598 &  3 &  -- & 103 \\
 44743 & $\beta$ CMa  & B1\,II/III & $\beta$ Cep &  -- &  -- &  11 &  5 &  700 &  -- &  -- &  108 \\ 
 46328 & $\xi^1$\,CMa	    & B1\,III             & $\beta$ Cep & 56 &  -- &  -- & 54 &  928 &  -- &  -- &  221 \\ 
 52089 & $\epsilon$ CMa  & B1.5\,II & 4 &  -- &  -- &  8 &  4 &  550 &  -- &  -- &  149 \\ 
 55522 & HR\,2718           & B2\,IV/V            & He-s &  9 &  -- &  -- &  9 &  596 &  -- &  -- & 224 \\
 58260 &                    & B3\,Vp              & He-s &  9 &  -- &  -- &  9 &  438 &  -- &  -- & 125 \\
 61556 & HR\,2949           & B5\,V               & He-w & 41 &  -- &  -- & 22 & 1227 &  -- &  6 & 282 \\
 63425 &                    & B0.5\,V             & SB1? & 11 &  -- &  -- & 11 &  648 &  -- &  -- & 111 \\
 64740 & HR\,3089	    & B1.5\,Vp            & He-s &  4 &  -- & 13 & 17 &  346 &  -- & 21 &  99 \\
 66522 &                    & B2\,III             & He-s &  1 &  -- &  4 &  5 &  296 &  -- &  -- & 117 \\
 66665 &                    & B0.5\,V             & -- &  5 & 16 &  -- & 21 &  414 &  -- &  -- &  49 \\
 66765 &                    & B1/B2\,V            & He-s &  1 &  -- & 10 &  8 &  449 &  -- & 11 & 117 \\
 67621 &                    & B2\,IV              & He-w &  1 &  -- &  8 &  6 &  497 &  -- &  -- & 133 \\
 96446 & V\,430\,Car	    & B1\,IVp/B2\,Vp      & $\beta$ Cep, He-s &  -- &  -- & 10 & 10 &  271 &  -- &  -- & 107 \\
105382 & HR\,4618           & B6\,III             & He-w &  -- &  -- &  3 &  3 &  939 &  -- &  -- & 175 \\
121743 & $\phi$\,Cen        & B2\,IV              & $\beta$ Cep, He-w &  8 &  -- &  7 & 15 &  629 &  -- &  -- &  76 \\
122451 & $\beta$\,Cen       & B1                  & SB2, $\beta$ Cep &  -- &  -- & 283 &  14 & 1802 &  -- &  -- & 119 \\
125823 & a\,Cen             & B7\,IIIp            & He-w & 17 &  -- &  -- & 17 &  708 &  -- &  -- & 101 \\
127381 & $\sigma$\,Lup      & B1\,V               & He-s & 20 &  -- &  -- & 20 &  919 &  -- &  -- & 158 \\
130807 & $o$\,Lup           & B5                  & He-w,SB1? &  2 &  -- & 10 & 12 &  564 &  -- &  -- &  56 \\
136504 & $\epsilon$\,Lup    & B2\,IV-V            & SB2, $\beta$ Cep & 91 &  -- &  -- & 14 & 2748 &  -- &  -- &  88 \\
136504 & $\epsilon$\,Lup B  & B2\,IV-V & SB2, $\beta$ Cep &  -- &  -- &  -- &  -- & -- &  -- &  -- &  -- \\ 
142184 & HR\,5907           & B2\,V               & He-s & 27 &  -- &  -- & 27 & 1080 &  -- &  -- &  83 \\
142990 & V\,913\,Sco        & B5\,V               & He-w &  16 &  -- &  -- &  15 & 1176 &  -- &  -- & 111 \\
149277 &                    & B2\,IV/V            & SB2,He-s & 23 &  -- &  -- & 23 &  320 &  -- &  -- &  65 \\
149438 & $\tau$\,Sco        & B0.2\,V             & -- &  -- &  -- &  -- &  0 &    0 &  -- &  -- &  -- \\ 
156324 &                    & B2\,V               & SB3,He-s & 21 &  -- &  3 & 21 &  314 &  -- & 11 &  -- \\ 
156424 &                    & B2\,V               & He-s, SB1$/\beta$ Cep &  9 &  -- &  3 & 11 &  265 &  -- & 12 &  -- \\ 
163472 & V\,2052\,Oph       & B1/B2\,V            & $\beta$ Cep &  -- & 13 &  -- & 13 &  631 &  -- &  -- &  66 \\
164492C & EM* LkHA 123      & B1.5\,V & SB3 & 17 &  -- &  6 & 23 &  461 &  -- &  8 &  -- \\ 
175362 & Wolff star         & B5\,V               & He-w & 24 &  -- &  -- & 23 &  765 &  -- &  -- &  89 \\
176582 & HR\,7185           & B5\,IV              & He-w &  -- & 44 &  -- & 43 &  865 &  19 &  -- & 112 \\
182180 & HR\,7355           & B2\,Vn              & He-s &  4 &  -- &  -- &  4 & 1426 &  -- &  -- &  46 \\
184927 & V\,1671\,Cyg       & B2\,Vp              & He-s & 24 &  -- &  -- & 21 &  562 & 11 &  -- & 161 \\
186205 &                    & B2\,Vp              & He-s & 15 &  -- &  -- & 10 &  221 &  7 &  -- & 226 \\
189775 & HR\,7651           & B5\,V               & He-w & 11 &  15  &  -- & 26 &  698 & 14 &  -- & 123 \\
205021 & $\beta$\,Cep       & B1\,IV              & SB2, $\beta$ Cep &  -- & 21 &  -- & 20 & 1439 &  -- &  -- & 120 \\
208057 & 16\,Peg            & B3\,V               & SPB & 13 &  8 &  -- & 21 & 1143 &  -- &  -- &  88 \\
       & ALS 3694 & B1      & He-s & 16 &  -- &  -- &  6 &  145 &  -- & 13 &  -- \\ 
\hline\hline
\end{tabular}
\end{table*}

\subsection{Sample Selection}

The sample consists of those magnetic main sequence B-type stars earlier than B5 identified by P13, for which sufficient spectroscopic and spectropolarimetric data are available to evaluate their rotational and magnetic properties. Three stars with reported spectral types later than B5 (HD 36526, HD 105382, and HD 125823) are also included, as their effective temperatures are above 15 kK. This is a consequence of surface chemical abundance peculiarities affecting spectral type determinations, with He-weak stars in particular having, as their name implies, weaker He lines than expected for their effective temperatures. Since He lines are the primary diagnostic for spectral typing amongst hot stars, the spectral types assigned to He-weak stars are systematically later than would be implied by their effective temperatures. We include 5 additional stars, discovered to be magnetic since the P13 sample was published: HD 23478 (B3 IV, \citealt{2015MNRAS.451.1928S,2015A&A...578L...3H}); the secondary star of the HD 136504 system ($\epsilon$ Lupi, B2 IV/V, \citealt{2015MNRAS.454L...1S}), in which the primary was already known to be magnetic \citep{2009AN....330..317H,shultz2012}; HD 164492C (B1.5 V), HD 44743 ($\beta$ CMa, B1 II/III) and HD 52089 ($\epsilon$ CMa, B1.5 II), discovered by the BOB collaboration \citep{2014AA...564L..10H,2015AA...574A..20F}. While HD 52089 was initially reported to be in the post main-sequence phase of its evolution, analysis of its stellar parameters by \cite{2017MNRAS.471.1926N} has shown that it is likely still in the core hydrogen burning phase. In total the initial sample consists of 52 early B-type stars reported to host magnetic fields. 

The sample is summarized in Table \ref{obstab}. Stars are listed in order of their HD number; ALS 3694, which does not appear in the Henry Draper catalogue, is listed last. The 2$^{nd}$ column gives alternate designations. The 3$^{rd}$ column gives the spectral type and luminosity class. Remarks as to chemical peculiarity (He-weak or -strong), binarity (SB1/2/3), and/or pulsation ($\beta$ Cep or Slowly Pulsating B-type star, SPB) are made in the 4$^{th}$ column. The remaining columns provide the number of spectropolarimetric, spectroscopic, and photometric observations available for each target. 

As a comprehensive sample drawn from the literature, this study is neither volume nor magnitude limited. The statistical properties of the MiMeS survey were summarized by \cite{2016MNRAS.456....2W}. The MiMeS survey was complete for all OB stars up to $V\sim1$, 50\% complete up to $V\sim3$, and overall observed 7\% of the OB stars with $V<8$. In spectral type, completeness was highest for the earliest stars (70\% of O4 stars with $V<8$), declining towards later spectral types. For the B-type stars, the sample is 30\% complete at B0, diminishing to about 15\% at B5: thus, the sample is most complete for the least common spectral types. However, the sample includes essentially all known and confirmed magnetic stars with spectral types between B5 and B0. 

\subsection{Observing programs}

The majority of the observations used in this paper were acquired under the auspices of the MiMeS Large Programs (LPs) at the 3.6~m Canada-France-Hawaii Telescope (CFHT) using ESPaDOnS, the 2~m Bernard Lyot Telescope (TBL) using Narval, and the ESO La Silla Observatory 3.6~m Telescope using HARPSpol. Five spectroscopic binaries in the sample (HD 35502, HD 136504, HD 149277, HD 156324, and HD 164492C) have also been observed by the BinaMIcS LPs at CFHT and TBL. The remainder of the high-resolution spectropolarimetric data were acquired by various programs using ESPaDOnS, Narval, and HARPSPol at CFHT\footnote{Program codes CFHT 13BC012, CFHT 14AC010, CFHT 17AC16, PI M. Shultz; CFHT 14BC011, PI J. Sikora.}, TBL, and the ESO La Silla 3.6~m telescope, as well some data collected via the BRITEpol LPs at CFHT, TBL, and ESO \citep{neiner2017ppas}. The total numbers of ESPaDonS, Narval, and HARPSpol observations for each star are given in the 5$^{th}$ to 7$^{th}$ columns of Table \ref{obstab}. 

The methodology, observing strategy, instrumentation, and scope of the MiMeS LPs were described in detail by \cite{2016MNRAS.456....2W}. The BinaMIcS LPs have largely adopted the strategies of the MiMeS LPs. 

While the MiMeS LPs ended in 2012, only limited data were available for several magnetic stars identified by the Survey Component. Therefore the Targeted Component was extended in two CFHT/ESPaDOnS PI programs in 2013 and 2014. In total 1159 spectropolarimetric observations were gathered (614 ESPaDOnS, 248 Narval, and 297 HARPSpol). After removal of low signal-to-noise ratio (S/N) observations and, where appropriate and necessary, binning of individual spectra obtained in close temporal proximity, the high-resolution magnetic dataset consists of 792 individual magnetic measurements with a mean of 15 measurements per star. 

In addition to the high-resolution circular spectropolarimetry that forms the core of the dataset, we have also utilized several supplementary datasets for certain stars. For 14 stars we have obtained low-resolution circular spectropolarimetry with the dimaPol instrument at the 1.8~m Plaskett Telescope (137 observations). For 11 stars we have high-resolution spectroscopy from FEROS at the MPG La Silla Observatory 2.2~m telescope, both gathered from the ESO archive, and (for 8 stars) presented here for the first time\footnote{Program codes MPIA 092.A-9018(A) and MPIA LSO22-P95-007, PI M. Shultz.}. Finally, we have downloaded epoch photometry from the Hipparcos archive, where available. 

\subsection{Spectropolarimetry}

ESPaDOnS (Echelle SpectroPolarimetric Device for the Observation of Stars) is a high-spectral resolution spectropolarimeter with a resolving power of $\lambda/\Delta\lambda \sim 65,000$ at 500~nm and a spectral range of 370 to 1050 nm across 40 spectral orders (e.g.\ \citealt{1993ASPC...40..136D}). Narval is an identical instrument in all important respects. ESPaDOnS and Narval data were reduced using Libre-ESPRIT \citep{d1997}. HARPSpol has a greater spectral resolving power than ESPaDOnS and Narval ($\lambda/\Delta\lambda \sim 100,000$), and a narrower spectral range, 378 to 691 nm, with a gap between 524 and 536 nm, across 71 spectral orders \citep{2003Msngr.114...20M,2011Msngr.143....7P}. HARPSpol data were reduced using a modified version of the {\sc reduce }reduction pipeline \citep{2002A&A...385.1095P,2011A&A...525A..97M}. The excellent agreement between observations obtained with ESPaDOnS and Narval was demonstrated by \cite{2016MNRAS.456....2W}, who also describe the reduction of ESPaDOnS, Narval, and HARPSpol data in detail. The new data acquired after 2013 have been processed in the same fashion as described by \cite{2016MNRAS.456....2W}. The agreement between ESPaDOnS/Narval and HARPSpol is less well studied, but comparisons have been performed for the magnetic B1 star HD 164492C by \cite{2017MNRAS.465.2517W}, and for the late-type Bp star HD 133880 by \cite{2017A&A...605A..13K}, both finding very good agreement between the instruments. 




The detection and measurement of stellar magnetic fields requires data with a very high S/N. Therefore, in some cases spectra were removed from the analysis when the S/N was insufficient to obtain a meaningful measurement (typically, when the maximum S/N per spectral pixel ${\rm S/N_{max}} \le 100$, below which the uncertainty in the longitudinal magnetic field is generally on the order of several kG). In other cases, if the time difference between observations could reasonably be expected to be small compared to either the known rotational period or the minimum rotational period inferred from the projected rotational velocity, spectra acquired on the same night were binned in order to increase the S/N (e.g. HD 136504, for which each measurement consists of at least 4 spectropolarimetric sequences). Especially noisy observations were not included in such binning. The final number of measurements used for magnetic analysis is listed in the 8$^{th}$ column of Table \ref{obstab}. The 9$^{th}$ column gives the median peak S/N per spectral pixel in the final dataset, $\langle{\rm S/N_{max}}\rangle$. The data quality is in general high, with a median S/N across all spectropolarimetric sequences of 612. The log of all spectropolarimetric observations is provided online.

\cite{2016MNRAS.456....2W} noted that data collected with Narval during the late summers of 2011 and 2012 were affected by occasional, random loss of control of a Fresnel rhomb. While this issue cannot produce spurious magnetic signatures, the accuracy of magnetic measurements conducted with these spectra cannot be trusted. Therefore, spectra from the time windows given by Wade et al. were excluded from magnetic analysis. Only two observations within these epochs, of HD 176582 acquired on 24 Aug 2011 and HD 35502 on 21 Sept 2012, show obvious inconsistencies with the expected \bz~variations. Another outlier, outside the windows given by Wade et al., was found for $\beta$ Cep on 04 Oct 2009; this measurement was also discarded. 

In addition to the high spectral resolution spectropolarimetry, 137 low spectral resolution measurements collected with the dimaPol spectropolarimeter mounted on the 1.8 m Dominion Astrophysical Observatory (DAO) Plaskett Telescope are available for 14 stars. The number of DAO observations are given in the 10$^{th}$ column of Table \ref{obstab}. dimaPol has a spectral resolution of approximately 10,000, covering a 25 nm region centred on the rest wavelength of the H$\beta$ line. Observations for HD 23478, HD 35502, HD 176582, and HD 184927 have been reported by \cite{2015MNRAS.451.1928S}, \cite{2016MNRAS.460.1811S}, \cite{bohl2011}, and \cite{2015MNRAS.447.1418Y}, respectively. The remainder are presented here for the first time. Unpublished measurements are also available online. The instrument and reduction pipeline are described in detail by \cite{2012PASP..124..329M}.

\subsection{Spectroscopy}


Eleven stars, mostly with optical emission originating in their magnetospheres were observed using the FEROS spectrograph at the MPG La Silla 2.2~m telescope. FEROS is a high-dispersion echelle spectrograph, with $\lambda/\Delta\lambda\sim$48,000 and a spectral range of 375--890 nm \citep{1998SPIE.3355..844K}. The data were reduced using the standard FEROS Data Reduction System MIDAS scripts\footnote{Available at https://www.eso.org/sci/facilities/lasilla/ \\ instruments/feros/tools/DRS.html}. The number of FEROS spectra are summarized in the 11$^{th}$ column of Table \ref{obstab}. 

\subsection{Photometry}

Hipparcos (High precision parallax collecting satellite) was an astrometric space telescope, whose mission lasted from 1989 to 1993. While the primary aim was to obtain high-precision trigonometric parallaxes, it also obtained photometry for a large number of stars. These data are available for 36 of the sample stars. As chemically peculiar stars exhibit photometric variability due to surface chemical abundance spots, in some cases rotational periods can be determined using Hipparcos photometry. Photometric variability may arise due to other physical mechanisms, e.g. pulsation, but in such cases the photometric and magnetic data should not phase coherently. These data were acquired from the online archive \citep{perry1997,vanleeuwen2007}. The number of Hipparcos measurements is given in the final column of Table \ref{obstab}.

\section{Magnetometry}\label{Magnetometry}

   \begin{figure}
   \centering
   \includegraphics[width=8.5cm]{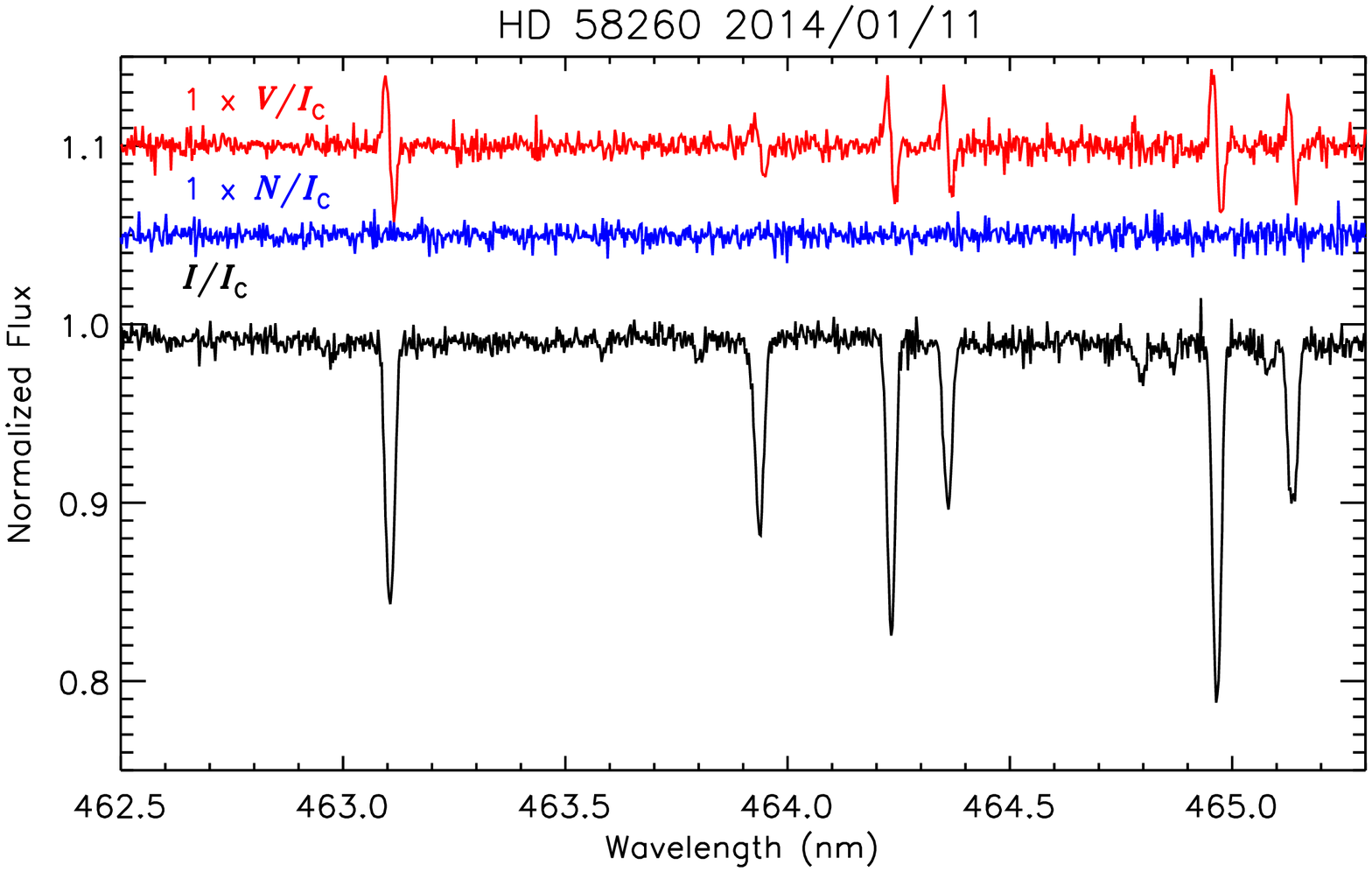}
   \includegraphics[width=8.5cm]{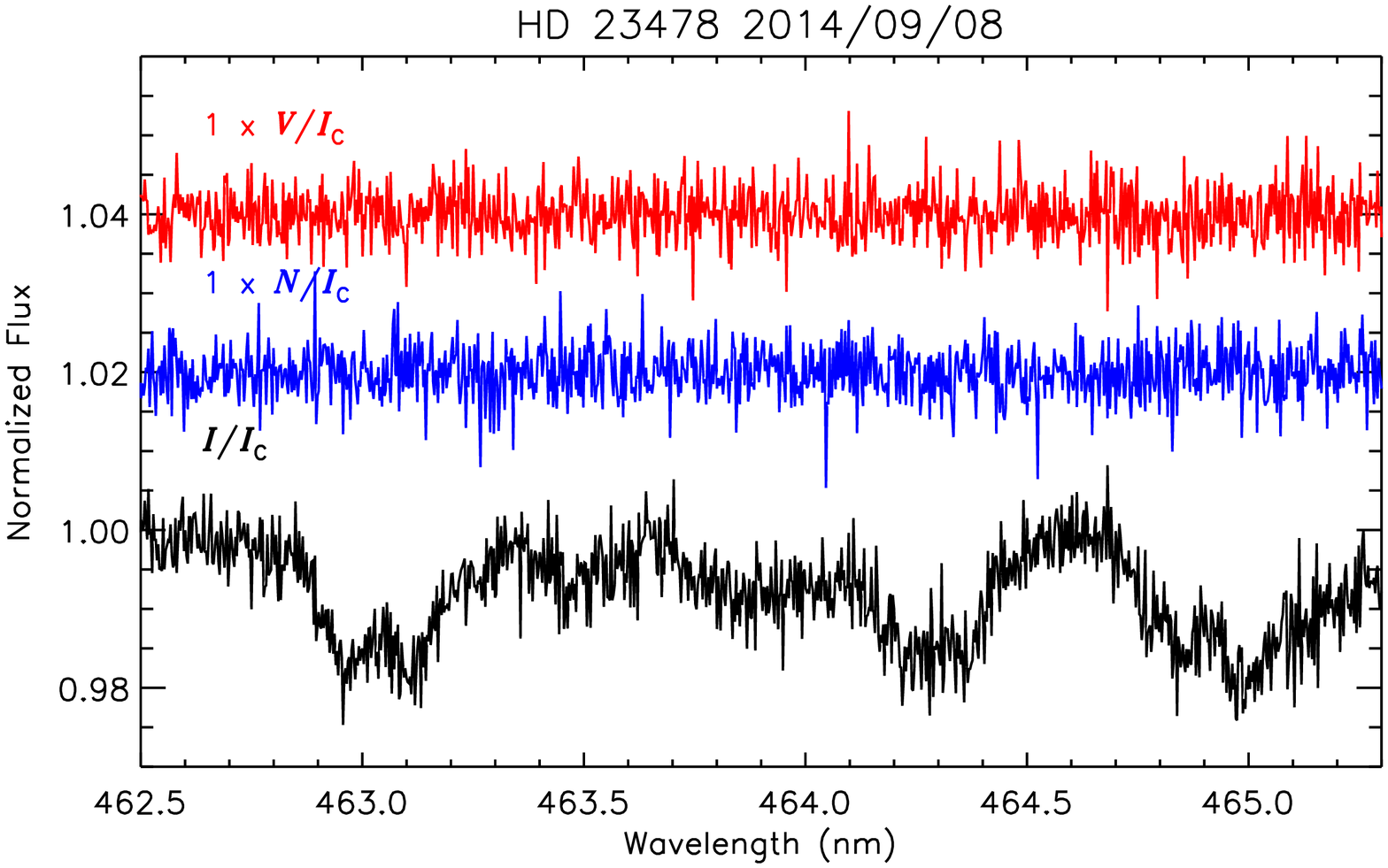}
      \caption[]{Stokes $I$, diagnostic null $N$, and Stokes $V$ in a representative spectral window for the B3Vp star HD 58260 ({\em top}) and for the B3V star HD 23478 ({\em bottom}). Zeeman signatures are clearly detectable in individual spectral lines of HD 58260, due to the star's sharp spectral lines and strong magnetic field. However, despite the high S/N Zeeman signatures are not detectable in the individual lines of HD 23478, which has a similarly strong magnetic field but broad spectral lines. Compare the Stokes $V$ profiles in this figure to the LSD Stokes $V$ profiles for the same stars in Fig.\ \ref{lsd_allplot_1}.}
         \label{vspecplot}
   \end{figure}

Zeeman signatures are visible in the Stokes $V$ profiles of individual spectral lines of many of the sample stars, as illustrated in Fig.\ \ref{vspecplot} for the case the HD 58260 (top). This is particularly the case for sharp-lined stars with strong magnetic fields, as is true for HD 58260. However, many of the stars in the sample have much broader spectral lines in which Zeeman signatures are intrinsically more difficult to detect, such as HD 23478, which has the same spectral type as HD 58260 (Fig.\ \ref{vspecplot}, bottom). Both stars have longitudinal magnetic fields $|\langle B_z\rangle|\sim 1$~kG \citep{1987ApJ...323..325B,2015MNRAS.451.1928S}. In such cases multi-line analysis methods such as least-squares deconvolution (LSD) are required in order to measure the star's magnetic field. 

   \begin{figure*}
   \centering
   \includegraphics[width=18.5cm]{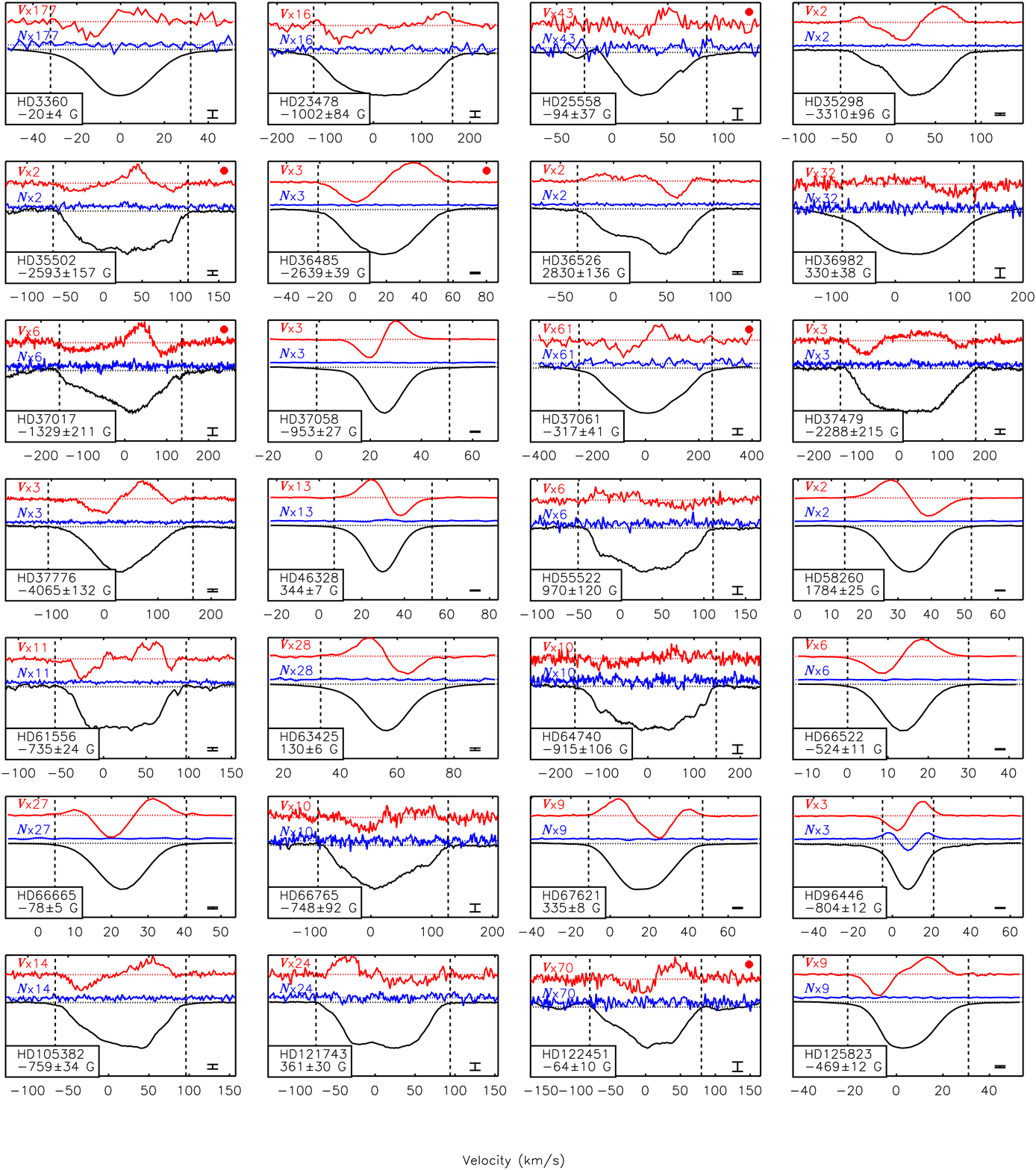}
      \caption[LSD profiles of individual stars.]{LSD profiles of individual stars. Stokes $I$ in black, Stokes $V$ in red, $N$ in blue. The amplification factor applied to $N$ and $V$ is indicated in the top left of each panel. Vertical dashed lines indicate the integration ranges for determination of the FAP and \bz. The mean error bar in $N$ and Stokes $V$, scaled to the amplification factor, is displayed in the lower right. The legend on the lower left gives the name of the star, and \bz: the LSD profiles shown here are those yielding the highest \bz~significance. Stokes $I$ profiles of binary stars corrected via disentangling are indicated by red circles in the upper right corner.}
         \label{lsd_allplot_1}
   \end{figure*}

   \begin{figure*}
   \centering
   \includegraphics[width=18.5cm]{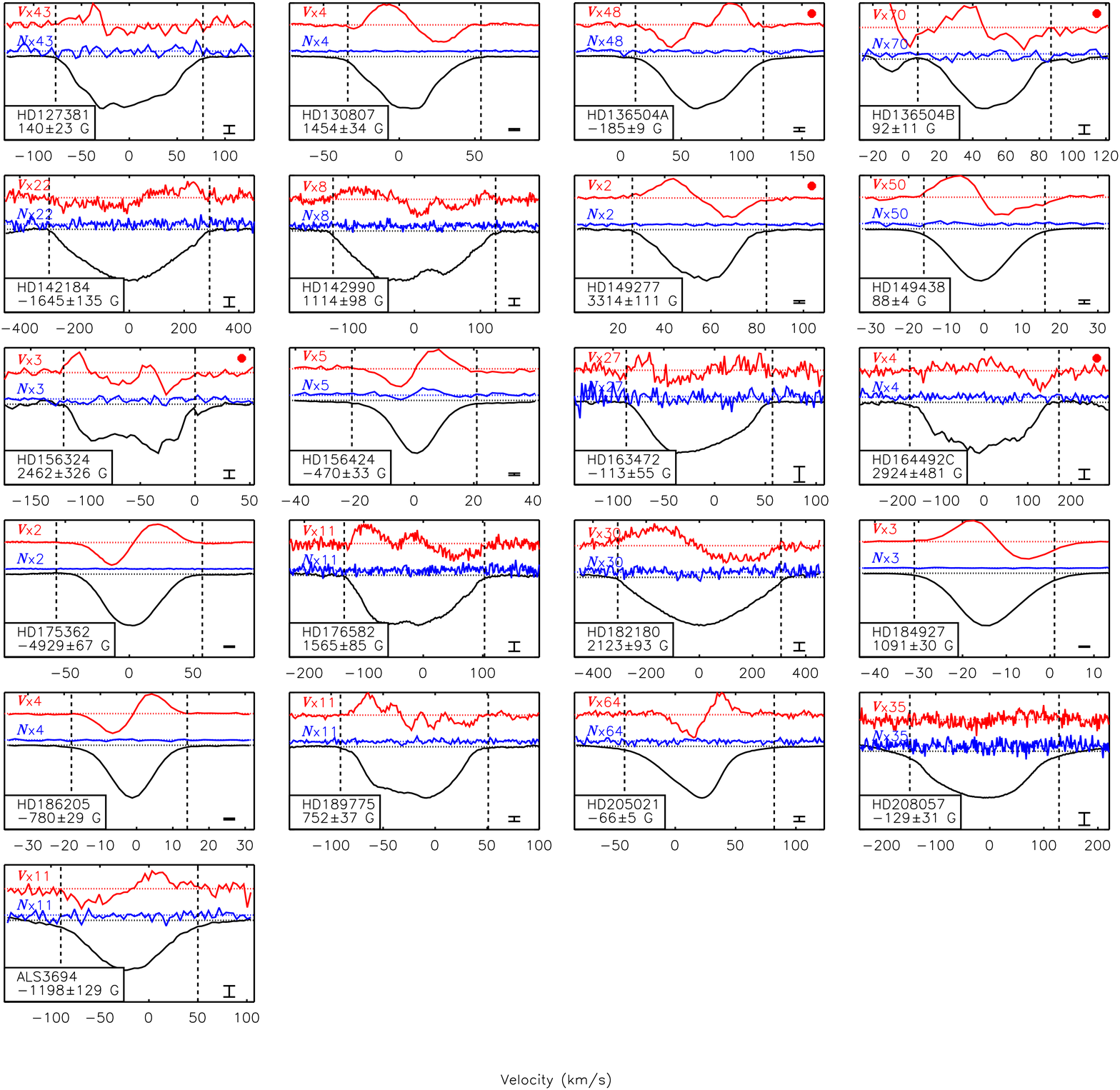}
      \caption[LSD profiles of individual stars, Cont'd.]{As Fig.\ \ref{lsd_allplot_1}. Note that the Stokes $V$ LSD profiles of HD 135604 were not disentangled, and thus a contribution from the Zeeman signature of HD 136504A is apparent in the blue edge of HD 136504B's Stokes $V$ profile.}
         \label{lsd_allplot_2}
   \end{figure*}


\subsection{Least squares deconvolution}\label{lsd}

We utilized the LSD procedure \citep{d1997} in order to maximize the per-pixel S/N of the polarization profiles. In particular, LSD was performed using the iLSD package \citep{koch2010}. Line lists were obtained from the Vienna Atomic Line Database\footnote{Available at http://vald.astro.uu.se/} (VALD3; \citealt{piskunov1995, ryabchikova1997, kupka1999, kupka2000}) using `extract stellar' requests, for effective temperatures from 15 kK to 30 kK in increments of 1 kK. 

To ensure a homogeneous analysis, the line lists were automatically cleaned by removing: H Balmer and Paschen lines; He lines with strongly pressure-broadened wings; lines blended with H or He lines; lines in spectral regions that are often significantly contaminated by telluric features, depending on the observing conditions; and lines in the spectral region affected by instrumental ripples between 618 and 628 nm. When the spectropolarimetric dataset included both HARPSpol and ESPaDOnS/Narval data, the line lists were further truncated by removing all lines outside of the smaller HARPSpol spectral range. Line masks were then customized to each star by adjusting the depths of the remaining lines (typically 200--300 lines) to the observed line depths (e.g.\ \citealt{2012PhDT.......265G}). While He lines were by default excluded as their pressure-broadened wings do not have the same shape as metallic lines, and therefore do not conform to the fundamental LSD hypothesis, in many cases (in particular the He-strong stars), including some He lines can greatly improve the S/N of the LSD profile and, hence, the detectability of the Stokes $V$ signature. Therefore we also extracted LSD profiles with line masks including the He~{\sc i} lines 416.9 nm, 443.8 nm, 471.3 nm, 501.6 nm, 504.8 nm, and 587.6 nm. Table \ref{masktab} gives the number of lines used to extract LSD profiles for each star, for masks including only metallic lines (Z), and masks including both metallic and He lines (YZ). Measurements using H lines (X) are also provided; these are described in further detail in \S~\ref{h_bz}. For HD 44743 and HD 52089 (for which we did not extract new LSD profiles, but rely on the data published by \citealt{2015AA...574A..20F} and \citealt{2017MNRAS.471.1926N}), Table \ref{masktab} gives the number of lines used by \cite{2015AA...574A..20F}. 


LSD profiles were extracted using wavelength, Land\'e factor, and line depth normalization constants of 500 nm, 1.2, and 0.1, respectively. Pixel velocity widths were set according to \vsini/40 rounded to the nearest 1.8 \kms~spectral pixel, with a minimum pixel width of 1.8 \kms~adopted for all ESPaDOnS, Narval, and HARPSpol spectra, in order to increase the S/N in Stokes $V$ in those stars with especially broad spectral lines. This pixel width was chosen based on the average pixel velocity width of ESPaDOnS and Narval data. The velocity range used for deconvolution was $\pm$2\vsini$+ v_{\rm sys}$, where $v_{\rm sys}$ is the systemic or central velocity, ensuring inclusion of the full spectral line while minimizing contamination in spectra of stars with narrow spectral features. For spectroscopic binary stars, velocity ranges were set using line width and radial velocity (RV) semi-amplitudes, which were determined from RV measurements performed using parametric fitting of individual line profiles using a proprietary {\sc idl} routine \citep{2017MNRAS.465.2432G}. Profiles were extracted for Stokes $I$, Stokes $V$, and both diagnostic null $N$ spectra. Representative LSD profiles for all stars except HD 44743, HD 52089 (for which we did not extract new LSD profiles), and HD 35912 (in which a magnetic field is not detected, see below) are shown in Figs.\ \ref{lsd_allplot_1} and \ref{lsd_allplot_2}. 

\begin{figure}
    \includegraphics[width=8.5cm]{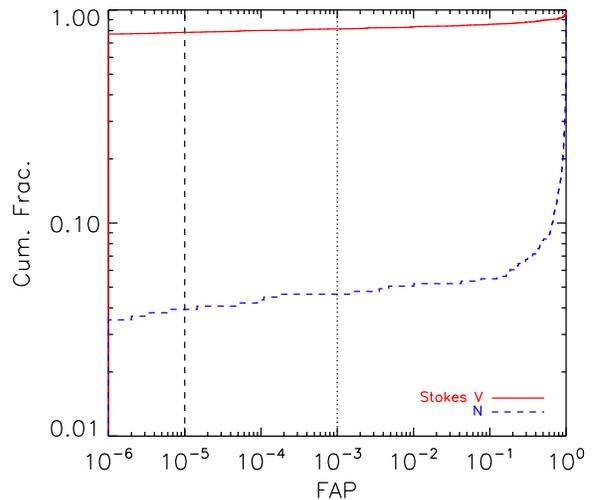}
    \caption[Cumulative distribution of False Alarm Probabilities]{Cumulative distribution of False Alarm Probabilities. The dashed (dotted) vertical line indicates the lower cutoff for definite (marginal) detections. $\sim$80\% of Stokes $V$ measurements are definite detections (DDs), with FAP$\le10^{-5}$, while $\sim$4\% of $N$ measurements are also DDs.}
 \label{fap_cum}
\end{figure}

As a first evaluation of the statistical significance of the magnetic signatures in the LSD Stokes $V$ profiles we calculated False Alarm Probabilities (FAPs) by comparing the signal inside the Stokes $V$ line profile to the signal in the wings \citep{1992AA...265..669D,d1997}, where the boundaries of the line profile were determined from $\pm$\vsini~adjusted to the rest frame of the star. Fig.\ \ref{fap_cum} shows the cumulative distribution function of FAPs for all Stokes $V$ profiles in the dataset, where for each star we used the LSD profiles extracted with the mask yielding the highest S/N. 78\% of the LSD profiles have ${\rm FAP}<10^{-5}$, the upper boundary for a formal definite detection (DD) \citep{d1997}. The remaining $\sim$20\% registering marginal detections (MD) with $10^{-5}<{\rm FAP}<10^{-3}$ or non-detections (ND) with ${\rm FAP}>10^{-3}$ are accounted for by the presence in the dataset of weak Zeeman signatures relative to the S/N. 




Approximately 4\% of $N$ LSD profiles also register a DD (Fig.\ \ref{fap_cum}, blue line). This is due to stars exhibiting rapid radial velocity variations over short time-spans compared to the exposure times. \cite{neiner2012a} noted the strong signal in the $N$ profile of the $\beta$ Cep star HD 96446, and we confirm its presence in all spectra in the dataset (Fig.\ \ref{lsd_allplot_1}). Also showing DDs in some $N$ spectra are HD 46328 (Fig.\ \ref{lsd_allplot_1}) and HD 205021 (Fig.\ \ref{lsd_allplot_2}), both $\beta$ Cep stars; HD 156324, a short-period binary (Fig.\ \ref{lsd_allplot_2}); and HD 156424, which is not listed as either a binary or a $\beta$ Cep variable, but does exhibit radial velocity variations (Fig.\ \ref{lsd_allplot_2}). The $N$ signatures in HD 156324 and HD 156424 were noted by \cite{alecian2014}. Such $N$ signatures can be reproduced by including RV variations between sub-exposures \citep{neiner2012a,2017MNRAS.471.2286S}. Correcting for rapid RV variations in the reduction stage does not result in changes to \bz~larger than the error bars \citep{neiner2012a,2017MNRAS.471.2286S}. Therefore RV variability should have no impact on \bz. 

\begin{figure}
    \includegraphics[width=8.5cm]{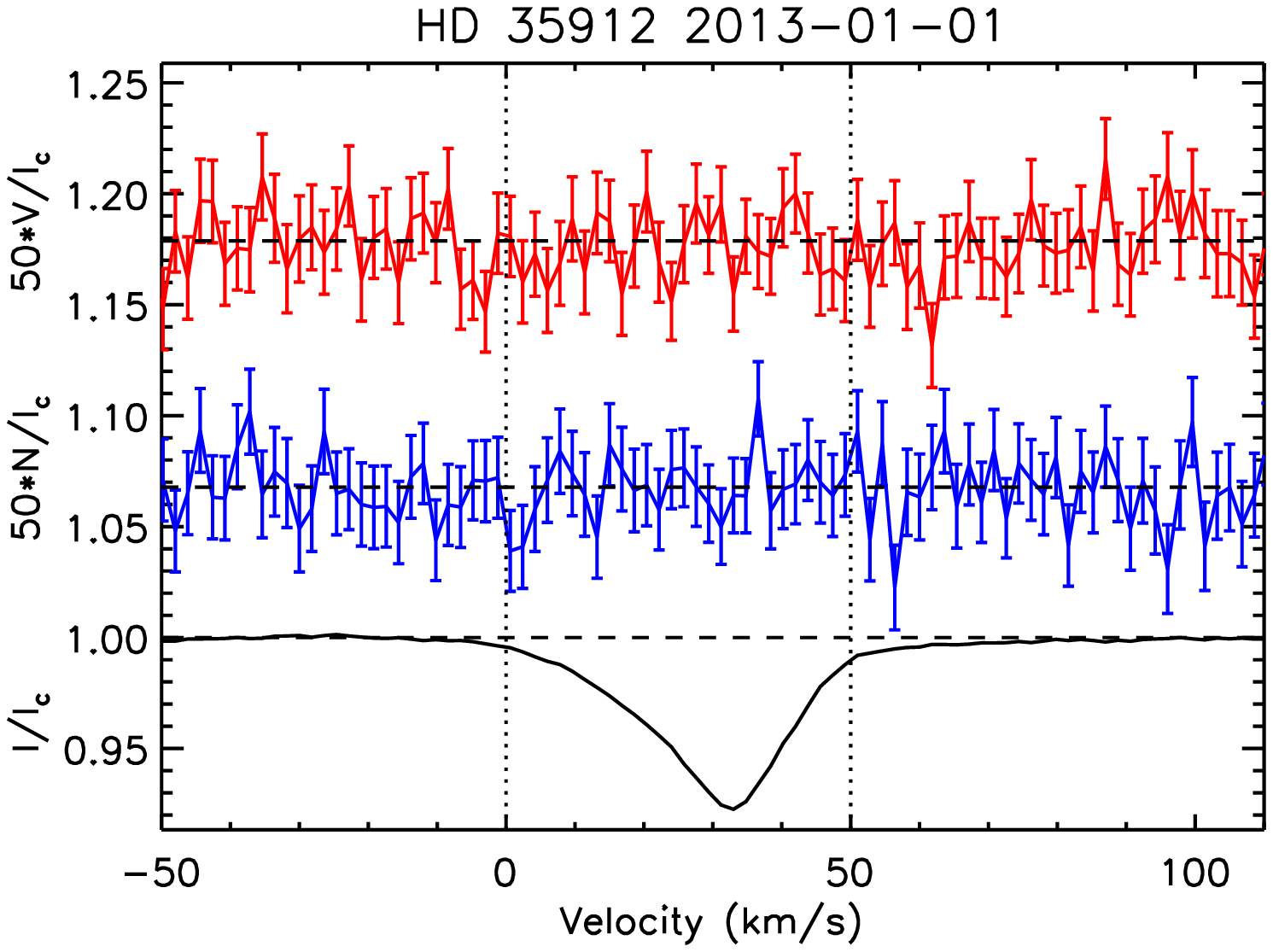}
    \caption[HD 35912 LSD profile.]{The highest S/N LSD profile of HD 35912. Dotted lines show the integration range used for calculation of FAP and measurement of \bz$=-13 \pm 36$ G. The Stokes $V$ profile (top) is indistinguishable from the $N$ profile (middle).}
         \label{HD35912}
\end{figure}

ESPaDOnS measurements did not confirm a magnetic field in one star, HD 35912, catalogued as a magnetic star by \cite{2005AA...430.1143B} based on their own low-resolution measurements, as well as measurements performed by \cite{1970ApJ...159..723C}. The LSD profile deconvolved from the spectrum with the highest S/N is shown in Fig.\ \ref{HD35912}. HD 35912 has sharp spectral lines, with \vsini~$= 12$~\kms. All 6 of the validated observations, with a median peak S/N per spectral pixel of 340, yielded non-detections in the LSD profile. The median error bar of the longitudinal magnetic field measurements of this star is $\sim$35 G, more than sufficient to detect the previously reported $\ge$6 kG magnetic dipole. DAO \bz~measurements, with a mean error bar of 200 G, also failed to detect a magnetic field (Table \ref{daotab}). We therefore removed this star from the analysis. 

\begin{table*}
\caption[Summary of magnetometry.]{Summary of magnetometry. For LSD profiles extracted with metallic lines (Z), metallic + He lines (YZ), and for measurements made using H lines (X), we provide the number of lines used in the mask (or the number of H lines to measure \bz), the maximum \bz~value obtained, and the mean significance level of the \bz~measurements $\Sigma_B$ (eqn. \ref{sigmab_eqn}) made using that set of lines. The final 3 columns give: the number of single-element masks yielding $\Sigma_B > 3$ as a fraction of the total number of single-element masks with which LSD profiles were extracted; the significance $A_{\rm e}/\sigma_{\rm e}$ of the elemental anomaly index (described in Sect.\ \ref{all_el}); and the set of measurements recommended for modelling.}
\resizebox{18 cm}{!}{
\label{masktab}
\centering
\begin{tabular}{l | r r r | r r r | r r r | r r r}  
 & \multicolumn{3}{c}{Z} & \multicolumn{3}{c}{YZ} & \multicolumn{3}{c}{X} & &  &  \\
Star  & No.   & \bz$_{,\rm max}$ & $\Sigma_B$ & No.   & \bz$_{,\rm max}$ & $\Sigma_B$ & No. H   & \bz$_{,\rm max}$ & $\Sigma_B$ & $N_{\rm 3}/N$ & $A_{\rm e}/\sigma_{\rm e}$ & Sel. \\
      & Lines & (kG)             &            & Lines & (kG)             &            & Lines & (kG)             &            &               & & \\ 
\hline
HD\,3360 & 253 &  $-0.029\pm 0.007$ &  1.4 & 271 &  $-0.020\pm  0.004$ &  1.2 &   3 &  $-0.05\pm  0.05$ &  0.9 &  0/13 &  2.2 & YZ \\
HD\,23478 & 140 & $-1.5\pm 0.2$ &  4.4 & 163 & $-1.2\pm 0.1$ &  8.0 &   2 & $-2.1\pm 0.2$ &  7.9 &  1/10 &  2.9 & X \\
HD\,25558 &  -- & -- & -- & 224 &  $-0.10\pm  0.04$ &  1.3 &  -- & -- & -- &  0/0 &  -- & YZ \\
HD\,35298 & 176 & $-3.31\pm 0.10$ & 16.8 & 189 & $-3.13\pm 0.09$ & 17.1 &   3 &  $3.2\pm 0.3$ & 17.3 &  7/ 9 &  2.3 & X \\
HD\,35502 & 189 & $-2.6\pm 0.2$ &  3.3 & 202 & $-2.6\pm 0.1$ &  7.4 &   2 & $-1.58\pm 0.09$ &  5.7 &  1/ 7 &  2.7 & X \\
HD\,36485 & 194 & $-2.57\pm 0.07$ & 34.4 & 207 & $-2.36\pm 0.07$ & 37.1 &   3 & $-2.17\pm 0.10$ & 24.1 &  9/11 &  5.8 & Z \\
HD\,36526 & 179 &  $3.2\pm 0.2$ &  8.1 & 192 &  $3.3\pm 0.3$ &  4.9 &   3 &  $3.4\pm 0.2$ & 12.9 &  2/ 5 &  1.8 & X \\
HD\,36982 & 230 &  $0.4\pm 0.8$ &  1.9 & 272 &  $0.34\pm 0.08$ &  1.4 &   3 &  $0.6\pm 0.2$ &  1.6 &  0/ 9 &  1.8 & YZ \\
HD\,37017 & 234 & $-1.9\pm 2.1$ &  2.9 & 253 & $-1.4\pm 0.3$ &  3.7 &   1 & $-1.8\pm 0.3$ &  7.3 &  1/ 6 &  1.3 & X \\
HD\,37058 & 166 & $-0.95\pm 0.03$ & 17.7 & 179 & $-0.56\pm 0.03$ &  9.4 &   3 & $-0.75\pm 0.05$ &  7.1 &  7/11 &  4.6 & X \\
HD\,37061 & 284 & $-0.4\pm 0.2$ &  1.5 & 326 & $-0.50\pm 0.08$ &  2.5 &   1 & $-0.42\pm 0.08$ &  2.6 &  0/ 9 &  0.9 & YZ \\
HD\,37479 & 222 & $-2.3\pm 0.3$ &  5.3 & 236 &  $2.0\pm 0.1$ &  9.0 &   2 &  $2.76\pm 0.09$ & 13.6 &  4/10 &  4.1 & X \\
HD\,37776 & 253 & $-4.1\pm 0.1$ & 15.2 & 268 & $-1.16\pm 0.07$ &  5.1 &   2 &  $1.7\pm 0.2$ &  5.0 &  9/10 & 10.6 & X \\
HD\,44743 & 161 & $-0.023\pm 0.004$ & 2.4 & 204 & $-0.026 \pm 0.003$ & 2.9 & -- & -- & -- & 0/0 & -- & YZ \\
HD\,46328 & 321 &  $0.35\pm 0.01$ & 54.4 & 343 &  $0.31\pm 0.01$ & 15.9 &   1 &  $0.51\pm 0.06$ &  8.0 &  8/10 &  1.9 & Z \\
HD\,52089 & 153 & $-0.011\pm 0.006$ & 1.4 & 119 & $-0.018\pm 0.006$ & 2.1 & -- & -- & -- & 0/0 & -- & YZ \\
HD\,55522 & 180 &  $1.0\pm 0.1$ &  4.1 & 192 & $-0.56\pm 0.10$ &  4.0 &   3 &  $0.9\pm 0.1$ &  4.2 &  0/11 &  2.1 & X \\
HD\,58260 & 141 &  $1.82\pm 0.03$ & 66.3 & 142 &  $1.77\pm 0.03$ & 66.3 &   3 &  $1.92\pm 0.06$ & 21.7 &  8/ 9 &  4.5 & Z \\
HD\,61556 & 213 & $-0.89\pm 0.06$ & 14.4 & 217 & $-0.89\pm 0.06$ & 14.4 &   3 & $-0.82\pm 0.03$ & 14.0 &  7/11 &  4.7 & X \\
HD\,63425 & 267 &  $0.130\pm 0.006$ &  9.8 & 283 &  $0.101\pm 0.005$ & 10.5 &   3 &  $0.2\pm 0.1$ &  1.1 &  3/10 &  1.0 & Z \\
HD\,64740 & 279 & $-0.9\pm 0.1$ &  3.1 & 298 & $-0.68\pm 0.06$ &  5.8 &   3 & $-0.82\pm 0.09$ &  8.6 &  1/10 &  2.4 & X \\
HD\,66522 & 207 & $-0.54\pm 0.01$ & 30.6 & 209 & $-0.50\pm 0.01$s & 30.8 &   3 &  $0.62\pm 0.07$ & 10.2 &  8/11 &  7.0 & X \\
HD\,66665 & 284 & $-0.126\pm 0.009$ &  4.6 & 299 & $-0.101\pm 0.009$ &  4.0 &   3 &  $0.24\pm 0.06$ &  1.0 &  1/10 &  1.3 & Z \\
HD\,66765 & 243 & $-1.1\pm 0.1$ &  6.8 & 288 & $-0.7\pm 0.2$ & 12.9 &   3 & $-0.81\pm 0.03$ & 15.7 &  3/10 &  2.0 & X \\
HD\,67621 & 250 &  $0.42\pm 0.01$ & 10.8 & 269 &  $0.26\pm 0.01$ &  8.8 &   3 &  $0.29\pm 0.03$ &  5.1 &  6/13 &  4.4 & X \\
HD\,96446 & 265 & $-1.00\pm 0.02$ & 47.5 & 308 & $-1.20\pm 0.06$ & 25.9 &   3 & $-1.40\pm 0.07$ & 25.6 &  8/11 &  4.3 & Z \\
HD\,105382 & 160 & $-0.76\pm 0.03$ & 14.2 & 173 & $-0.70\pm 0.03$ & 11.4 & 3 & $-0.74\pm 0.03$ & 16.3 & 4/ 9 &  2.1 & X \\
HD\,121743 & 225 &  $0.39\pm 0.06$ &  3.7 & 250 &  $0.3\pm 0.1$ &  5.0 &   3 &  $0.33\pm 0.09$ &  4.2 &  1/ 6 &  1.4 & X \\
HD\,122451 & 276 &  $-0.03\pm  0.01$ &  1.4 & 303 &  $-0.065\pm  0.010$ &  3.1 &   3 &  $-0.05\pm  0.03$ &  2.2 &  0/ 3 &  0.9 & Z \\
HD\,125823 & 230 & $-0.47\pm 0.01$ & 22.9 & 243 &  $0.30\pm 0.02$ & 13.3 &   3 &  $0.53\pm 0.07$ &  7.4 &  6/11 &  4.0 & X \\
HD\,127381 & 269 &  $0.17\pm 0.05$ &  2.8 & 292 &  $0.14\pm 0.02$ &  2.8 &   3 & $-0.14\pm 0.10$ &  1.6 &  0/10 &  2.2 & Z \\
HD\,130807 & 194 &  $1.45\pm 0.03$ & 23.8 & 207 &  $1.00\pm 0.02$ & 24.4 &   3 &  $1.00\pm 0.03$ & 17.1 &  4/10 &  9.9 & X \\
HD\,136504\,A & 193 & $-0.23\pm 0.02$ &  9.5 &  329 & $-0.16\pm 0.01$ & 16.3 &  -- & -- & -- &  0/0 &  -- & Z \\
HD\,136504\,B & 193 &  $0.14\pm 0.05$ &  2.6 &  329 & $0.10\pm 0.03$ & 4.9 &  -- & -- & -- &  0/0 &  -- & Z \\
HD\,142184 & 148 & $-2.1\pm 0.4$ &  2.3 & 161 & $-1.6\pm 0.1$ &  7.9 &   2 & $-2.01\pm 0.10$ & 14.8 &  1/ 8 &  2.7 & X \\
HD\,142990 & 193 &  $1.11\pm 0.10$ &  4.2 & 206 & $-0.9\pm 0.1$ &  2.7 &   3 & $-1.2\pm 0.1$ &  6.1 &  1/10 &  2.7 & X \\
HD\,149277 & 226 &  $3.3\pm 0.1$ &  7.7 & 239 &  $2.34\pm 0.06$ & 11.3 &   3 &  $2.9\pm 0.2$ &  5.2 &  6/ 9 &  3.7 & X \\
HD\,149438 & 267 &   $0.088\pm  0.004$ &  7.5 & 283 &   $0.062\pm  0.005$ &  5.5 &   2 &  $0.18\pm 0.04$ &  0.8 &  3/12 &  2.8 & Z \\
HD\,156324 & 218 &  $2.7\pm 0.4$ &  4.2 & 286 &  $2.3\pm 0.1$ &  9.3 &   2 &  $2.2\pm 0.2$ &  4.2 &  1/ 4 &  1.3 & Z \\
HD\,156424 & 233 & $-0.50\pm 0.05$ &  8.9 & 238 & $-0.65\pm 0.03$ & 15.3 &   3 &  $0.27\pm 0.08$ &  1.0 &  7/12 &  1.3 & Z \\
HD\,163472 & 266 & $-0.3\pm 0.1$ &  1.2 & 306 &   $0.09\pm  0.04$ &  0.7 &   3 & $-0.19\pm 0.09$ &  0.7 &  0/10 &  1.7 & Z \\
HD\,164492\,C & 307 &  $2.8\pm 0.5$ &  2.6 & 353 &  $1.8\pm 0.1$ &  6.5 &   1 &  $2.3\pm 0.2$ &  7.2 &  1/10 &  0.8 & YZ \\
HD\,175362 & 177 &  $-5.19\pm  0.07$ & 45.1 & 190 & $-4.44\pm 0.06$ & 41.8 &   3 &  $-5.13\pm  0.03$ & 63.3 &  9/10 &  7.7 & X \\
HD\,176582 & 173 &  $1.56\pm 0.09$ &  9.3 & 186 &  $1.40\pm 0.07$ &  9.4 &   3 &  $1.64\pm 0.06$ & 17.4 &  4/ 6 &  2.0 & X \\
HD\,182180 & 144 &  $2.4\pm 0.3$ &  8.9 & 157 &  $2.3\pm 0.1$ & 16.4 &  2 & $2.7\pm 0.1$ & 21.6 &  3/ 8 &  2.8 & X \\
HD\,184927 & 242 &  $1.18\pm 0.03$ & 22.5 & 255 &  $0.65\pm 0.01$ & 17.3 &   3 &  $1.82\pm 0.07$ & 25.7 & 10/13 & 14.1 & X \\
HD\,186205 & 213 & $-0.91\pm 0.04$ & 21.8 & 233 & $-0.82\pm 0.03$ & 28.2 &   3 & $-0.63\pm 0.04$ &  7.9 &  9/10 &  2.8 & YZ \\
HD\,189775 & 152 &  $1.21\pm 0.08$ &  7.8 & 165 &  $0.99\pm 0.06$ &  11.0 &   3 &  $1.29\pm 0.07$ &  9.4 &  3/ 8 &  4.5 & X \\
HD\,205021 & 311 &  $0.10\pm 0.01$ &  4.2 & 344 &   $0.076\pm  0.006$ &  6.6 &   3 &  $0.15\pm 0.03$ &  2.0 &  3/ 9 &  1.3 & YZ \\
HD\,208057 & 196 &  $0.2\pm 0.1$ &  1.2 & 236 & $-0.13\pm 0.03$ &  2.5 &   3 & $-0.23\pm 0.04$ &  1.8 &  0/ 9 &  1.6 & YZ \\
ALS\,3694 & 141 &  $0.9\pm 0.7$ &  1.2 & 154 & $-1.1\pm 0.1$ &  5.7 &   2 & $-3.7\pm 1.0$ &  3.7 &  1/12 &  3.0 & X \\
\hline\hline
\end{tabular}
}
\end{table*}

\subsection{Longitudinal magnetic field}\label{section_bz}

   \begin{figure}
   \centering
   \includegraphics[width=8.5cm]{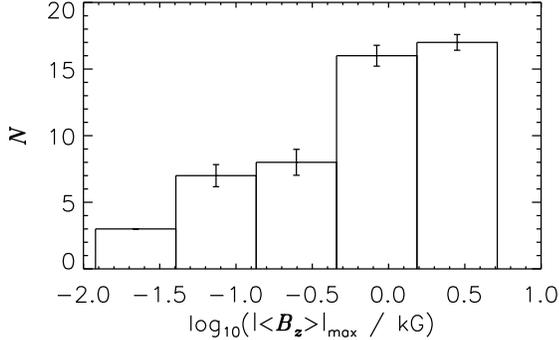}
      \caption[]{Histogram of maximum $|\langle B_z \rangle|$ values. Uncertainties were computed from the standard deviation in each bin via Monte Carlo simulations of synthetic datasets of the same size, with synthetic datapoints randomly shifted from the original within the measured error bars.}
         \label{bzmax_hist}
   \end{figure}

To evaluate the strength of the magnetic field, we measured the line-of-sight or longitudinal magnetic field \bz~in G, averaged over the stellar disk, by measuring the first-order moment of Stokes $V$ normalized to the equivalent width of Stokes $I$ (e.g.\ \citealt{mat1989}). Integration ranges were set by $\pm$\vsini$+ v_{\rm sys}$. The same measurement can be applied to the diagnostic $N$ profile, yielding an analagous `null longitudinal magnetic field' \nz, with which \bz~can be compared \citep{wade2000}. The error bars $\sigma_B$ and $\sigma_N$ were obtained via error propagation of single-pixel photon noise uncertainties in the LSD profiles. Table \ref{masktab} gives the value of the strongest \bz~extremum, \bz$_{\rm max}$, in order to give an indication iof the strength of the stars' magnetic fields. These range from $\sim$10 G (HD 52089) to $\sim$5 kG (HD 175362). For some stars the positive and negative \bz~extrema are of a similar magnitude, leading to an apparent sign inversion when comparing \bz~maxima obtained from different line lists. The histogram of $|\langle B_z \rangle|_{\rm max}$ is shown in Fig.\ \ref{bzmax_hist}. The distribution peaks between 1.5 and 5 kG, with a cut-off above, and a low \bz~tail extending to 10~G. Since strong magnetic fields are intrinsically easier to detect, the high-\bz~cutoff likely reflects the intrinsic rarity of early B-type stars with higher values of \bz. It is not so obvious that the distribution at low values reflects a real rarity of stars with relatively weak magnetic fields, as such fields are much harder to detect (see also the discussion of this possibility by \citealt{2015AA...574A..20F}).  



   \begin{figure}
   \centering
   \includegraphics[width=8.5cm]{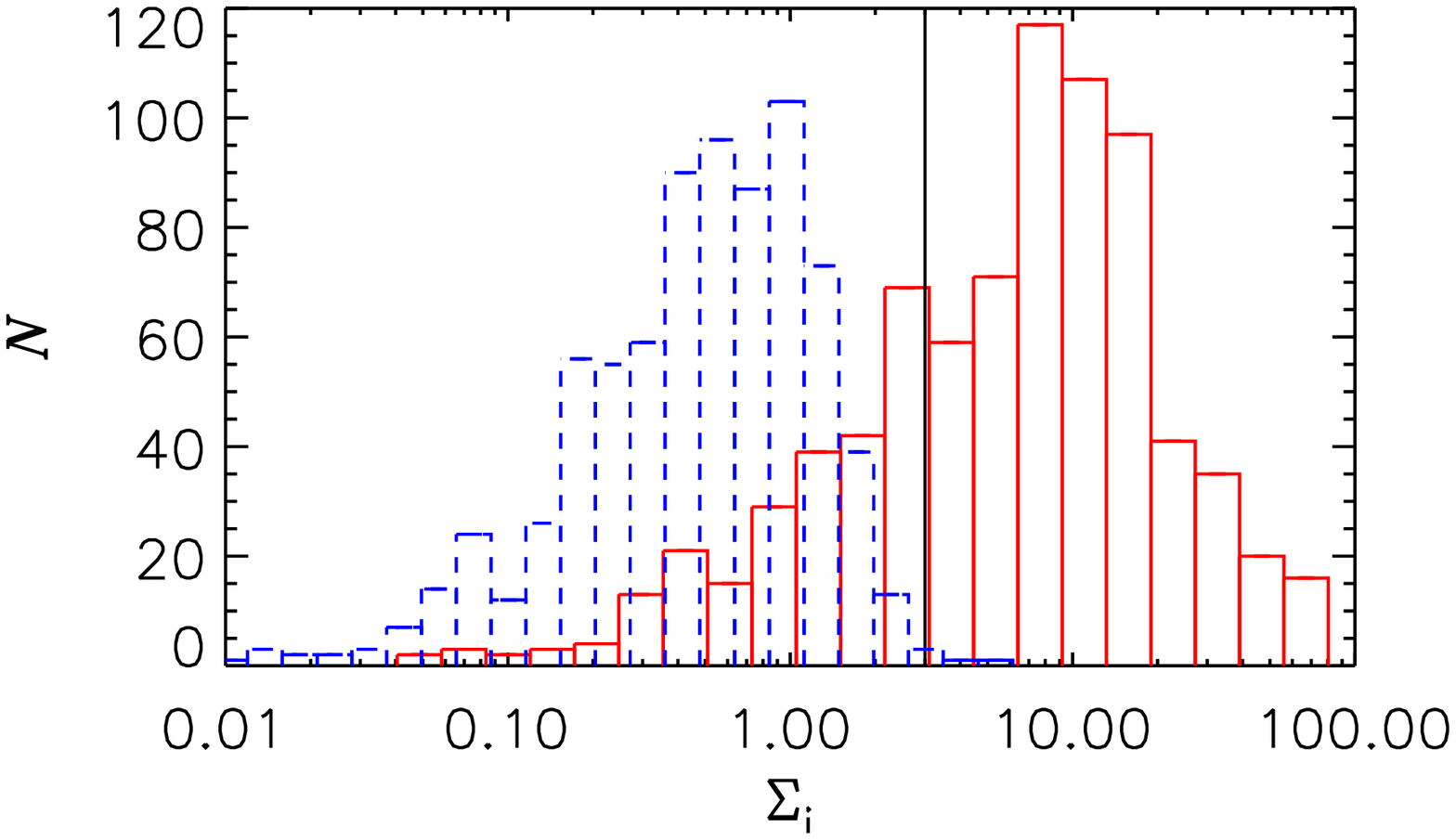}
   \includegraphics[width=8.5cm]{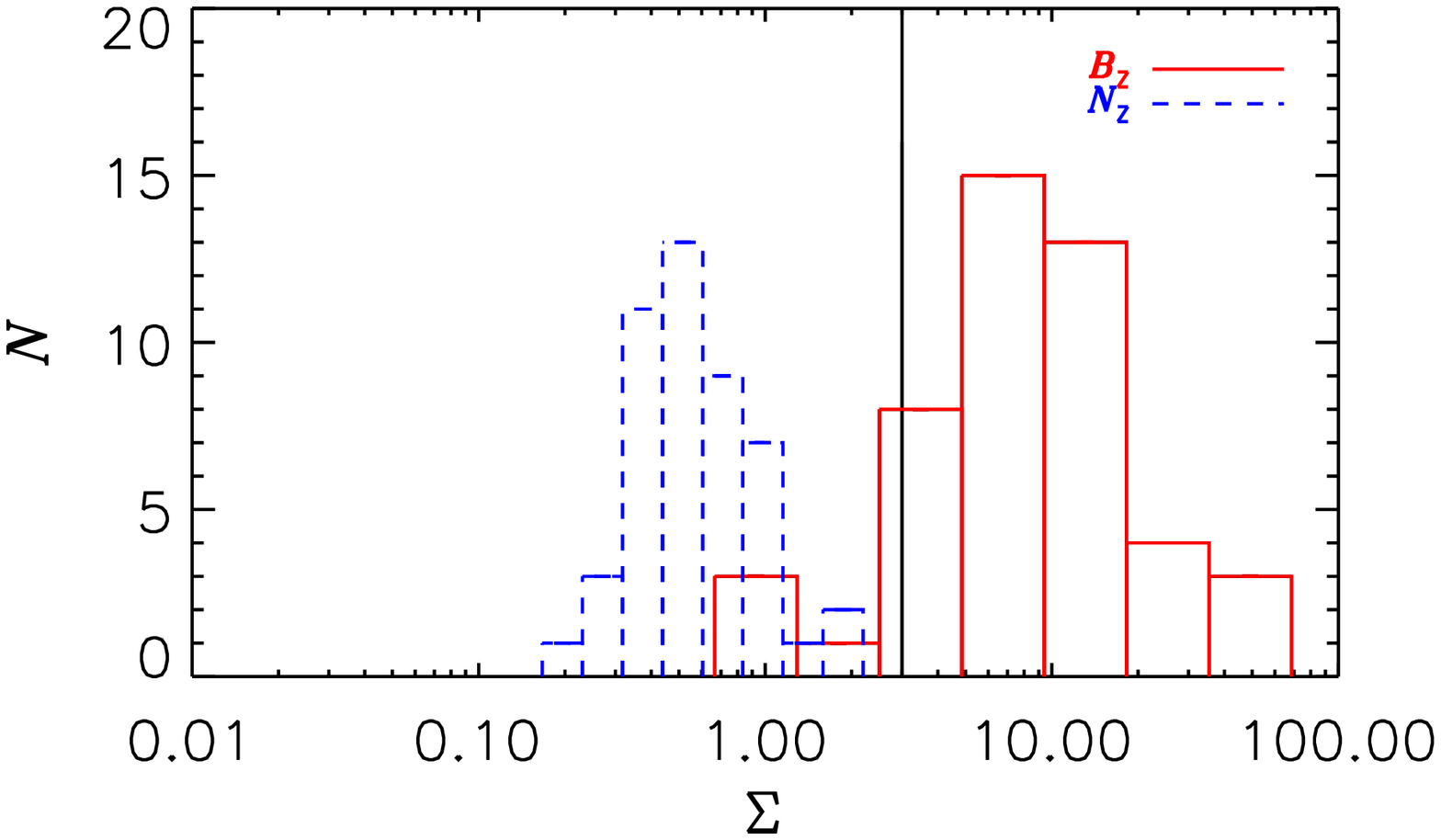}
      \caption[Histogram of \bz~significance.]{Histograms of $\Sigma_i = |\langle B_z \rangle|/\sigma_B$ and $|\langle N_{\rm z} \rangle|/\sigma_B$ for all individual measurements (above), and the mean value of $\Sigma$ for individual stars (below). The vertical solid lines indicate 3$\sigma$. The \bz~$\Sigma$ distribution peaks at 10$\sigma$, while the \nz~distribution peaks below 1$\sigma$. 67\% of \bz~measurements are above 3$\sigma$, and only 0.2\% of \nz~measurements. There are no stars for which the mean $\Sigma_N$ is above 3$\sigma$ significance, while 83\% of stars have $\Sigma_B$ above the 3$\sigma$ threshold.}
         \label{bzbs_hist}
   \end{figure}

To evaluate the data quality, we calculated the mean significance $\Sigma_B$ of the \bz~measurements, which we define as:

\begin{equation}\label{sigmab_eqn}
\Sigma_B = \frac{1}{n}\sum\limits_{i=1}^n \frac{|\langle B_z \rangle|_i}{\sigma_{B,i}},
\end{equation}

\noindent where $i$ denotes an individual measurement out of the total number $n$ of observations for a given star. For an individual measurement, $|\langle B_{\rm z}\rangle|/\sigma_B$ is a measurement of the significance of the \bz~measurement. Eqn.\ \ref{sigmab_eqn} then gives the mean of this significance across all measurements, with $\Sigma_B \le 1$ indicating that the dataset is dominated by noise. $\Sigma_B$ is given in Table \ref{masktab} for LSD profiles extracted using only metallic lines, and metallic plus He lines. For HD 44743 and HD 52089 $\Sigma_B$ was evaluated from the \bz~measurements presented by \cite{2015AA...574A..20F} and \cite{2017MNRAS.471.1926N}. Fig.\ \ref{bzbs_hist} shows a histogram of these values, along with the corresponding histogram for $\Sigma_N$, the analaogue of $\Sigma_B$ for \nz. The significance of \bz~peaks at $\sim 10\sigma$, with a tail extending to 70$\sigma$. Approximately 10\% of the sample has $\Sigma_{\rm B} \le 3$: these are all stars with weak magnetic fields (\bz$_{\rm max} \le$ 300 G). \nz~is tightly clustered around $0.5\sigma$, and is below 3$\sigma$ for all stars. The DDs obtained from $N$ profiles, examined in the previous sub-section, do not yield spurious signals in \nz.


For DAO data, the longitudinal magnetic field is measured via the Zeeman shift between two spectra of opposite circular polarizations in the core of the H$\beta$ line, as well as nearby lines such as He~{\sc i} 492.2 nm or Fe~{\sc ii} 492.3 nm. \bz~is proportional to the Zeeman shift with a per-pixel scaling factor of 6.8 kG in H$\beta$, 6.6 kG in He~{\sc i}, and 14.2 kG in Fe~{\sc ii}. The DAO \bz~measurements are summarized in Table \ref{daotab}. 

\begin{table}
\caption[Summary of DAO \bz~measurements.]{Summary of DAO \bz~measurements.}
\resizebox{8.5 cm}{!}{
\centering
\label{daotab}
\begin{tabular}{l | r r | r r | r r}
\hline
\hline
 & \multicolumn{2}{c}{H$\beta$} & \multicolumn{2}{c}{Fe~{\sc ii} 492.3 nm} & \multicolumn{2}{c}{He~{\sc i} 492.2 nm} \\
HD & \bz$_{\rm max}$ & $\Sigma_B$  & \bz$_{\rm max}$ & $\Sigma_B$  & \bz$_{\rm max}$  & $\Sigma_B$ \\
No. & (kG) & & (kG) & & (kG) & \\
\hline
3360  & $-0.17 \pm 0.09$ & 1.8 & -- & -- & -- & -- \\
23478 & $-2.9\pm 0.4$ & 4.8 & -- & -- & -- & -- \\
35298 & $5.7\pm0.5$ & 5.3 & $-6.8\pm 0.5$ & 4.8 & -- & -- \\
35502 & $-4.0\pm1.0$ & 4.6 & -- & -- & -- & -- \\
35912 & $-0.5\pm 0.2$ & 1.6 &  -- & -- & -- & -- \\
36485 & $-3.1\pm 0.4$ & 10 & -- & -- & -- & -- \\
36526 & $4.0\pm 0.3$ & 6.5  & $7\pm 2$ & 3.2  & -- & -- \\
37058 & $-0.6\pm0.1$ & 2.4 & -- & -- & -- & -- \\
37061 & $-0.02 \pm 0.26$ & 0.06 & -- & -- & -- & -- \\
37776 & $-3.9 \pm 0.8$ & 3.5 & -- & -- & -- & -- \\
176582 & $-2.1\pm 0.2$ & 7.6 &  -- & -- & -- & -- \\
184927 & $2.1\pm 0.3$ & 6.0  & -- & --  & $1.5\pm 0.1$ & 4.3 \\
186205 & $-1.0 \pm 0.2$ & 2.1  & -- & --  & -- & -- \\
189775 & $1.7 \pm 0.3$ & 3.3  & $1.1 \pm 0.2$ & 1.6  & -- & -- \\
\hline
\hline
\end{tabular}
}
\end{table}

Ten of the sample stars are multi-lined spectroscopic binaries. The LSD profiles of these stars are indicated in Figs.\ \ref{lsd_allplot_1} and \ref{lsd_allplot_2} with red circles in the top right corner. Since \bz~is calculated using the centre-of-gravity of the Stokes $I$ and $V$ profiles, if the components are blended (as is generally the case), \bz~can be affected by the contribution of binary companions to the Stokes $I$ spectrum. In order to remove this influence, disentangled LSD profiles were obtained via an iterative algorithm similar to that described by \cite{2006AA...448..283G}, and \bz~was measured from the resulting Stokes $I$ profiles of the magnetic component. Line profile modelling, radial velocity measurement, and disentangling of Stokes $I$ will be discussed in detail in the context of an analysis of the binary sub-population by Shultz et al.\ (in prep.). This correction is negligible for HD 36485 and HD 37061, due to the insignificant contribution of the non-magnetic star to Stokes $I$. For HD 35502, \bz~is increased by $\sim$25\% after correction. For HD 149277, the correction is non-existent for most observations, as the RV amplitude is much larger than the line widths and thus the line profiles are blended in only a few observations; however, for blended observations the correction for this star can be up to 40\%. For HD 122451 (in which the secondary is the magnetic star) and HD 156324, the correction is important. In the former case, \bz~is approximately twice as high when measured using disentangled line profiles, since the components are blended in all observations and contribute approximately equally to the line profile. In the latter case, while the contributions of the non-magnetic stars are not large compared to the magnetic component, scatter in \bz~is greatly reduced due to the strongly variable blending. 

Iterative disentangling assumes that the Zeeman signatures in Stokes $V$ are entirely due to one star. This assumption does not hold for HD 136504, in which both components are magnetic \citep{2015MNRAS.454L...1S}. Therefore, for this star the only measurements used were those obtained when the separation of the stellar line profiles in velocity space was greater than the summed \vsini~of the components. This left only 5/14 observations available for analysis. Iterative disentangling was also not adopted for HD 25558: as both stars are Slowly Pulsating B-type (SPB) stars \citep{2014MNRAS.438.3535S}, line profile variability in both components makes spectral disentangling unreliable. Therefore we limited the HD 25558 dataset to only those observations in which the line profiles are separately distinguishable (leaving 11/31 measurements), and used model fits to remove the flux of the non-magnetic primary using the binary line profile fitting program described by \cite{2017MNRAS.465.2432G}. As the EW ratio is a free parameter in model fits, this method is able to compensate for the changing EWs of the two components due to pulsations. 

\subsubsection{Measurements with different elements}\label{all_el}

   \begin{figure}
   \centering
   \includegraphics[width=8.5cm]{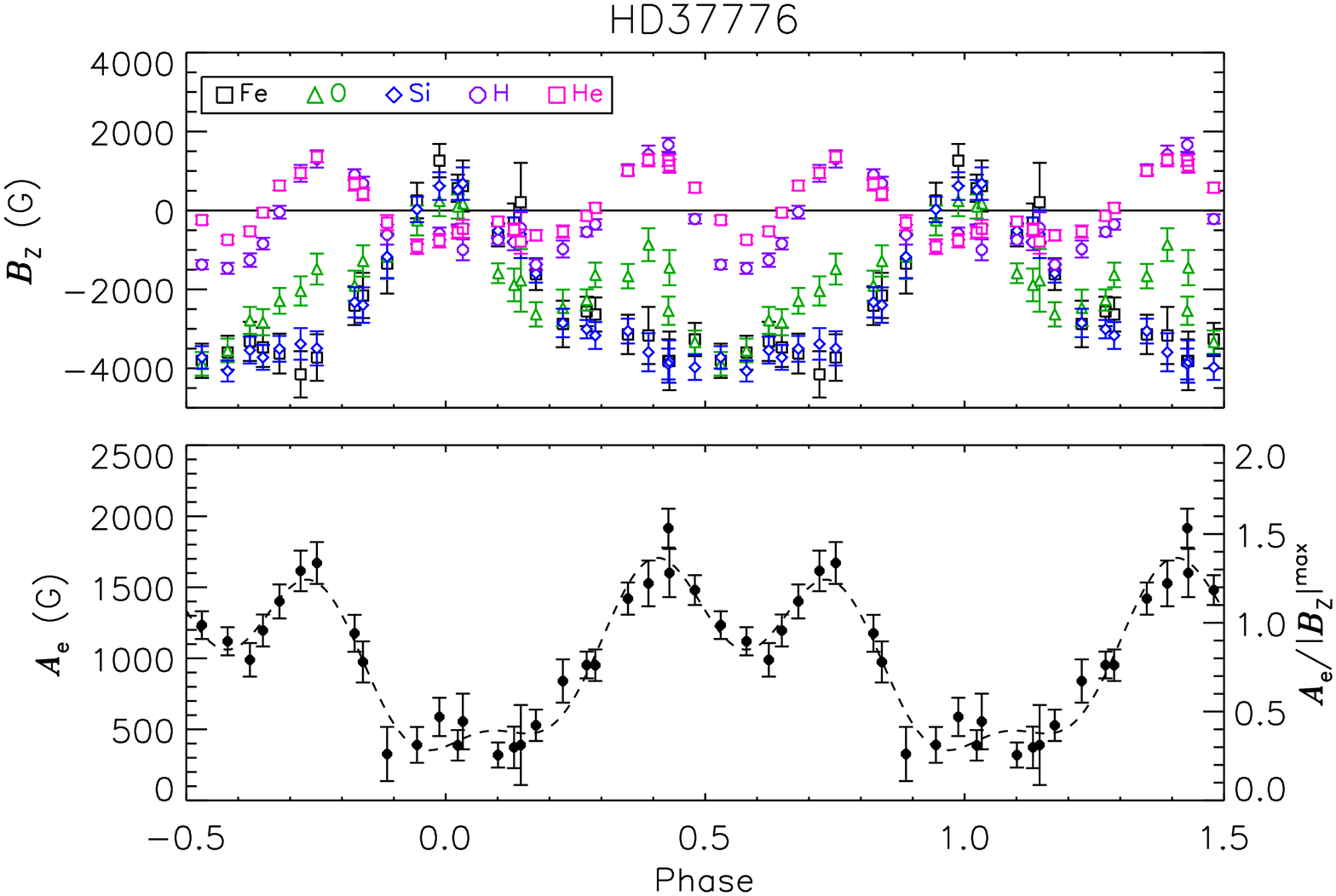} 
   \includegraphics[width=8.5cm]{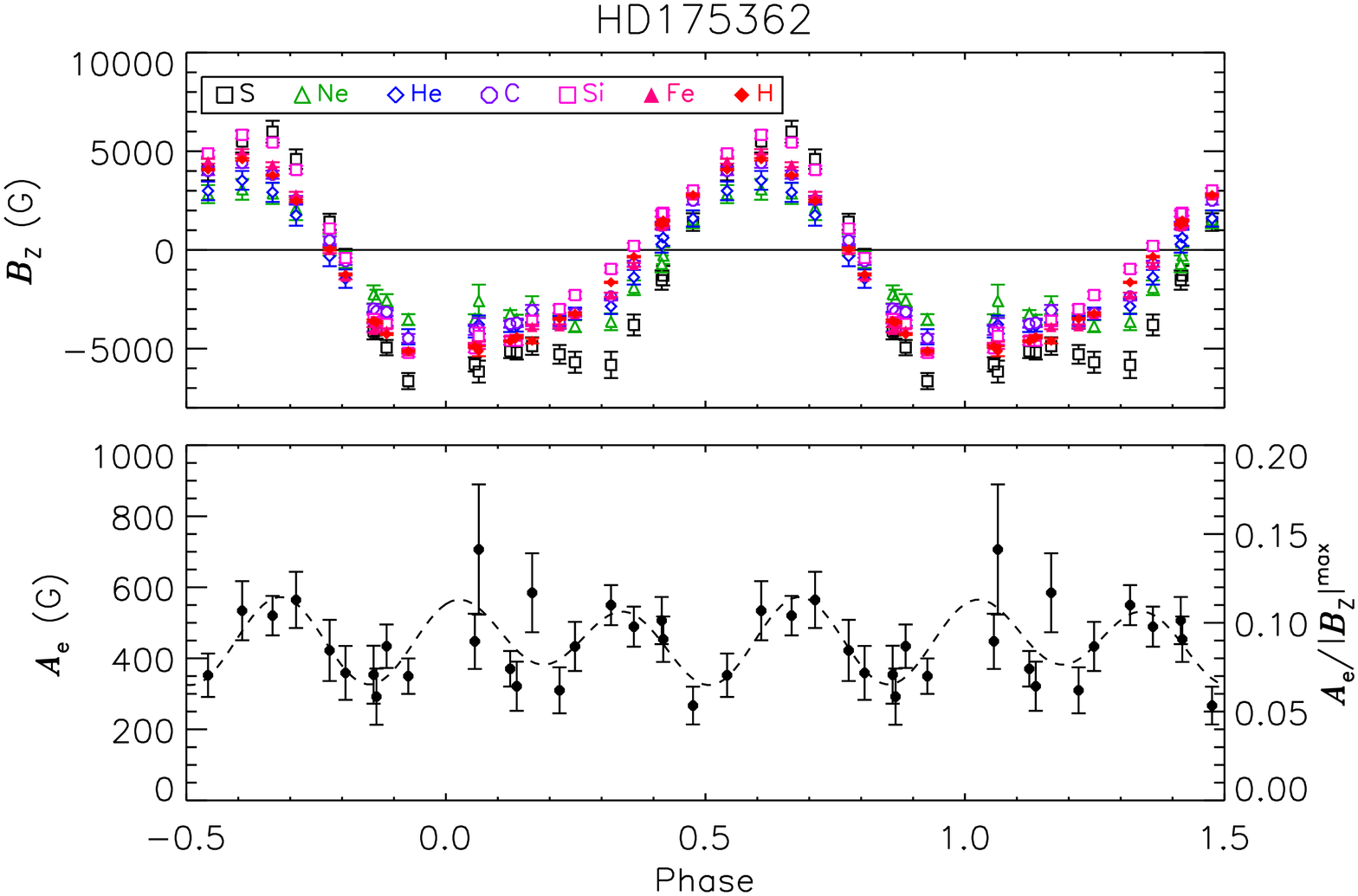} 
   \includegraphics[width=8.5cm]{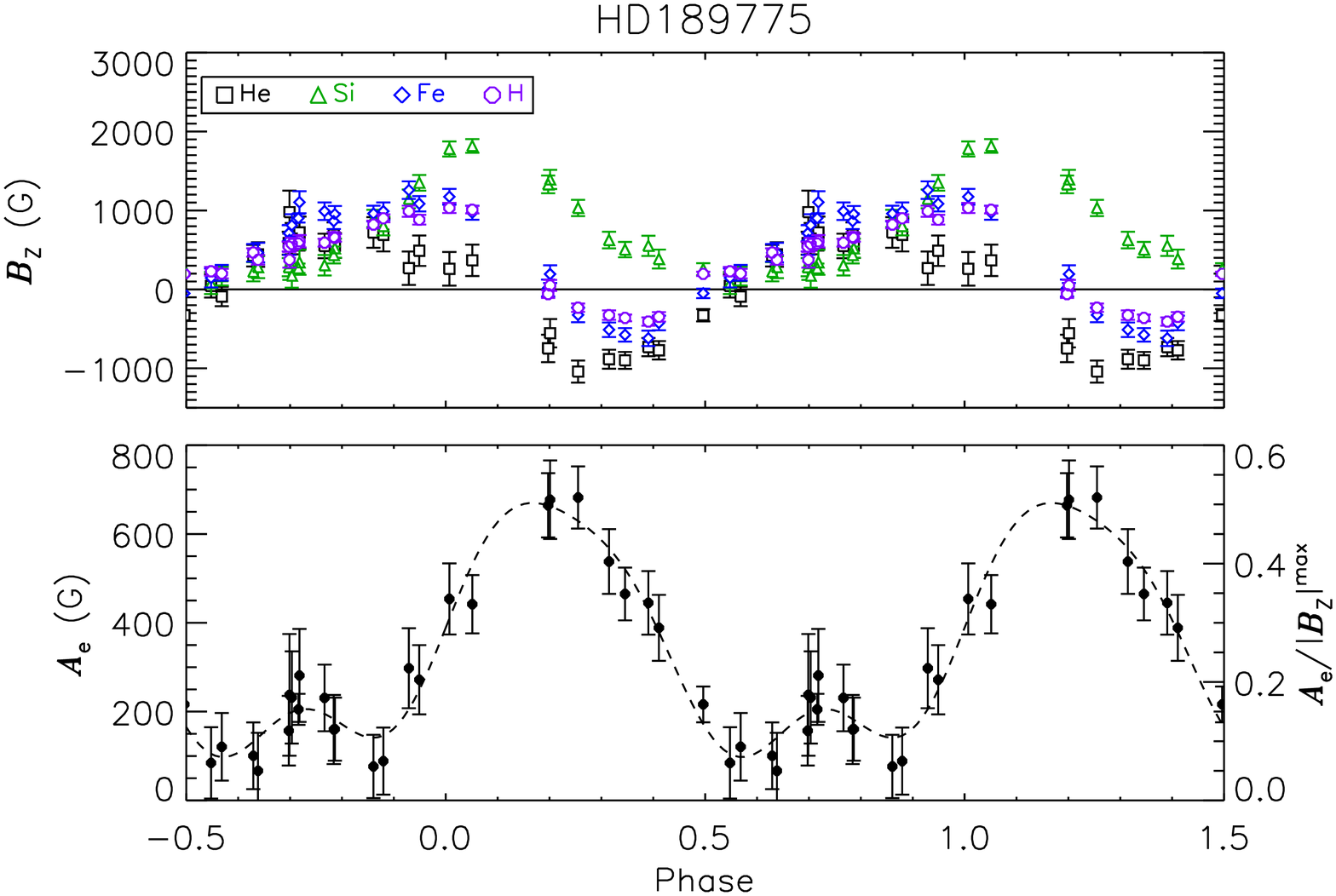} 
      \caption[Variation of \bz~with chemical elements for HD 37776.]{Variation in \bz~when measured using different chemical elements for HD 37776, HD 175362, and HD 189775. {\em Top panels}: Single-element \bz~measurements. {\em Bottom panes}: elemental anomaly index $A_{\rm e}$ (Eqn.~\ref{ae}). The left axis gives the absolute $A_{\rm e}$ in G, the right axis the relative $A_{\rm e}$ obtained via normalization to $|\langle B_{z} \rangle|_{\rm max}$. The dashed line indicates a 3$^{rd}$-order sinusoidal fit, used to determine the integrated value of $A_{\rm e}$ across all rotational phases (see text). Note the variation between stars in both the magnitude and variability of $A_{\rm e}$.}
         \label{bz_multi_el_hd37776}
   \end{figure}







While the relative precision of \bz~is quite high, as evaluated by $\Sigma_B$, there is also the question of accuracy. \cite{bl1977} showed that \bz~measurements of Ap stars displayed systematic differences when measured using spectral lines from different elements. This phenomenon has subsequently been reported for some Bp stars \citep{2005AA...430.1143B, yakunin2011, 2015MNRAS.447.1418Y, 2015MNRAS.449.3945S}. The physical origin of these discrepancies is thought to be the same as that leading to the photometric variability of Ap/Bp stars, namely, surface chemical abundance spots. Spots lead to differential line formation such that there is a greater contribution to Stokes $V$ in regions of enhanced abundance, causing the polarized flux to be enhanced in some regions relative to others and, hence, warping the Stokes $V$ profile \citep{2015MNRAS.447.1418Y}.

To explore the prevalence and influence of this effect in our sample, we extracted LSD profiles using single-element line masks, i.e.\ line masks in which lines of only a single chemical element were included. These were obtained from the cleaned and tweaked masks described in Sect.\ \ref{lsd}, with the criterion that a given element have at least 3 isolated lines in the analysis region. Table \ref{masktab} gives the fraction of masks $N_{\rm 3}/N$, where $N$ is the total number of masks for which LSD profiles could be extracted for each star, and $N_3$ is the number of masks for which $\Sigma_B>3$. We also measured \bz~using H lines (see below, \S~\ref{h_bz}). \bz~measurements using different elements are shown in Fig.\ \ref{bz_multi_el_hd37776} for the examples of HD 37776, HD 175362, and HD 189775, phased according to the ephemerides given below in \S~\ref{rotation_periods}. All 3 stars show a large variance in \bz.


To quantify the degree to which \bz~differs when measured using the spectral lines of different elements, we calculated the {\em elemental anomaly} $A_{\rm e}$. This is the weighted standard deviation across all $n$ single-element measurements obtained from a given spectrum $i$:

\begin{equation}\label{ae}
A_{\rm e} = \sqrt{\frac{1}{\sum\limits_{i=1}^n (\sigma_{B,i} - \overline{\sigma_B})^{-2}} \sum\limits_{i=1}^n\frac{(\langle B_{z,i}\rangle - \overline{\langle B_z\rangle})^2}{(\sigma_{B,i} - \overline{\sigma_B})^2}}
\end{equation}

\noindent where $\sigma_{B,i}$ is the uncertainty in \bz~for element $i$, $\overline{\sigma_B}$ is the mean uncertainty across all $n$ elements, and $\overline{\langle B_z\rangle}$ is likewise the mean \bz~over all $n$ elements. An error-bar weighted standard deviation is used because measurements obtained from different single-element line masks have systematic differences in uncertainty, due to the large differences in the number of lines available for different elements. Normalization to $|\langle B_z\rangle|_{\rm max}$ is then performed in order to determine the fractional variation in \bz~across different elements, rather than the absolute variation, which will be higher for stars with intrinsically stronger magnetic fields. The uncertainty $\sigma_{\rm e}$ in $A_{\rm e}$ was calculated from the weighted mean error bar across all elements. 

Examples of $A_{\rm e}$ curves are shown in the bottom panels of Fig.\ \ref{bz_multi_el_hd37776}, with the left axes giving $A_{\rm e}$ in G, and the right axes giving the fractional $A_{\rm e}$. The behaviour of $A_{\rm e}$ is different with each star. In the case of HD 37776, $A_{\rm e}$ changes very significantly with rotation phase, peaking near one or more of the magnetic extrema, and reaching a minimum near \bz$=0$. HD 189775 also shows large changes with phase, but the largest difference is seen near the magnetic equator (\bz$\sim$0) due to a phase shift between Si \bz~measurements and other elements. By contrast, HD 175362 shows little variation of $A_{\rm e}$ with rotation. The peak values are also quite different: 150\% for HD 37776, 50\% for HD 189775, and only 10\% for HD 175362. Stars with highly variable $A_{\rm e}$ likely exhibit greater surface abundance anistropies than stars in which $A_{\rm e}$ is less variable. 

\begin{figure}
\centering
\includegraphics[width=8.5cm]{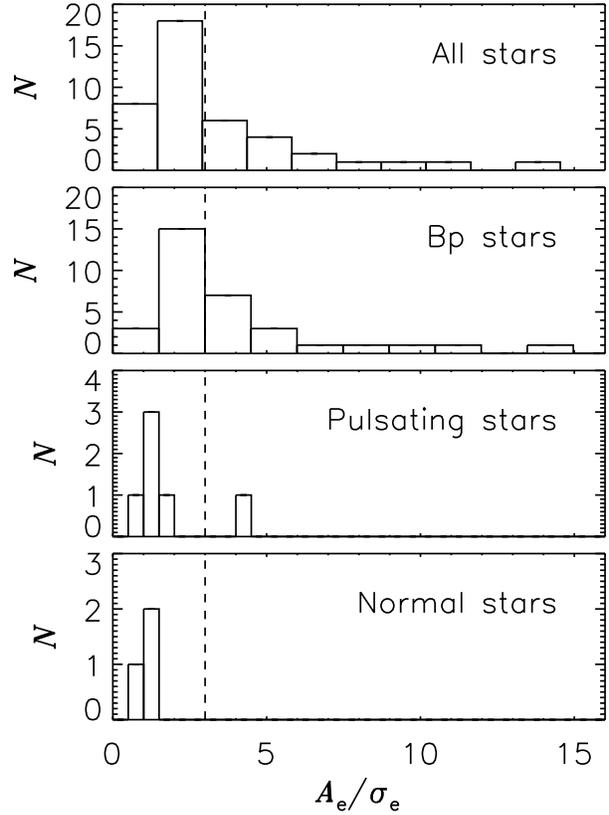}
\caption[Significance of elemental \bz~variation.]{Signifiance histogram for $A_{\rm e}$ for ({\em top -- bottom}) all stars in the sample for which $A_{\rm e}$ could be measured, Bp stars, pulsating stars, and chemically normal, non-pulsating stars. The dashed line marks 3$\sigma$ significance. Approximately half the sample shows at least a 3$\sigma$ spread in \bz~measurements conducted with different elements. All of these are Bp stars: the only pulsator with $A_{\rm e}/\sigma_{\rm e} \ge 3$ is HD 96446, which is also a He-strong star.}
\label{ae_hist}
\end{figure}

In order to obtain a single number with which to characterize the average strength of deviations between the magnetic curves derived for different elements for a given star, we used the {\sc idl} {\sc curvefit} routine to generate 2$^{nd}$- or 3$^{rd}$-order harmonic best-fits, which were then integrated as functions of rotational phase. The significance of this integrated value is then $A_{\rm e}/\sigma_{\rm e}$, which is given for each star in the 2$^{\rm nd}$-last column of Table \ref{masktab}, where $\sigma_{\rm e}$ is the mean uncertainty in the individual $A_{\rm e}$ measurements. If $A_{\rm e}/\sigma_{\rm e} < 1$, it can be concluded that differences in \bz~between different elements are a consequence of noise rather than real variations. It is this significance in which we are interested, as it will help to decide whether accurate measurements are available using metallic lines, or whether H lines should be used instead. Fig.\ \ref{ae_hist} shows the histogram of $A_{\rm e}/\sigma_{\rm e}$ for all stars. For the full sample (top panel) the median significance is 2.7$\sigma$. For the sub-sample of He-weak and He-strong Bp stars, the median of the distribution is at 2.9$\sigma$ (2$^{nd}$ panel from the top). Amongst pulsating $\beta$ Cep and SPB stars (3$^{rd}$ panel from the top) and the remainder of the sample with no apparent peculiarities (bottom panel), only one star, HD 96446, has $A_{\rm e}/\sigma_{\rm e} \ge 3$: as HD 96446 is also a He-strong star, this likely reflects the star's chemical peculiarities. The medians of these distributions are 1.4$\sigma$ and 1$\sigma$, respectively. While there are only 8 non-chemically peculiar stars for which $A_{\rm e}$ could be measured, these results are consistent with an origin of the effect in distortion due to chemical spots, and confirm that when chemical peculiarities are not present there is no significant difference in \bz~when measured using different chemical elements. Even amongst the Bp stars, although there are several stars for which $A_{\rm e}/\sigma_{\rm e} \gg 3$, differences in \bz~measured from different elements are negligible for many of the stars.

It should be noted that while the mathematical treatment of $A_{\rm e}$ is based upon {\em random} uncertainties, these variations are in fact {\em systematic}. Unfortunately, it is not obvious what simple mathematical tools might be utilized in order to capture the systematic variations between different sets of measurements. A rigorous exploration of this effect will require Zeeman Doppler Imaging (ZDI) of the surface magnetic fields together with Doppler Imaging (DI) of the chemical abundance patterns. As ZDI and DI cartography is outside of the scope of this work, we have limited ourselves to determining $A_{\rm e}$ as a somewhat crude indicator for comparing the magnitude of the effect between different stars. However, it should be kept in mind that as of yet no pattern in the distribution of chemical elements relative to the surface magnetic field has been detected, hence, a treatment of these variations in terms of random error might not be entirely unwarranted. 

\subsubsection{H line \bz~measurements}\label{h_bz}

\begin{figure}
   \centering
   \includegraphics[width=8.5cm]{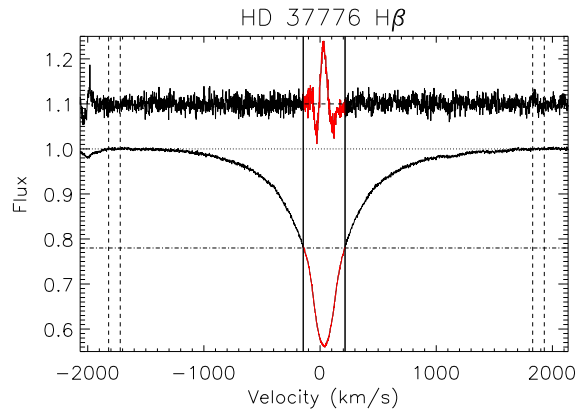}
      \caption[Measuring \bz~from H$\beta$.]{Measuring \bz~from the H$\beta$ line of HD 37776. Stokes $I$ (bottom) is normalized to the continuum. Normalization regions are shown by vertical dashed lines, with the `true continuum' indicated by the horizontal dotted line. The integration limits are indicated by vertical solid lines. Note that the Stokes $V$ (top) signature is entirely within the rotationally broadened non-LTE core of the line. The dot-dashed line indicates the `line continuum', used to evaluate \bz~(see text).}
         \label{hd37776_halpha_bz}
   \end{figure}

In addition to yielding different \bz~measurements, chemical spots can also lead to anharmonic \bz~variations that can be mistaken for contributions from higher-order multipolar components to the photospheric magnetic field. \cite{bl1977} showed that measurements performed using H lines avoid this problem, as H is in general distributed relatively uniformly over the photosphere. Historically, such measurements were performed with photopolarimeters using the wings of the H$\beta$ line. However, \cite{2015AA...580A.120L} have shown that, when high-resolution spectropolarimetry is available, \bz~can be measured using the rotationally broadened non-LTE core of H$\alpha$ rather than using the line wings. Fig.\ \ref{hd37776_halpha_bz} shows Stokes $I$ and $V$ for the H$\beta$ line of HD 37776, demonstrating that the magnetic signature is also clearly detectable in the Stokes $V$ profile of this line. 

We measured \bz~using H$\alpha$ through $H\gamma$, except for stars with significant emission (as is the case for e.g.\ HD\,37776, \citealt{petit2013}), in which case H$\alpha$ measurements were not used. Lines at shorter wavelengths than H$\gamma$ were not used as the S/N of ESPaDOnS/Narval spectra is typically much lower in this region. The final \bz~was calculated as the error bar-weighted mean across all Balmer lines used. Table \ref{masktab} gives the number of H lines used for each star, as well as $\Sigma_B$ for the H line measurements. We used the laboratory wavelengths of the lines and a Land\'e factor $g = 1$. The EW was measured using a continuum $I_{\rm c}$ taken at the boundary of the rotationally broadened line core, rather than the `true' continuum bounding the pressure-broadened wings: this ensures the centroid of the line, and thus the separation of the circularly polarized components, is evaluated at the highest possible S/N and thus the maximum precision \citep{2015AA...580A.120L}. 

\begin{figure}
   \centering
   \includegraphics[width=8.5cm]{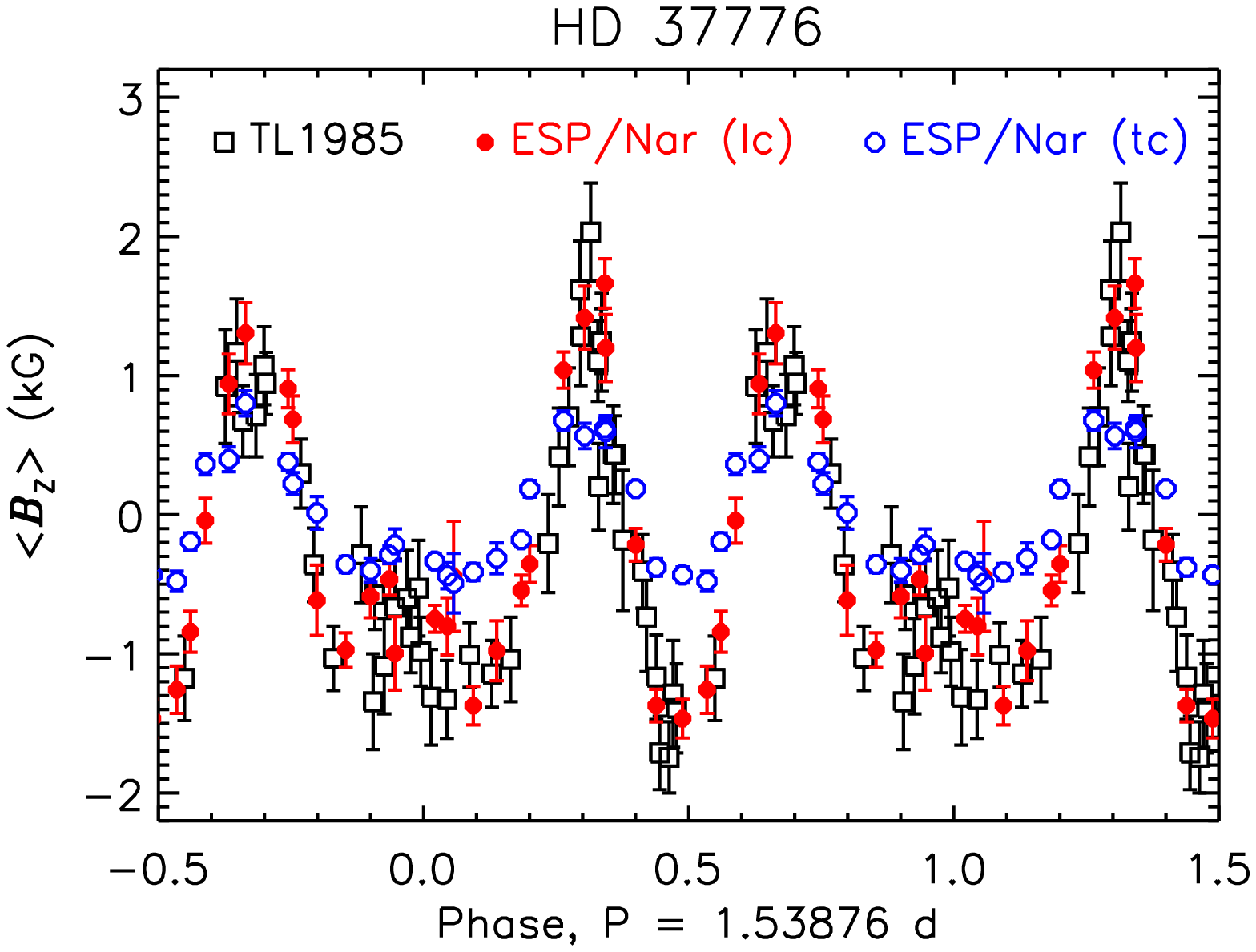}
      \caption[Comparisons of H \bz~measurements to historical data]{Photopolarimetric H$\beta$ wing measurements of HD 37776 reported by \protect\cite{thom1985} compared to ESPaDOnS and Narval data. The data are phased using the non-linear ephemeris determined by \protect\cite{miku2008}. The high-resolution spectropolarimetric measurements were renormalized either to the `true continuum' (tc) or `line continuum' (lc) (see text and Fig.\ \ref{hd37776_halpha_bz}). The latter measurements provide a much better agreement with the historical data.}
         \label{hyd_bz}
   \end{figure}

An example of the different result obtained using `true' and `line' continua is shown in Fig.\ \ref{hd37776_halpha_bz} for HD 37776. If H-line \bz~measurements are evaluated with $I_{\rm c}$ at the `true' continuum rather than the boundaries of the non-LTE core (as determined by eye), the amplitude of the \bz~curve is greatly reduced such that the measurements no longer agree with literature values, as demonstrated in Fig.\ \ref{hyd_bz} for HD 37776. We compare to photopolarimetric H$\beta$ wing measurements reported in the literature \citep{1983ApJS...53..151B,thom1985,1987ApJ...323..325B}. \cite{2000A&A...358.1151M} showed that, with H profiles calculated with a more accurate treatment of limb darkening and Stark broadening, photopolarimetric \bz~measurements made using H$\beta$ lines should be corrected to $\sim$80\% of their published values. The HD 37776 data have been phased with the non-linear ephemeris calculated by \cite{miku2008}, which accounts for the spindown of the star. The modern data are more precise, but the general features of the \bz~curve are essentially identical. Further comparisons of high-resolution H line \bz~measurements with photopolarimetric data are shown in Appendix \ref{bz_ind} for HD 36485, HD 37017, HD 37058, HD 64740, HD 125823, HD 142990, and HD 175362: in all cases the agreement between modern and historical data is very good. H line \bz~measurements are also compared to \bz~measurements from LSD profiles extracted from single-element line masks in Fig.\ \ref{bz_multi_el_hd37776}: the reduced amplitude obtained when measuring \bz~using the `true' continuum would leave the amplitude of H line \bz~measurements in disagreement with those obtained from other metallic lines. 

We have also found that $I_{\rm c}$ should be evaluated in the same fashion for He lines with strong pressure-broadened wings. This introduces some ambiguity into \bz~measurement for LSD profiles extracted using line masks dominated by He lines: to minimize this problem, He lines with broad wings were excluded from the line masks whenever possible.

A further consideration relates to the nature of echelle spectra. Before extracting LSD profiles from ESPaDOnS or Narval data, the spectra are in general normalized using a polynomial spline fit to the continuum of each echelle order. However, the broad wings of the H Balmer lines, especially H$\beta$ and H$\gamma$, overlap with the edges of their respective echelle orders: thus, polynomial normalization may distort the line profiles. This will then change the EW, and lead to an incorrect measurement of \bz. To avoid this we used spectra that had not been normalized using polynomial splines, instead normalizing using a linear fit between the edges of the line cores. Experimentation with different stars indicated that this strategy minimizes scatter in the measurements. As the {\sc reduce} pipeline does not perform global normalization after merging the subexposures, HARPSpol data should not be affected by this issue. 

\section{Rotation}\label{rotation}



\subsection{Velocity broadening}\label{section_vsini}

Line-profile fitting was utilized to measure \vsini. ESPaDOnS, Narval, and HARPSpol spectra combine a high spectral resolution with a large spectral range, and offer numerous resolved metallic absorption lines with which to measure line broadening. In order to identify an optimal set of spectral lines, we first searched the VALD3 line lists described in \S~\ref{lsd} for isolated metallic lines. The final list for all stars includes: C~{\sc ii} 426.7 nm and 658.2 nm; N~{\sc ii} 404.4 nm and 568.0 nm; O~{\sc ii} 418.5 nm and 445.2 nm; Ne~{\sc i} 640.2 nm; Ne~{\sc ii} 439.2 nm; Si~{\sc ii} 412.8 nm, 504.1 nm, 637.1 nm, and 567.0 nm; Si~{\sc iii} 455.3 nm and 457.5 nm; Si~{\sc iv} 411.6 nm; S~{\sc ii} 543.3 nm and 566.5 nm; S~{\sc iii} 425.4 nm\footnote{While this line is blended with an O~{\sc ii} line, the strength of this O~{\sc ii} line is negligible below about 20 kK; this line was not used for stars above this \teff.}; and Fe~{\sc ii} 526.0 nm and 538.7 nm. For each star, the list was curated to remove lines that were absent (due to chemical peculiarities or effective temperature), or blended with other lines (due to high \vsini). For HD 37061, for which none of the given lines were detected, we selected He~{\sc i} 501.6 nm (pressure broadening being fairly low in this line), He~{\sc ii} 468.6 nm, and Si~{\sc iii} 456.8 nm. 

We used mean spectra created from all available spectra for each star, so as to minimize the impact of line profile variability. For binary stars, using the same method as in \S~\ref{section_bz}, we first decomposed the spectra into their stellar components using an iterative algorithm similar to that described by \cite{2006AA...448..283G}. 

\begin{figure}
  \centering
   \includegraphics[width=8.5cm]{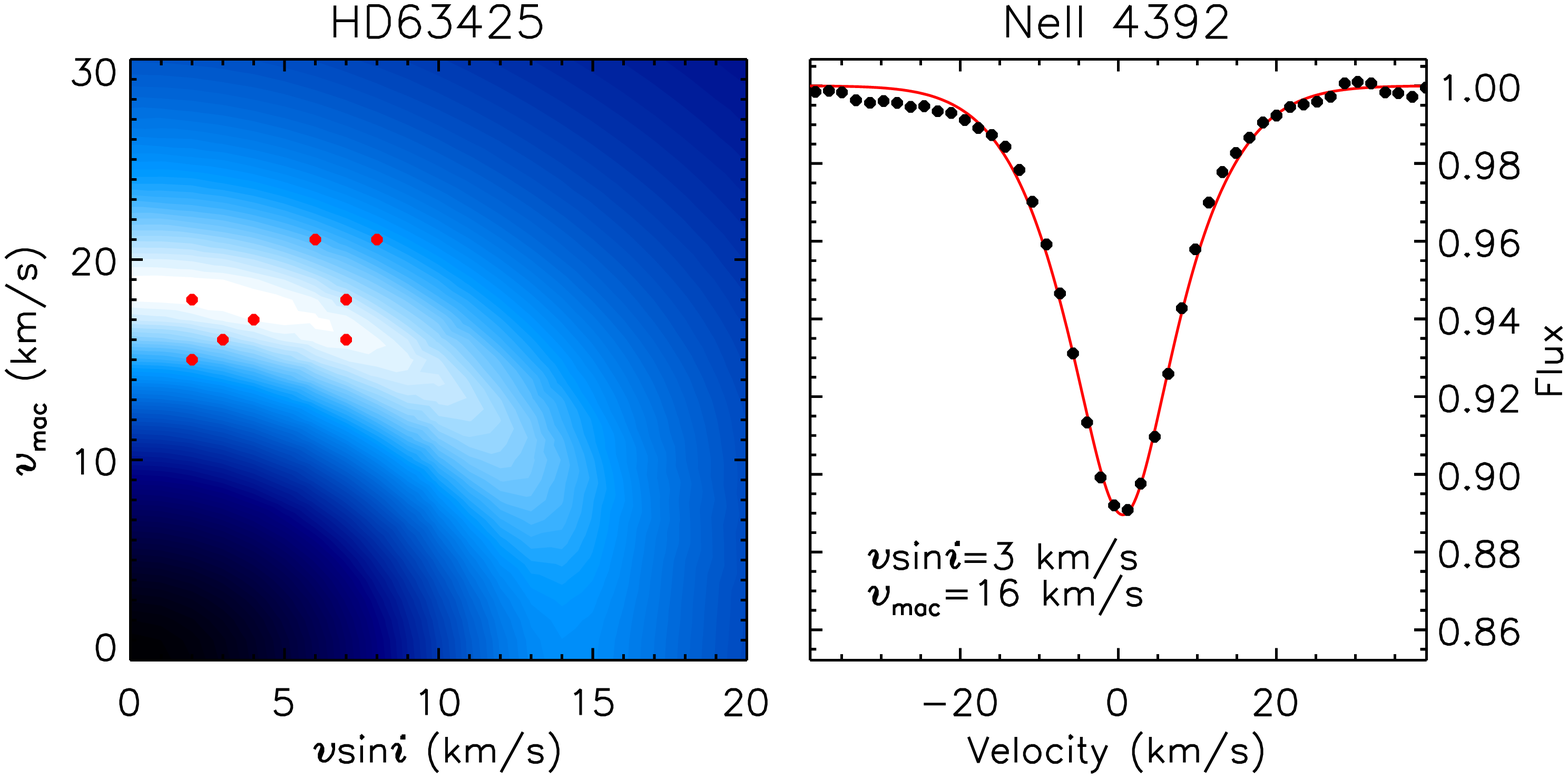} 
   \includegraphics[width=8.5cm]{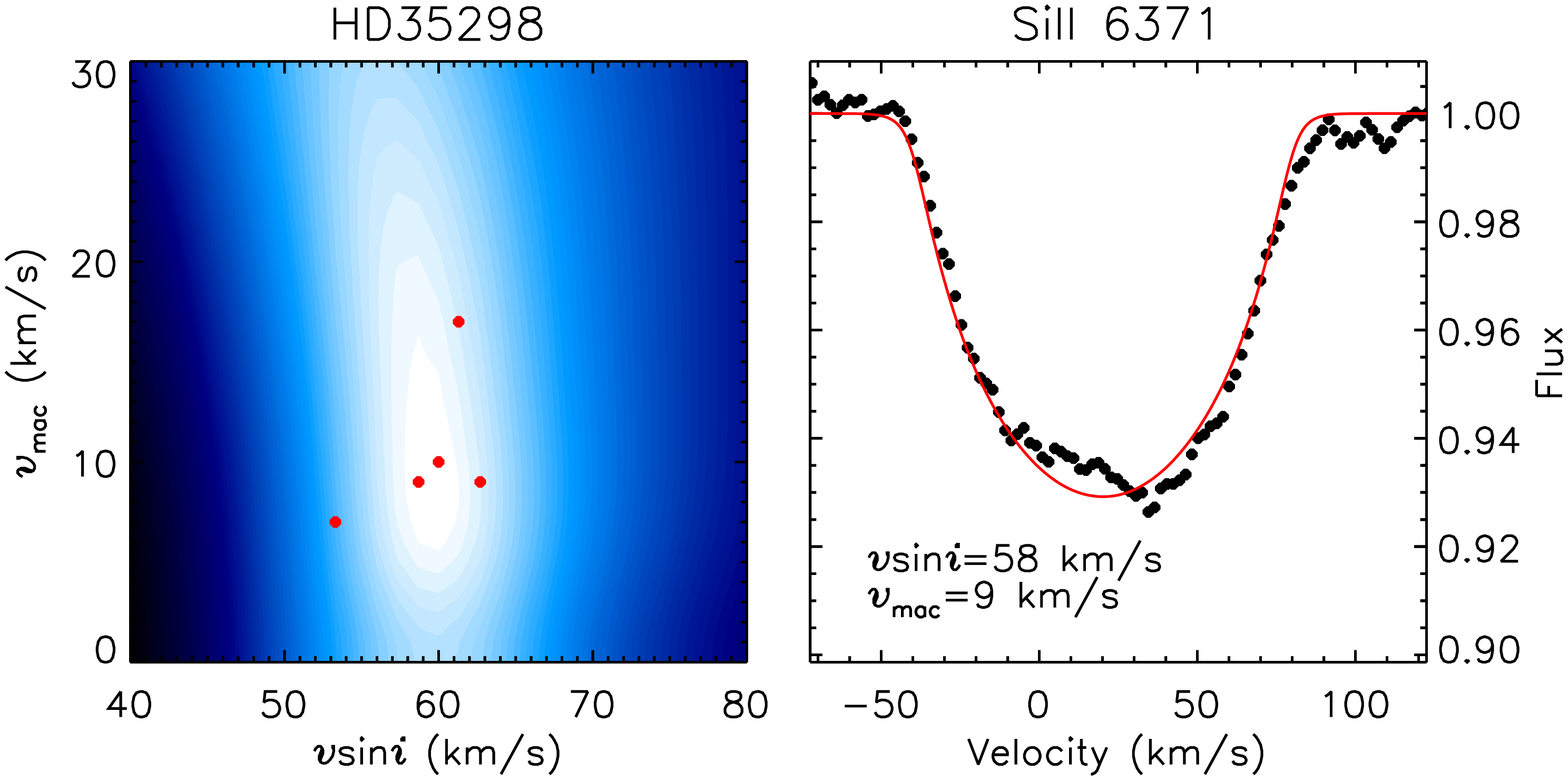} 
   \includegraphics[width=8.5cm]{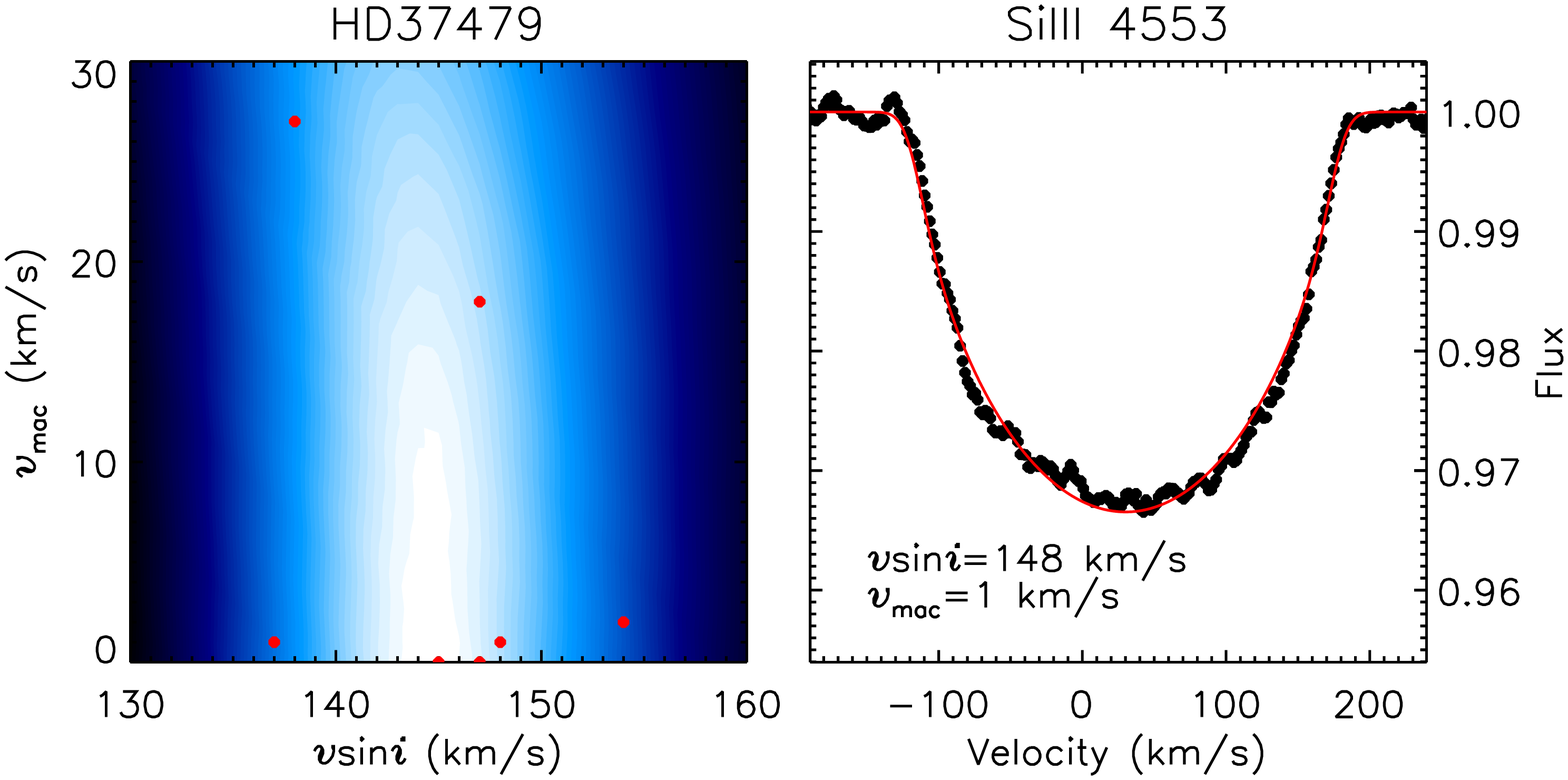} 
      \caption[\vsini~vs.~$v_{\rm mac}$ $\chi^2$ landscape for HD 63425.]{{\em Left}: \vsini~vs.~$v_{\rm mac}$ $\chi^2$ landscapes. Shading is proportional to $\log{(\chi^2)}$, with lighter colours indicating lower $\chi^2$ values. The map reflects the total $\chi^2$ across all lines tested. The best-fit models for individual lines are indicated by red circles. {\em Right}: representative model fits. Observed and synthetic line profiles are indicated by black circles and the red line. {\em Top}: The slow rotator HD 63425. Lines used for analysis were: (O~{\sc ii} 418.5 nm and 445.2 nm; N~{\sc ii} 567.9 nm; Ne~{\sc ii} 439.2 nm; Al~{\sc iii} 451.2 nm; Si~{\sc iii} 455.3 nm and 457.5 nm; Si~{\sc iv} 411.6 nm). {\em Middle}: The intermediate rotator HD 35298. Due to strong blending in this He-weak Bp star, only 5 suitable lines could be found: Ne~{\sc i} 640.2 nm, Si~{\sc ii} 412.8 nm, 504.1 nm, 634.7 nm, and 637.1 nm. {\em Bottom}: the rapid rotator HD 37479. The lines used were C~{\sc ii} 426.7 nm, N~{\sc ii} 568.0 nm, O~{\sc ii} 466.2 nm, Ne~{\sc i} 640.2 nm, Si~{\sc ii} 637.1 nm, and Si~{\sc iii} 455.3 nm and 456.8 nm.}
         \label{vsini_vmac_chi}
   \end{figure}

Line broadening was evaluated using a $\chi^2$ goodness of fit test, comparing each line to a grid of synthetic line profiles covering a range of \vsini~and $v_{\rm mac}$ values. The synthetic profiles were convolved with a gaussian with a FWHM corresponding to the instrumental resolution of ESPaDOnS and Narval spectra, providing a lower limit to line broadening measurements of about 2~\kms. Examples of the $\chi^2$ landscapes are shown in Fig.\ \ref{vsini_vmac_chi} for 3 stars: the sharp-lined star HD 63425; the He-weak star HD 35298, with an intermediate rotational period; and HD 37479, a rapidly rotating star. 

Disk integration was performed with an engine similar to that described by \cite{petit2012a}, with some modifications. First, local profile widths were calculated using Maxwellian velocity distributions appropriate to the stellar \teff~and the atomic weight. Second, radial-tangential macroturbulence rather than isotropic turbulence was implemented \citep{1975ApJ...202..148G}. This was motivated by the inclusion of higher-mass stars in the sample, especially the pulsating $\beta$ Cep stars. For Bp stars $v_{\rm mac}$ values may be fictitious in that they do not likely reflect actual velocity fields within the stellar atmosphere, but can be taken as standing in for distortions to the line profile introduced by chemical spots or Zeeman splitting (see below). In general, macroturbulence may arise from a variety of physical processes, and can be taken as a stand-in for non-rotational broadening \citep{2017A&A...597A..22S}.

For slow rotators (e.g., HD 63425, Fig.\ \ref{vsini_vmac_chi}, top), solutions with high $v_{\rm mac}$ and \vsini~$\sim 0$~\kms~produce much better fits. For intermediate rotators such as HD 35298 (Fig.\ \ref{vsini_vmac_chi}, middle), the quality of the fit is improved by inclusion of non-zero $v_{\rm mac}$, although the lowest $\chi^2$ solutions with $v_{\rm mac}=0$ \kms~yield essentially the same \vsini~as the best-fit solution with higher values of $v_{\rm mac}$. For rapid rotators, inclusion of $v_{\rm mac}$ makes very little difference, as shown for the example of HD 37479 in Fig.\ \ref{vsini_vmac_chi} (bottom). 

The final values of \vsini~and $v_{\rm mac}$ were taken as the mean of the best-fit values across all analyzed spectral lines. Uncertainties were determined based on the standard deviation of the best-fit parameters across all lines, and are typically on the order of 5 \kms. \vsini~and $v_{\rm mac}$ are given in Table \ref{rottab}. 

\cite{2013AA...559L..10S} examined \vsini~diagnostics for magnetic O-type stars known to have extremely long rotation periods, such that the true \vsini~should be essentially zero, and found that in such cases \vsini~was often drastically over-estimated, by up to about 50 \kms~(i.e.\ a similar value to that found for $v_{\rm mac}$). They concluded that macroturbulence was likely contaminating the measurements. \cite{2014AA...569A.118A} found that measurements of \vsini~could, at least in the case of the Fast Fourier Transform (FFT) method, be significantly affected by both pulsation and chemical spots, as \vsini~was strongly modulated with known pulsation and rotation periods. Both of these studies suggest that uncertainties in \vsini~may in some cases be underestimated.  Therefore, in the case of stars for which we determine extremely long ($>1$ yr) rotation periods, we consider our \vsini~measurements to be upper limits based upon the spectral resolution of the data, as the true projected rotational velocities are necessarily much lower, and as the measurements are similar in magnitude to the instrumental profile. 

Another mechanism that may affect line broadening is gravity darkening. \cite{2004MNRAS.350..189T} showed that, for stars with surface equatorial rotational velocities above 80\% of their critical velocities, line broadening ceases to be a sensitive measure of the projected rotational velocity. This is because gravity darkening reduces the contribution of equatorial regions to the integrated flux. Accounting for this requires detailed spectral modelling including meridional temperature and surface gravity variations, together with knowledge of the inclination of the rotational axis from the line of sight. Only two stars in this sample are rotating in this regime, HD 142184 \citep{grun2012}, and HD 182180 \citep{rivi2013}. In both cases careful spectral modelling accounting for oblateness as well as the meridional spectral differences arising from gravity darkening has been performed \citep{grun2012,rivi2013}, and the \vsini~values so obtained are adopted here. 

   \begin{figure}
  \centering
   \includegraphics[width=8.5cm]{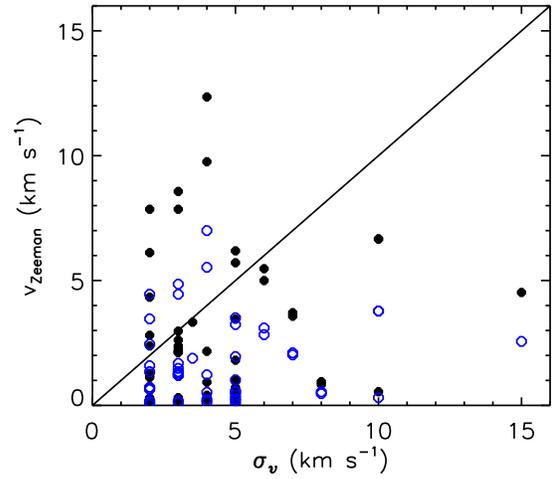}
      \caption[]{Zeeman broadening as a function of the uncertainty in \vsini, $\sigma_v$. Filled black circles show the Zeeman broadening expected for a fictitious line with $\lambda_0=500$~nm and $g_{\rm eff}=1$. Open blue circles show the Zeeman broadening for the S~{\sc ii} 566.5~nm line, with $g_{\rm eff}=0.5$.}
         \label{sigv_vzeeman}
   \end{figure}





The line profiles of magnetic stars are subject to additional line broadening due to Zeeman splitting. For the majority of stars in the sample, this does not significantly affect line broadening: a 1 kG field will cause a line with an effective Land\'e factor $g_{\rm eff}=1$ to split by $\Delta\lambda \approx 0.0012$ nm~at 500 nm, which is equivalent to about 0.7 \kms. However, for stars with strong surface magnetic fields (5-10~kG) and sharp spectral lines, Zeeman splitting can be a significant source of additional broadening as it is comparable to the broadening due to rotation and turbulence. Fig.\ \ref{sigv_vzeeman} shows the predicted Zeeman broadening $v_{\rm Zeeman}$ as a function of the uncertainty in \vsini, $\sigma_v$. The degree of Zeeman splitting was computed for a fictitious line with $\lambda_0 = 500$~nm and $g_{\rm eff} = 1$, taking the surface strength of the magnetic field to be $3.4$\bz$_{\rm max}$ (the minimum surface magnetic field strength at the magnetic pole, assuming a dipolar magnetic field). For the majority of the sample $\sigma_v \ge v_{\rm Zeeman}$, and in these cases Zeeman splitting can be neglected. However, there are 11 stars for which $\sigma_v \le v_{\rm Zeeman}$. These are: HD~35298, HD~35502, HD~36485, HD~36526, HD~37776, HD~58260, HD~96446, HD~149277, HD~175362, HD~182180, and HD~184927. 

For HD~36485, HD~58260, HD~96446, HD~149277, HD~175362, and HD~184927 the S~{\sc ii}~566.5~nm line is detectable. This line has the lowest effective Land\'e factor in the VALD line lists, $g_{\rm eff} = 0.5$, and is therefore much less strongly affected by Zeeman splitting. Fig.\ \ref{sigv_vzeeman} also shows the amount of Zeeman splitting expected for S~{\sc ii}~566.5~nm, for which $v_{\rm Zeeman}$ is much closer to $\sigma_v$. Therefore, for these stars we used this line to measure line broadening, obtaining uncertainties from the $1\sigma$ $\chi^2$ contours. For HD~35298 and HD~36526, the Fe~{\sc ii}~450.8~nm line, which also has $g_{\rm eff} = 0.5$, was used instead. In general using these lines reduced \vsini~by 3 to 5 \kms, consistent with the predicted degree of Zeeman splitting. For HD~37776, no low-$g_{\rm eff}$ lines could be detected, but the amount of Land\'e splitting ($\sim 10$~\kms) predicted for this star according to the approximation described in the previous paragraph is close to the difference between the value determined above ($101 \pm 4$~\kms), and the value reported by \cite{koch2011}, 91~\kms, who performed a detailed spectrum synthesis accounting for the star's strong magnetic field, which we adopt here. For HD~35502 and HD~182180, the lines with $g_{\rm eff} \le 0.5$ are all too weak to detect, even in the mean disentangled spectrum, but $v_{\rm Zeeman} \sim 6$~\kms~is only slightly higher than $\sigma_v = 5$~\kms, therefore we do not expect Zeeman splitting to significantly affect the \vsini~measurements of these two stars. The \vsini~values for these stars listed in Table \ref{rottab} reflect the considerations described here.

   \begin{figure}
  \centering
   \includegraphics[width=8.5cm]{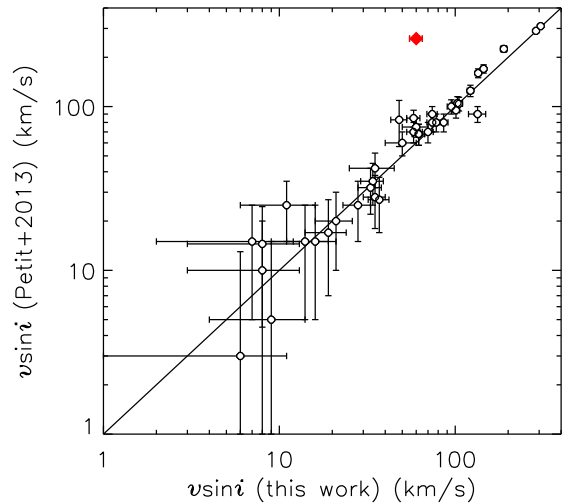} \\
      \caption[Comparison of \vsini~measurements with literature values.]{Comparison of \vsini~measurements with literature values. Note that the axes are logarithmic. The solid red diamond indicates HD 35298, discussed further in the text.}
         \label{vsini_compare}
   \end{figure}

Fig.\ \ref{vsini_compare} compares these values of \vsini~to those collected by \cite{petit2013}. There is only one 3$\sigma$ outlier, HD 35298, which is highlighted. The value given by P13, 260 \kms, is much higher than that found here, 58$\pm$2 \kms, as illustrated in Fig.\ \ref{vsini_vmac_chi} (middle). The original value appears to have been measured by \cite{1963ApJ...137..316M}, based on a 10~\AA/mm spectrogram. It is not clear how such a high \vsini~was obtained. In any case, it is inconsistent with the value determined here; our results are however consistent with the measurement published by \cite{1975mazb.conf...75K}, 57~\kms. 

\begin{table*}
\caption[\vsini~and Rotation periods.]{Projected rotation velocities, macroturbulent velocities, and rotational periods. An asterisk next to the HD number indicates that the star's \bz~curve is included in the Appendix. In most cases JD0 (in HJD) is defined at $|\langle B_z\rangle|_{\rm max}$, as discussed in the text; ephemerides for which JD0 is defined otherwise are indicated by a superscript in the 5$^{th}$ column: maximum light: $p$; minimum EW: $s$. The 6$^{th}$ column gives the method by which the period was determined: $m$: magnetic; $u$ ultraviolet spectroscopy; $s$: optical spectroscopy; $p$: photometry; $sp$: a non-linear ephemeris accounting for the spindown. Where more than one method was used to determine $P_{\rm rot}$, the S/N in the 7$^{th}$ column is given in the same order as Method. The final column gives the reference for the period; where `This work' is given together with another reference, we have refined the period determined in a previous study.}
\label{rottab}
\begin{tabular}[l]{l | r r r r l r l}
\hline
\hline
Star  & \vsini & $v_{\rm mac}$ & $P_{\rm rot}$ & JD0             & Method & S/N & Reference\\
      & (\kms) & (\kms)        & (d)           & -2400000 (d)    &        &     &          \\
\\
HD   3360 &  $19\pm 2$ &  $13\pm 1$ &  5.37045(7) & 45227.2(25)$^{s}$ & {\em u} & 8.6  & \cite{neiner2003a}\\ 
HD  23478 & $136\pm 7$ &   $8\pm 8$ & 1.0498(2) & 47933.7(2)$^{p}$ & {\em p} & 5.2  & \cite{2015MNRAS.451.1928S}\\ 
HD  25558$^*$ &  $35\pm 4$ &  $20\pm 10$ & 1.233(1) & 55400.0(2) & {\em m} & 4.7  & \cite{2014MNRAS.438.3535S}, This work\\ 
HD  35298$^*$ &  $58\pm 2$ &  $10\pm 8$ & 1.85458(3) & 54486.91(7) & {\em p, m} & 8.0, 32.7  & \cite{1984AA...141..328N,2013AstBu..68..214Y}, This work\\ 
HD  35502 &  $78\pm 5$ &  $10\pm 9$ & 0.853807(3) & 56295.812850(3) & {\em p, s, m} & 7.1, 20.7, 32.8  & \cite{2016MNRAS.460.1811S}\\ 
HD  36485$^*$ &  $26\pm 4$ &   $6\pm 5$ & 1.47775(3) & 48298.86(3)$^s$ & {\em s} & --  & \cite{leone2010}\\ 
HD  36526$^*$ &  $55\pm 5$ &   $5\pm 5$ & 1.54185(4) & 55611.93(6) & {\em p, m} & 27.4  & \cite{1984AA...141..328N}, This work\\ 
HD  36982$^*$ &  $86\pm 5$ &  $17\pm 12$ & 1.8551(5) & 54412.5(3) & {\em m} & 5.9  & This work\\ 
HD  37017$^*$ & $134\pm 15$ &  $12\pm 22$ & 0.901186(2) & 43441.20(9) & {\em m} & 11.0  & \cite{1987ApJ...323..325B}, This work\\ 
HD  37058$^*$ &  $11 \pm 2$ &  $15 \pm 2$ & 14.581(2) & 56522.0(4) & {\em m} & 28.0  & \cite{pederson1979}, This work\\ 
HD  37061$^*$ & $189\pm 8$ &  $61\pm 30$ & 1.0950(4) & 55223.2(2) & {\em m} & 11.0  & This work\\ 
HD  37479 & $145\pm 5$ &  $13\pm 16$ & 1.1908100(9) & 42778.829(1)$^{p}$ & {\em sp} & --  & \cite{town2010}\\ 
HD  37776 & $91\pm 4$ &  $35\pm 4$ & 1.5387115(9) & 48857.124(3)$^{p}$ & {\em sp} & --  & \cite{miku2008}\\ 
HD  44743 &  $20\pm 7$ &  $41\pm 4$ & -- & -- & -- & -- & \cite{2015AA...574A..20F}, This work\\ 
HD  46328 &   $\le 8$ &  $15\pm 5$ & $>$30 (yr) & 44296(304) & {\em s, m} & 29.0, 26.0  &   \cite{2017MNRAS.471.2286S}\\ 
HD  52089 &  $21\pm 2$ &  $47\pm 2$ &  -- & -- & -- & -- & \cite{2015AA...574A..20F}\\
HD  55522$^*$ &  $70\pm 2$ &  $11\pm 9$ & 2.7292(3) & 53000.5(2)$^{p}$ & {\em p, m} & 18.9, 11.8  & \cite{2004AA...413..273B}, This work\\ 
HD  58260$^*$ &   $3\pm 2$ &  $16\pm 2$ &  -- & -- & -- & -- & -- \\
HD  61556 &  $58\pm 3$ &  $19\pm 10$ & 1.9087(6) & 55198.05(5) & {\em p, s, m} & 6.4, 13.8, 17.1  & \cite{2015MNRAS.449.3945S}\\ 
HD  63425$^*$ &   $3\pm 3$ &  $19\pm 2$ & 163(4) & 55285(13) & {\em m} & 9.4  & This work\\ 
HD  64740$^*$ & $135\pm 2$ &  $18\pm 11$ & 1.330205(3) & 43498.43(7) & {\em m} & 14.3  & \cite{1987ApJ...323..325B}, This work\\ 
HD  66522$^*$ &   $3\pm 2$ &  $12 \pm 2$ &  909(75) & 47365(76)$^{p}$ & {\em p} &  7.3  & This work\\ 
HD  66665$^*$ &   $8\pm 2$ &  $10\pm 1$ & 24.5(1) & 55274.4(8) & {\em m} & 14.7  & \cite{petit2013}, This work\\ 
HD  66765$^*$ &  $95\pm 3$ &  $12\pm 16$ & 1.6079(5) & 55907.9(2) & {\em p, s, m} & 4.02, 11.8, 12.5  & \cite{alecian2014}, This work\\ 
HD  67621$^*$ &  $21\pm 5$ &   $8\pm 5$ & 3.593(1) & 55904.7(1) & {\em m} & 34.3  & \cite{alecian2014}, This work\\ 
HD  96446 &   $6\pm 2$ &   $15\pm 15$ & 23.38(3) & 55734(5)$^{s}$ & {\em s, m} & --, 28.4  & \cite{2017MNRAS.464L..85J}\\ 
HD 105382 &  $74\pm 5$ &   $7\pm 9$ & 1.295(1) & 47866.0(1)$^{p}$ & {\em p} & 19.2  & \cite{2001AA...366..121B}\\ 
HD 121743 &  $74\pm 4$ &  $10\pm 8$ & 1.130170(9) & 56760.855 & {\em m} & --  &  Briquet et al.\ (in prep.)\\ 
HD 122451$^*$ &  $70\pm 10$ &  $20\pm 20$ & 2.885(1) & 55706.6(4) & {\em m} & 5.4  & This work\\ 
HD 125823$^*$ &  $16\pm 2$ &  $10\pm 5$ &  8.8169(1) & 43733.6(35) & {\em p, m} & 17.7, 32.4  & \cite{1996AA...311..230C}, This work\\ 
HD 127381 &  $62\pm 4$ &  $15\pm 6$ & 3.0194(2) & 47620.5(6)$^{p}$ & {\em p, m} & 4.9, 7.4  & \cite{henrichs2012}\\ 
HD 130807 &  $28\pm 5$ &  $16\pm 6$ & 2.9533(1) & 55707.06(9) & {\em p, m} & 4.4, 25.3  & \cite{2017omiLup_inprep} \\ 
HD 136504A &  $35\pm 10$ &  $18\pm 6$ & -- & -- & -- & -- & --\\ 
HD 136504B &  $35 \pm 5$ &   $5\pm 5$ & -- & -- & -- & -- & -- \\ 
HD 142184 & $288\pm 6$ &   $5\pm 5$ & 0.508276(1) & 47913.694(1)$^{s}$ & {\em p, s, m} & 7.6, 10.4, 4.4  & \cite{grun2012}\\ 
HD 142990$^*$ & $122\pm 2$ &   $1\pm 1$ & 0.978832(2) & 43563.00(5)$^{p}$ & {\em p m} & 15.5  & \cite{2005AA...430.1143B}, This work\\ 
HD 149277$^*$ &  $8\pm 3$ &  $15\pm 2$ & 25.380(7) & 56119.5(3) & {\em m} & 34.8  & This work\\ 
HD 149438 &   $7\pm 3$ &   $2\pm 2$ & 41.033(2) & 53482(2) & {\em m} & 66.7  & \cite{2006MNRAS.370..629D}\\ 
HD 156324 &  $50\pm 10$ &  $10\pm 6$ & 1.5805(3) & 56128.89(9)$^{s}$ & {\em s, m} & 22.6, 14.4  & \cite{shultz_hd156324_2017}\\ 
HD 156424$^*$ &   $7\pm 2$ &  $14\pm 5$ & 2.8721(6) & 56126.3(3) & {\em m} & 15.8  & This work\\ 
HD 163472 &  $62\pm 5$ &  $10\pm 5$ & 3.6388330(9) & 44807.71(6)$^{s}$ & {\em u} & 18.8  & \cite{neiner2003b}\\ 
HD 164492 & $138\pm 10$ &  $34\pm 10$ & 1.36986(7) & 56770.53(7) & {\em s, m} & 11.6, 68.9  &  \cite{2017MNRAS.465.2517W}\\ 
HD 175362$^*$ &  $34\pm 4$ &  $13\pm 8$ & 3.67381(1) & 43733.04(8) & {\em p, m} & 34.8, 38.3   & \cite{1987ApJ...323..325B}, This work\\ 
HD 176582 & $103\pm 7$ &  $15\pm 15$ & 1.581984(3) & 54496.694(2)$^{s}$ & {\em s, m} & 6.9, 20.7  & \cite{bohl2011}\\ 
HD 182180 & $306\pm 5$ &   $1\pm 1$ & 0.5214404(6) & 54940.83(5)$^{s}$ & {\em p, s} & 57.2, 37.7  & \cite{2010MNRAS.405L..51O,2010MNRAS.405L..46R}\\ 
HD 184927 &   $8\pm 2$ &   $8\pm 1$ &  9.53102(7) & 55706.517(5) & {\em s, m} & --, 15.8  & \cite{2015MNRAS.447.1418Y}\\ 
HD 186205$^*$ &   $4\pm 4$ &  $12\pm 3$ & 37.21(6) & 55640(3) & {\em m} & 10.4  & This work\\ 
HD 189775$^*$ &  $58\pm 2$ &  $13\pm 7$ & 2.6071(3) & 56262.01(8) & {\em p, m} & 32.7  & \cite{petit2013}, This work\\ 
HD 205021 &  $37\pm 2$ &  $25\pm 5$ & 12.000750(9) & 52366.3(1) & {\em u, m} & 37.5, 13.0  & \cite{henrichs2013}\\ 
HD 208057$^*$ & $105\pm 5$ &  $33\pm 11$ & 1.3678(4) & 55028.2(1) & {\em m} & 11.6  & This work\\ 
ALS 3694$^*$  &  $48\pm 3$ &  $22\pm 12$ & 1.678(3) & 56819.5(2)$^{s}$ & {\em s} & 5.4  & This work\\ 
\hline\hline
\end{tabular}
\end{table*}

\subsection{Rotation periods}\label{rotation_periods}

Amongst magnetic stars with radiative envelopes, magnetic measurements are in general modulated entirely by a star's rotation, since their fossil magnetic fields do not exhibit cyclic variability of the kind observed in cool stars with dynamo-generated surface magnetic fields. The same is true of their photometric and spectroscopic variability, since these arise in photospheric chemical spots and, in some cases, magnetospheric emission lines: in neither case has any variability been detected that could not be ascribed to rotation. The only exception to this rule of relevance to the magnetic early B-type stars are the $\beta$ Cep and SPB pulsators, whose photometric and spectroscopic variability is dominated by pulsation; however, in these cases magnetic diagnostics are still primarily sensitive to rotation, e.g.\ \cite{neiner2003a,neiner2003b,henrichs2013}, and the timescales of variability are also very different (hours for $\beta$ Cep pulsation, days for rotation)\footnote{In the case of the $\beta$ Cep pulsator $\xi^1$ CMa, a weak modulation of \bz~with pulsation phase has been detected, however the amplitude of this modulation is much smaller than that of the rotational modulation, and this difference, combined with the very different timescales, hours vs.\ decades for this star, make their respective influences easily distinguishable \citep{2017MNRAS.471.2286S}.}. Therefore, to determine rotation periods, in most cases we have relied on the \bz~measurements described in \S~\ref{Magnetometry}. In two cases, ALS 3694 and HD 156324, we obtained periods using H$\alpha$ EWs, as both of these stars display magnetospheric H$\alpha$ emission \citep{alecian2014,2016ASPC..506..305S}, and the spectroscopic datasets are larger than the magnetic datasets. For HD 66765 we refined $P_{\rm rot}$ using Hipparcos photometry and He EWs, where we assumed  both the photometric and the line profile variability in this He-strong star to be due to chemical spots. In one case, HD 66522, $P_{\rm rot}$ was determined using archival Hipparcos photometry, where once again we assume the origin of the photometric variability to be rotational modulation (the very long period determined for this star, $\sim$900 d, is too long to be compatible with pulsation, while orbital modulation is unlikely as there is no evidence of RV variability). Hipparcos photometry was also used to refine $P_{\rm rot}$ in two further cases, HD 142990 and HD 35298. 

Period analysis was performed using Lomb-Scargle statistics \citep{1976ApSS..39..447L, 1982ApJ...263..835S} as implemented in the {\sc idl} program {\sc periodogram.pro}\footnote{Available at https://hesperia.gsfc.nasa.gov/ssw/gen/idl/\\util/periodogram.pro}, which normalizes the periodogram to the total variance \citep{1986ApJ...302..757H}.  The uncertainty in each frequency was determined using the formula from \cite{1976fats.book.....B}, $\sigma_{\rm F}=\sqrt{6}\sigma_{\rm obs}/(\pi\sqrt{N_{\rm obs}}A\Delta T)$, where $\sigma_{\rm obs}$ is the mean uncertainty in the measurements, $N_{\rm obs}$ is the number of measurements, $A$ is the amplitude of the RV curve, and $\Delta T$ is the timespan of observations.  

In order to check that the rotation periods are physically plausible, period windows were bounded from above by

\begin{equation}\label{prot_upper}
P_{\rm rot} \le \frac{2\pi R_{\rm eq}}{v\sin{i}},
\end{equation}

\noindent and from below by the breakup velocity $v_{\rm br}$ \citep{1928asco.book.....J}, where $R_{\rm eq}$ is the equatorial radius. For the upper bound of the period window, the polar radius $R_{\rm p} = R_{\rm eq}$, while for the breakup velocity $R_{\rm eq} = 1.5 R_{\rm p}$ due to rotationally induced oblateness. The majority of the stellar masses and radii used to determine $v_{\rm br}$ were obtained from \cite{petit2013}, with the exception of those stars in which magnetic fields were discovered subsequent to the publication of their catalogue. In these cases masses and radii were obtained from \cite{2015MNRAS.451.1928S} for HD 23478, \cite{2015AA...574A..20F} for HD 44743 and HD 52089, \cite{uytterhoeven2005} for HD 136504, and \cite{2017MNRAS.465.2517W} for HD 164492C. 







Within a given period window, it is frequently the case that there are multiple peaks in the periodogram that phase the data more or less equally well, all of which meet the formal criterion for signifiance (a S/N$>$4, \citealt{1993A&A...271..482B,1997A&A...328..544K}). In order to check for spurious peaks, we used the \nz~measurements discussed in \S~\ref{section_bz}. Peaks which appear in the \nz~period spectrum are likely a consequence of the window function, and can be ignored. When \nz~is not available (for historical \bz~measurements as well as spectroscopic and photometric data), we used synthetic null measurements obtained via random Gaussian noise normalized to the mean uncertainty in order to derive the window contribution to the periodogram within the sampling window. 

The statistical significance of a given peak in the periodogram can be quantified by means of the false alarm probability (FAP), where we use Eqn.~22 from \cite{1986ApJ...302..757H} which gives the FAP as a function of the number of data points and the amplitude of the period spectrum. Similarly to the FAPs used in \S~\ref{Magnetometry} to evaluate the statistical significance of the signal within Stokes $V$, smaller FAPs indicate that a signal is less likely to be a consequence of white noise. We also calculated the S/N of each period, after prewhitening with the most significant period (and the harmonics, for variations with higher-order terms). No periods in Table \ref{rottab} have a S/N below 4. The 13.6~d period for HD 44743 found by \cite{2015AA...574A..20F} has a S/N of only 1.9, is thus likely to be spurious, and is therefore not included (and indeed, unpublished HARPSpol \bz~measurements collected via the BRITEpol LP are not coherently phased with this period; Bram Buysschaert, priv. comm.). 

Depending on the size of a given dataset, the time-sampling, and the amplitude of the signal, in some cases there may be multiple peaks in the periodogram that meet the formal criterion for significance, and are of similar power, i.e.\ the rotational period given in Table \ref{rottab} may not be unique. When this is the case, this is directly addressed in the Appendix, where for each star with a new or refined rotational period we show both periodograms, \bz~curves, and where appropriate or available light curves phased with the adopted ephemeris. In general we preferred solutions that implied higher rotational axis inclinations (i.e.\ longer periods), since higher inclinations are intrinsically more likely than lower inclinations. 

In the end, new rotation periods have been determined for 10 stars. For a further 14 stars, comparison of the new magnetic data to \bz~measurements in the literature has enabled refinement of the rotation periods. Rotation periods are given in Table \ref{rottab}. In general JD0 is taken to be the heliocentric Julian date at which \bz=$|\langle B_z\rangle|_{\rm max}$ in the rotational cycle preceding the first observation in the time series, based upon a sinusoidal fit to the data. In some cases maximum light or minimum equivalent width were used to define JD0; in these cases, this is indicated with a superscript in the 5$^{th}$ column of Table \ref{rottab}. The uncertainty in JD0 was determined from the uncertainty in the phase of the sinusoidal fit. Also given in Table \ref{rottab} are references for rotational periods obtained from the literature, and the method by which the period was obtained: via ultraviolet spectroscopy, optical spectroscopy, photometry, or magnetometry. When comparison of our measurements to historical data has allowed refinement of $P_{\rm rot}$, this work is also given as a reference. 



Rotation periods could not be determined for five stars: HD 44743, HD 52089, HD 58260, and HD 136504A and B. In the cases of HD 44743 and HD 136504, period analysis is hampered by both the small number of high-resolution \bz~measurements, and by the low levels of variability relative to the median error bar in the datasets. For HD 52089, there are 8 high-resolution \bz~measurements available, in addition to Hipparcos photometry, but due to the very low level of variability in both datasets relative to the observational uncertainties no period above the S/N threshold of 4 could be identified. All of these stars have relatively weak \bz~in comparison to the majority of the stars in the sample, around 10-100~G. In the case of HD 58260, despite the mean \bz~of $\sim$1.8~kG being significant at the 30$\sigma$ level, the standard deviation of \bz~is less than the mean 1$\sigma$ error bar. 

In two cases, HD 37479 and HD 37776, non-linear ephemerides are available that account for the observed spin-down of the star \citep{town2010, miku2008}. \cite{oks2012} demonstrated the agreement achieved by this ephemeris between historical and modern \bz~measurements of HD 37479. On their own, ESPaDOnS and Narval data are unable to distuinguish between the spin-down ephemeris provided by \cite{miku2008} for HD 37776 and the newer ephemeris, in which spin-{\em up} of the star is reported \citep{miku2011}. We therefore adopt the earlier ephemeris as being the more conservative option, as shown in Fig.\ \ref{hyd_bz}. This small ambiguity in ephemeris has no impact on the magnetic modelling.

The \bz~curves of stars for which \bz~curves have not already been published, or for which detailed individual studies are not currently in preparation, are show in Appendix \ref{bz_ind}. These are indicated with asterisks next to the star name in Table \ref{rottab}. 

\section{Discussion}

\subsection{Selection of \bz~datasets for modelling}\label{bz_selection}

The final column of Table \ref{masktab} gives the type of measurement to be favoured for future modelling, based on the arguments below: `Z' (LSD profiles extracted using line masks with metallic lines), `YZ' (LSD profiles extracted using line masks with metallic and He lines), or `X' (H lines). 

With an ideal dataset in which S/N is not a limitation, H line measurements would be used in all cases in order to avoid distortions due to chemical spots. However, for most stars the mean \bz~significance $\Sigma_B$ is lower in H lines than for LSD profiles, i.e.\ H line measurements are generally less precise than those obtained with LSD profiles. Furthermore, in the majority of cases differences in \bz~measured from different elements are negligible (Fig.\ \ref{ae_hist}). H line measurements were thus selected only when $A_{\rm e}/\sigma_{\rm e} \ge 2$. 

For stars with $A_{\rm e}/\sigma_{\rm e} \le 2$, measurements obtained from LSD profiles extracted using metallic line masks are in general preferred, as these are unaffected by the extra broadening introduced by He lines. However, for stars with $\Sigma_{\rm B} \sim 1$, \bz~measurements using all available spectral lines were selected, as in these cases meaningful measurements are only possible when the maximum possible precision is achieved. 

\subsection{Fits to \bz~curves}\label{curve_shape}

\begin{table*}
\caption[]{Curve-fitting of \bz~curves. Stars for which FORS data were used to constrain the fit are indicated with an asterisk next to the HD number in the 1$^{st}$ column. The $2^{nd}$ column gives the significance $\Sigma_{\rm Amp}$ of the \bz~amplitude, as explained in the text. The $3^{rd}$ column gives the reduced $\chi^2$ of a single-order sinusoidal fit. The $4^{th}$ through $7^{th}$ columns gives the amplitudes $B_{n}$ of the $n^{th}$-order least-squares sinuoisdal fits.}
\label{bzfittab}
\centering
\begin{tabular}{l | r r r | r r r r}  
\hline
\hline
HD       & $\Sigma_{\rm Amp}$ & $\chi^2_1/\nu$ & $B_0$ & $B_1$ & $B_2$ & $B_3$ \\ 
No.      &                    &                & (kG)  & (kG)  & (kG)  & (kG)  \\ 
\hline
  3360   & 6.5 & 1.0 & $-0.0037\pm 0.0006$ & $0.0098\pm 0.0009$ &  --  &  --  \\
 23478   & 4.3 & 1.5 & $-0.98\pm 0.03$ & $0.14\pm 0.05$ &  --  &  --  \\
 25558   & 3.4 & 0.5 & $-0.045\pm 0.006$ & $0.044\pm 0.008$ &  --  &  --  \\
 35298   &  43 & 4.8 & $0.40\pm 0.04$ & $3.40\pm 0.06$ & $0.17\pm 0.05$ & $0.22\pm 0.06$ \\
 35502   &  17 & 0.6 & $-1.22\pm 0.03$ & $1.66\pm 0.04$ &  --  &  --  \\
 36485   & 3.9 & 1.4 & $-2.48\pm 0.02$ & $0.06\pm 0.03$ & $0.08\pm 0.02$ &  --  \\
 36526   &  27 & 1.3 & $1.34\pm 0.05$ & $2.25\pm 0.08$ &  --  &  --  \\
 36982   & 3.8 & 0.5 & $0.129\pm 0.008$ & $0.09\pm 0.01$ &  --  &  --  \\
 37017   &  12 & 0.9 & $-0.80\pm 0.03$ & $0.98\pm 0.05$ &  --  &  --  \\
 37058   &  33 & 1.5 & $-0.19\pm 0.02$ & $0.77\pm 0.03$ &  --  &  --  \\
 37061   & 9.6 & 1.7 & $-0.11\pm 0.02$ & $0.17\pm 0.02$ &  --  &  --  \\
 37479   &  37 &  19 & $0.67\pm 0.05$ & $2.37\pm 0.07$ & $0.68\pm 0.07$ &  --  \\
 37776   &  18 &  39 & $-0.19\pm 0.04$ & $0.35\pm 0.05$ & $0.96\pm 0.06$ & $0.96\pm 0.06$ \\
 46328   &  99 & 1.1 & $-0.21\pm 0.01$ & $0.54\pm 0.01$ &  --  &  --  \\
 55522   &  13 & 1.0 & $-0.01\pm 0.04$ & $0.83\pm 0.06$ &  --  &  --  \\
 61556   &  33 & 8.3 & $-0.37\pm 0.01$ & $0.45\pm 0.01$ & $0.07\pm 0.02$ & $0.06\pm 0.02$ \\
 63425   &  11 & 1.1 & $0.095\pm 0.002$ & $0.050\pm 0.003$ &  --  &  --  \\
 64740   &  27 & 3.0 & $-0.21\pm 0.01$ & $0.57\pm 0.02$ & $0.07\pm 0.02$ &  --  \\
 66522   &  23 & 0.9 & $-0.02\pm 0.02$ & $0.74\pm 0.01$ &  --  &  --  \\
 66665   &  14 & 0.7 & $-0.040\pm 0.002$ & $0.081\pm 0.002$ &  --  &  --  \\
 66765   &  18 & 1.3 & $-0.17\pm 0.01$ & $0.56\pm 0.02$ &  --  &  --  \\
 67621   &  13 & 1.0 & $0.160\pm 0.004$ & $0.250\pm 0.006$ &  --  &  --  \\
 96446   &  20 & 0.9 & $-0.74\pm 0.04$ & $0.31\pm 0.02$ &  --  &  --  \\
105382$^*$   & 8.5 & 4.5 & $-0.33\pm 0.04$ & $0.34\pm 0.06$ &  --  &  --  \\
121743   & 3.1 & 1.0 & $0.26\pm 0.01$ & $0.06\pm 0.03$ &  --  &  --  \\
122451   & 9.3 & 3.0 & $-0.024\pm 0.002$ & $0.010\pm 0.004$ &  --  &  --  \\
125823   &  18 & 1.2 & $0.03\pm 0.01$ & $0.48\pm 0.02$ &  --  &  --  \\
127381   & 7.6 & 1.4 & $0.016\pm 0.009$ & $0.12\pm 0.01$ &  --  &  --  \\
130807   &  48 & 2.6 & $0.384\pm 0.006$ & $0.626\pm 0.008$ & $0.050\pm 0.008$ &  --  \\
142184   &  11 & 3.6 & $-1.55\pm 0.03$ & $0.54\pm 0.05$ & $0.24\pm 0.04$ &  --  \\
142990   &  20 & 3.2 & $-0.12\pm 0.03$ & $1.16\pm 0.04$ & $0.20\pm 0.05$ &  --  \\
149277   &  24 & 1.2 & $0.41\pm 0.05$ & $2.35\pm 0.07$ &  --  &  --  \\
149438   &  34 &  38 & $0.0094\pm 0.0009$ & $0.049\pm 0.001$ & $0.032\pm 0.001$ & $0.019\pm 0.001$ \\
156324   &  10 & 1.4 & $0.90\pm 0.07$ & $1.18\pm 0.09$ &  --  &  --  \\
156424   & 6.7 & 1.4 & $-0.34\pm 0.02$ & $0.17\pm 0.03$ &  --  &  --  \\
163472   & 8.7 & 1.2 & $-0.029\pm 0.004$ & $0.080\pm 0.006$ &  --  &  --  \\
164492   &  13 & 0.8 & $0.75\pm 0.03$ & $1.07\pm 0.04$ &  --  &  --  \\
175362   & 155 & 307 & $-1.02\pm 0.02$ & $4.89\pm 0.03$ & $0.93\pm 0.03$ & $0.22\pm 0.03$ \\
176582   &  28 & 3.2 & $-0.00\pm 0.02$ & $1.49\pm 0.03$ & $0.05\pm 0.03$ & $0.15\pm 0.03$ \\
182180   &  44 & 1.8 & $0.24\pm 0.08$ & $2.44\pm 0.09$ &  --  &  --  \\
184927   &  37 & 1.3 & $0.92\pm 0.01$ & $0.80\pm 0.02$ &  --  &  --  \\
186205   & 9.3 & 1.1 & $-0.717\pm 0.009$ & $0.12\pm 0.02$ &  --  &  --  \\
189775   &  25 &  14 & $0.38\pm 0.02$ & $0.61\pm 0.02$ & $0.23\pm 0.02$ &  --  \\
205021   &  19 & 0.4 & $0.003\pm 0.002$ & $0.102\pm 0.003$ &  --  &  --  \\
208057   & 8.0 & 0.8 & $-0.001\pm 0.005$ & $0.090\pm 0.008$ &  --  &  --  \\
ALS 3694$^*$ & 1.8 & 0.8 & $-1.85\pm 0.05$ & $0.2\pm 0.1$ &  --  &  --  \\
\hline\hline
\end{tabular}
\end{table*}

   \begin{figure}
  \centering
   \includegraphics[width=8.5cm]{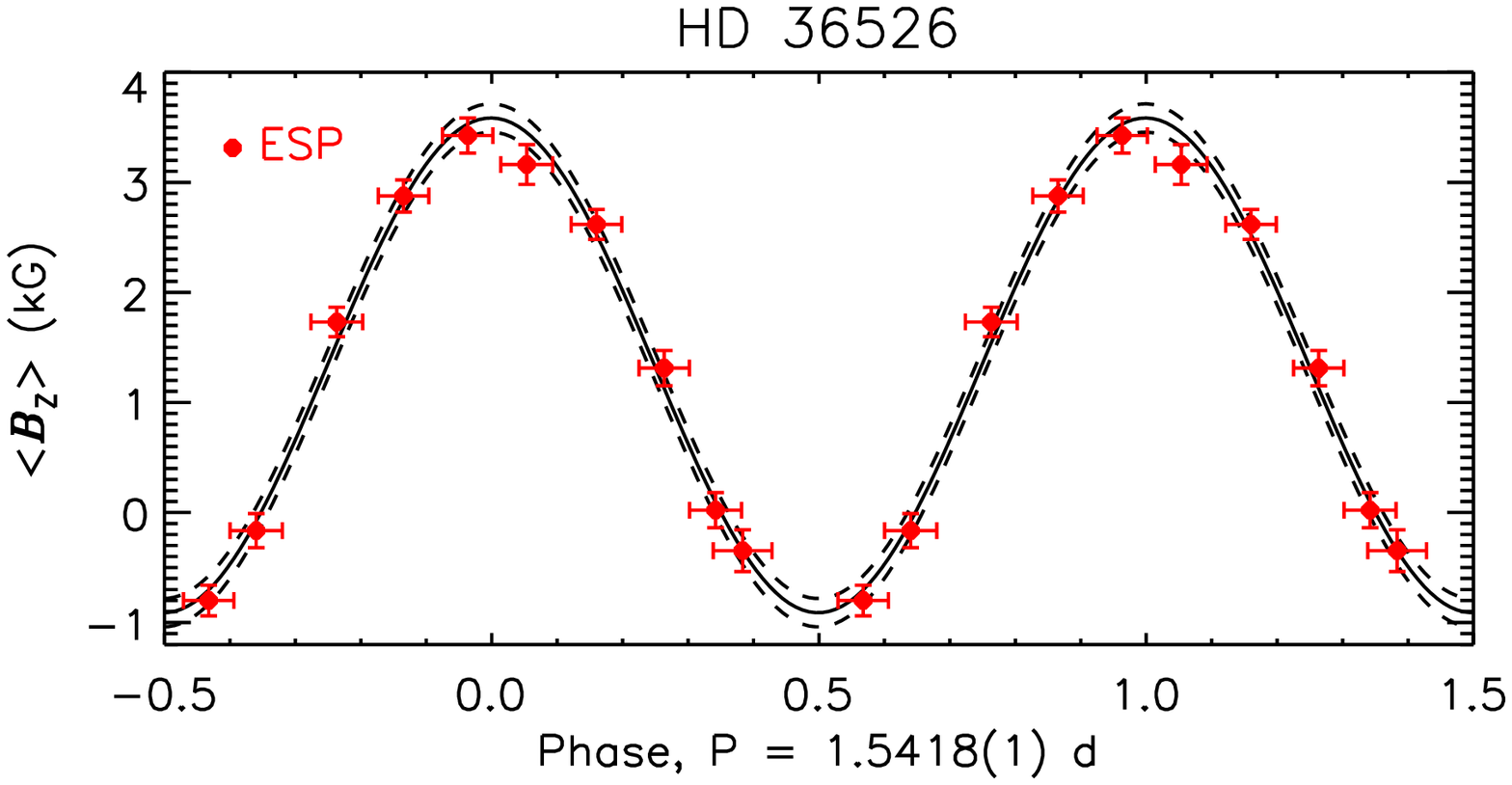}
   \includegraphics[width=8.5cm]{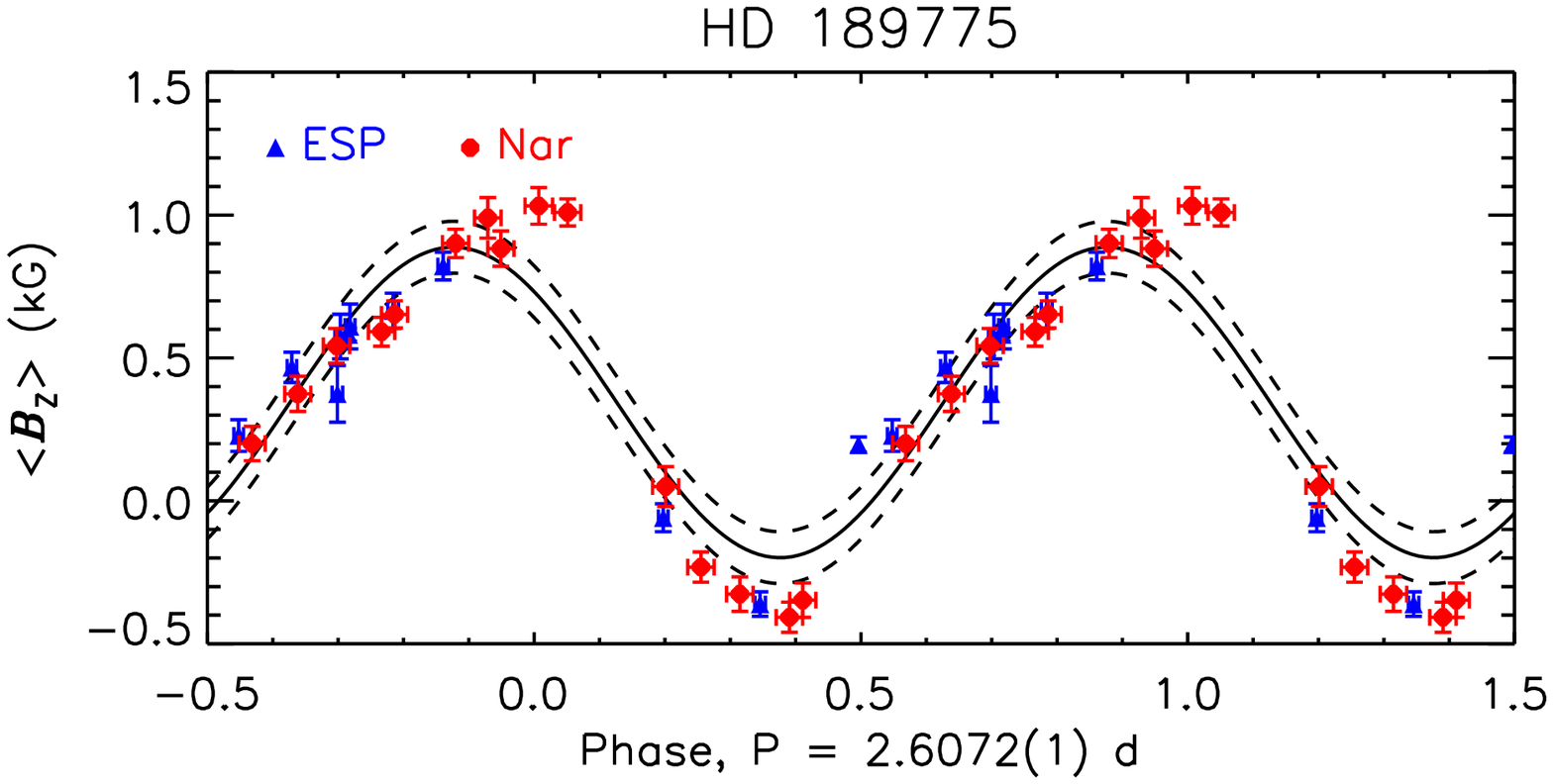}
      \caption[]{$1^{st}$-order sinusoidal fits for HD 36526 (which yields $\chi^2/\nu \sim 1$) and HD 189775 (with $\chi^2/\nu \sim 20$). Solid curves show the fits; dashed curves show the $1\sigma$ uncertainty in the fits. The legend indicates the source of the data, ESP(aDOnS) or Nar(val).}
         \label{good_bad_bz_fit}
   \end{figure}

\begin{figure}
\centering
\includegraphics[width=8.5cm]{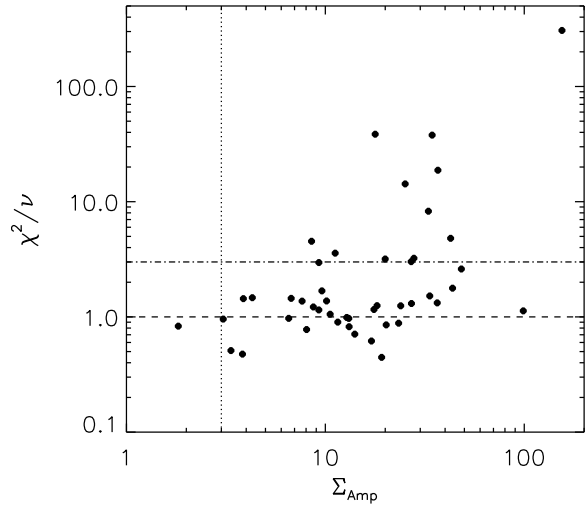}
\caption[Goodness-of-fit $\chi^2/\nu$ of 1$^{st}$-order fits to \bz~as a function of the significance of \bz~variation.]{Goodness-of-fit $\chi^2/\nu$ of 1$^{st}$-order fits to \bz~as a function of the significance of the \bz~variation $\Sigma_{\rm Amp}$. Points to the right of the dotted line possess a \bz~variation of at least 3$\sigma$ significance. The dashed line indicates $\chi^2/\nu=1$, the formal definition of a `good' fit. The dot-dashed line indicates $\chi^2/\nu=3$. Points above the dot-dashed line and to the right of the dotted line show the strongest evidence for higher-order multipole contributions to \bz.}
\label{bzchi}
\end{figure}



\begin{figure}
\centering
\includegraphics[width=8.5cm]{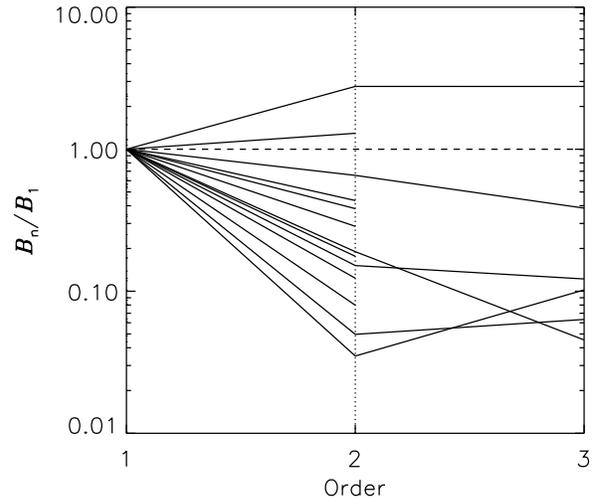}
\caption[]{Coefficients $B_{n}$ normalized to $B_1$, for the stars for which $1^{st}$ order sinusoids yield $\chi^2/\nu > 3$.}
\label{bn}
\end{figure}

Several of the Stokes $V$ profiles in Figs.\ \ref{lsd_allplot_1} and \ref{lsd_allplot_2} show distortions that might be ascribed to non-dipolar surface magnetic fields, e.g.\ HD~35298, HD~36526, HD~61556, HD~127381, HD~142990, HD~156324, HD~176582, and HD~189775. However, such distortions might also easily arise from chemical spots or pulsations. As a pure centred dipole should produce a sinusoidal \bz~variation, in order to determine the dominant magnetic topologies of the sample, we have modelled the \bz~curves by fitting least-squares sinusoids, using the equation 

\begin{equation}\label{sinfit}
\langle B_z\rangle(\phi) = \sum_{n=0}^{3} B_n \sin{(n\phi + \Phi_n)},
\end{equation}

\noindent where $\phi$ is the rotational phase determined using the ephemerides in Table \ref{rottab}, $B_n$ is the amplitude of the $n^{th}$-order sinusoid, and $\Phi_n$ is the phase offset. We utilized the {\sc idl} routine {\sc curvefit}, which computes the least-squares solution and its 1$\sigma$ uncertainties to an arbitrary non-linear function so long as its partial derivatives are known. We kept $B_n$ and $\Phi_n$ as free parameters, initially setting $B_0$ to the mean \bz, $B_1$ to $B_3$ as the standard deviation of \bz~divided by the order, and $\Phi_n = 0$. This simple model is meant to reproduce approximately the expected \bz~behaviour of dipolar, quadrupolar, and octupolar magnetic field components. The coefficients $B_n$ are given in Table \ref{bzfittab}. Including a phase offset between the components accounts for possibility that they do not all share the same tilt angle with respect to the rotational axis. 

We utilized the same \bz~measurements chosen for modelling in \S~\ref{bz_selection} (see also Table \ref{masktab}). In the case of HD 105382 there are too few high-resolution \bz~measurements (only 3) to constrain even a first-order fit, therefore for this star the fit was calculated together with the FORS1 \bz~measurements performed by \cite{2015AA...583A.115B}. The FORS1 \bz~measurements published by \cite{bagn2006} were also used to help constrain the fit for ALS 3694, due to the very large error bars of the ESPaDOnS \bz~measurements relative to the variation in \bz.

The most straightforward way to look for the presence of non-dipolar surface magnetic fields is to evaluate the goodness-of-fit $\chi^2/\nu$ of the best-fitting 1$^{st}$-order sinusoidal curves, where $\nu$ is the number of degrees of freedom. If $\chi^2/\nu \sim 1$, the model is a reasonable fit to the data, indicating that \bz~is consistent with a centred dipole. Examples are shown in Fig.\ \ref{good_bad_bz_fit} for HD 36526, for which a $1^{st}$-order sinusoid is a good fit, and HD 189775, which is not fit well by this model. For those stars for which $\chi^2/\nu \sim 1$, only $B_0$ and $B_1$ are given in Table \ref{bzfittab}, since fitting higher-order terms would tend to reduce $\chi^2/\nu$ below 1 by over-fitting to the noise in the data. Since the reduced $\chi^2$ is sensitive to the precision of the dataset, we compared the goodness of fit to the significance of the amplitude of the \bz~curve, $\Sigma_{\rm Amp} = |\langle B_z\rangle_{\rm max} - \langle B_z\rangle_{\rm min}|/\sigma_B$, where $\sigma_B$ is the median \bz~error bar. It should be recalled that the mean signifiance of \nz~$\Sigma_N < 1$ for all stars, indicating that the \bz~error bars are consistent with the scatter in the data, hence values of $\chi^2/\nu > 1$ should reflect real departures from sinusoidal variations. $\chi^2/\nu$ and $\Sigma_{\rm Amp}$ are given in Table \ref{bzfittab}.

Fig.\ \ref{bzchi} shows $\chi^2/\nu$ as a function of $\Sigma_{\rm Amp}$. The comparison is performed for 46/51 stars, with the remainder left out due to the absence of a firmly established $P_{\rm rot}$. There is an approximate correlation between the two measures, suggesting that, as expected, departures from purely sinusoidal behaviour are more easily detected in datasets with a higher precision relative to the strength of the magnetic field. Only 1 star, ALS 3694, has $\Sigma_{\rm Amp} \le 3$: for this star a departure from sinusoidal behaviour cannot be evaluated, since the quality of the data is too low to reliably constrain a higher-order fit. Of the 45 stars with $\Sigma_{\rm Amp} \ge 3$, 33 have $\chi^2/\nu \sim 1$, indicating that a single-order sinusoid provides a reasonably good fit to the data. 

The remaining 12 stars have $\chi^2/\nu \ge 3$, suggesting a dipolar model may not be appropriate in these cases. In all cases H line measurements were used to evaluate $\chi^2/\nu$, therefore it is unlikely that this result is merely a consequence of distortion of \bz~by chemical spots. Of these, 6 stars are already well known to exhibit significant departures from the sinusoidal behaviour expected of a centred dipole: HD 175362 \citep{1976ApJ...203..171W, 1983ApJS...53..151B}, HD 37776 \citep{thom1985}, HD 37479 \citep{town2005b}; HD 149438 \citep{2006MNRAS.370..629D}; HD 142184 \citep{grun2012}; and HD 61556 \citep{2015MNRAS.449.3945S}. Indeed, HD 37776, HD 37479, and HD 149438 have been verified to possess multipolar magnetic fields via Zeeman Doppler Imaging (ZDI) \citep{koch2011, 2015MNRAS.451.2015O,2006MNRAS.370..629D}. The remainder, in increasing order of $\chi^2/\nu$, are: HD 64740, HD 142990, HD 176582, HD 105382, HD 35298, and HD 189775. Since the fit to HD 105382 was constrained using FORS1 data, the high value of $\chi^2/\nu$ likely reflects systematic differences between FORS1 and HARPSpol, rather than instrinsic magnetic complexity \citep{2014AA...572A.113L}.

Examination of $\chi^2/\nu$ can identify poor fits, but it does not necessarily indicate that the departure from purely sinusoidal behaviour is either large or systematic. In Fig.\ \ref{bn} the coefficients $B_{n}$ are shown as a function of $n$, normalized to $B_1$. In all but 3 cases $B_2 \le 0.5 B_1$, and it is often much smaller than this. In only two cases is $B_2 > B_1$: HD 37776, and HD 36485 (which, although it is well-fit by a $1^{st}$-order sinusoid, we fit with a $2^{nd}$ order sinusoid for reasons explained below). $B_3$ is typically less than or similar to $B_2$, and is larger than $B_1$ only for HD 37776, for which star ZDI has found the magnetic topology to be dominated by the $\ell=2$ mode \citep{koch2011}. This suggests that even for the majority of the stars with \bz~curves suggestive of some degree of magnetic complexity, their surface magnetic fields are still dominated by the dipolar component. 



The \bz~curve of one star, HD 36485, yields relatively low values of the above measures ($\Sigma_{\rm Amp} = 3.9$ and $\chi^2/\nu = 1.4$). While the low level of variation, likely a consequence of its rotational pole being nearly aligned with the line of sight \citep{leone2010}, means that the signature of a complex surface magnetic field is not formally significant according to the criteria adopted here, close examination of its \bz~curve suggests a double-wave variation that would indicate significant contribution from multipolar components (see Fig.\ \ref{HD36485_prot}). This is the only star other than HD 37776 and HD 149438 whose \bz~curve shows more than two extrema.

We conclude that, out of the 46 stars in the sample with rotation periods, we are not able to detect any evidence in \bz~of magnetic field topologies more complex than a centred dipole in 34 of them. This is despite the quality of the data being sufficient to do so for 33 of the stars with \bz~curves consistent with a dipolar magnetic field. Amongst the 12 stars showing evidence of some degree of magnetic complexity, in all but 3 cases the dipolar component is still dominant. Thus, the majority of the sample are well-described by dipolar models. Therefore, non-dipole components of the magnetic field of sufficient magnitude to strongly distort \bz~are relatively uncommon, and even when detectable, generally modify \bz~to only a small degree. However, we have detected evidence in the \bz~curves of multipolar magnetic field components in 6 stars for which this has not previously been reported: HD 35298, HD 36485, HD 64740, HD 142990, HD 176582, and HD 189775. 


\subsection{Rotational and turbulent broadening in magnetic vs.\ non-magnetic stars}

   \begin{figure}
  \centering
   \includegraphics[width=8.5cm]{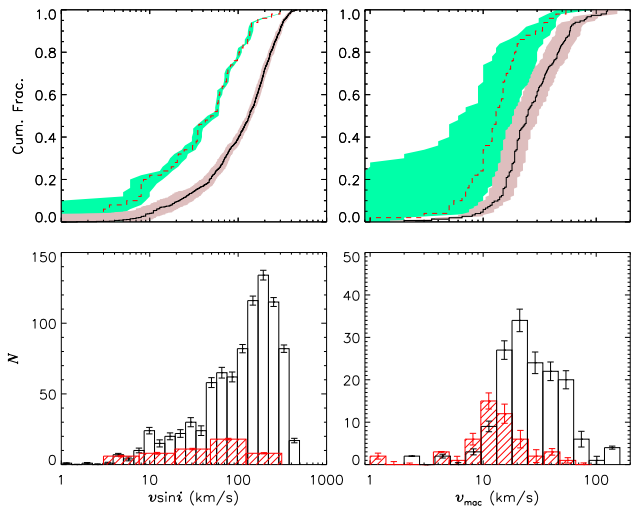}
      \caption[Histograms of \vsini~and $v_{\rm mac}$.]{Cumulative distributions ({\em top}) and histograms ({\em bottom}) of \vsini~({\em left}) and $v_{\rm mac}$~({\em right}) for the magnetic B-type stars in the present study (dashed and hatched red) and main-sequence B-type stars from the literature (black). Shaded regions in the cumulative distributions indicate uncertainties.}
         \label{vsini_hist}
   \end{figure}

   \begin{figure}
  \centering
   \includegraphics[width=8.5cm]{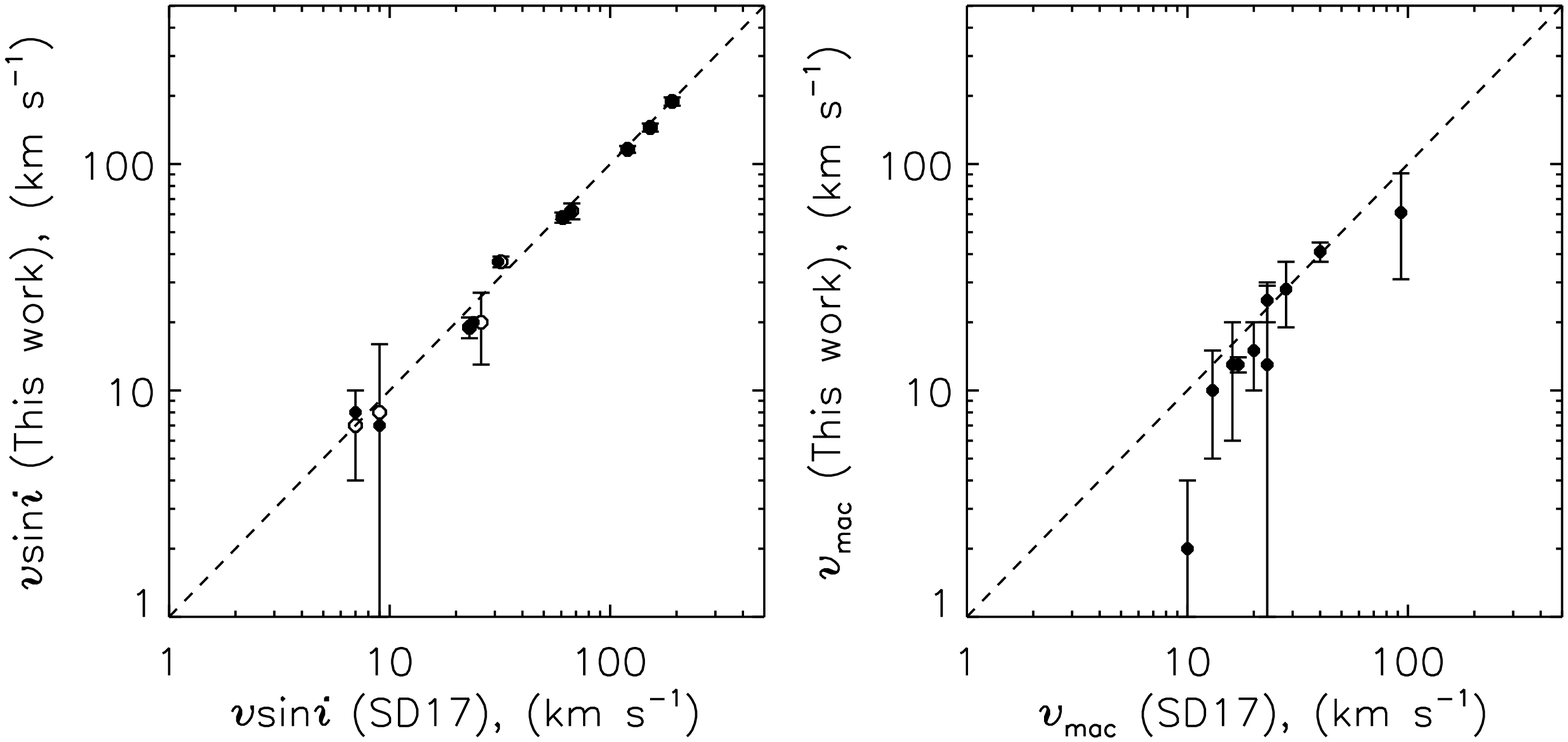}
      \caption[]{Comparison of \vsini~({\em left}) and $v_{\rm mac}$ ({\em right}) for the stars appearing in both the present sample and in that of \protect\cite{2017A&A...597A..22S}. Open circles indicate Fourier transform measurements, filled circles indicate goodness-of-fit measurements.}
         \label{vsini_vmac_sd17_compare}
   \end{figure}

Fig.\ \ref{vsini_hist} compares the \vsini~and $v_{\rm mac}$ measurements of the magnetic B-type stars in the present sample to those reported for non-magnetic, main-sequence B-type stars in the literature, where for the literature stars \vsini~measurements were taken from \cite{2013A&A...550A.109D}, \cite{2014A&A...562A.135S}, \cite{2015AJ....150...41G}, and \cite{2017A&A...597A..22S} (a total of 890 \vsini~measurements of non-magnetic B-type stars), while $v_{\rm mac}$ measurements were provided only by \cite{2014A&A...562A.135S} and \cite{2017A&A...597A..22S} (for a total of 154 $v_{\rm mac}$ measurements of non-magnetic B-type stars). Bin sizes were determined using the Freedman-Diaconis rule, which optimizes the bin size for the variance and size of the data set \citep{Freedman1981}. Error bars arise from the 1$\sigma$ uncertainties in the individual measurements. 10000 simulated datasets were created, of the same size as the original dataset, with values of individual datapoints equal to the original measurement plus a random number generated from a Gaussian disribution normalized to the 1$\sigma$ uncertainty in the original measurement. Histograms were then generated from each simulated dataset, and the error bar in each histogram bin was obtained from the standard deviation across the histograms for all synthetic datasets. Cumulative distribution uncertainties were obtained from the cumulative distributions of the measurements plus or minus their 1$\sigma$ error bars.

The literature \vsini~measurements combine extragalactic stars (the VLT-FLAMES Tarantula Survey, \citealt{2013A&A...550A.109D}) and Galactic stars \citep{2014A&A...562A.135S,2015AJ....150...41G,2017A&A...597A..22S}. The samples presented by \cite{2014A&A...562A.135S} and \cite{2017A&A...597A..22S} overlap slightly with one another and with the present sample; stars appearing in both were removed from the \citealt{2014A&A...562A.135S} and \citealt{2017A&A...597A..22S} data, and for stars appearing in both \cite{2014A&A...562A.135S} and \cite{2017A&A...597A..22S} the more recent values were used. Fig.\ \ref{vsini_vmac_sd17_compare} compares the \vsini~and $v_{\rm mac}$ values determined here to those determined for the same stars by \cite{2017A&A...597A..22S}. There is good agreement for \vsini, with no appearance of systematic differences between the two samples. However, our values of $v_{\rm mac}$ are systematically lower by about 20\%. This is likely due to the use of a $\delta$-function for the intrinsic profile by \cite{2014A&A...562A.135S} and \cite{2017A&A...597A..22S}, whereas we elected to use Gaussian intrinsic profiles with full-width at half-maximum values determined by the thermal broadening of the line, typically around a few \kms. Experiments for a few stars in which $\delta$-functions were utilized instead recovered values of $v_{\rm mac}$ comparable with those found by \cite{2014A&A...562A.135S} and \cite{2017A&A...597A..22S}.

The two other samples presumably contain some number of magnetic stars, but the visual magnitudes of these stars are too faint for high-resolution magnetometry to distinguish magnetic from non-magnetic stars; in any case, the fraction of magnetic stars should be no higher than 10\%. Comparing \vsini~values, there is a clear difference between magnetic and non-magnetic B-type stars, with the distribution of the former peaking at under 100 \kms~while the latter peaks at about 200 \kms. A two-sample Kolmogorov-Smirnov (K-S) test yields a probability of $7\times10^{-8}$ that the two samples belong to the same distribution. This is in line with expectations that magnetic stars should be systematically more slowly rotating than non-magnetic stars due to braking by the magnetized stellar wind. It should be noted that this difference may be even more pronounced than suggested by Fig.\ \ref{vsini_hist}: \cite{2013A&A...550A.109D} noted the distribution of \vsini~in the Large Magellanic Cloud B-type stars is bimodal, with a broad-lined component ($\sim$75\% of the sample) with \vsini~$\sim$~250 \kms, and a narrow-lined component with \vsini~$<100$ \kms. \citeauthor{2013A&A...550A.109D} speculated that the narrow-lined population is their sample might be a consequence of magnetic braking, noting that the fraction expected to be spun down by tidal interactions with a binary companion (the other leading mechanism for producing slow rotation) is much smaller than the observed fraction of slow rotators \citep{2013ApJ...764..166D}. Nevertheless, given that \cite{2013A&A...550A.109D} had no evidence beyond rotation to infer magnetic fields in the slowly rotating sub-sample, and that removing these stars from the comparison would artificially increase the difference between stars with and without confirmed magnetic fields, we have made the conservative assumption that the sub-sample of narrow-lined stars are not magnetic, and so have included them in the comparison. It is worth noting that the velocity peak of their narrow-lined component is similar to that found here for magnetic stars, suggesting that these stars may be worthy of follow-up observations to determine if they exhibit the spectroscopic peculiarites and periodic photometric modulations expected for magnetic stars. However, both MiMeS and BOB have observed a large number of sharp-lined Galactic stars, many of which are not detectably magnetic \citep{2015A&A...582A..45F,2016MNRAS.456....2W,2017MNRAS.465.2432G}.

   \begin{figure}
  \centering
   \includegraphics[width=8.5cm]{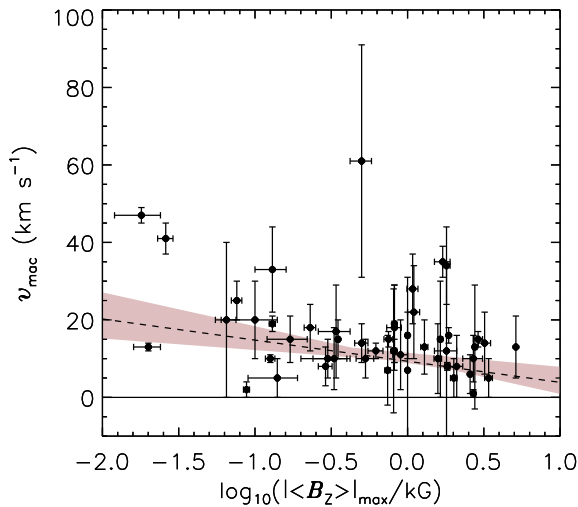}
      \caption[.]{$v_{\rm mac}$ as a function of $\log{(|\langle B_z\rangle|_{\rm max})}$. The dashed line indicates the least-squares linear fit, and the shaded region shows the uncertainty in the fit derived from iterative removal of individual data-points.}
         \label{vmac_bzmax}
   \end{figure}

Comparison of $v_{\rm mac}$ values determined here to those presented for non-magnetic B-type stars by \cite{2014A&A...562A.135S} (right panels of Fig.\ \ref{vsini_hist}) yields a somewhat more ambiguous result. The probability that the two samples belong to the same distribution according to the K-S test is $10^{-10}$, a low enough value to suggest the difference is real. The magnetic sample peaks at a lower value ($\sim$10 \kms) than the non-magnetic sample ($\sim$20 \kms), with the non-magnetic stars having generally higher values of $v_{\rm mac}$: no magnetic star has $v_{\rm mac} > 60$~\kms, while the non-magnetic sample extends up to 100~\kms; conversely, the lowest value in the non-magnetic sample is around 10 \kms, while several of the magnetic stars have $v_{\rm mac}$ consistent with 0~\kms. It should be noted that in the sample of sharp-lined early B-type stars studied by \cite{2012A&A...539A.143N}, several stars with $v_{\rm mac} = 0$~\kms~were found. Only one of the stars ($\beta$ Cep) studied by \cite{2012A&A...539A.143N} is known to possess a magnetic field, and the value of $v_{\rm mac}$ they found ($20 \pm 7$~\kms) is compatible with the value found here ($25 \pm 5$~\kms). However, the \cite{2012A&A...539A.143N} sample was selected for stars with sharp spectral lines, is thus intrinsically biased towards small line-broadening parameters, and therefore may not give an accurate representation of the true distribution of macroturbulent velocities in main-sequence early B-type stars. 

While the physical origin of macroturbulence is still debated, the leading possibility is that it arises from the super-position of multiple high-order non-radial $g$-mode pulsations excited in the sub-surface Fe convection zone \citep{2009A&A...508..409A,2017A&A...597A..23G}. Strong magnetic fields are expected to suppress convection and, hence, inhibit macroturbulence, an effect which has been observed in the strongly magnetized Of?p star NGC~1624-2 \citep{2013MNRAS.433.2497S}. If the magnetic field plays a role in supressing macroturbulence in the present sample, we should expect to see that more strongly magnetized stars tend to have lower values of $v_{\rm mac}$. This comparison is performed in Fig.\ \ref{vmac_bzmax}. A least-squares linear fit finds a trend of decreasing $v_{\rm mac}$ with increasing $|\langle B_z\rangle|_{\rm max}$, which is robust against iterative removal of individual data-points as indicated by the shaded region. The Pearson's correlation coefficient for $v_{\rm mac}$ vs.\ $\log{|\langle B_z\rangle|_{\rm max}}$ is $-0.36 \pm 0.03$. 

The much smaller difference in $v_{\rm mac}$ distributions as compared to \vsini~could also be a consequence of subtle differences in the line profile fitting routines used here and employed by \citeauthor{2014A&A...562A.135S}. There is evidence for such a systematic discrepancy in Fig.\ \ref{vsini_vmac_sd17_compare}, which could explain part or all of the difference. The median ratio of $v_{\rm mac}$ values from \citeauthor{2014A&A...562A.135S} to those found here (Fig.\ \ref{vsini_vmac_sd17_compare}) is $1.3$, while the median ratio between the cumulative distributions in Fig.\ \ref{vsini_hist} is 2.2, suggesting that the different choices of intrinsic profiles may not explain all of the differences in the distributions. Uncertainties in $v_{\rm mac}$ tend to be higher than those in \vsini~(around 25\% as compared to around 7\%), and as can be seen the from the uncertainties in the cumulative distributions of the magnetic and non-magnetic samples, they overlap at around 1$\sigma$. The results of \cite{2012A&A...539A.143N}, who found $v_{\rm mac} = 0$~\kms~for several early B-type stars not known to possess a magnetic field, further suggests that methodological differences (non-LTE spectral fitting vs.\ line profile fitting and Fourier transforms) may underlie the different results obtained in their study, vs.\ that of \citeauthor{2014A&A...562A.135S}. Verifying whether or not this difference in macroturbulent velocities is real will require a self-consistent analysis of the combined sample of magnetic and non-magnetic main sequence B-type stars. It may also benefit from a more sophisticated treatment of the line profile variability of the chemically peculiar stars in this sample. In particular, accounting for the effect of chemical spots on line profile shapes may be necessary, as it is possible these distortions may be misinterpreted as macroturbulence by the method used here (as was found by \cite{2014AA...569A.118A} for HD 105382, for which $v_{\rm mac}$ was a strong function of rotational phase). If this is the case, the values of $v_{\rm mac}$ in Table \ref{rottab} will tend to be higher than the true $v_{\rm mac}$, thus accentuating the difference between the magnetic and non-magnetic stars. Such a comparison should also utilize predictions based upon stellar structure models of the depth to which the surface magnetic fields are expected to inhibit convection in any given star. 

\subsection{Comparison of rotational periods in magnetic early B-type stars to Ap/Bp stars and magnetic O-type stars}

\begin{figure}
   \centering
   \includegraphics[width=8.5cm]{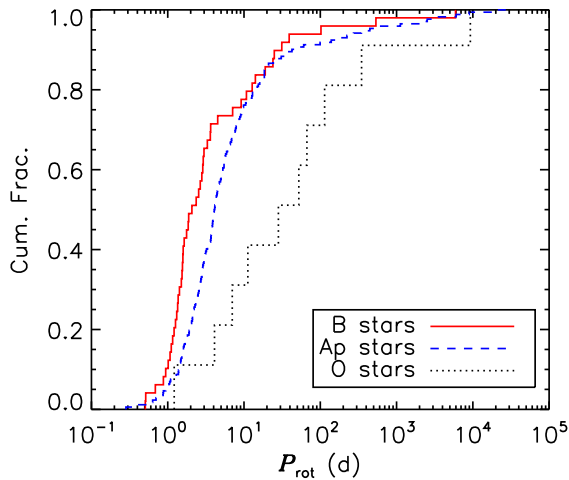} 
      \caption[]{Cumulative distributions of the rotational periods of magnetic early B-type stars, Ap and late-type Bp stars, and magnetic O-type stars.}
         \label{prot_ks}
   \end{figure}

Fig.\ \ref{prot_ks} compares the cumulative distribution of rotation periods for the early B-type stars, to the cumulative distribution of the rotational periods for Ap stars and late-type Bp stars, where the Ap/Bp stars have been compiled from the catalogues published by \cite{landmat2000}, \cite{2005AA...430.1143B}, \cite{2007AA...475.1053A}, and \cite{2007MsT..........1P}, for a total comparison sample of 172 stars. As there is some overlap between the samples, particularly with the comprehensive \citeauthor{2005AA...430.1143B} catalogue, each sample was cleaned by HD number to ensure no duplication. The overall distributions are quite similar: the median rotation period of both the present sample and the cooler stars is on the order of days, and both possess tails extending to around 10$^4$ d. However, there appears to be a higher percentage of rapidly rotating stars amongst the early-type stars. 

To evaluate whether or not the difference between the two populations is significant, we performed a K-S test. The K-S test significance is 0.0017, which is quite small and suggests the two distributions may indeed be different. This is a somewhat unexpected result, as given the stronger stellar winds and consequently more rapid magnetic braking experienced by early-type stars, they should presumably be, in general, slower rotators than the late-type stars. The difference may be indicative that the present sample is generally younger than the late-type stars, and/or that they were born with substantially more angular momentum than the Ap/Bp stars. Alternatively, this could be an observational bias arising from the inclusion in this sample of stars that were first identified as spectropolarimetric targets based upon emission signatures suggestive of a centrifugal magnetosphere \citep{town2005c, petit2013}, e.g. HD~23478 \citep{2015MNRAS.451.1928S,2015A&A...578L...3H}. 

Fig.\ \ref{prot_ks} also compares the distribution of rotational periods in the magnetic B-type stars to those of the magnetic O-type stars, which we have obtained from those collected by \cite{petit2013}, with the exception of HD 47129 which was provided by J.\ Grunhut (priv. comm.). While the extrema of the O-type star period distribution are similar to those of the other two distributions, the magnetic O-type stars are systematically more slowly rotating than either the Ap/Bp stars or the early B-type stars. The K-S test yields a significance of 0.0011 that the B-type stars and O-type stars belong to different distributions. The slower rotation of the magnetic O-type stars is an expected result since, due to their powerful stellar winds, they should lose angular momentum much more rapidly than cooler stars. 

\section{Conclusions and Future Work}


We measured \bz~using both LSD profiles and H lines. Comparison of \bz~measurements obtained from H lines and LSD profiles extracted using single-element masks has revealed that many of the Bp stars display systematic element-dependent differences in the shapes and amplitudes of their \bz~curves, confirming this phenomenon as essentially ubiquitous in this class of stars. Chemically normal stars, while a minority of the sample, show no evidence of this effect. \bz~measurements obtained using H lines show excellent agreement with historical data, further confirming the reliability of ESPaDOnS, Narval, and HARPSpol magnetometry, and in several cases enabling more precise rotational periods to be determined via comparison of the new \bz~curves to older measurements. 


No magnetic field was detected in HD 35912, indicating that the star has been incorrectly identified as magnetic in the literature, and reducing the size of the sample from 52 to 51 stars. Using magnetic, spectroscopic, and archival photometric data, we have determined new rotational periods for 10 of 51 stars. No period could be determined for 5 stars: HD~44743, HD~52089, HD~58260, and the two components of the doubly-magnetic close binary HD~136504. The remarkable stability of the \bz~measurements of HD~58260 over the 35 years of observations suggests either that the star's rotational period is extremely long, and/or that its rotational pole is almost perfectly aligned with the line of sight, and/or that its magnetic field is axisymmetric with the magnetic axis almost perfectly aligned with the rotational axis. For the remaining stars without rotational periods, the available datasets are simply too small and/or imprecise to identify a period, but all show some degree of variability and there is no reason to think that their rotational periods cannot be determined if additional, higher-S/N data is obtained. 

The distributions of rotational periods appear to be broadly similar between magnetic early B-type stars, and Ap/late-type Bp stars. There is some suggestion that the hotter stars possess an excess of rapid rotators compared to the cooler stars, a discrepancy that could be reflective either of observational bias or a real difference in the populations arising from e.g.\ differing angular momentum distributions at birth. 

Projected rotational broadening \vsini~was measured together with a radial/tangential turbulent broadening parameter $v_{\rm mac}$, with multiple spectral lines used for each star in order to obtain statistical uncertainties in the line broadening parameters. Zeeman splitting is able to account for the majority of the line broadening in HD~58260 and HD~96446, both of which have strong magnetic fields and sharp spectral lines. Comparison to previous \vsini~measurements revealed only one 3$\sigma$ outlier, HD~35298, for which we revise \vsini~sharply downwards. Comparing the distribution of \vsini~to non-magnetic stars shows that \vsini~peaks about 100 \kms~below the peak for non-magnetic stars, consistent with the magnetic stars having lost angular momentum more quickly due to magnetic braking. A similar comparison of $v_{\rm mac}$ suggests that turbulent broadening may also be systematically lower in magnetic stars, although this conclusion is only tentative and should be investigated with a larger sample measured with a self-consistent methodology. We find a trend of decreasing $v_{\rm mac}$ with increasing $|\langle B_z\rangle_{\rm max}|$, which tends to support the conclusion that strong magnetic fields suppress turbulence. If this proves to be the case, this may give some insight into the origin of the macroturbulence phenomenon in B-type stars in general. 


We have searched for the signature of multipolar magnetic fields via comparison of \bz~curves to the $1^{st}$-order sinusoidal variation expected for a centred, tilted dipole. The \bz~curves of 74\% of the sample (34 of 46 stars with rotational periods) are consistent with dipolar magnetic fields. Of the remaining 12 stars (26\% of the sample), the contributions to \bz~from higher-order multipoles are typically minor, with amplitude coefficients generally less than 20\% of the magnitude of the dipolar component. In addition to the 6 stars previously identified as having multipolar surface fields, we have found evidence for magnetic complexity in 6 stars for which it was not previously reported: HD 35298, HD 36485, HD 64740, HD 142990, and 176582, and HD 189775. While the reduced $\chi^2$ of a first-order fit to HD 36485's \bz~curve is formally consistent with a centred dipole, there is some evidence of a quadrupolar component, which requires verification at higher S/N given the very small level of variability exhibited by \bz.

This is Paper I in a series that will explore the magnetic and magnetospheric properties of the magnetic early B-type stars. In Paper II, we will use the rotational periods and \bz~measurements presented here to obtain dipolar oblique rotator model parameters together with magnetospheric parameters, with which we will revisit the rotation-magnetic confinement diagram introduced by \cite{petit2013} with the aim of exploring 1) the boundary between stars with H$\alpha$ in emission vs.\ absorption, and 2) the rotational evolution of these stars. 
\section*{Acknowledgements} This work has made use of the VALD database, operated at Uppsala University, the Institute of Astronomy RAS in Moscow, and the University of Vienna. MS acknowledges the financial support provided by the European Southern Observatory studentship program in Santiago, Chile. MS and GAW acknowledge support from the Natural Sciences and Engineering Research Council of Canada (NSERC). EA, CN, and the MiMeS collaboration acknowledge financial support from the Programme National de Physique Stellaire (PNPS) of INSU/CNRS. We acknowledge the Canadian Astronomy Data Centre (CADC). 

\bibliography{bib_dat.bib}{}


\appendix
\renewcommand{\thetable}{\Alph{section}\arabic{table}}


\section{Rotation periods and longitudinal magnetic field measurements of individual stars}\label{bz_ind}

In this appendix we present \bz~curves for stars with existing periods that we have refined using new data (14 stars), the stars for which we have determined new rotational periods using magnetic, spectroscopic, or photometric data (10 stars), 1 star for which we could not uniquely determine $P_{\rm rot}$ (HD 58260), and 1 star for which we have new \bz~measurements but could not improve $P_{\rm rot}$ (HD 36485). We also present periodograms and, in some cases, photometric or spectroscopic data together with \bz. The abscissae of the periodograms are restricted to the period windows discussed in \S~\ref{rotation_periods}. For stars with newly determined rotation periods, we state in the text whether we consider the period to be confidently established, or either tentative or only partially constrained. 

Stars for which the existing magnetic data have already been published, or for which detailed individual studies are in preparation, are not included in this appendix. These are HD 3360 \citep{2016AA...587A.126B}, HD 23478 \citep{2015MNRAS.451.1928S}; HD 35502 \citep{2016MNRAS.460.1811S}; HD 37479 \citep{oks2012}; HD 44743 and HD 52089 \citep{2015AA...574A..20F, 2017MNRAS.471.1926N}; HD 46328 \citep{2017MNRAS.471.2286S}; HD 61556 \citep{2015MNRAS.449.3945S}; HD 105382 \citep{alecian2011}; HD 121743 (Briquet et al., in prep); HD 127381 \citep{henrichs2012}; HD 130807 \citep{2017omiLup_inprep}; HD 136504 \citep{2015MNRAS.454L...1S,pablo_epslup}; HD 142184 \citep{grun2012}; HD 149438 (Kochukhov et al., in prep.); HD 156324 \citep{shultz_hd156324_2017}; HD 163472 \citep{neiner2012b}; HD 164492C \citep{2017MNRAS.465.2517W}; HD 176582 (Neiner et al., in prep.); HD 182180 \citep{2010MNRAS.405L..51O,rivi2013}; HD 184927 \citep{2015MNRAS.447.1418Y}; and HD 205021 (Neiner et al., in prep).



Measurements from ESPaDOnS, Narval, and HARPSpol are plotted with separate symbols and abbreviated in the legends as ESP, Nar, and HAR, respectively. The excellent agreement between ESPaDOnS and Narval measurements has been demonstrated by \cite{2016MNRAS.456....2W}. Those stars for which ESPaDOnS and HARPSpol measurements of similar quality are both available (e.g., HD 130807, HD 133880, HD 164492C; \citealt{2017omiLup_inprep}; \citealt{2017A&A...605A..13K}; \citealt{2017MNRAS.465.2517W}) demonstrate that longitudinal magnetic field measurements collected with these instruments also agree within 1$\sigma$, a result which we confirm here for the cases of HD 64740, HD 67621, and HD 156424.

For comparisons to \bz~measurements from the literature, the following abbreviations for the references are used in the legends: BL77 \citep{bl1977}; BL79 \citep{1979ApJ...228..809B}; BLT83 \citep{1983ApJS...53..151B}; B87 \citep{1987ApJ...323..325B}; BLT93 \citep{1993AA...269..355B}; M91 \citep{1991AAS...89..121M}; MH97 \citep{1997AAS..124..475M}; B04 \citep{2004AA...413..273B}; B06 \citep{bagn2006}; B07 \citep{2007AN....328...41B}; Y11 \citep{yakunin2011}; BM11 \citep{bohl2011}; Y13 \citep{2013AstBu..68..214Y}; B15 \citep{2015AA...583A.115B}; and F15 \citep{2015A&A...582A..45F}. Photopolarimetric \bz~measurements conducted using the wings of the H$\beta$ line (BL77, BL79, BLT83, B87) have been corrected to 80\% of the published values as suggested by \cite{2000A&A...358.1151M} (see \S~\ref{h_bz}).





   \begin{figure}
   \centering
   \includegraphics[width=8.5cm]{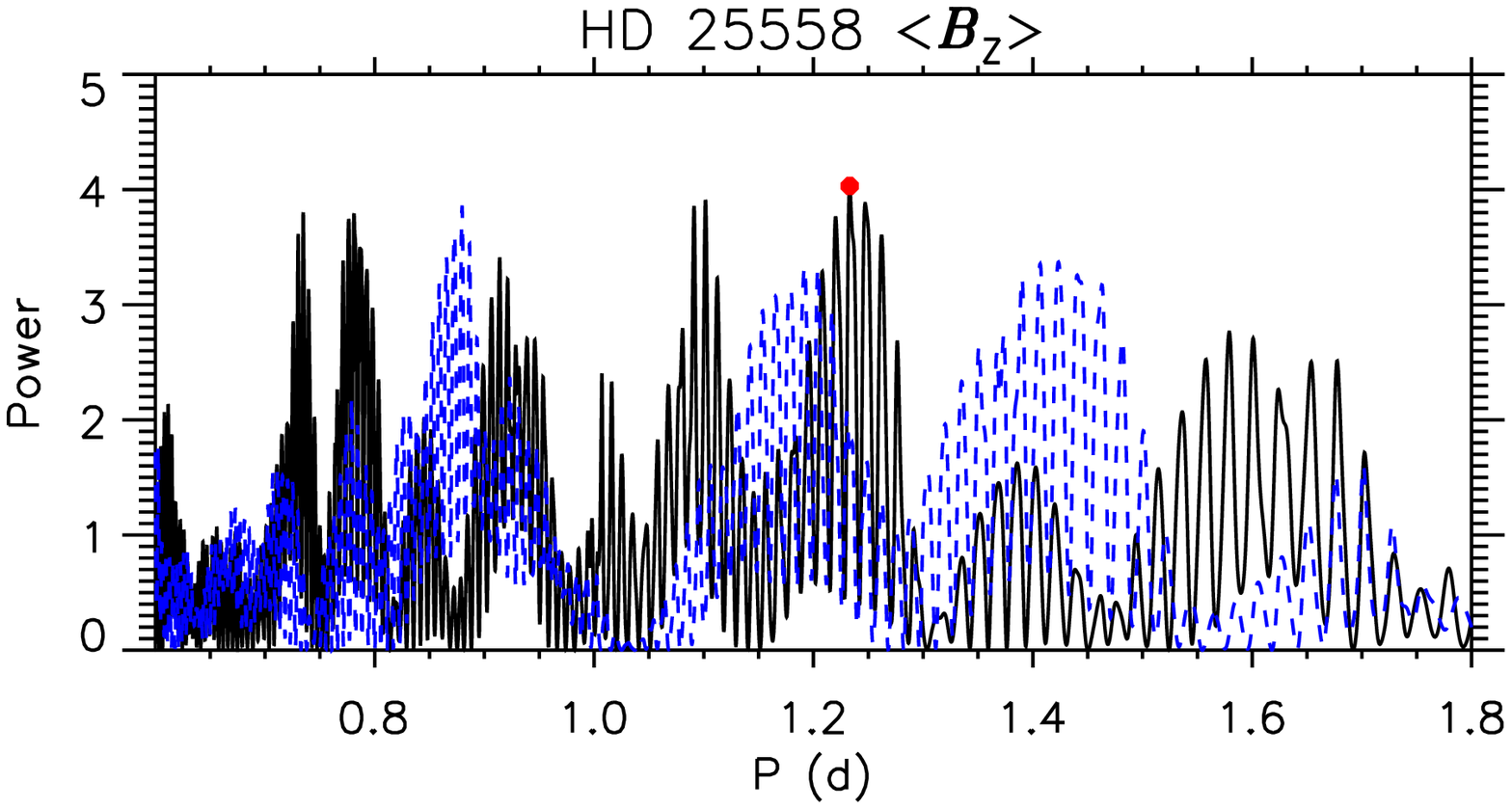} 
   \includegraphics[width=8.5cm]{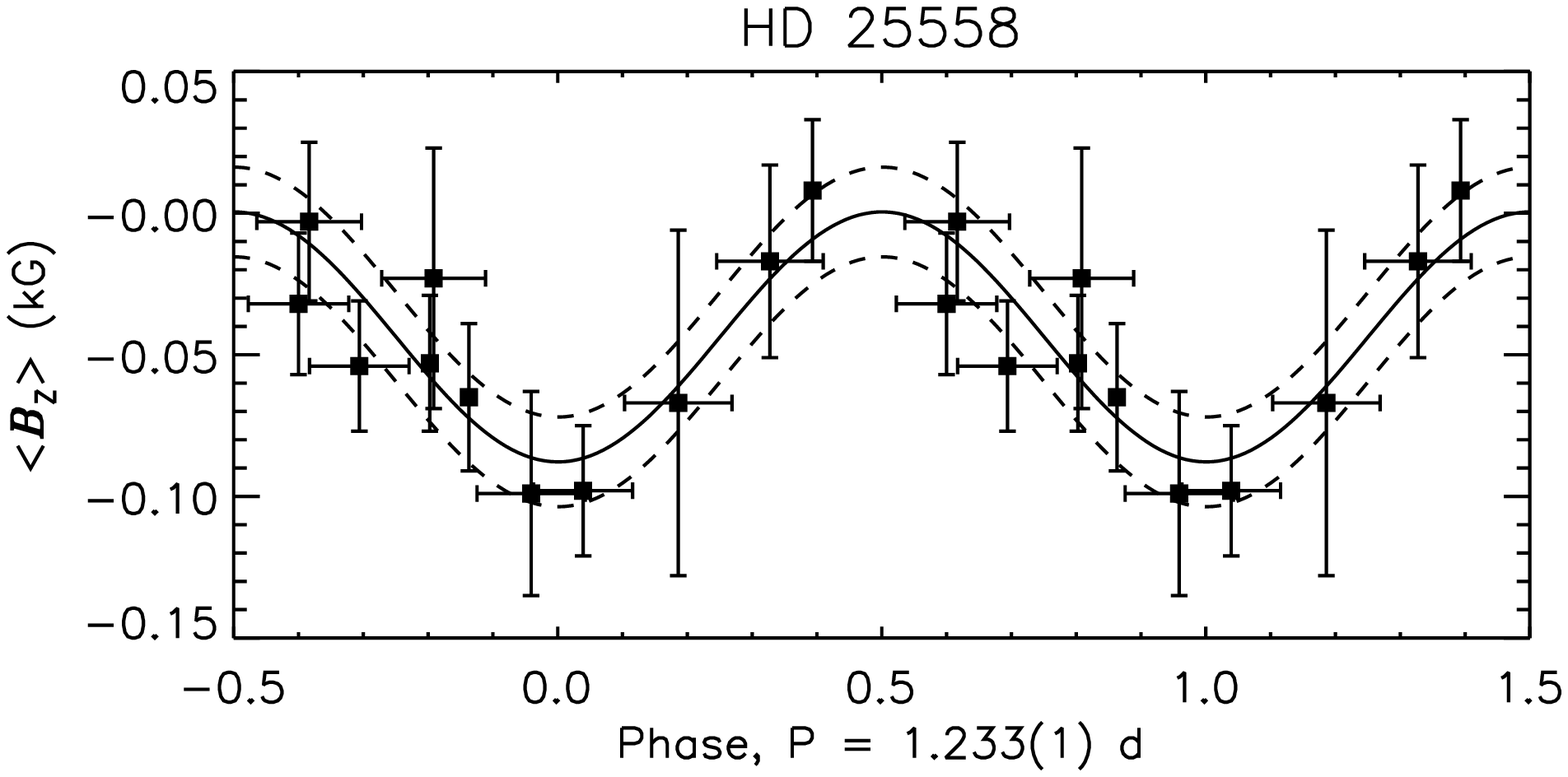} 
      \caption[Periodogram and \bz~for HD 25558.]{{\em Top}: periodogram for \bz~measurements of HD 25558. The abscissa is limited to the period window. Solid (black) line indicates the periodogram for \bz; dashed (blue) line the periodogram for \nz. The red circle indicates the adopted rotation period. {\em Bottom}: \bz~phased with $P_{\rm rot}$. Solid lines indicate the best-fit sinusoidal curves; dashed lines, the 1$\sigma$ uncertainty in the fit. In this and in other figures, the best-fit sinusoidal curve is unless otherwise stated of $1^{st}$ order.}
         \label{HD25558_prot}
   \end{figure}

\noindent {\bf HD 25558}: \cite{2014MNRAS.438.3535S} determined that the rotational period of the magnetic secondary of this SB2 system should be about 1.2$\pm$0.6 d. Using the restricted data-set described in \S~\ref{section_bz}, and restricting the period window to the range given by \cite{2014MNRAS.438.3535S}, we find maximum amplitude at 1.233(1) d, although there are numerous nearby peaks that provide a comparable phasing of the data (Fig.\ \ref{HD25558_prot}, top). The FAP of the maximum amplitude peak in the \bz~period spectrum is 0.12, only slightly lower than the maximum FAP in the \nz~spectrum, 0.17. \bz~is shown phased with this period in the bottom panel of Fig.\ \ref{HD25558_prot}.

\begin{figure}
    \includegraphics[width=8.5cm]{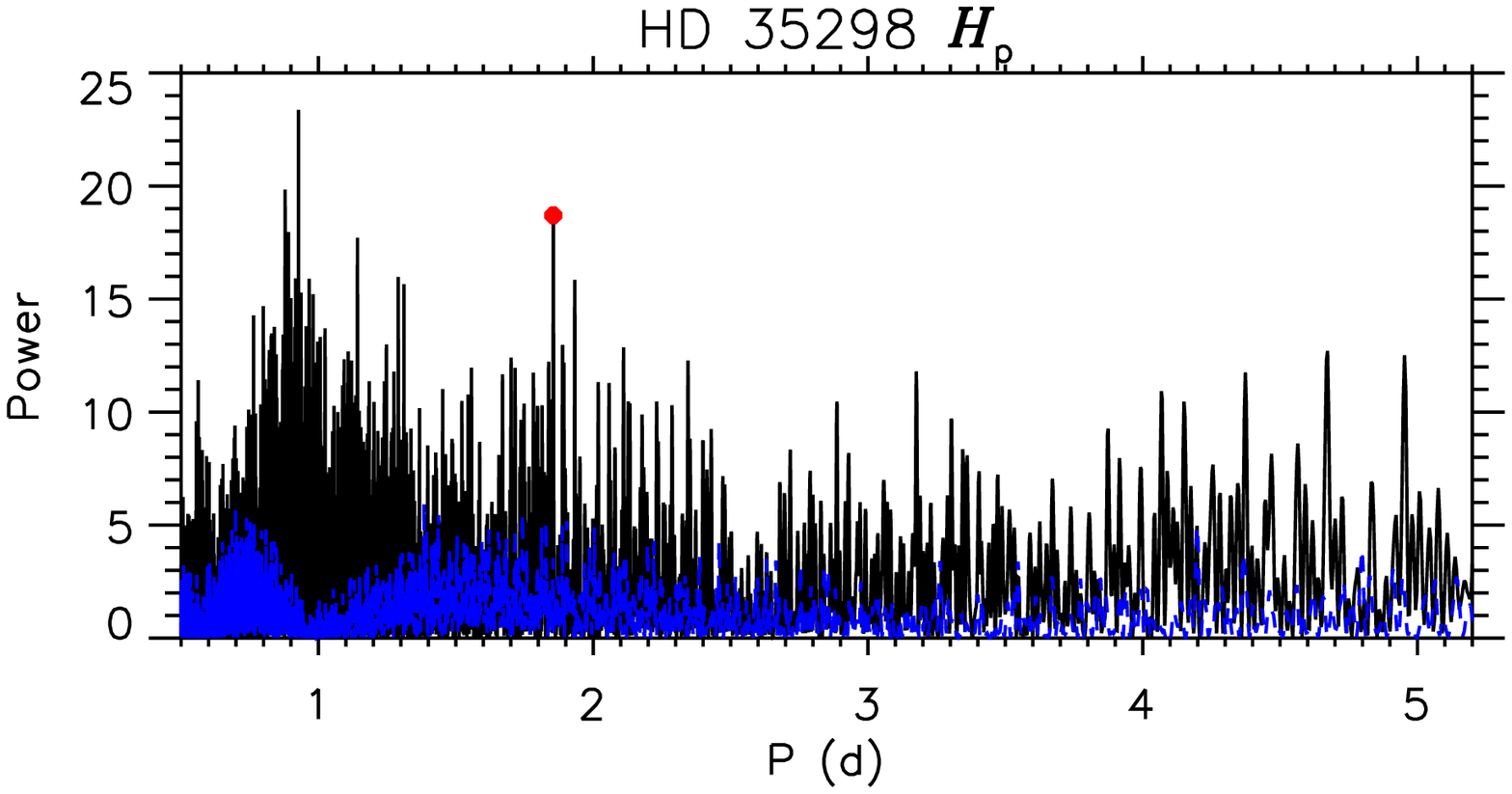}
    \includegraphics[width=8.5cm]{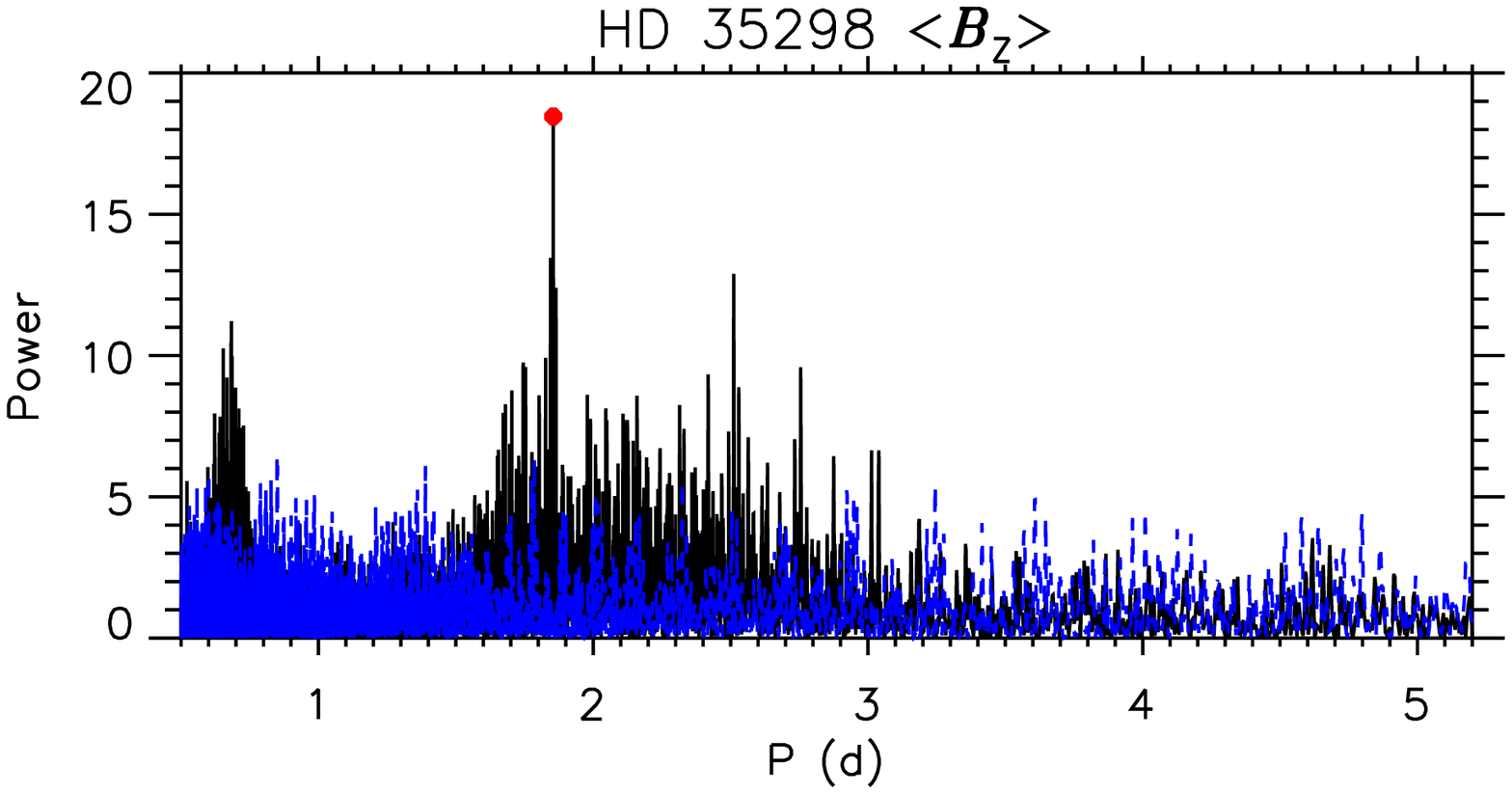}
    \includegraphics[width=8.5cm]{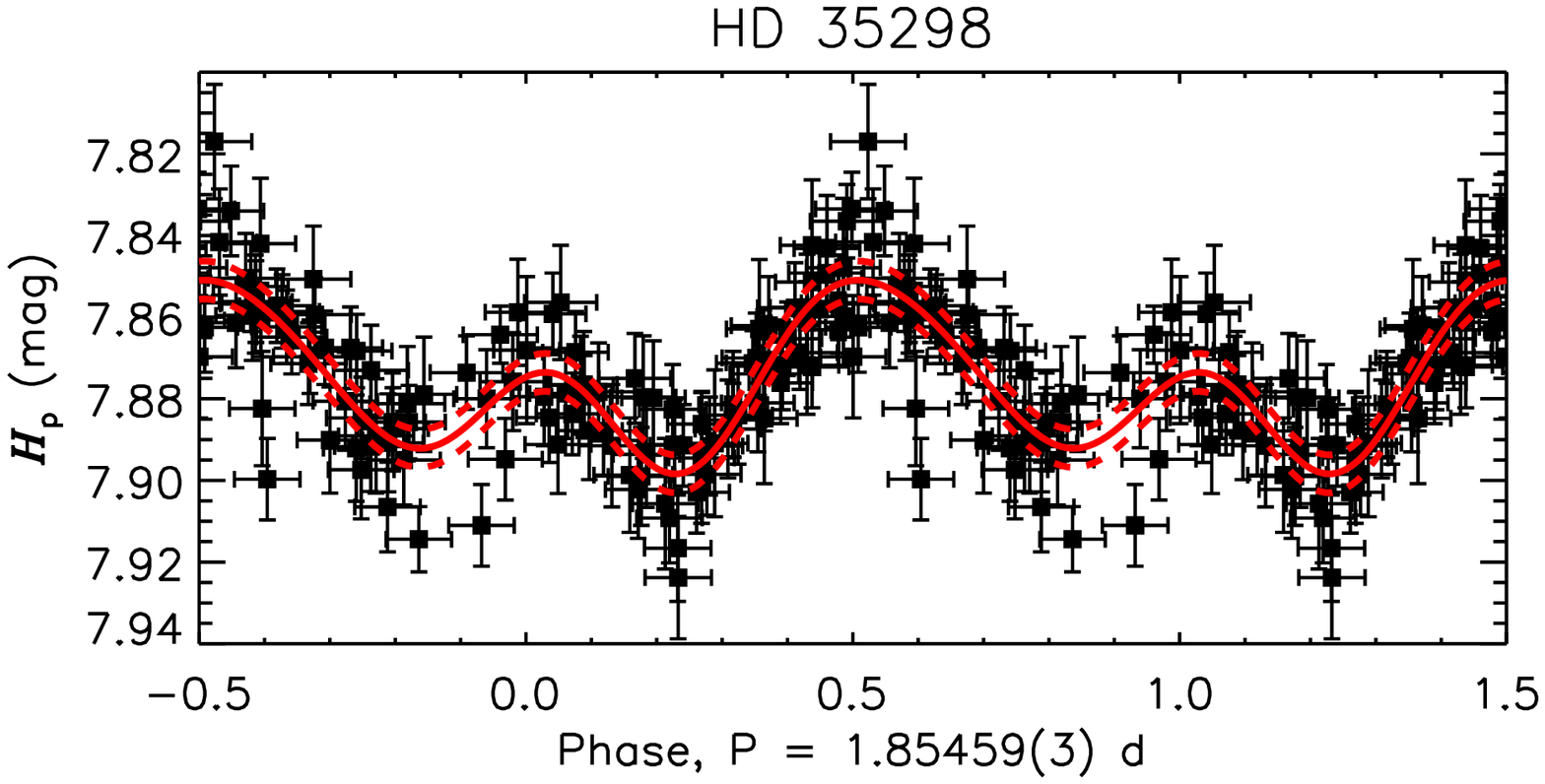}
    \includegraphics[width=8.5cm]{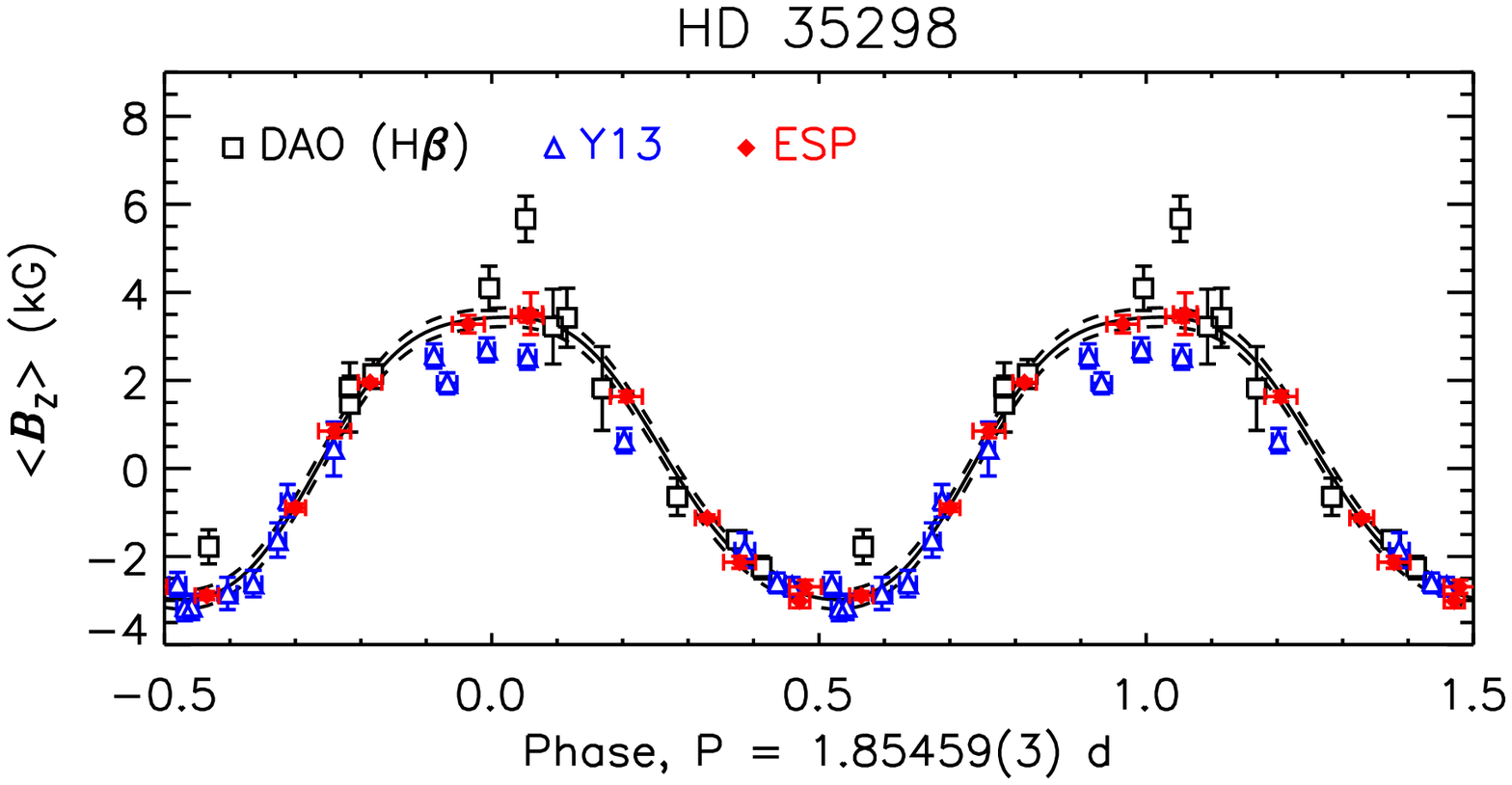}
    \caption[Periodogram, Hipparcos photometry, and \bz~for HD 35298.]{
       Rotational period of HD 35298. {\em Top }: periodogram constructed for $H_{\rm p}$ photometry (solid black line) and synthetic null measurements (dashed blue line). The adopted period is indicated by the red circle. {\em Second panel}: periodogram from ESPaDOnS \bz~measurements. {\em Third panel}: $H_{\rm p}$ phased with $P_{\rm rot}$. The red curve is 2$^{nd}$-order sinusoidal fit; dashed curves indicate uncertainties in the fit. {\em Bottom}: \bz~phased with $P_{\rm rot}$. The sinusoidal fit is of $3^{rd}$ order. 
    } \label{HD35298_prot}
\end{figure}

\noindent {\bf HD 35298}: \cite{1984AA...141..328N} found a photometric period of $P_{\rm rot}$ = 1.85336(1) d. The periodogram constructed from archival Hipparcos photometry shows maximum power at $\sim$0.927 d, which is almost exactly half of the period reported by \cite{1984AA...141..328N} (Fig.\ \ref{HD35298_prot}, top). This period does not provide an acceptable phasing of the magnetic data, however. The 2$^{nd}$-highest peak in the Hipparcos photometry periodogram is at 1.8548(2) d. The FAP of this peak is $\sim10^{-3}$, much lower than the FAP of the highest peak in the null spectrum, $\sim0.999$.  While the FAP of the highest peak in the ESPaDOnS \bz~periodogram is only 0.03, maximum power in the \bz~periodogram is at 1.8546(1)~d, with no power in the \nz~periodogram at this period. This period also provides an acceptable phasing of the dimaPol H$\beta$ \bz~measurements. Including the best dimaPol measurements (i.e.\ those that do not fall outside the range of the ESPaDOnS measurements), along with the \bz~measurements published by \cite{2013AstBu..68..214Y}, increases the precision of the period to 1.85459(3)~d, with a FAP of $10^{-8}$. This is compatible with $P_{\rm rot}=1.85457~{\rm d}$ found by \cite{2013AstBu..68..214Y}. \bz~and $H_{\rm p}$ are shown phased with this period in Fig.\ \ref{HD35298_prot}. $H_{\rm p}$ shows a double-wave variation, accounting for the appearance of the 0.927 d peak. The \bz~curve is fit best with a 3$^{rd}$-order sinusoid, which may be indicative of departures from a pure centred dipole. 

   \begin{figure}
   \centering
   \includegraphics[width=8.5cm]{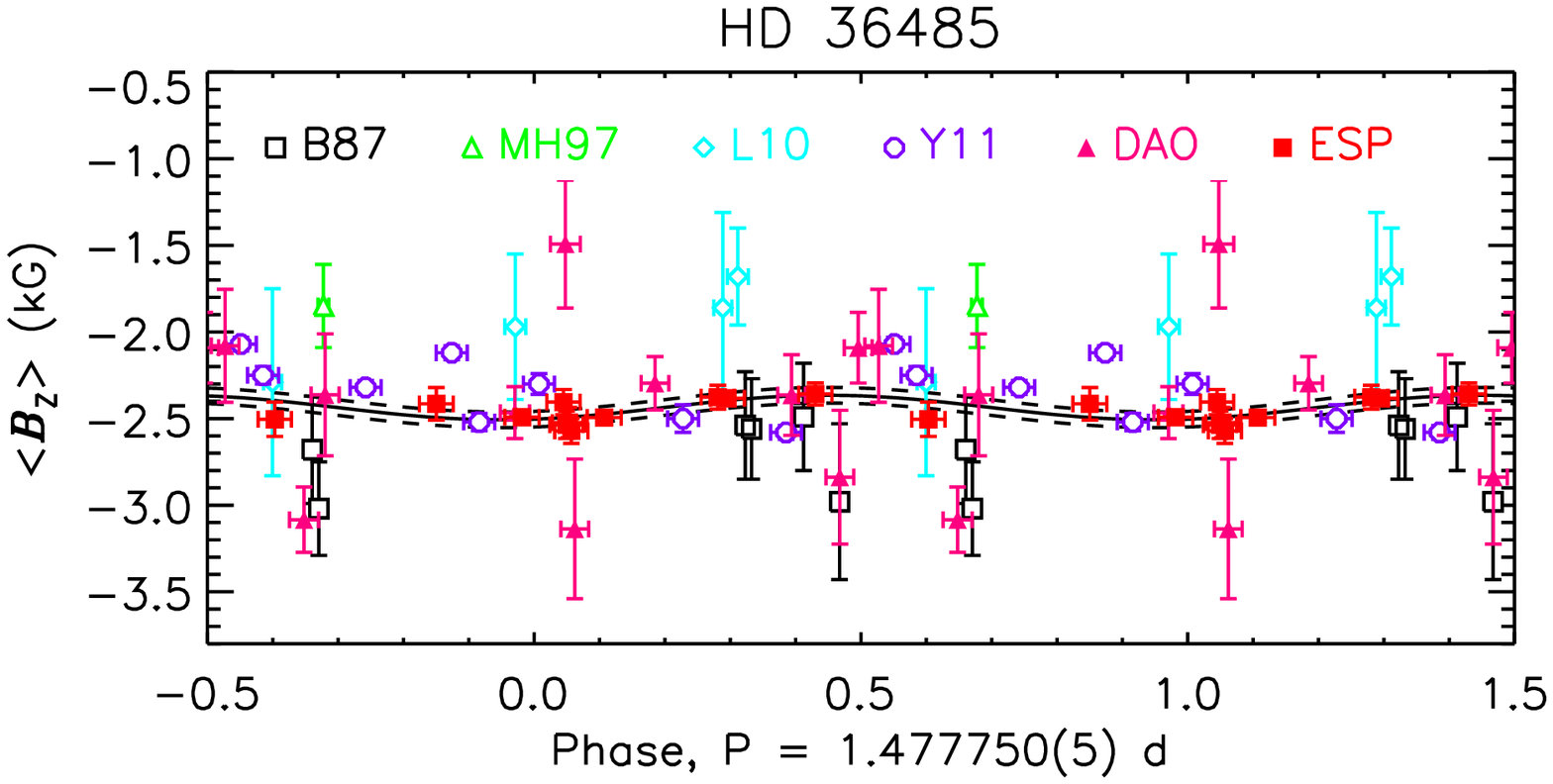}
   \includegraphics[width=8.5cm]{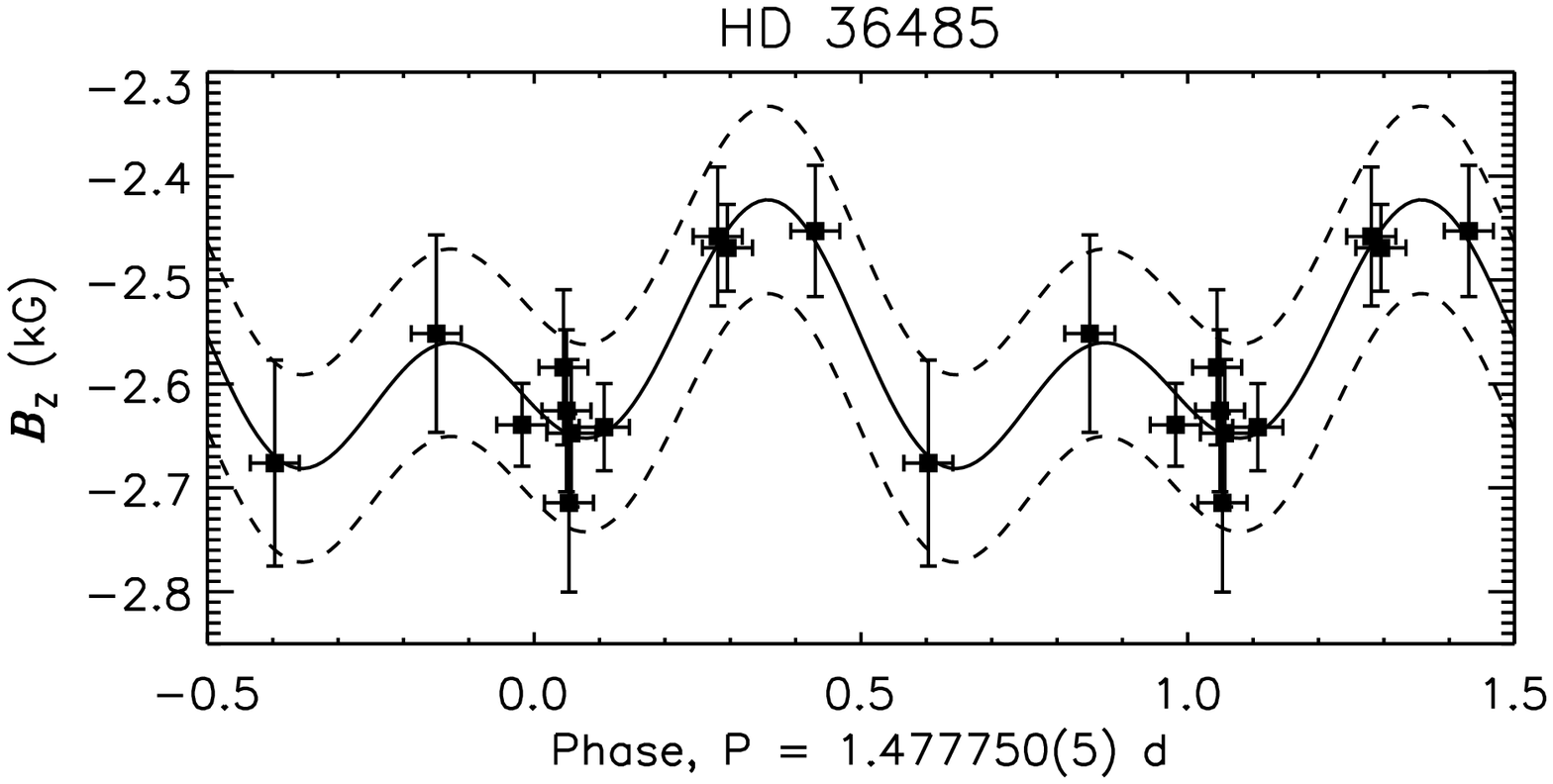}
      \caption[\bz~for HD 36485.]{\bz~curves for HD 36485 ($\delta$ Ori C). The top panel shows a comparison to historical data, demonstrating the much higher precision of the ESPaDOnS measurements. The bottom panel shows ESPaDOnS measurements only. Note that the \bz~curve is apparently better fit by a second-order sinusoid.}
         \label{HD36485_prot}
   \end{figure}

\noindent {\bf HD 36485}: \bz~measurements of HD 36485 ($\delta$ Ori C) have been published by \cite{1987ApJ...323..325B}, \cite{1997AAS..124..475M}, \cite{leone2010}, and \cite{yakunin2011}. ESPaDOnS data is unable to improve the period, which was determined spectroscopically using the magnetic primary's variable H$\alpha$ emission \citep{leone2010}. They are consistent in magnitude with previous data (see Fig.\ \ref{HD36485_prot}), but much more precise: in contrast to earlier measurements, a weak variation with a semi-amplitude of 0.06$\pm$0.03 kG can be discerned in the high-resolution measurements. The ESPaDOnS data agree in magnitude with the Special Astrophysical Observatory observations reported by \cite{yakunin2011}, but there is less scatter in the new data. The bottom panel shows the ESPaDOnS measurements only, fit with a second-order sinusoid, which provides a somewhat better fit to the data. The possibility that the star's magnetic field contains significant quadrupolar contributions was earlier suggested on the basis of a pronounced assymetry in the magnetospheric H$\alpha$ emission (\citealt{leone2010}). While the ESPaDOnS measurements are compatible with this hypothesis, and indeed represent the best test of it to date, the extremely low level of variation means that it cannot yet be unambiguously confirmed. Measurements with a precision of approximately 3$\times$ that of the existing dataset should be sufficient to definitively test the hypothesis that HD 36485 possesses a complex surface magnetic field.

   \begin{figure}
   \centering
\begin{tabular}{c}
   \includegraphics[width=8.5cm]{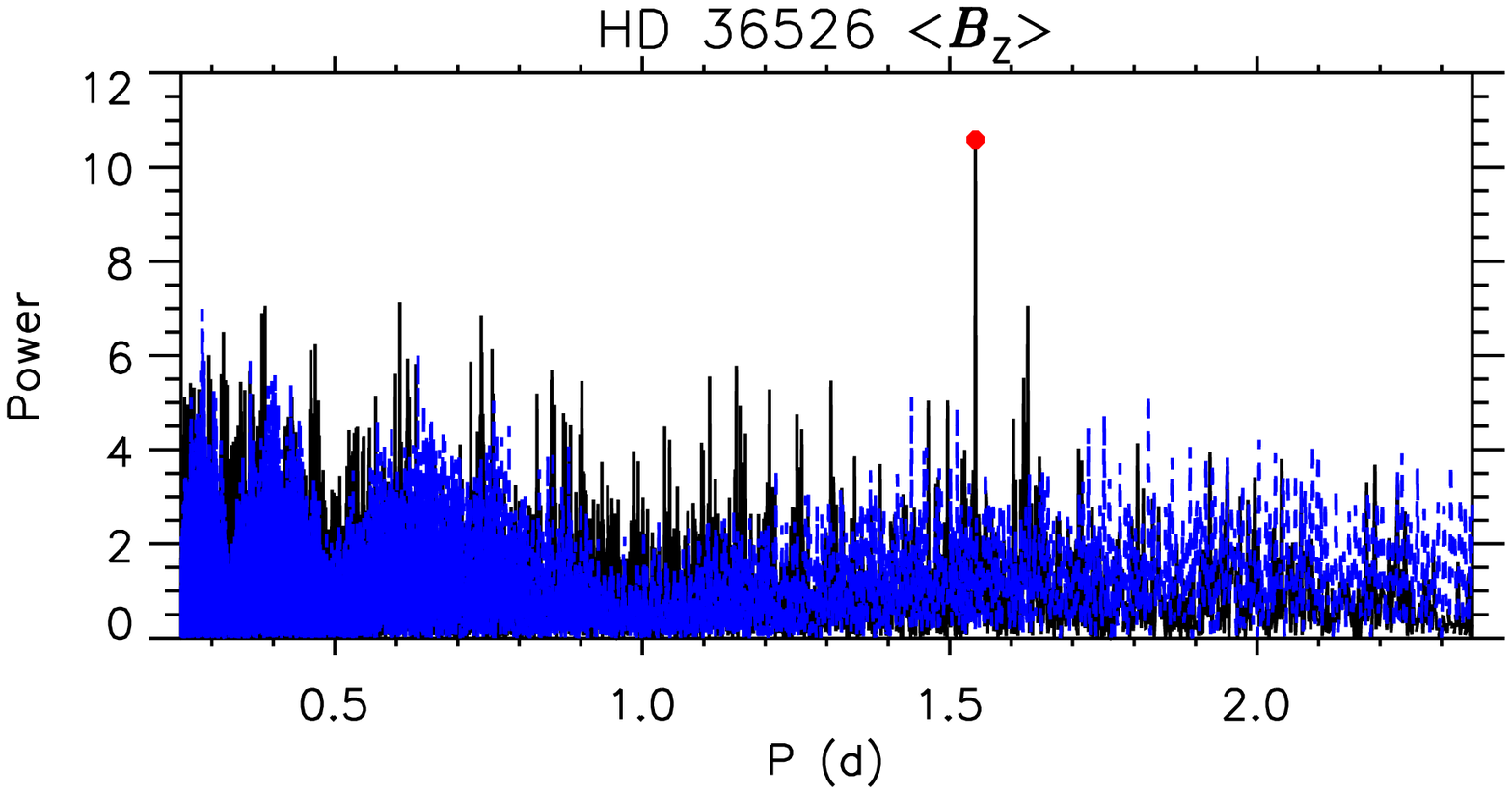} \\
   \includegraphics[width=8.5cm]{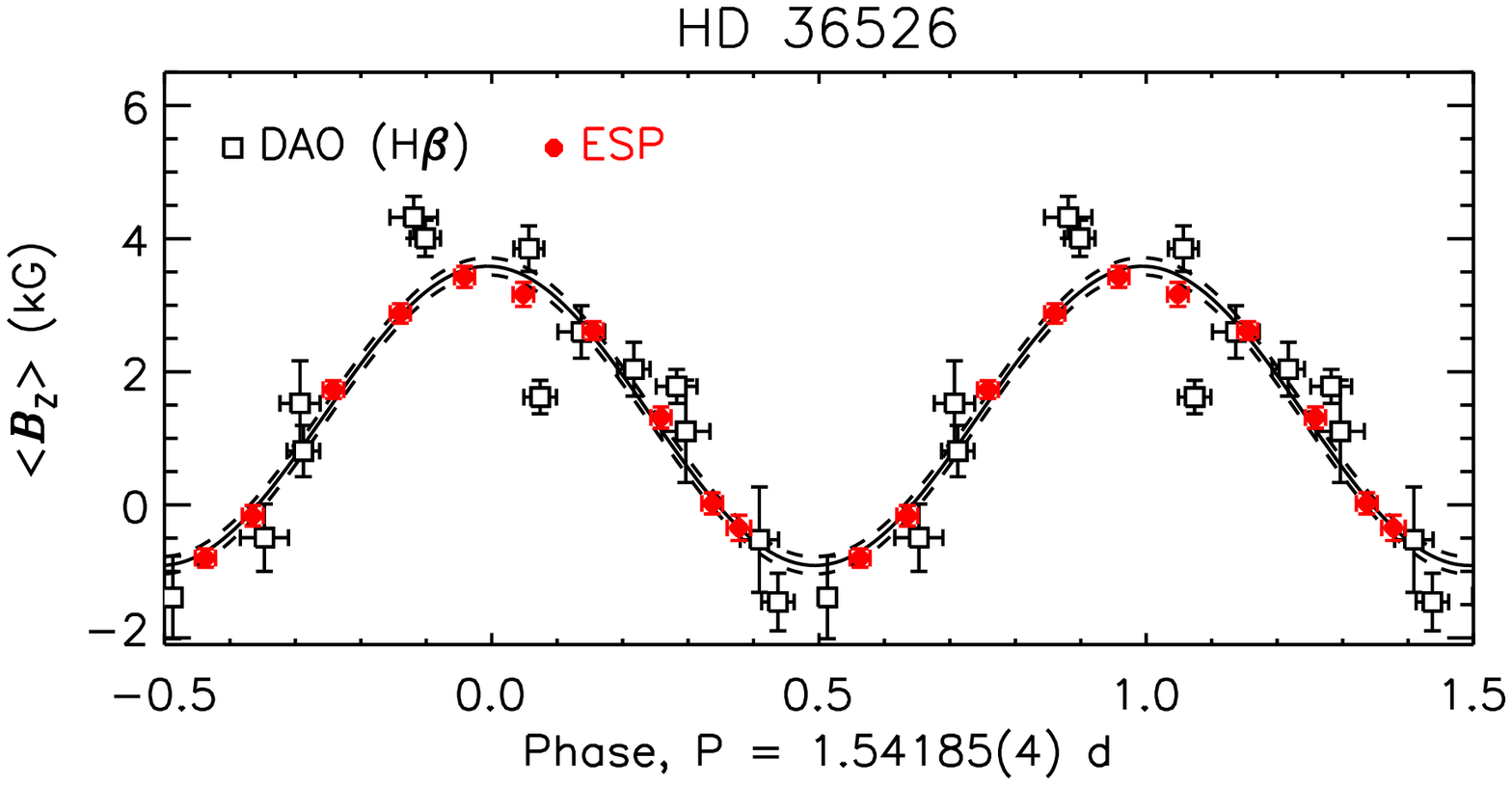} \\
\end{tabular} 
      \caption[Periodogram and \bz~for HD 36526.]{As Fig.\ \ref{HD25558_prot}, for HD 36526.}
         \label{HD36526_prot}
   \end{figure}

\noindent {\bf HD36526}: \cite{1984AA...141..328N} found a photometric period of 1.5405(1) d. This period provides a somewhat imperfect phasing of the ESPaDONS \bz~measurements; period analysis of the new ESPaDOnS magnetic data indicate a slightly longer $P_{\rm rot} = 1.542(1)$ d, while the dimaPol H$\beta$ \bz~measurements yield 1.5419(2)~d. Combining the datasets yields 1.54185(4)~d (Fig.\ \ref{HD36526_prot}).  The FAP of this peak is $2\times10^{-4}$, while the highest peak in the \nz~periodogram has a FAP of 0.01. \bz~‎is shown phased with this period in the bottom panel of Fig.\ \ref{HD36526_prot}.

   \begin{figure}
   \centering
   \includegraphics[width=8.5cm]{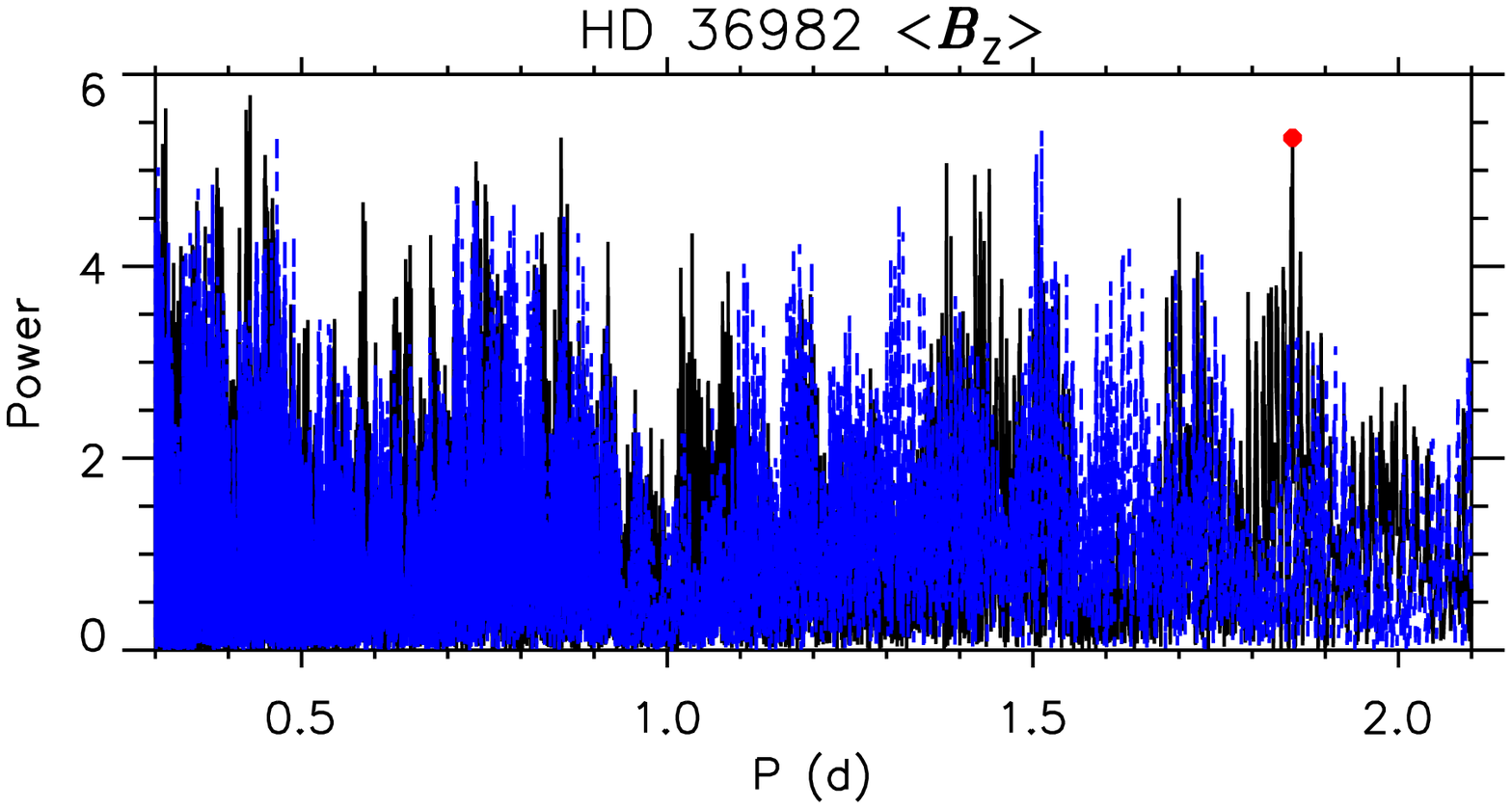} 
   \includegraphics[width=8.5cm]{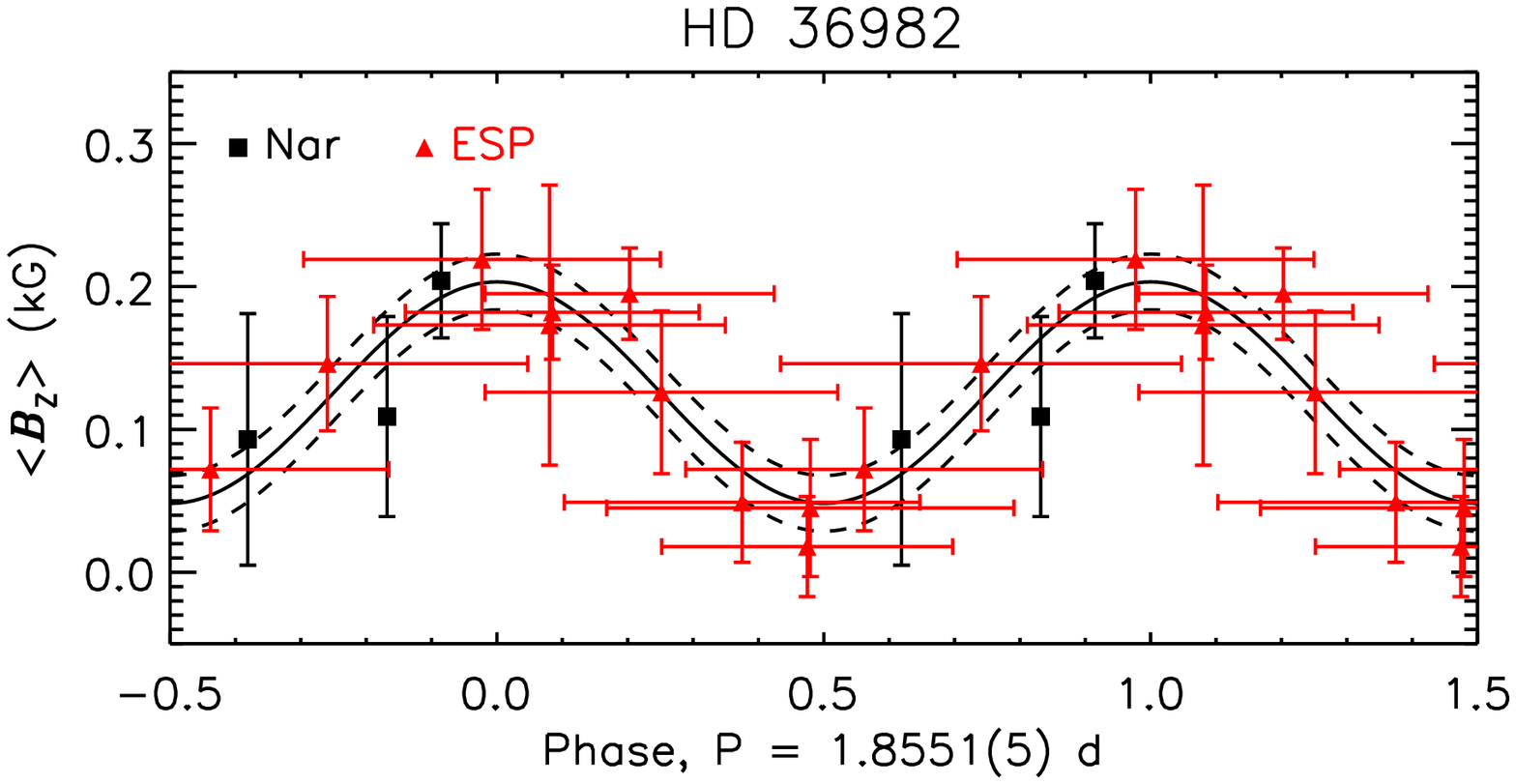} 
      \caption[Periodogram and \bz~for HD 36982.]{As Fig.\ \ref{HD25558_prot}, for HD 36982.}
         \label{LPOri_bz}
   \end{figure}

\noindent {\bf HD 36982}: spectroscopic variability is likely associated with this Herbig Be star's accretion disk, and therefore cannot be used to constrain $P_{\rm rot}$. Within the period window, the maximum power in the \bz~periodogram is at 0.425 d. However, there is a strong peak in the \nz~periodogram at this period (Fig.\ \ref{LPOri_bz}, top), and this is furthermore below the Nyquist frequency of 0.8~d$^{-1}$. We therefore adopt the next-strongest peak, 1.8551(5) d. However, it should be noted that the FAPs of the highest peaks in \bz~and \nz~are both similar, about 0.03. \bz~is shown phased with this period in Fig.\ \ref{LPOri_bz} (bottom).

   \begin{figure}
   \centering
\begin{tabular}{c}
   \includegraphics[width=8.5cm]{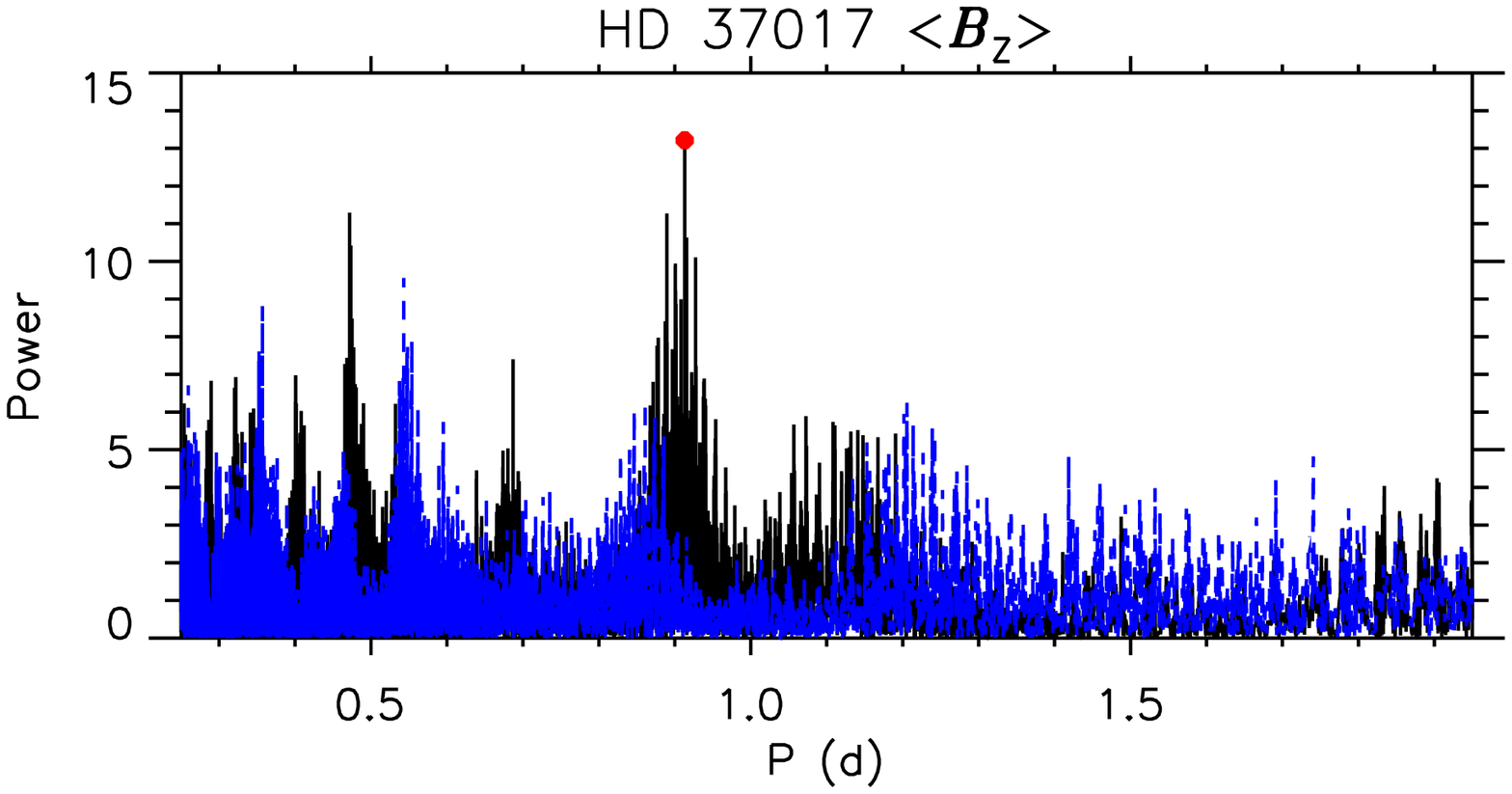} \\
   \includegraphics[width=8.5cm]{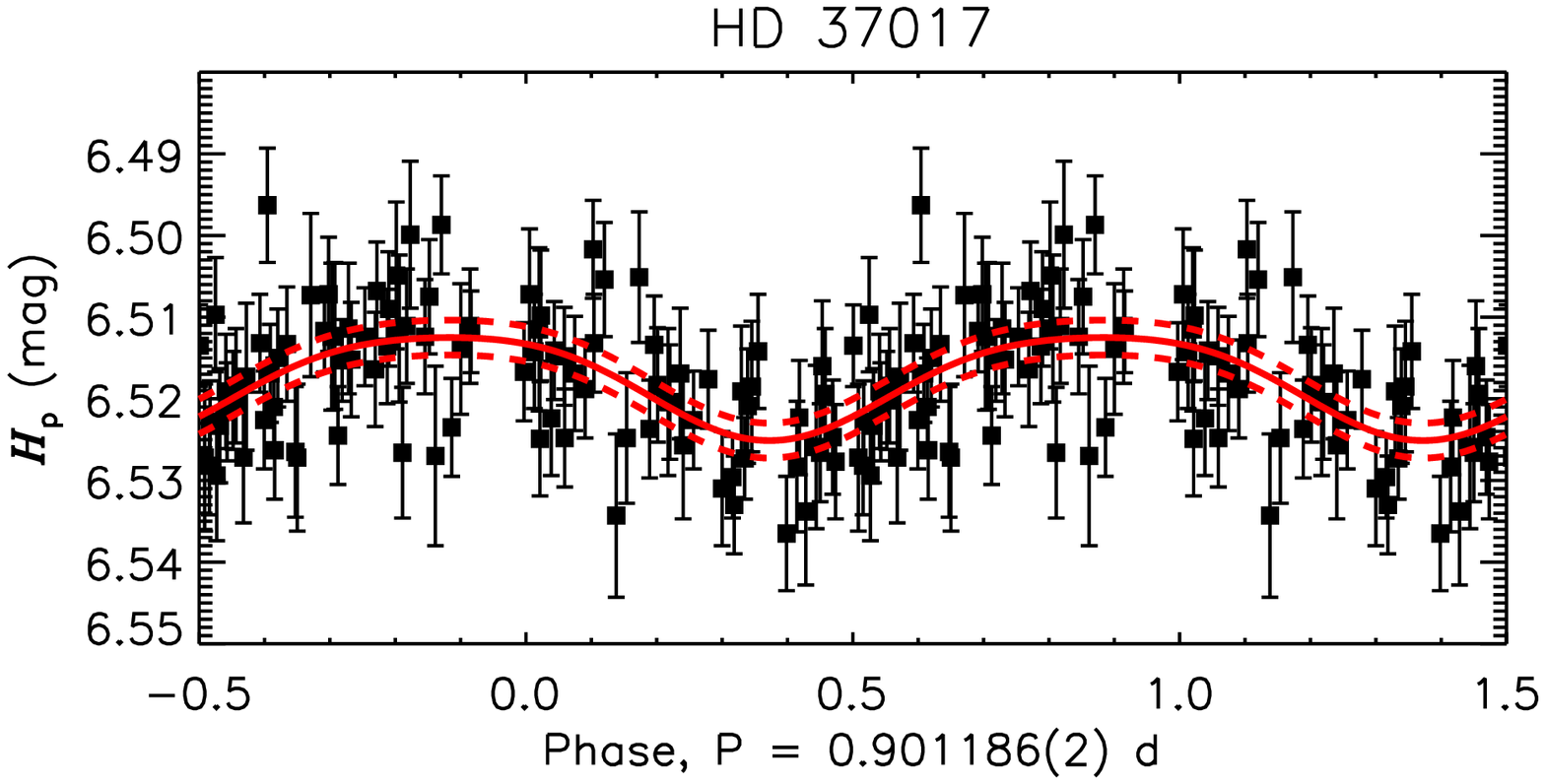} \\
   \includegraphics[width=8.5cm]{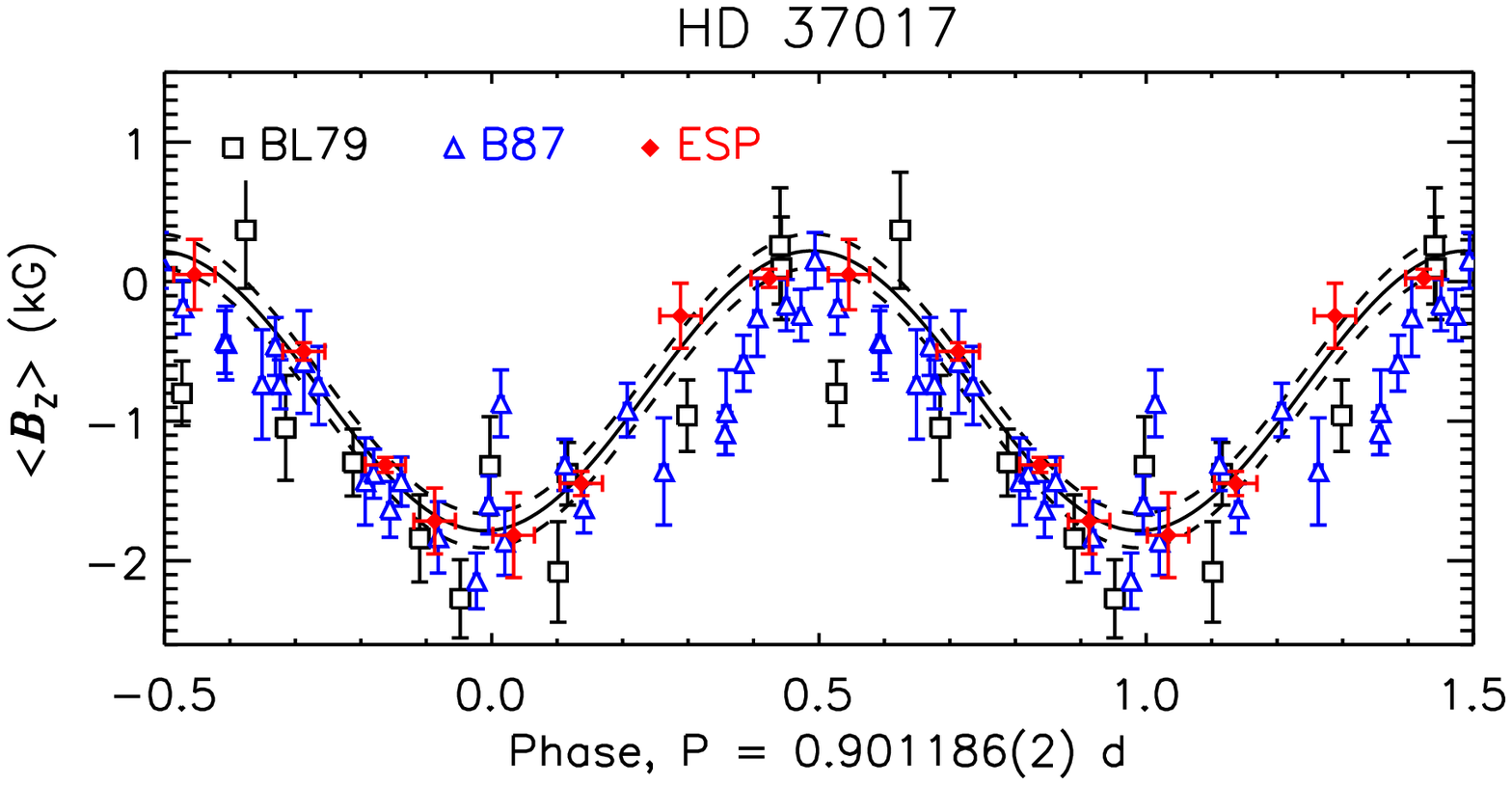} \\
\end{tabular} 
      \caption[Periodogram and \bz~for HD 37017.]{{\em Top}: \bz~periodogram for HD 37017. {\em Middle}: Hipparcos photometry phased with $P_{\rm rot}$. The red curve is 2$^{nd}$-order sinusoidal fit; dashed curves indicate uncertainties in the fit. {\em Bottom}: \bz~phased with $P_{\rm rot}$.}
         \label{HD37017_prot}
   \end{figure}

\noindent {\bf HD 37017}: \cite{1987ApJ...323..325B} combined periodograms for photometric and magnetic data and found $P_{\rm rot} = $0.901195(5) d. This period does not provide a reasonable phasing of the ESPaDONS data. Combining historical measurements with our own, we find 0.901186(2) d. The maximum amplitude period has a FAP of about $10^{-7}$, while the FAP of the highest-amplitude peak in the \nz~spectrum is 0.12. The periodogram and \bz~measurements are shown in the top and middle panels of Fig.\ \ref{HD37017_prot}. The Hipparcos photometry is also coherently phased with this period (bottom panel), although the low level of photometric variation means that it cannot be used to determine a more precise period. 

   \begin{figure}
   \centering
   \includegraphics[width=8.5cm]{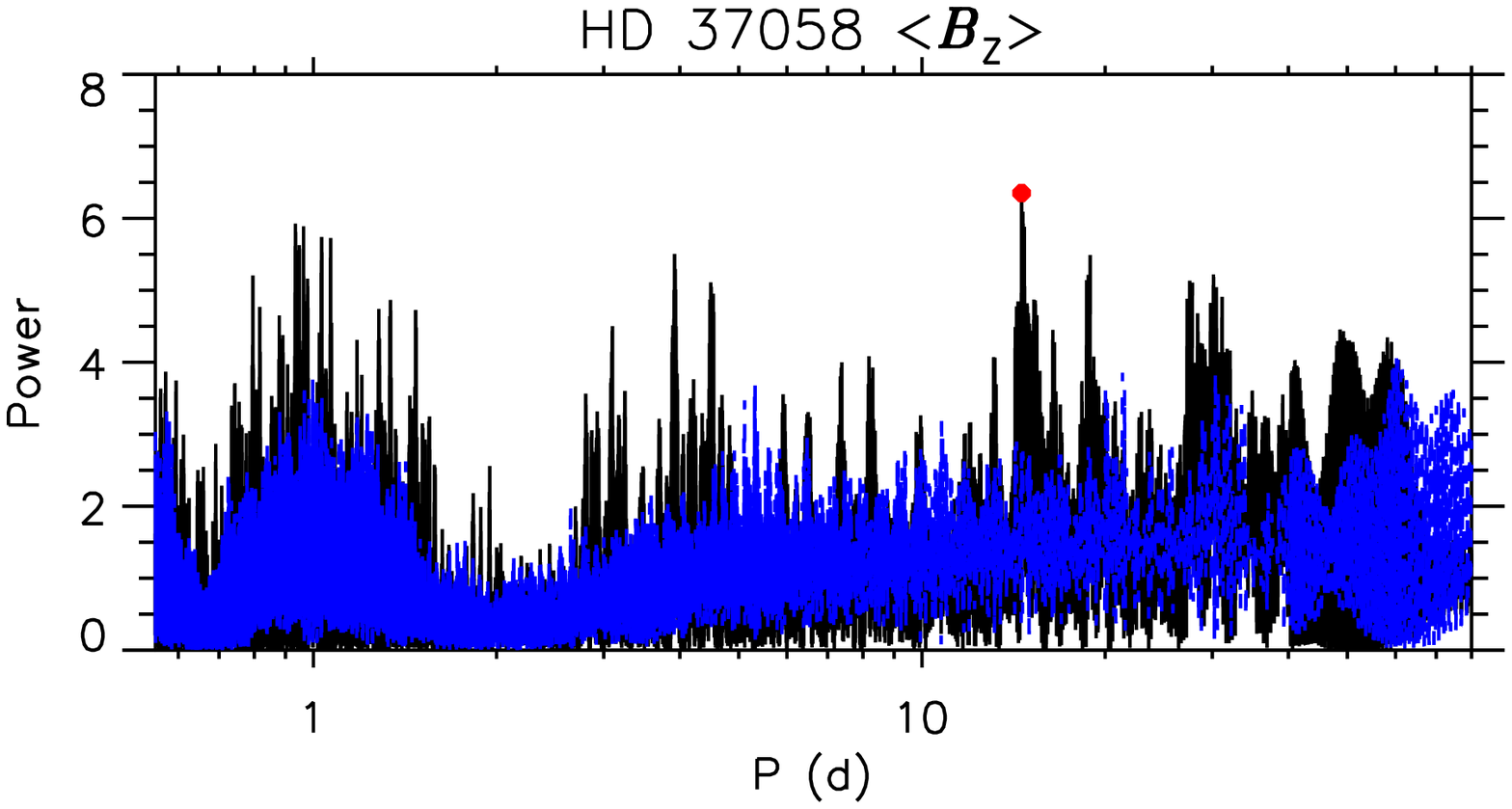} 
   \includegraphics[width=8.5cm]{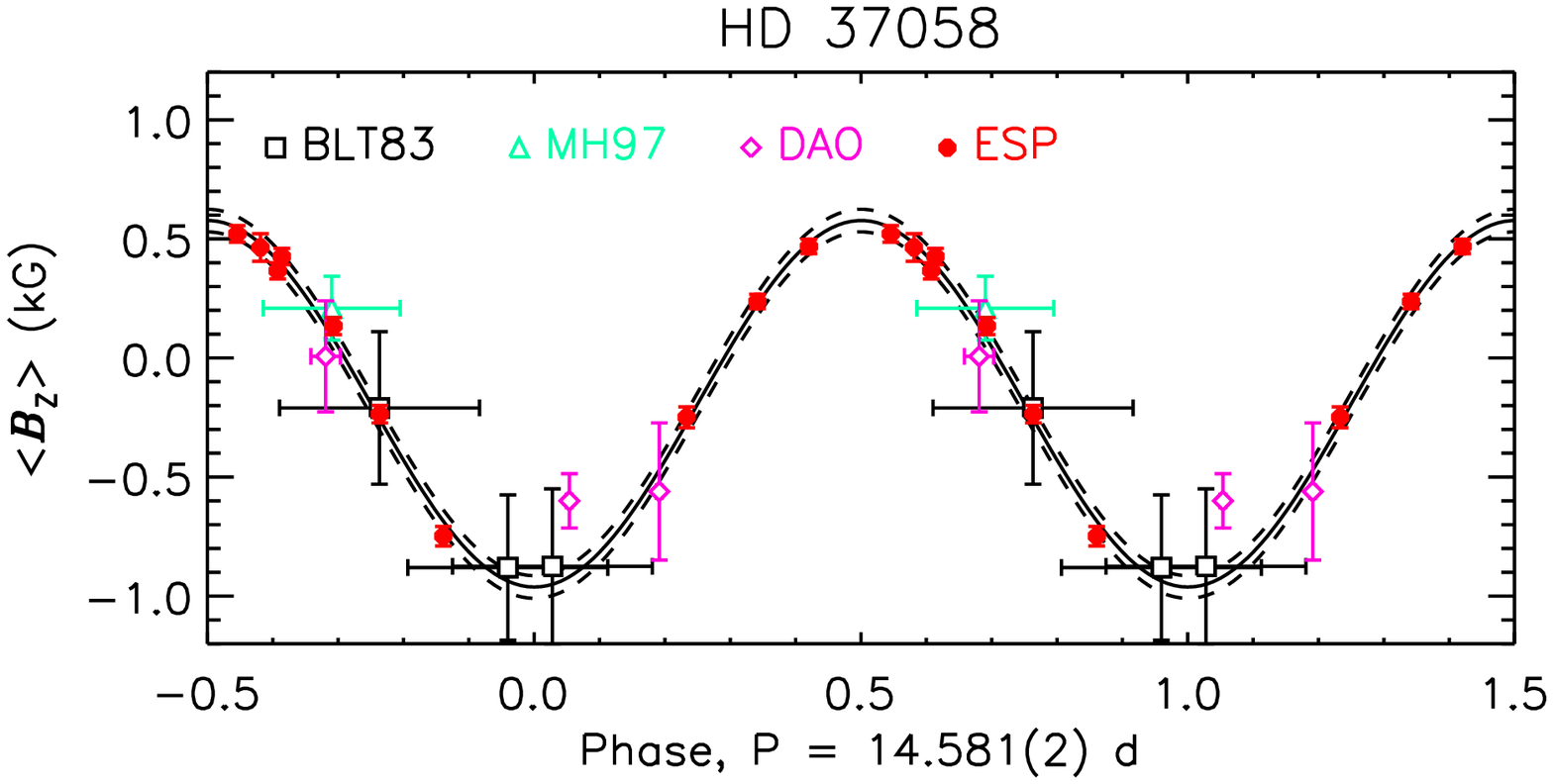} 
      \caption[Periodogram and \bz~for HD 37058.]{As Fig.\ \ref{HD25558_prot}, for HD 37058.}
         \label{HD37058_prot}
   \end{figure}

\noindent {\bf HD 37058}: \cite{pederson1979} reported a photometric period of $\sim$14 d, while \cite{2005AA...430.1143B} determined a 1.022 d period. In both cases the available datasets were too small to firmly establish $P_{\rm rot}$. The ESPaDOnS measurements confirm \cite{pederson1979}'s period. By combining with \bz~measurements from the literature, we find $P_{\rm rot}$=14.581(2)~d, with a FAP of 0.03, much lower than the FAP of the maximum amplitude peak in the null spectrum, $\sim$0.3. \bz~is shown phased with this period in Fig.\ \ref{HD37058_prot}.

   \begin{figure}
   \centering
   \includegraphics[width=8.5cm]{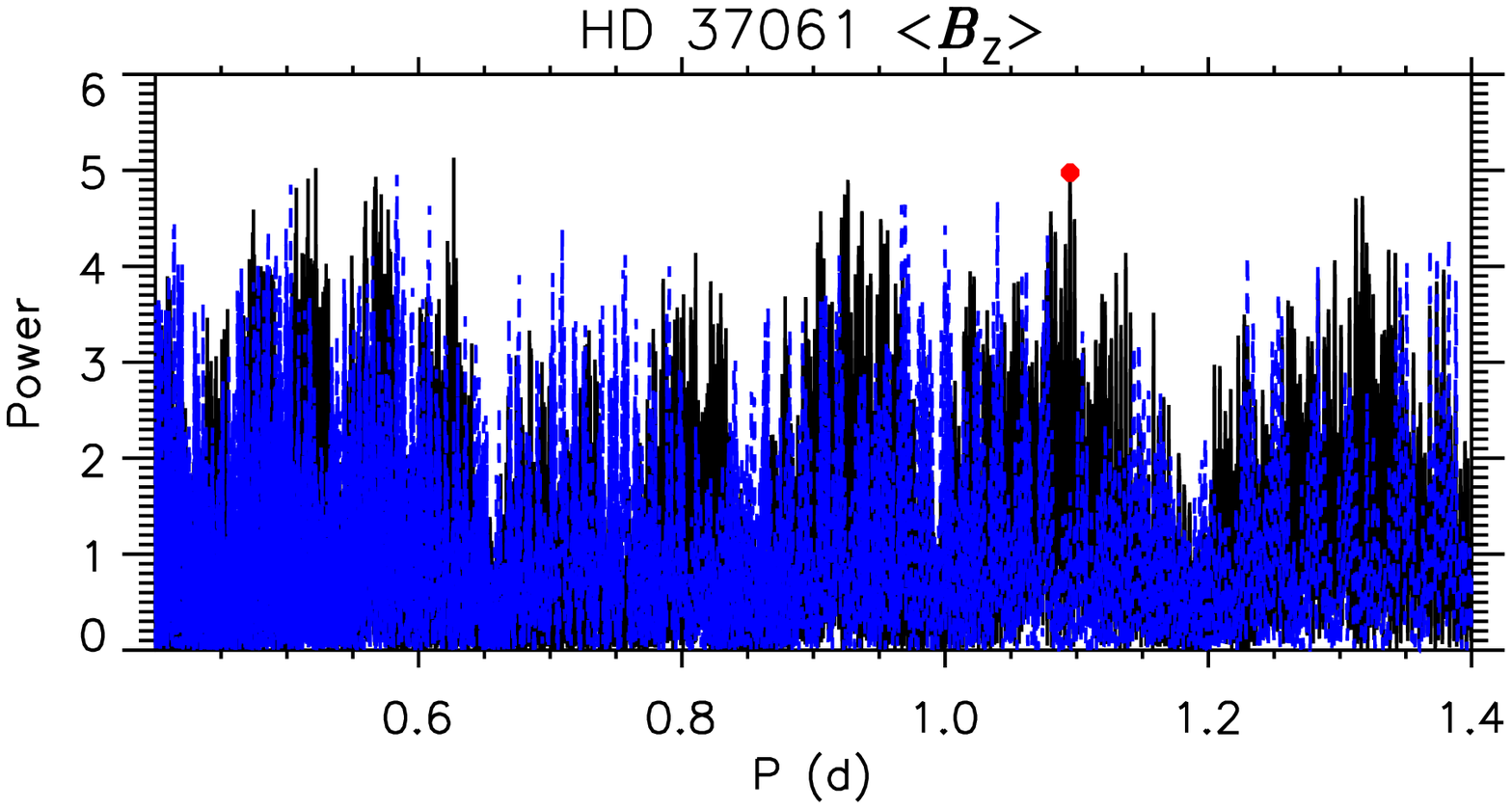} 
   \includegraphics[width=8.5cm]{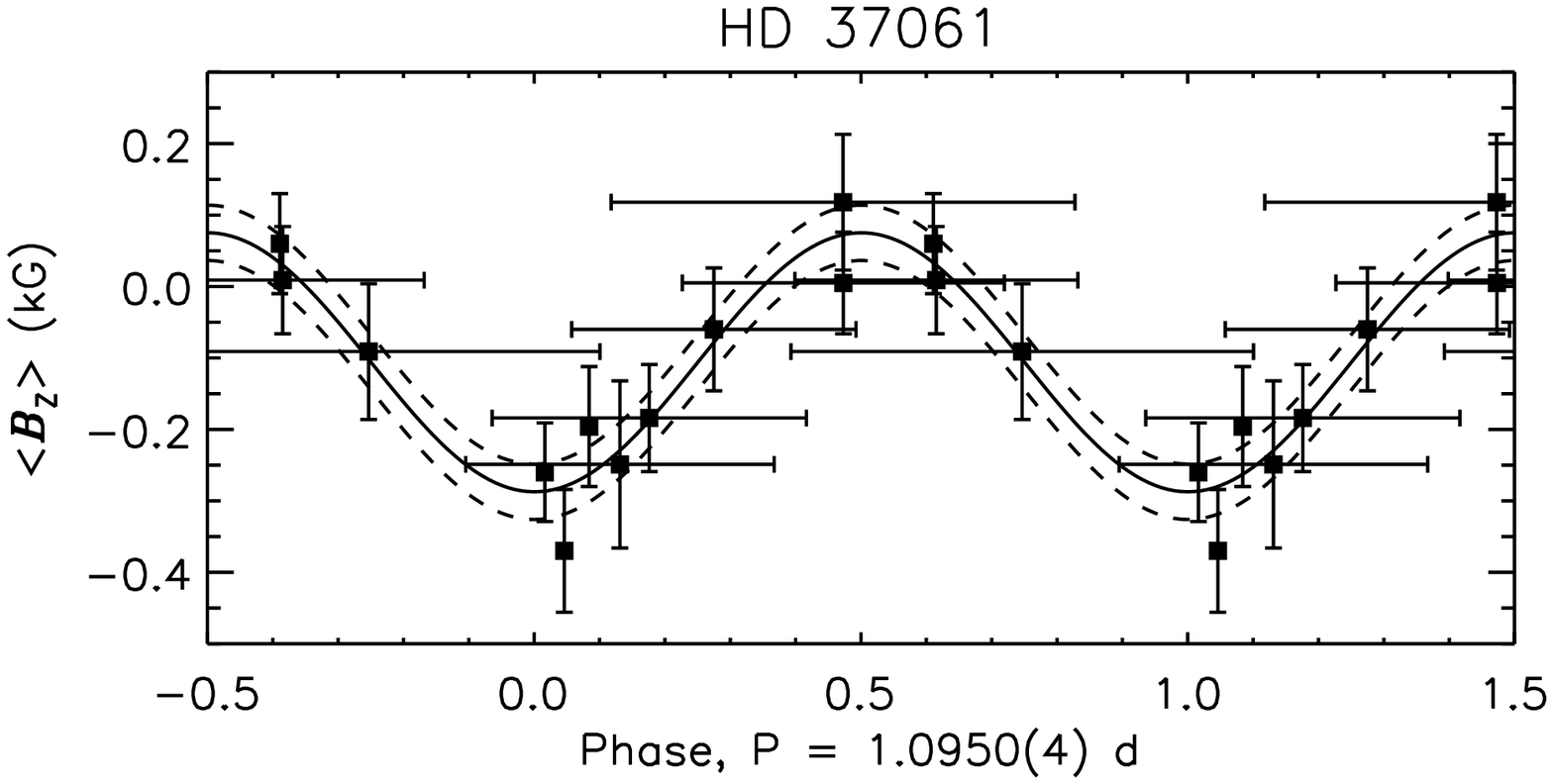} 
      \caption[Periodogram and \bz~for HD 37061.]{As Fig.\ \ref{HD25558_prot}, for HD 37061.}
         \label{HD37061_prot}
   \end{figure}

\noindent {\bf HD 37061}: while HD 37061 (NU Ori) displays spectroscopic variability, it is not clear that this is purely due to rotational modulation as periodograms created for these data yield contradictory results. Therefore we base our period search on the magnetic data. The star's high \vsini~(225$\pm$8 \kms) means the period must be no longer than $\sim$1.4 d. There are two peaks of comparable amplitude within the period window, at $\sim$0.6 d and $\sim$1.1 d (Fig.\ \ref{HD37061_prot}, top). The shortest period would indicate $v_{\rm eq} \sim v_{\rm br}$, necessitating very tight limits on the stellar parameters. The FAPs of either peak are both about 0.08, only slightly below the \nz~maximum amplitude FAP of about 0.1. As the next-highest peak yields less extreme rotational parameters, we adopt the peak at 1.0950(4)~d, as the more conservative option. \bz~is shown phased with this period in Fig.\ \ref{HD37061_prot}.

   \begin{figure}
   \centering
      \includegraphics[width=8.5cm]{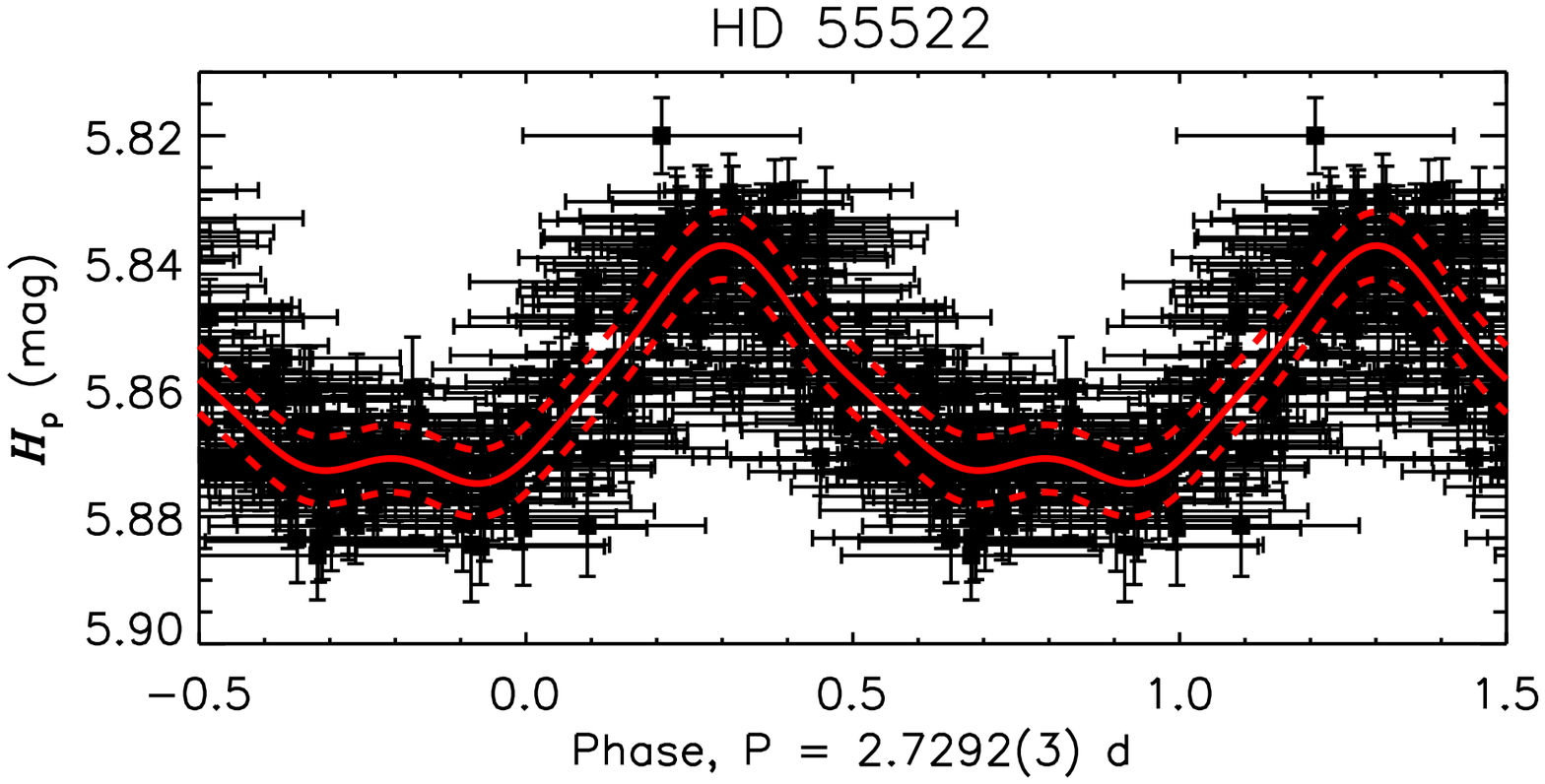} 
      \includegraphics[width=8.5cm]{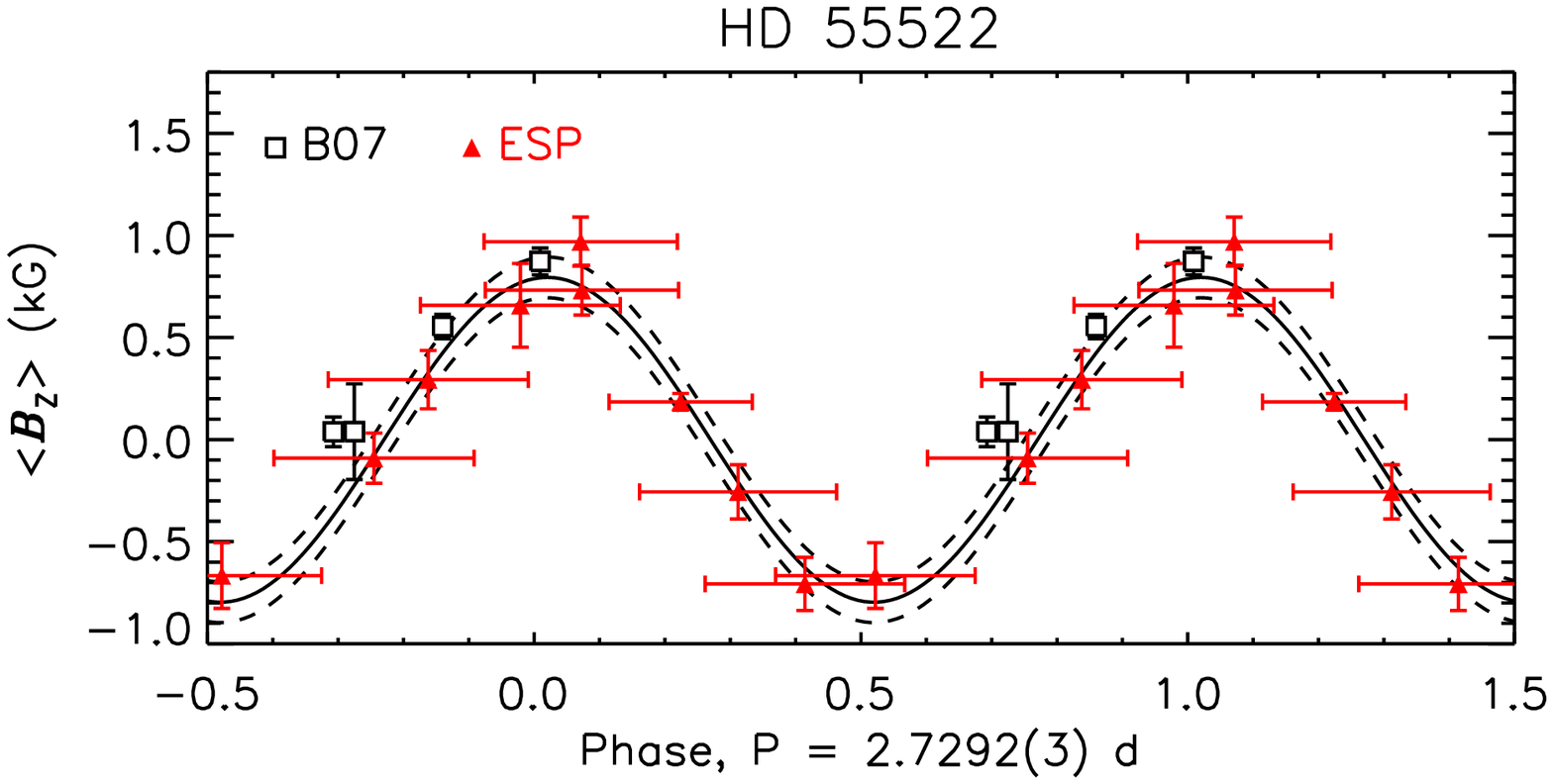} 
      \caption[Periodogram and \bz~for HD 55522.]{ {\em Top}: Hipparcos photometry phased with the revised period for HD 55522. The red curve is a 2$^{nd}$-order sinusoidal fit; dashed curves indicate the uncertainty in the fit. {\em Bottom}: \bz, phased with the revised period. The solid curve shows the best-fit sinusoid; the dashed curve shows the 1$\sigma$ uncertainty.}
         \label{HD55522_prot}
   \end{figure}

\noindent {\bf HD 55522}: \cite{2004AA...413..273B} used spectroscopy and ground-based as well as archival Hipparcos photometry to find $P_{\rm rot} = 2.729(1)$ d. This period phases the ESPaDOnS data well when examined in isolation, but comparison of the \bz~measurements presented by \cite{2007AN....328...41B} to our own reveals a small phase offset. While there are numerous closely spaced peaks in the periodogram constructed for the combined magnetic data, the only peak consistent with the photometric data is at 2.7292(3) d. \bz~and $H_{\rm p}$ are shown phased with this period in Fig.\ \ref{HD55522_prot}. While the phase uncertainties are somewhat large, maximum light appears to occur at about \bz$\sim 0$. 

   \begin{figure}
   \centering
      \includegraphics[width=8.5cm]{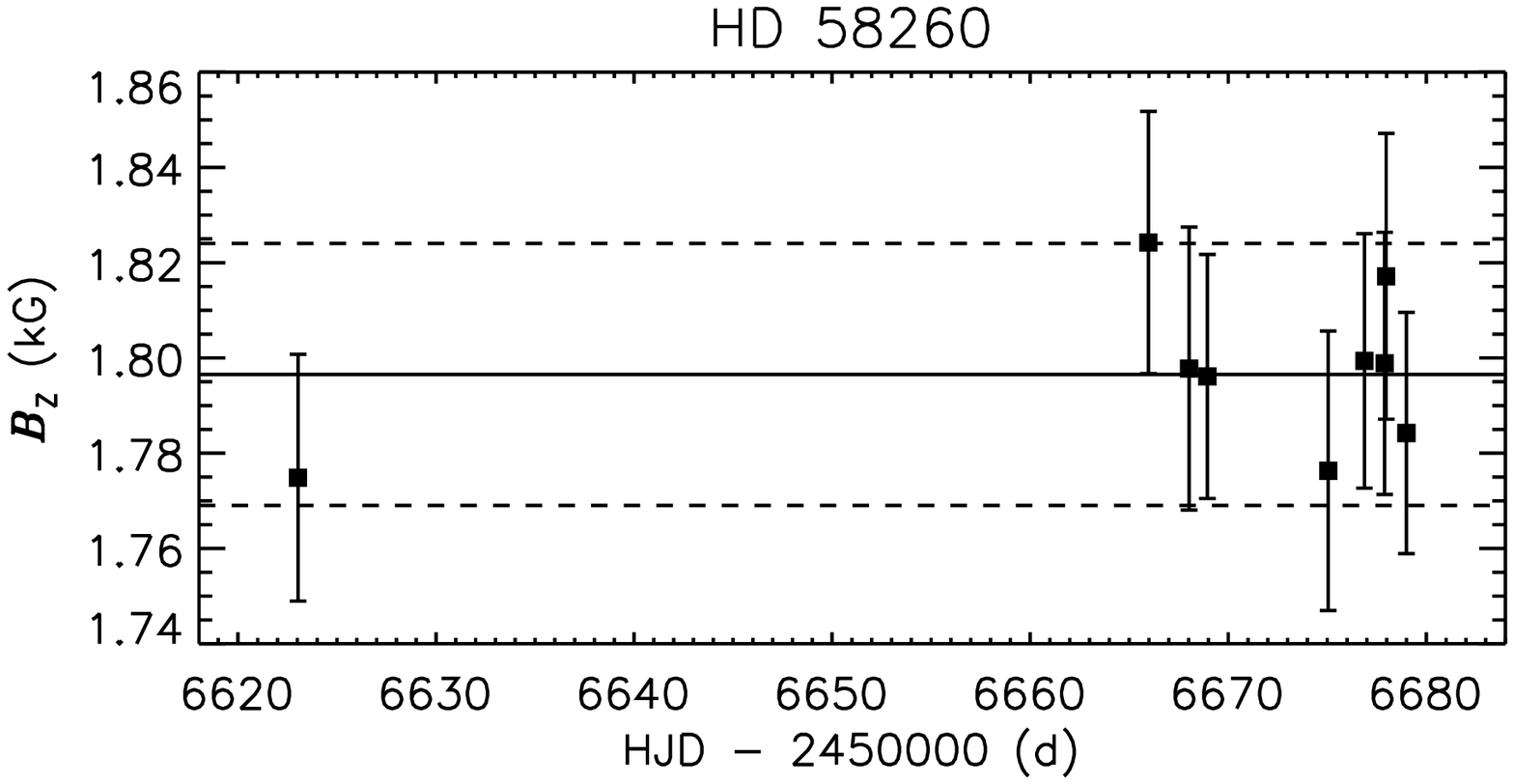} 
      \includegraphics[width=8.5cm]{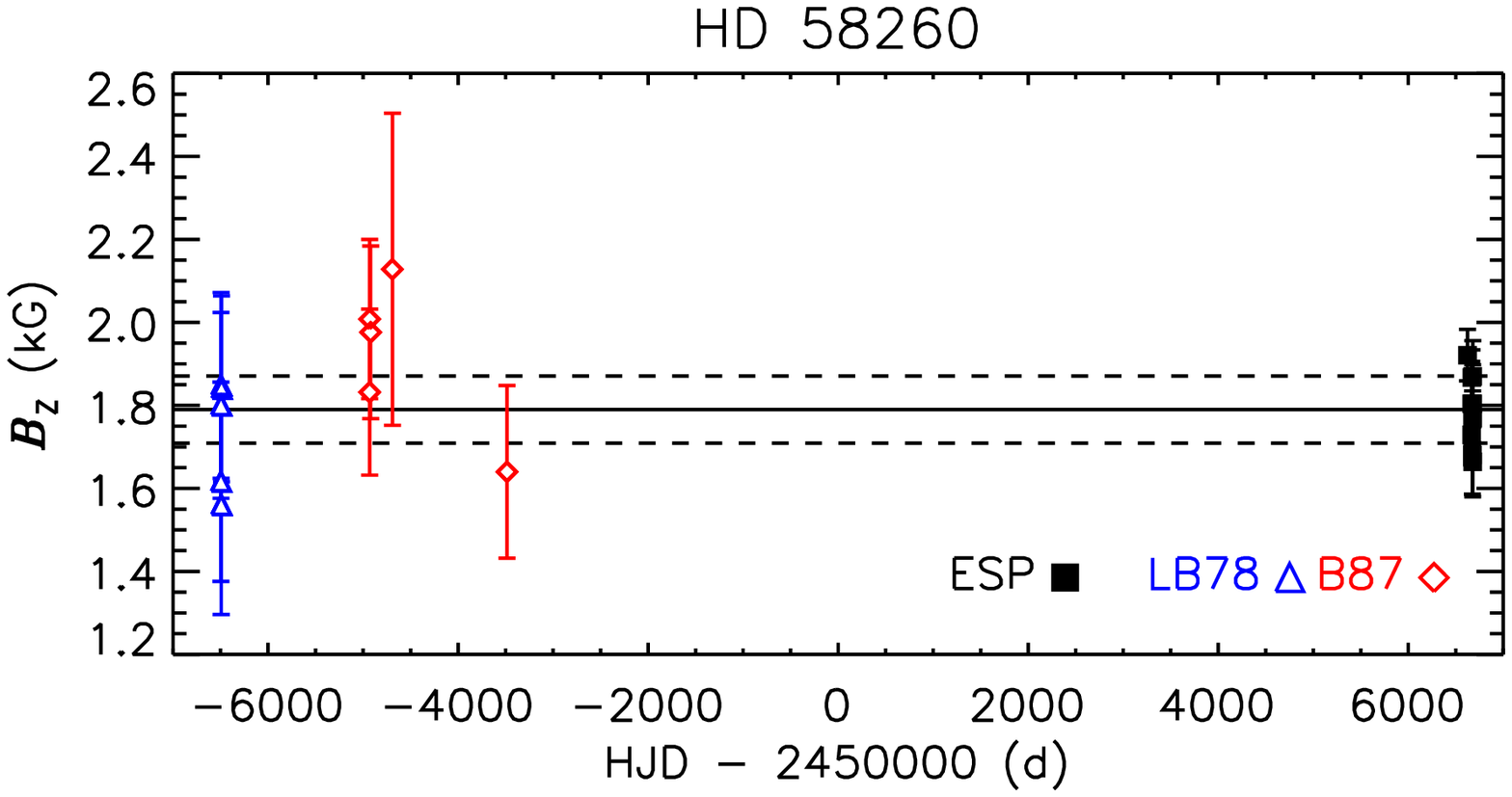} 
      \caption[\bz~for HD 58260.]{\bz~measurements of HD 58260 as a function of time. The top panel shows the ESPaDOnS data, the bottom panel a comparison to historical data. Solid lines indicate the mean ESPaDOnS \bz, dashed lines the mean ESPaDOnS \bz~error bar.}
         \label{HD58260_prot}
   \end{figure}

\noindent {\bf HD 58260}: \cite{pederson1979} suggested $P_{\rm rot}=1.657$ d based on the star's photometric variation. While this period is compatible with the period window of 1.42 to 164 d, we cannot confirm this period using Hipparcos photometry. \cite{1987ApJ...323..325B} were unable to determine a period for this star due to the negligible variation in \bz. Even at the much higher precision of the ESPaDOnS data, the standard deviation of \bz~is only 17 G, with a mean error bar of 28 G. While there are some higher-amplitude peaks at short periods, and very little power at periods longer than $\sim$10~d, this is likely an artifact caused by the relatively short temporal baseline (56 days, with all but 1 observation acquired within 13 days). Analysis of Hipparcos photometry yields no peaks with a S/N in excess of 4 at periods of greater than 1~d. Comparison to historical data (bottom panel) demonstrates that \bz~is consistent with no variation over a timescale of $\sim$35 years: all measurements are within 1$\sigma$ of the mean ESPaDOnS \bz.


   \begin{figure}
   \centering
   \includegraphics[width=8.5cm]{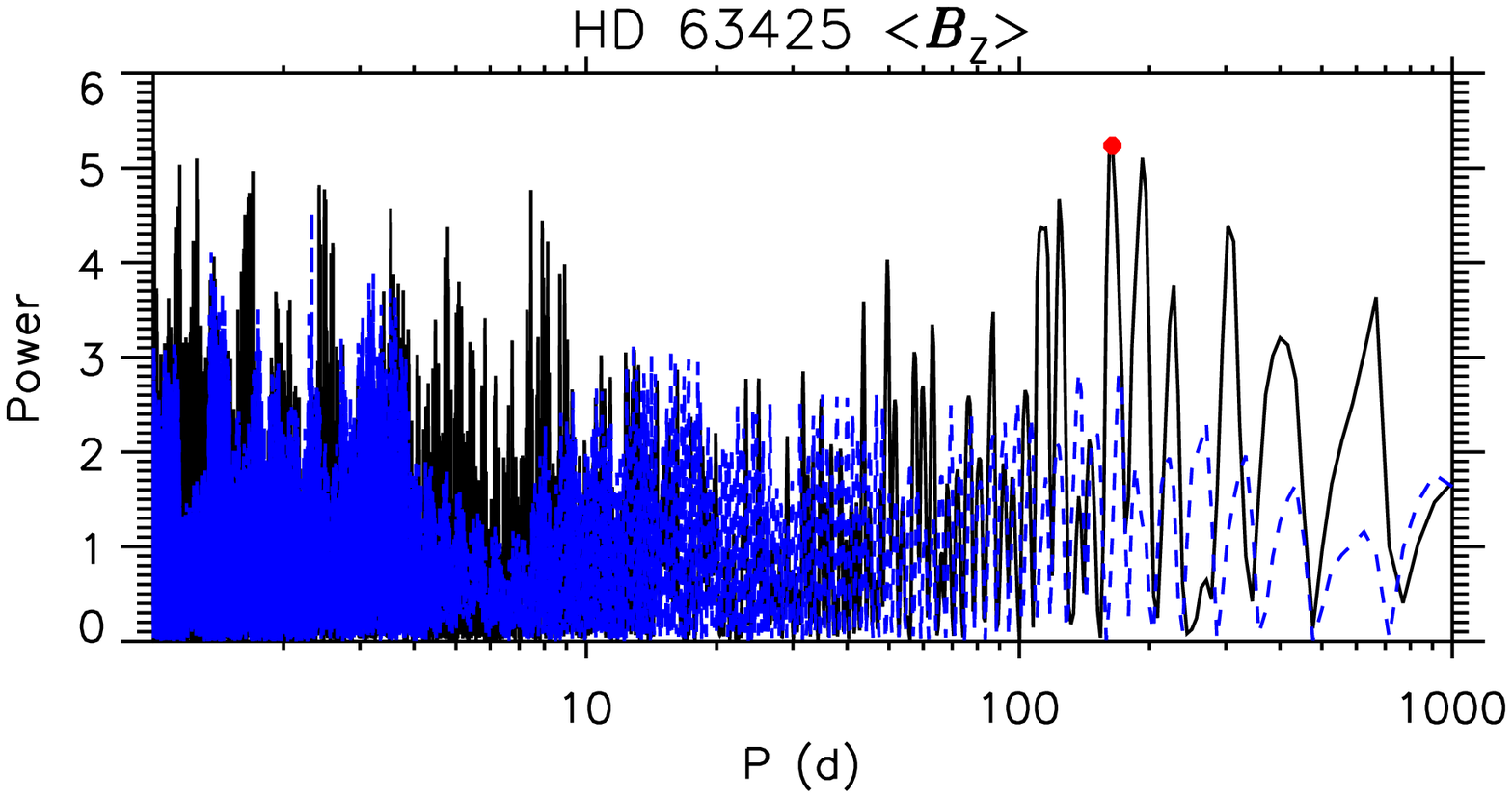} 
   \includegraphics[width=8.5cm]{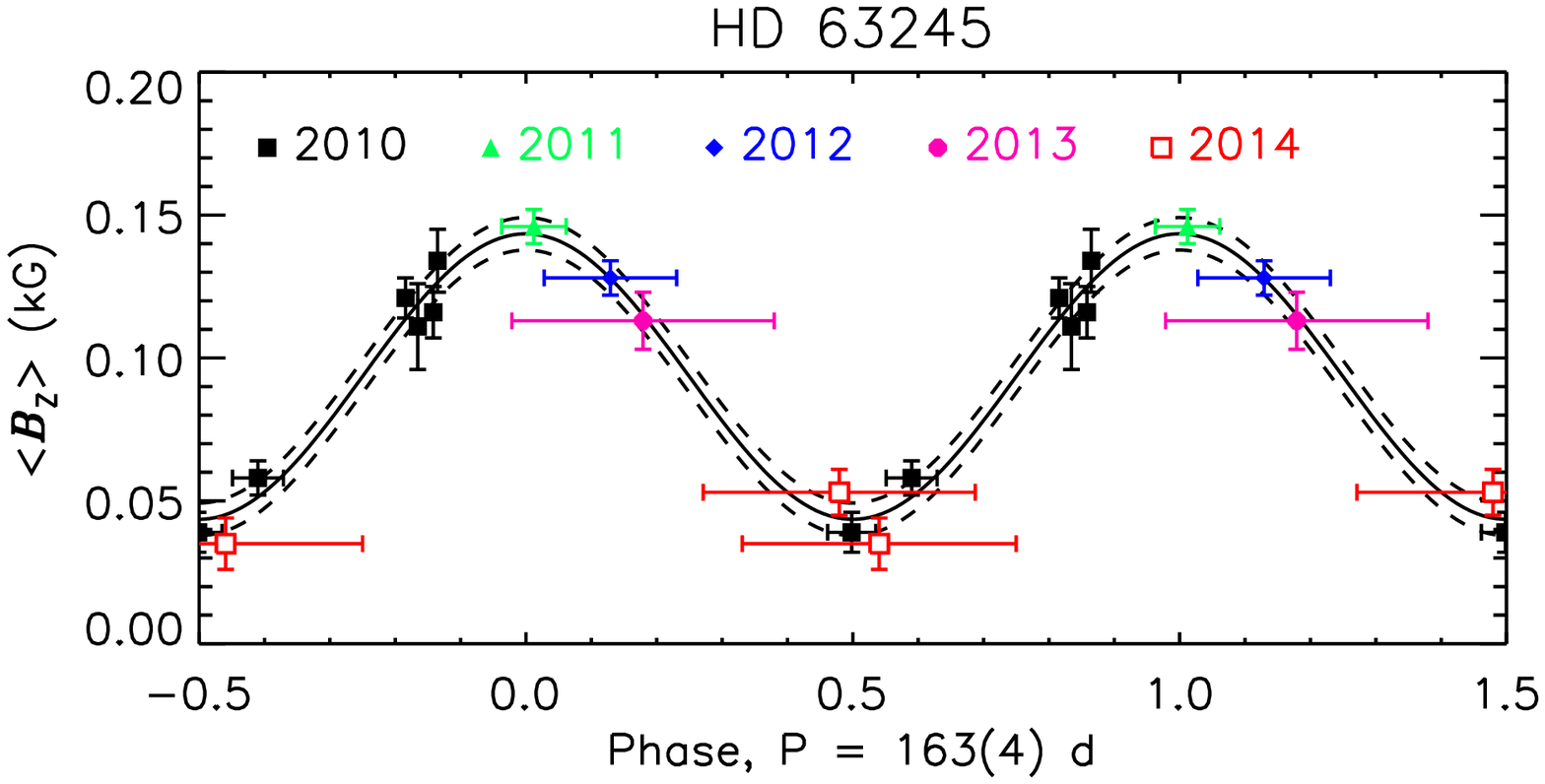} 
      \caption[Periodogram and \bz~for HD 63425.]{As Fig.\ \ref{HD25558_prot}, for HD 63425.}
         \label{HD63425_prot}
   \end{figure}

\noindent {\bf HD 63425}: two periods are compatible with the existing \bz~measurements. Maximum power is at 163(4)~d, which is compatible with the low \vsini. The other is very short, $\sim$0.55~d, which would make HD 63425 amongst the most rapidly rotating magnetic massive stars. The FAPs of the two periods are similar, about 0.08, and only slightly lower than the minimum \nz~FAP of 0.1. Given the very small \vsini, this would require the rotational pole to be almost perfectly aligned with the line-of-sight, which is {\em a priori} less likely. Furthermore, the \bz~measurements from 2010 (the year with the largest number of observations) show statistically significant differences only between those measurements separated by $\sim$250 d, while the variation of those collected within $\sim$10 d is of the same order as the error bars. We therefore adopt the longer period. \bz~is shown phased with this period in Fig.\ \ref{HD63425_prot}.

   \begin{figure}
   \centering
   \includegraphics[width=8.5cm]{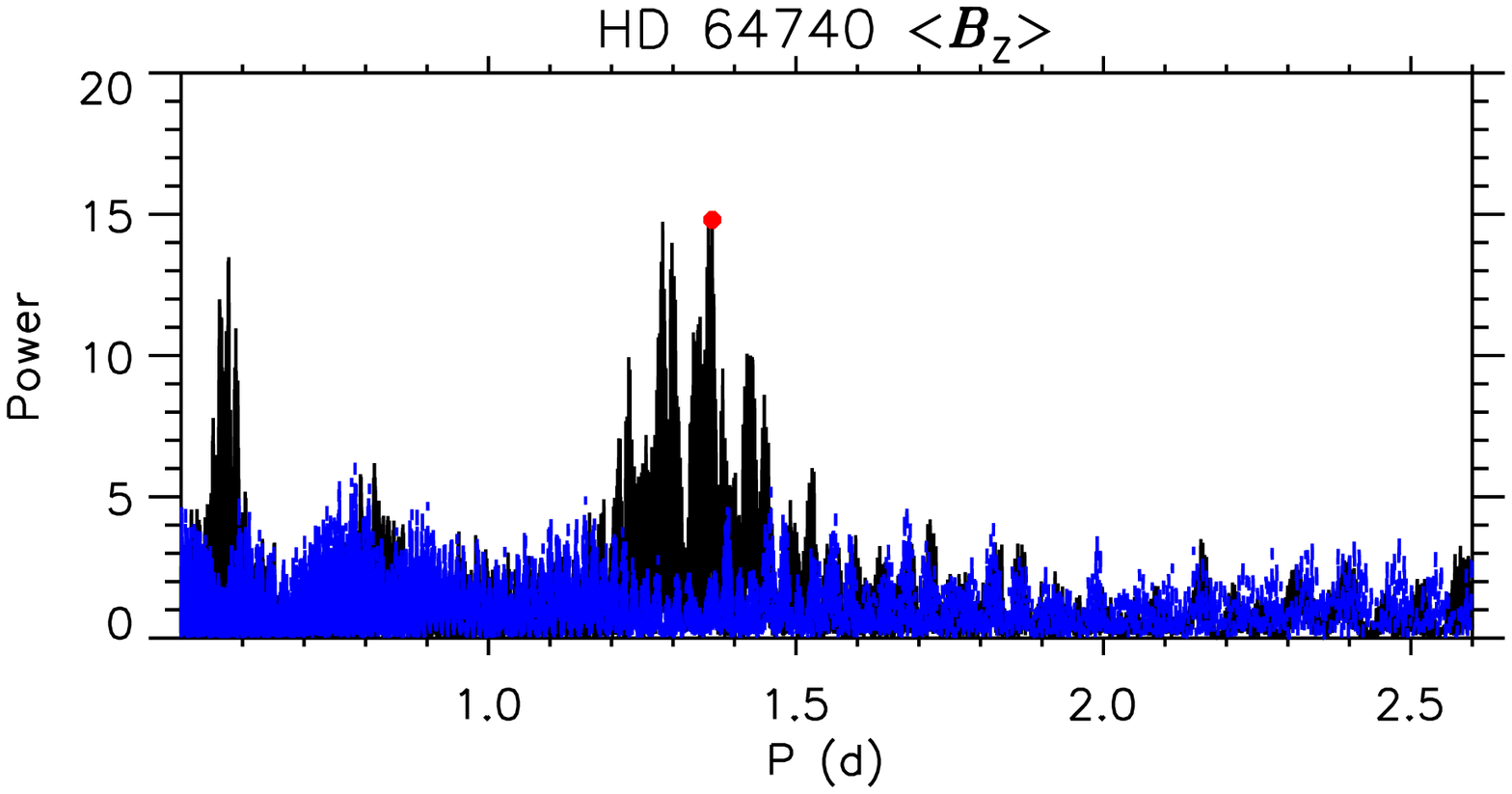} 
   \includegraphics[width=8.5cm]{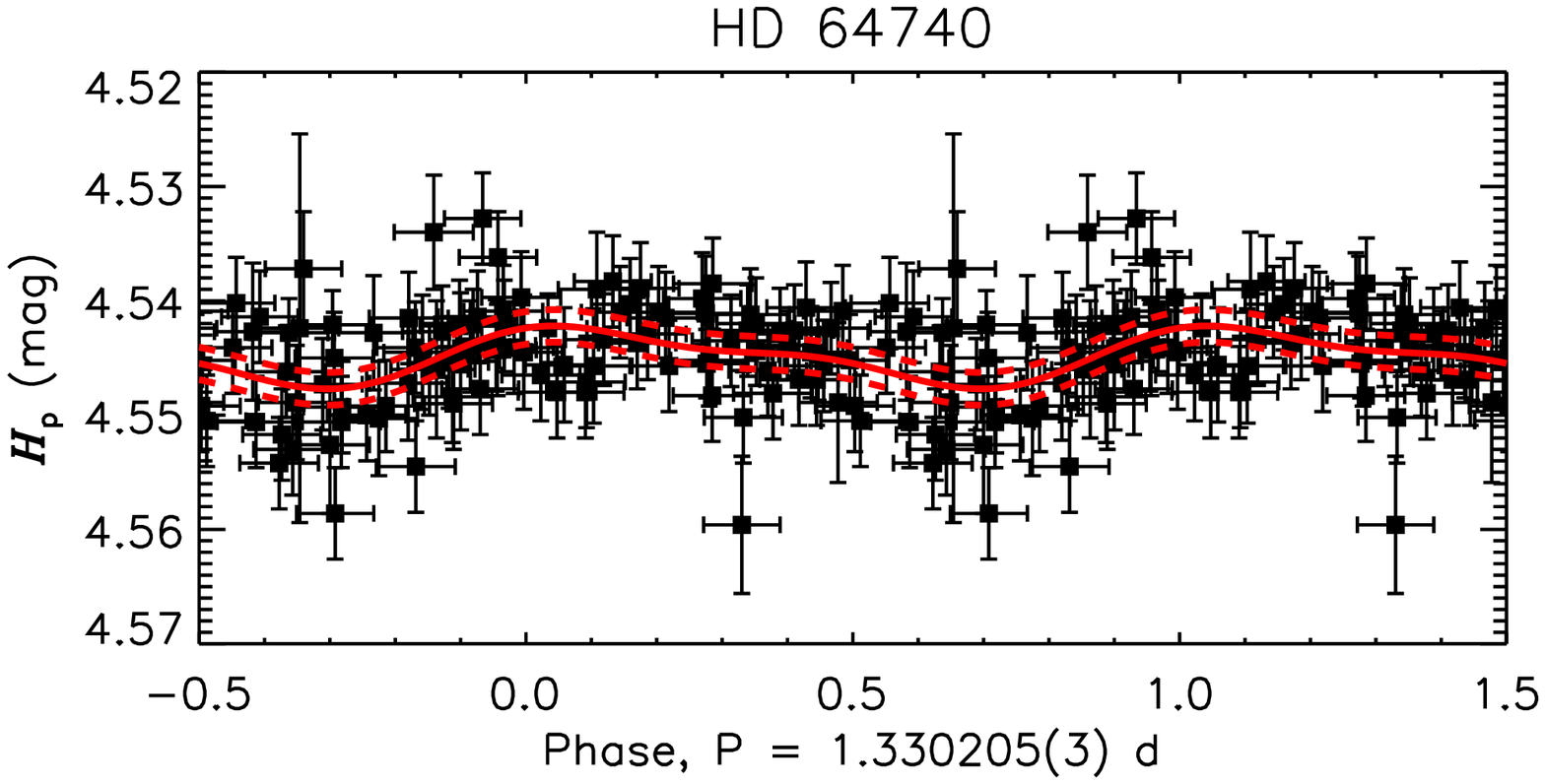}
   \includegraphics[width=8.5cm]{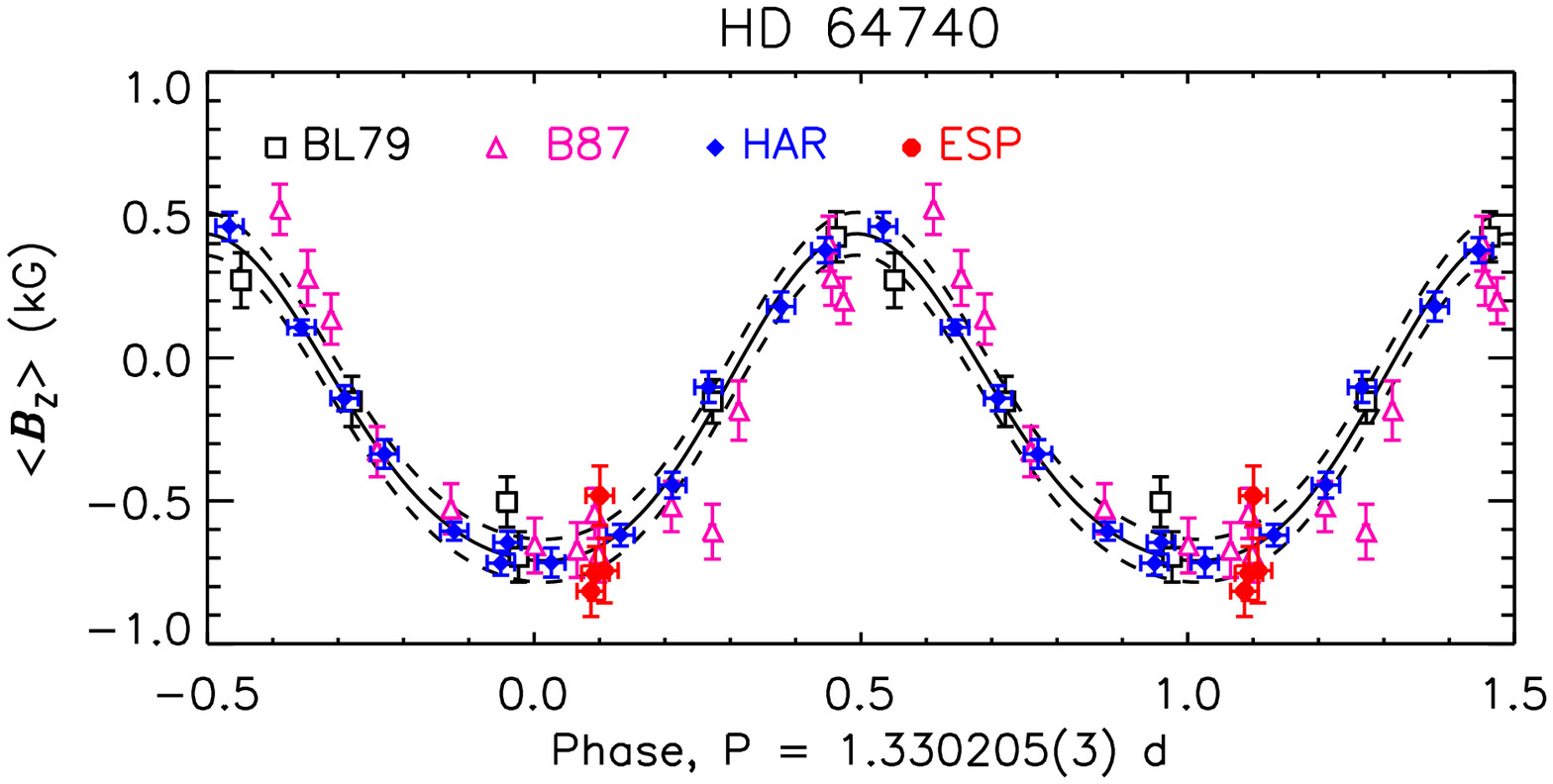} 
      \caption[Periodogram and \bz~for HD 64740.]{As Fig.\ \ref{HD37017_prot} for HD 64740. The sinusoidal fit is of $2^{nd}$ order.}
         \label{HD64740_prot}
   \end{figure}

\noindent {\bf HD 64740}: combining photometric and magnetic data, \cite{1987ApJ...323..325B} found 1.33026(6) d. Combining the magnetic measurements presented by \cite{1979ApJ...228..809B} and \cite{1987ApJ...323..325B} with the new ESPaDOnS and HARPSpol data, we find $P_{\rm rot}=$1.330205(3)~d, with a FAP of $4\times10^{-7}$, much lower than the minimum \nz~FAP of 0.28. The periodogram, \bz~measurements, and $H_{\rm p}$ light curve are shown in Fig.\ \ref{HD64740_prot}. Despite the relatively strong magnetic field, the light curve is only weakly variable, and could not be used to constrain $P_{\rm rot}$. A 2$^{nd}$-order sinusoidal fit has been used for \bz, which may indicate that the surface magnetic field is slightly more complex that a centred dipole.

\begin{figure}
    \includegraphics[width=8.5cm]{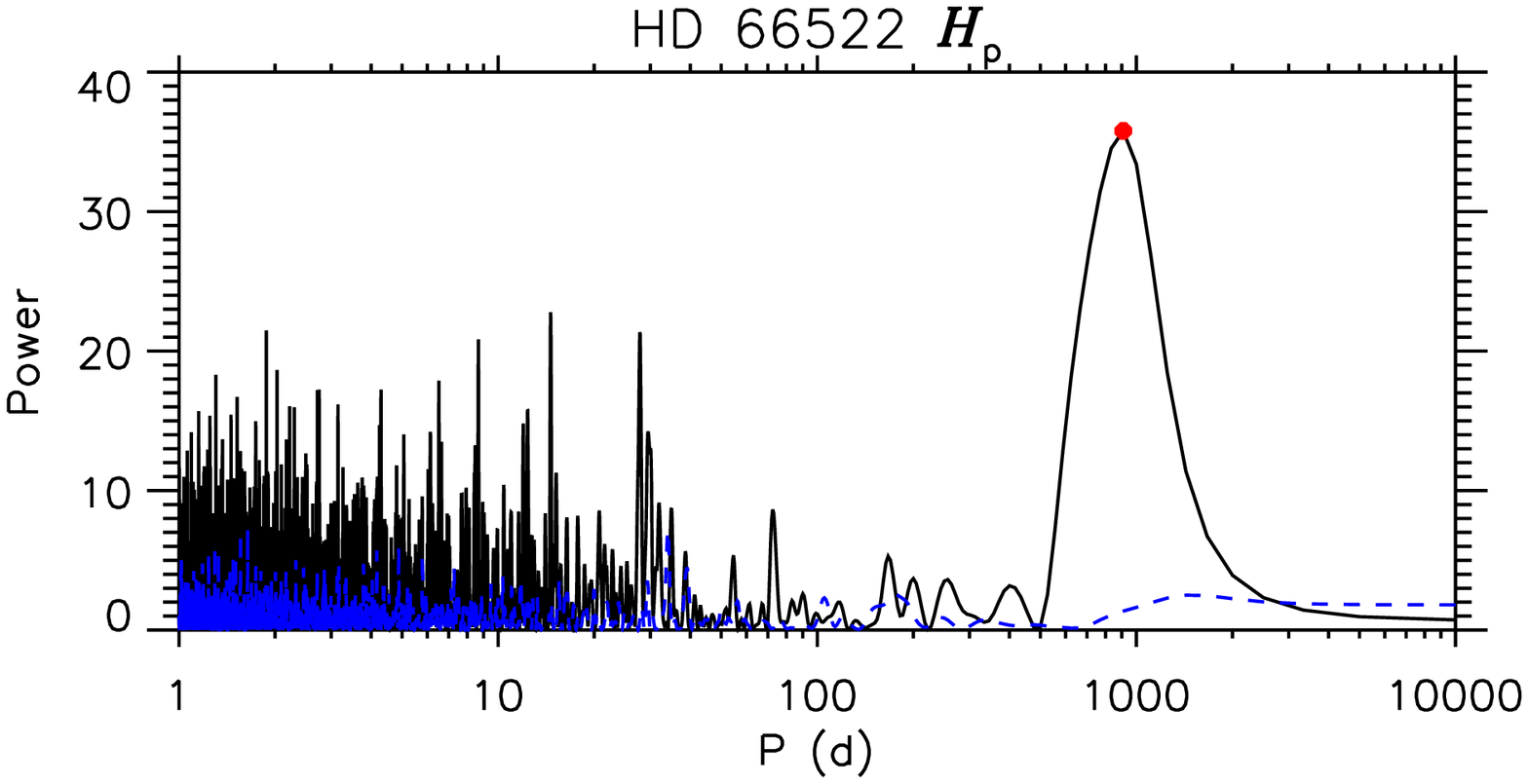}
    \includegraphics[width=8.5cm]{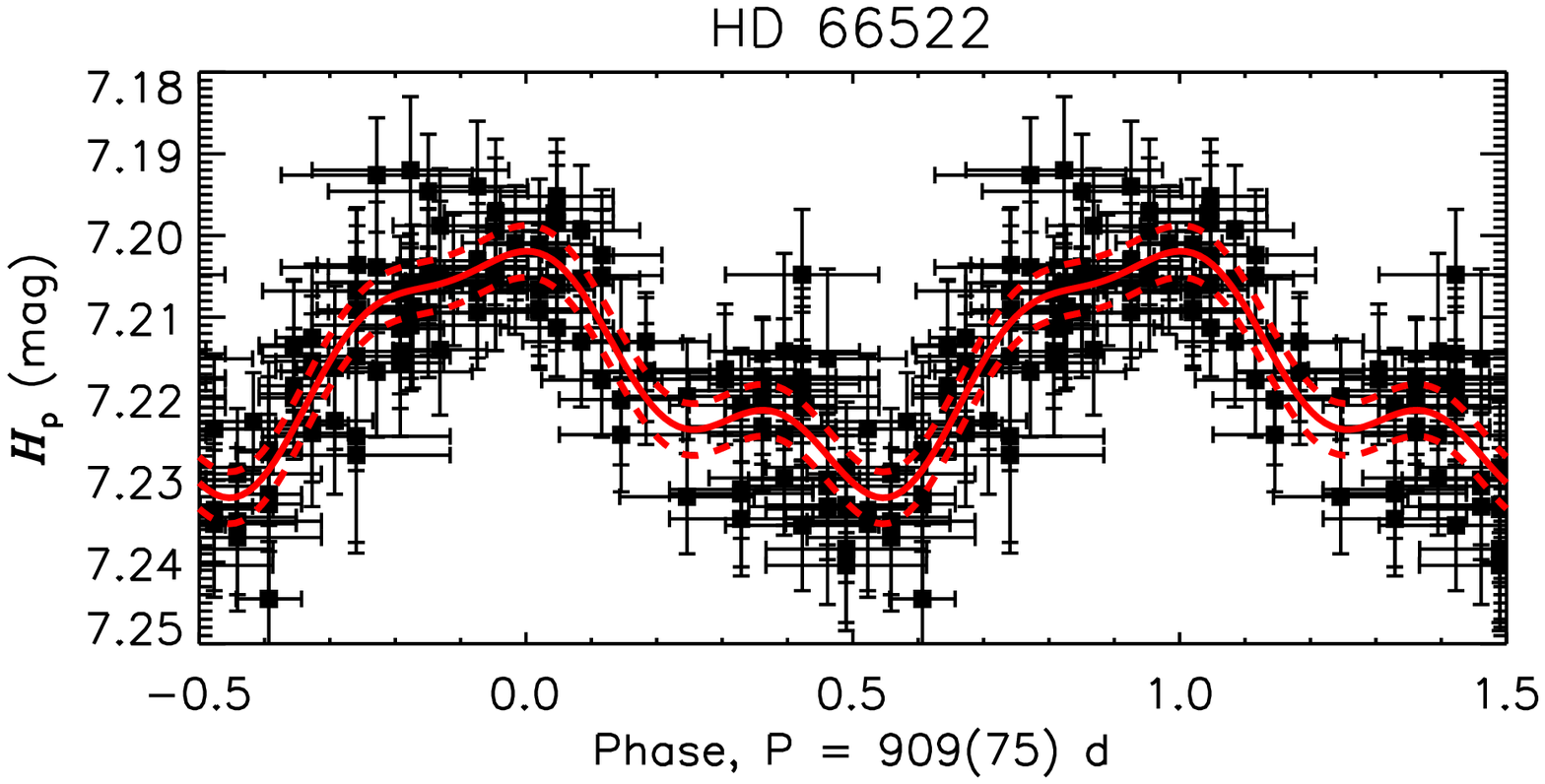}
    \includegraphics[width=8.5cm]{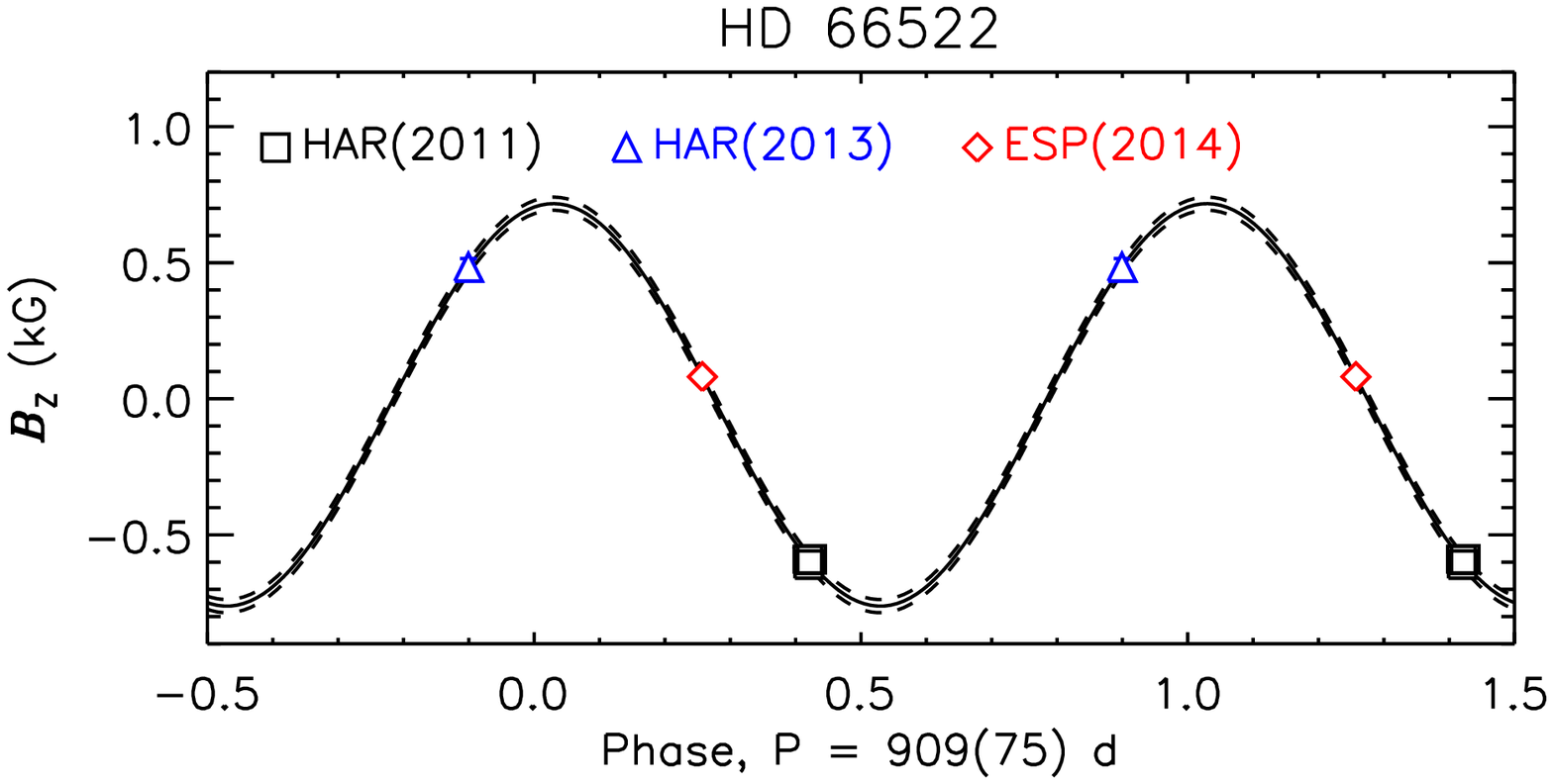}
      \caption[Periodogram, Hipparcos photometry, and \bz~for HD 66522.]{{\em Top}: Hipparcos photometry periodogram for HD 66522. {\em Middle}: $H_{\rm p}$ light curve. {\em Bottom}: \bz~curve. Numbers in brackets in the legend indicate the year of observation. In the bottom panel phase errors are larger than the x-axis and are not shown.}
 \label{HD66522_prot}
\end{figure}


\noindent {\bf HD 66522}: a period cannot be determined uniquely from the sparse magnetic data, but the Hipparcos photometry shows a clear modulation at 909(75)~d. The FAP is 2$\times10^{-14}$, while the minimum FAP in the null period spectrum is 0.99. This extremely long period is compatible with the star's negligible \vsini. The $H_{\rm p}$ periodogram is shown in the top panel of Fig.\ \ref{HD66522_prot}; $H_{\rm p}$ and \bz~are shown phased with the 909~d period in the middle and bottom panels. We chose the date of maximum light as JD0, which seems to correlate well to $|\langle B_z\rangle|_{\rm max}$ (although it is worth noting that the phase uncertainties are very large for the magnetic data, about 0.8 cycles). The star also shows spectroscopic variation; minimum line strength corresponds to maximum light, and vice versa, suggesting that both spectroscopic and photometric variations are due to the rotational modulation of chemical spots.  

   \begin{figure}
   \centering
\begin{tabular}{c}
   \includegraphics[width=8.5cm]{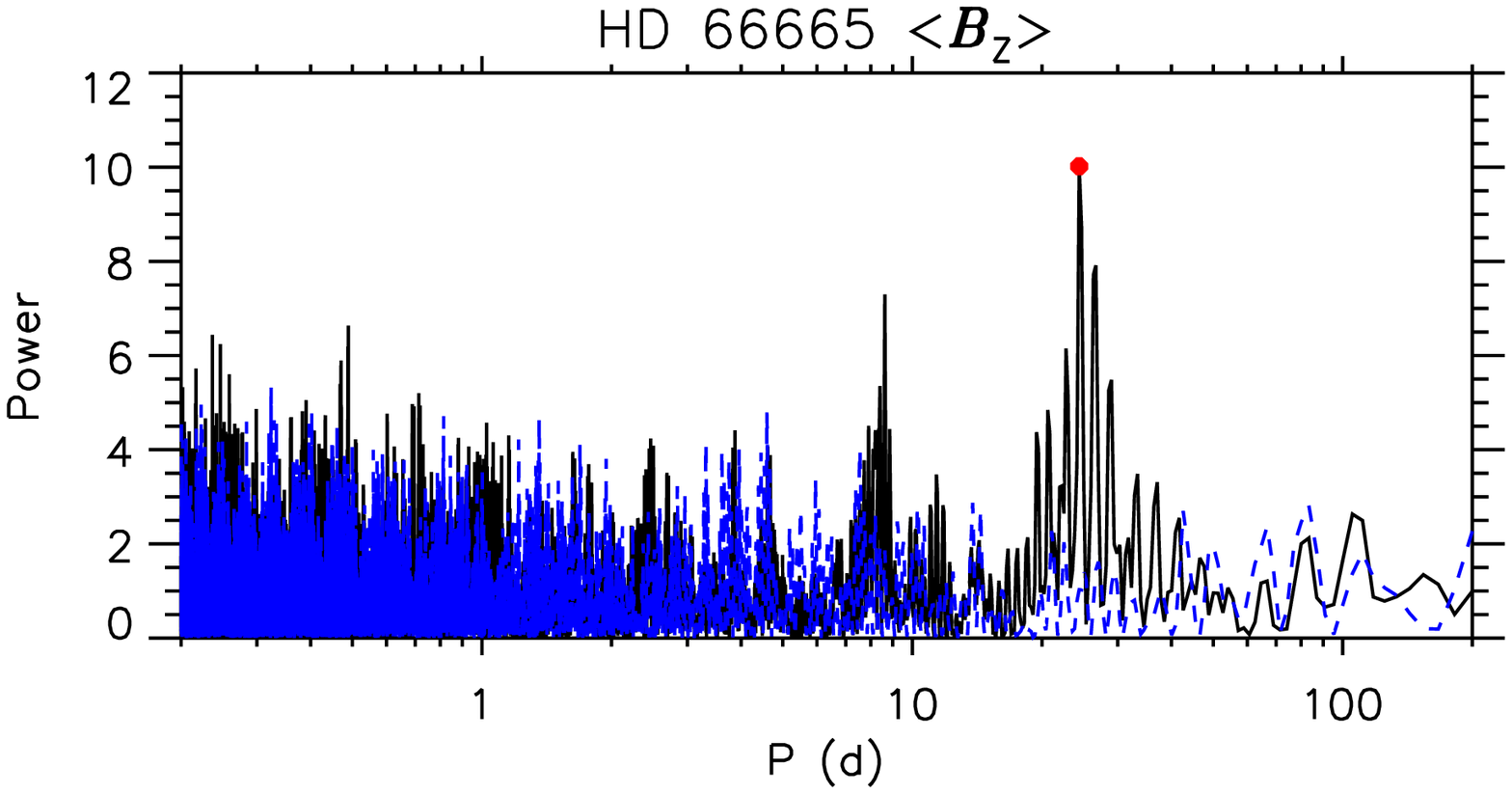} \\
   \includegraphics[width=8.5cm]{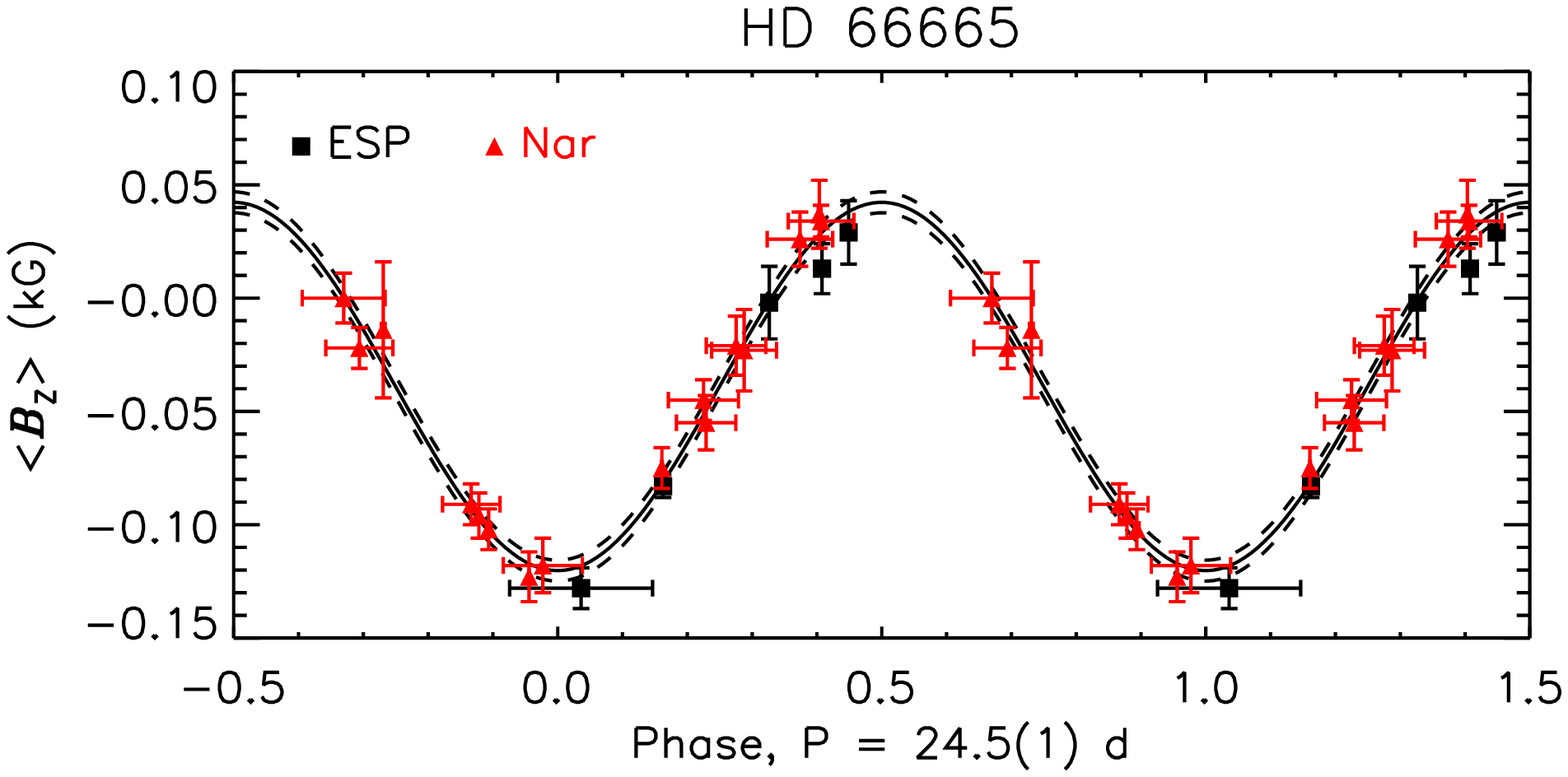} \\
\end{tabular}
      \caption[Periodogram and \bz~for HD 66665.]{As Fig.\ \ref{HD25558_prot}, for HD 66665.}
         \label{HD66665_prot}
   \end{figure}

\noindent {\bf HD 66665}: \cite{petit2013} reported a period of 21 d based upon a preliminary analysis of ESPaDOnS and Narval magnetic measurements. We find 24.5(1)~d using the same data, with a FAP of $2\times10^{-4}$. The 21 d period is definitely ruled out. There is very little power in the \nz~periodogram near 25 d, and the minimum FAP of the \nz~period spectrum is 0.08. The periodogram is shown in the top panel of Fig.\ \ref{HD66665_prot}; \bz~is shown phased with this period in the bottom panel.

\begin{figure}
    \includegraphics[width=8.5cm]{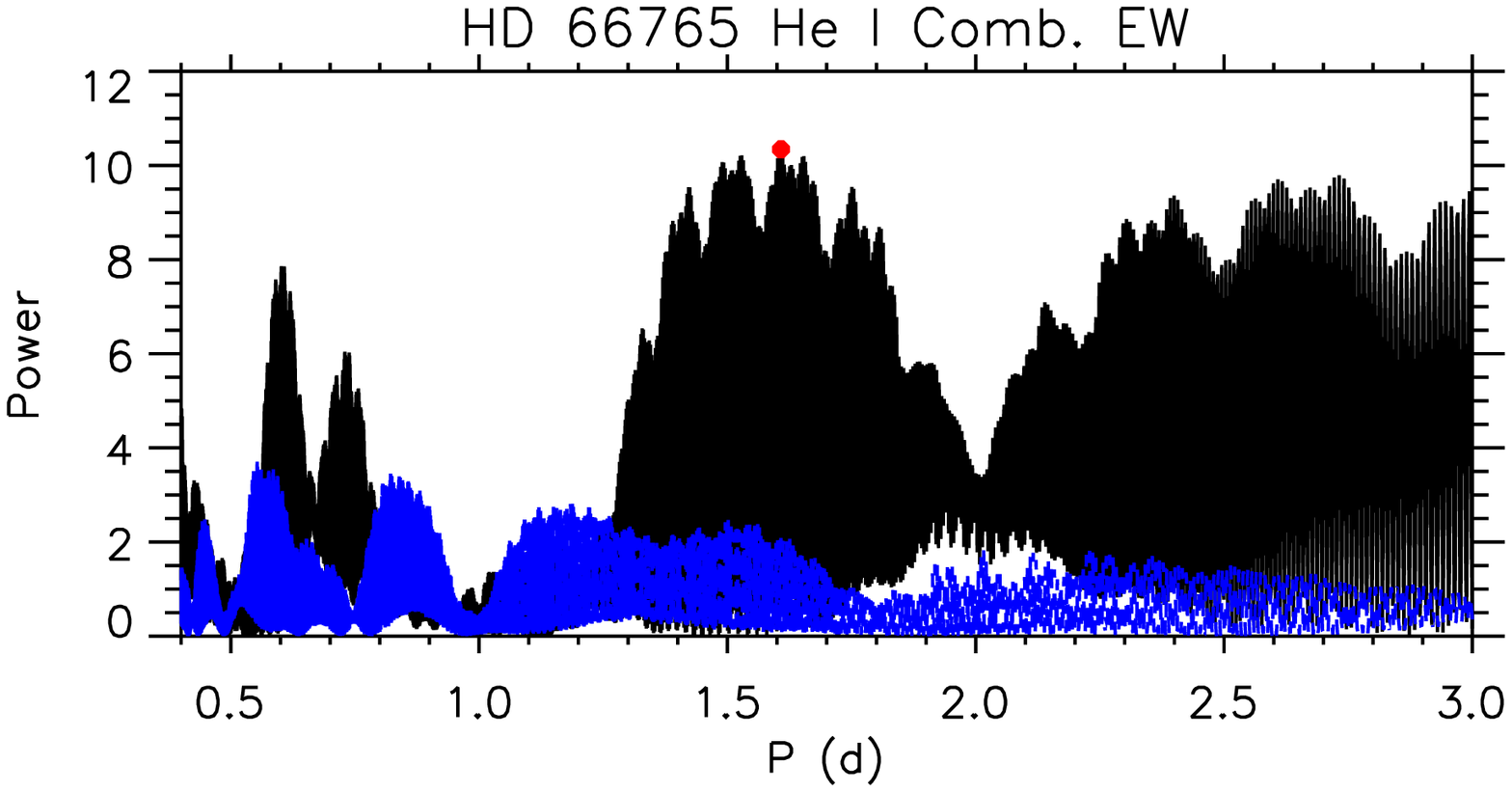}
    \includegraphics[width=8.5cm]{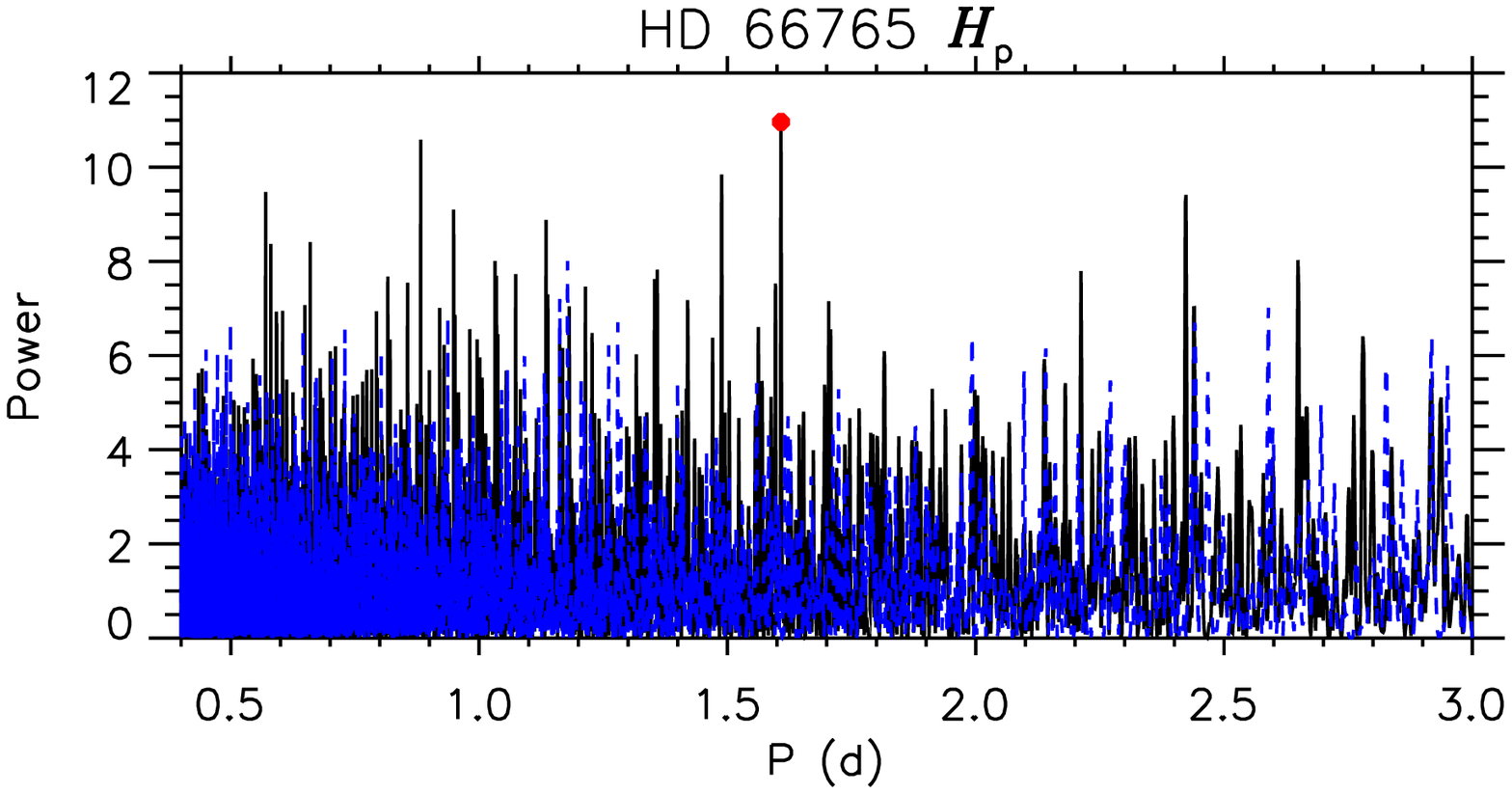}
    \includegraphics[width=8.5cm]{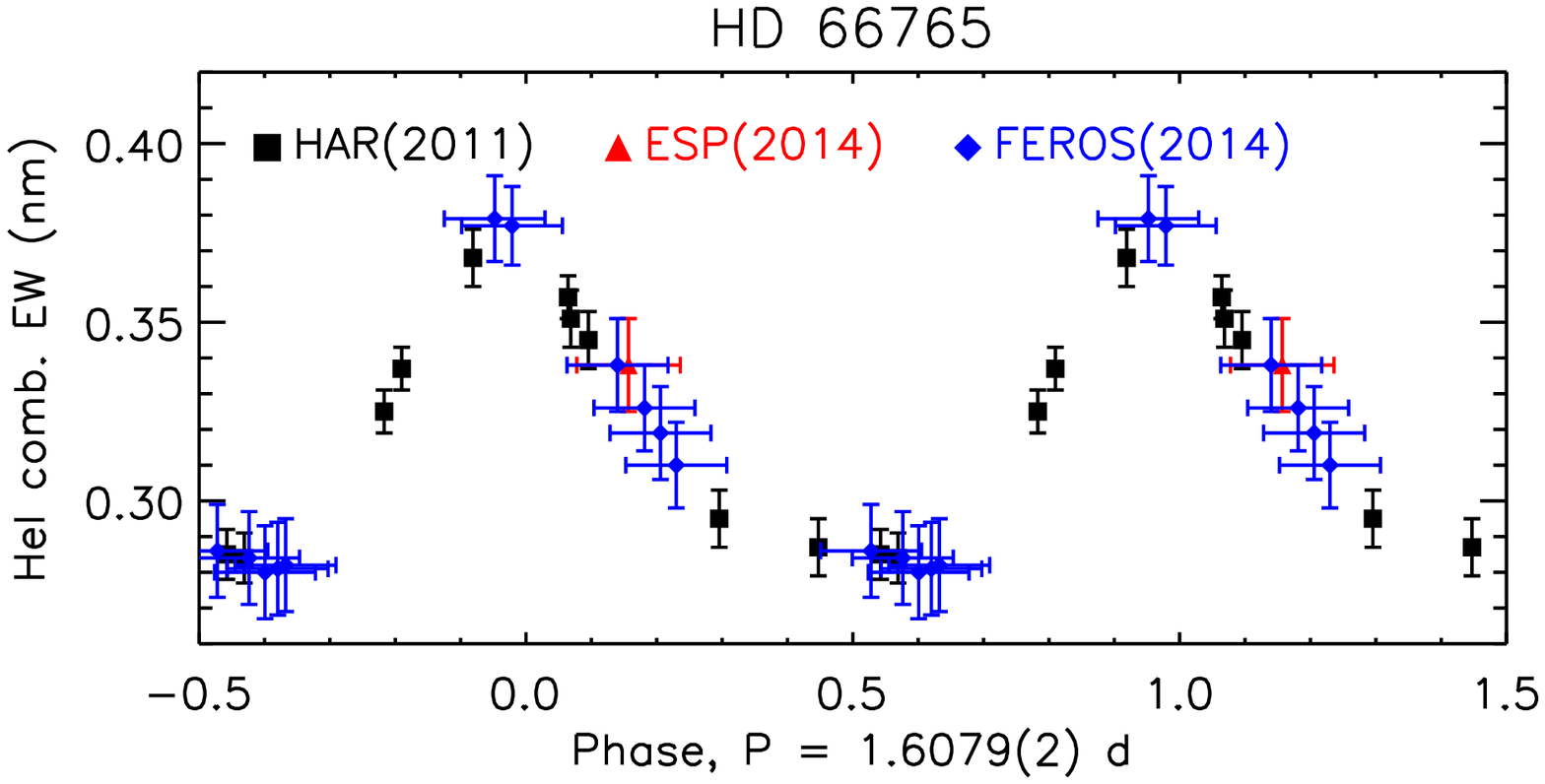}
    \includegraphics[width=8.5cm]{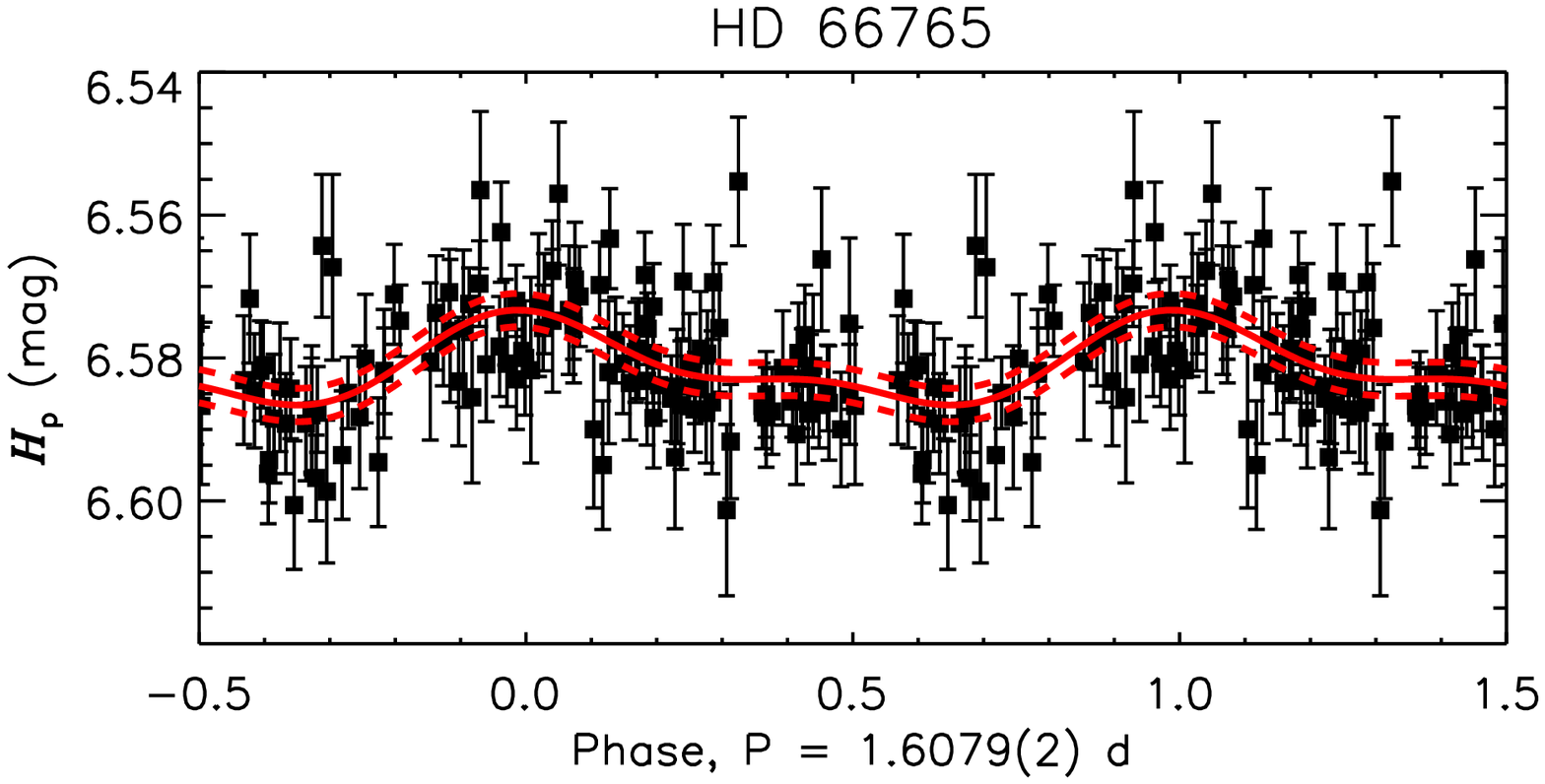}
    \includegraphics[width=8.5cm]{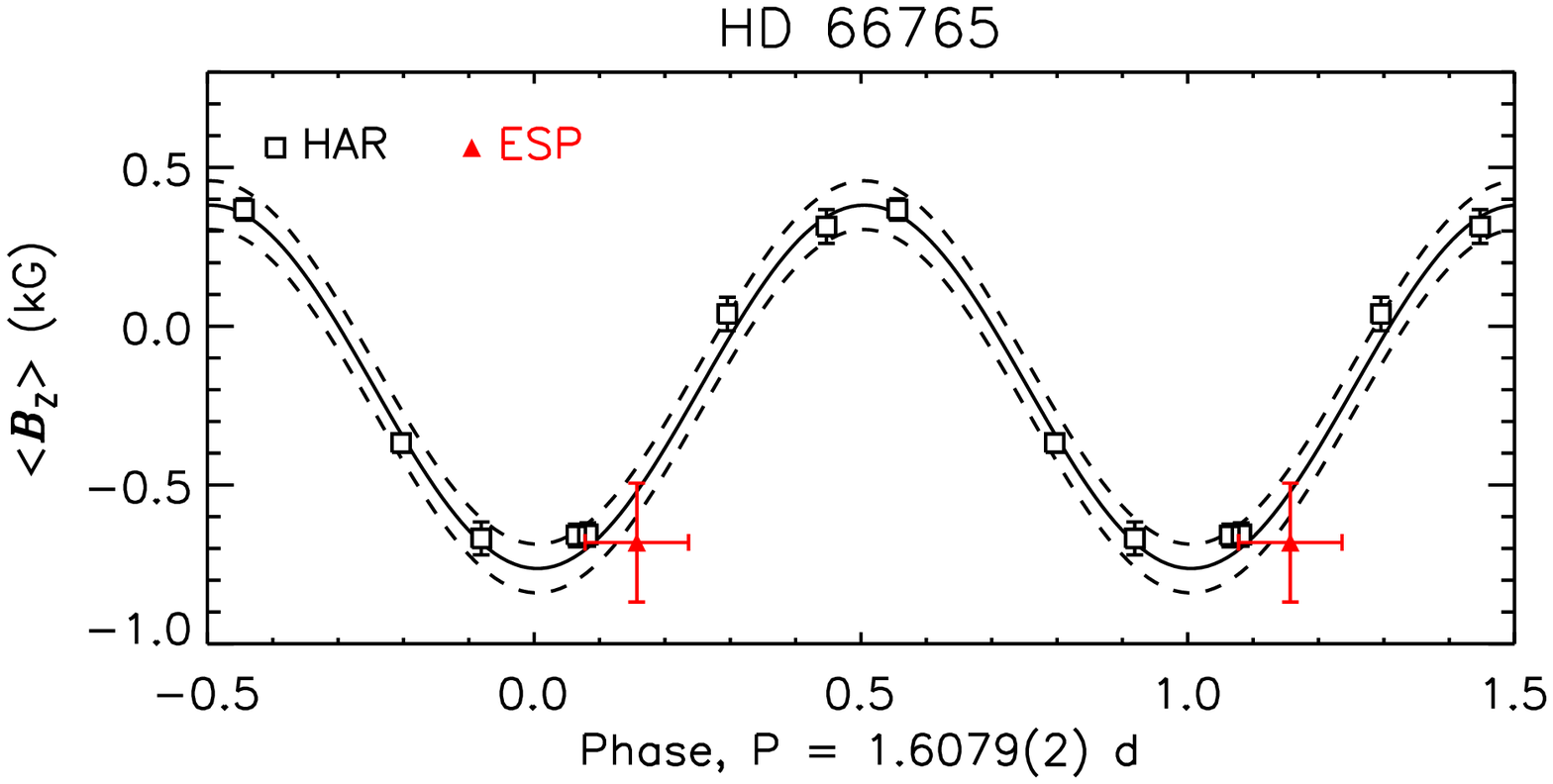}
      \caption[Periodogram, He~{\sc i} EWs, and \bz~for HD 66765.]{{\em Top-Bottom}: Periodogram for combined He~{\sc i} EWs; periodogram for Hipparcos photometry; combined He~{\sc i} EWs; Hipparcos photometry; \bz~measurements. $P_{\rm rot}$ was determined by combining $H_{\rm p}$ and He~{\sc i} EW periodograms.}
 \label{HD66765_prot}
\end{figure}

\noindent {\bf HD 66765}: \cite{alecian2014} found $P_{\rm rot}=1.62(15)$ d based on 4 days of observations with HARPSpol. We acquired a single new ESPaDOnS observation approximately 2 years later. The quality of the measurement is not as high as that of the HARPSpol data, and does not enable improvement of the period. However, we also acquired 11 FEROS spectra. As HD 66765 is a He-variable star, we measured the EWs of the He~{\sc i} 447.1 nm, 587.6 nm, and 667.8 nm lines. Within the uncertainty in $P_{\rm rot}$~determined by \cite{alecian2011}, the highest peak in the periodogram for the combined EWs is at 1.6079(5)~d, with a FAP of 2$\times10^{-4}$ and a minimum FAP in the null spectrum of 0.58. While the star is only weakly photometrically variable, the highest peak in the Hipparcos periodogram is at 1.6079(2)~d, somewhat more precise than the spectroscopic determination, although the FAP is only 0.54 (Fig.\ \ref{HD66765_prot}, top). The combined He~{\sc i} EWs, $H_{\rm p}$, and the \bz~measurements are shown phased with this period in the bottom three panels of Fig.\ \ref{HD66765_prot}. Note that phase uncertainties are not shown for the light curve as, with the precision of this period, they are larger than the x-axis.

   \begin{figure}
   \centering
\begin{tabular}{c}
      \includegraphics[width=8.5cm]{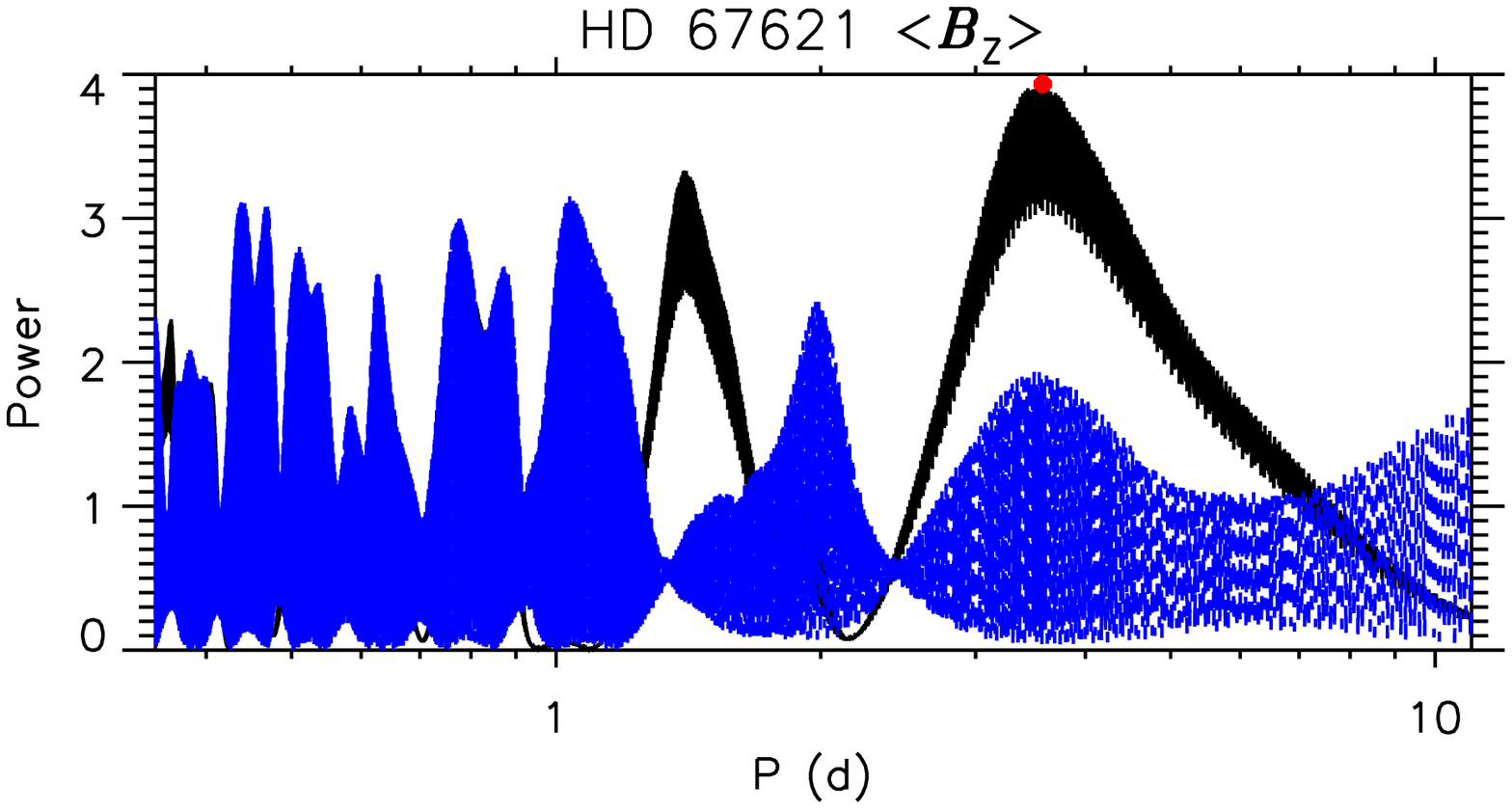} \\
      \includegraphics[width=8.5cm]{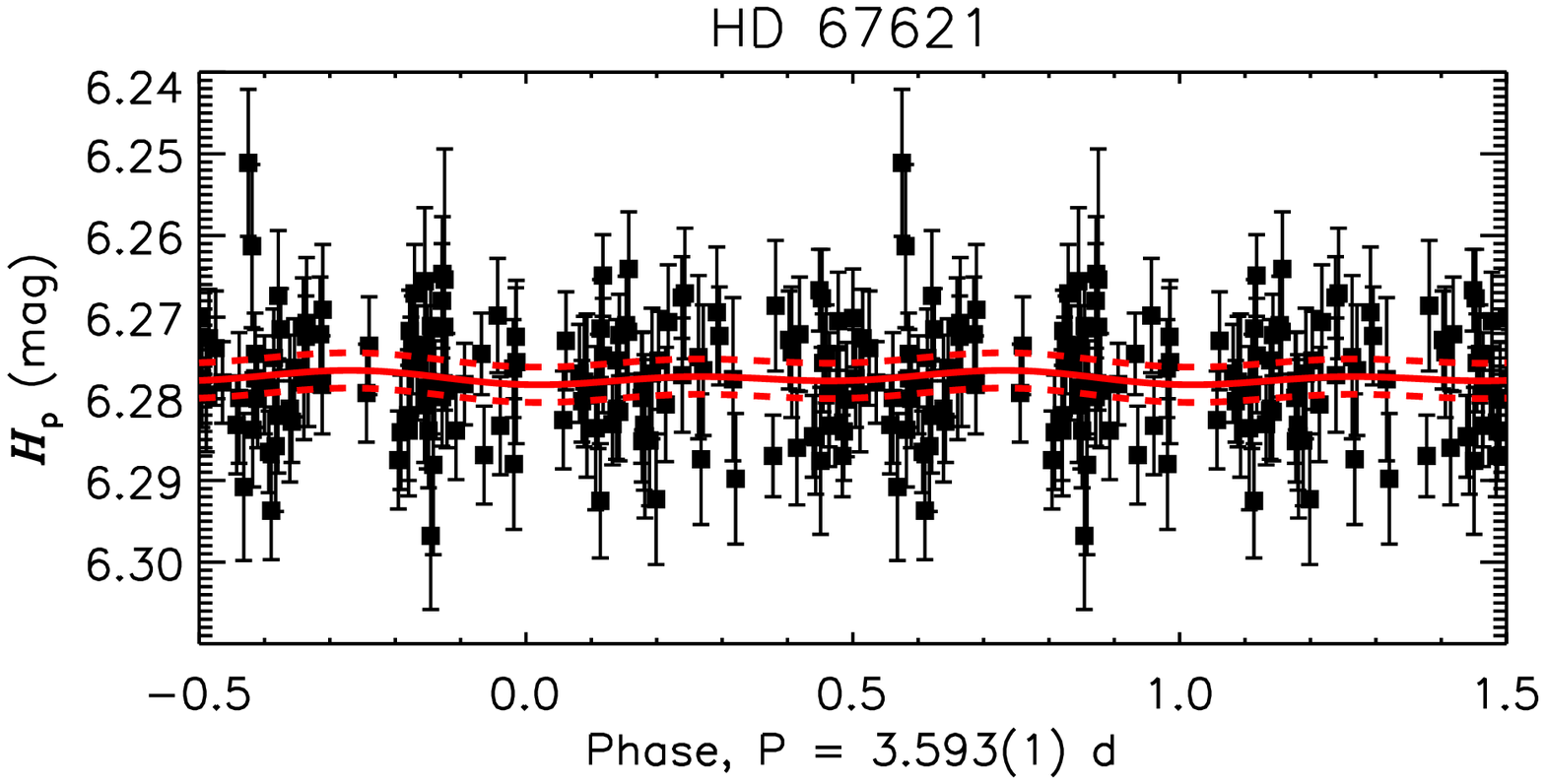} \\
      \includegraphics[width=8.5cm]{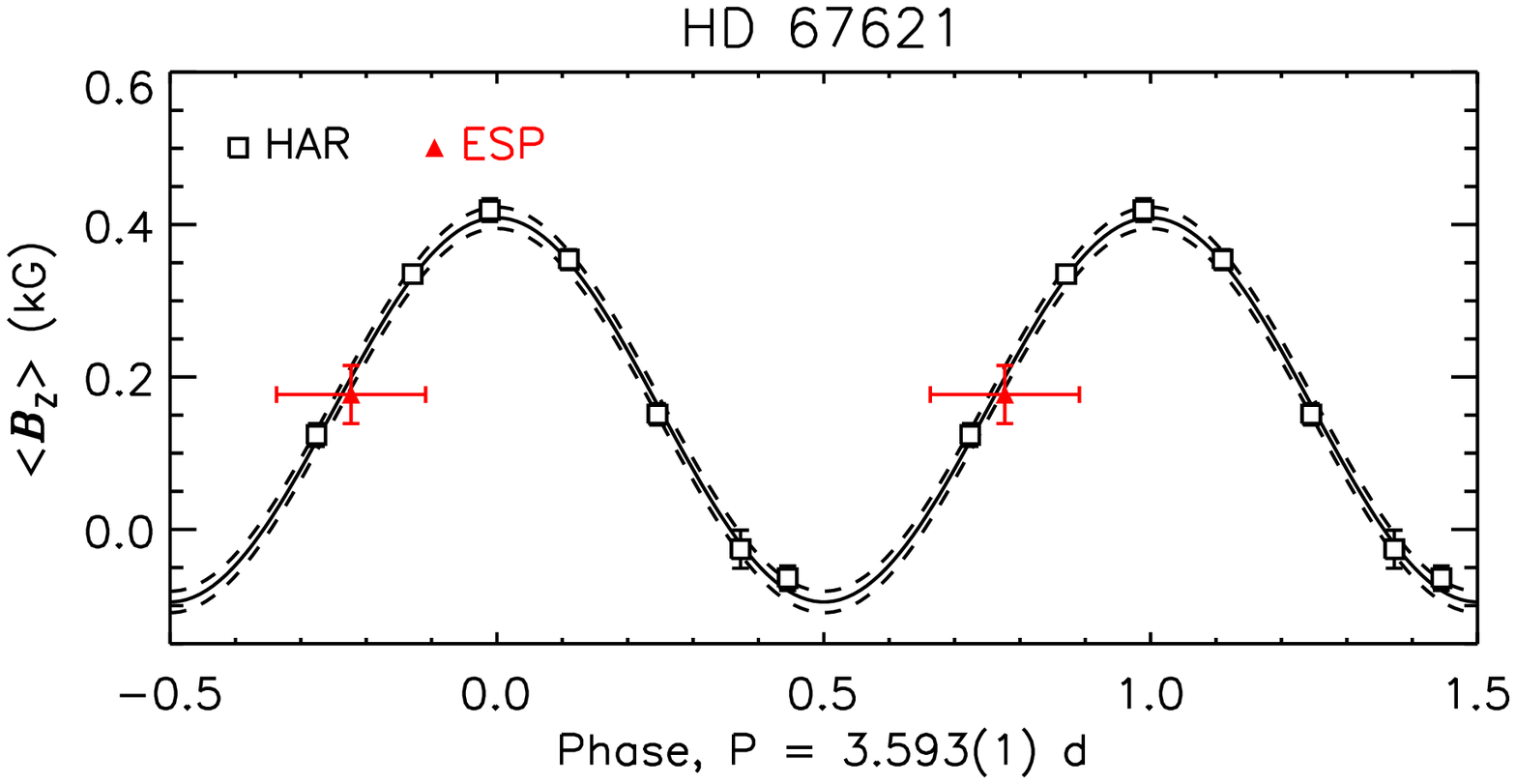} \\
\end{tabular}
      \caption[Periodogram and \bz~for HD 67621.]{As Fig.\ \ref{HD37017_prot} for HD 67621.}
         \label{HD67621_prot}
   \end{figure}

\noindent {\bf HD 67621}: \cite{alecian2014} determined $P_{\rm rot}=3.6(2)$ d. As with HD 66765, we observed the star once more with ESPaDOnS, and refine the period to 3.593(1) d, although we note that the periodogram has numerous nearby peaks within the original range (Fig.\ \ref{HD67621_prot}, top). While the minimum FAPs of the \bz~and \nz~period spectra are similar (0.27 and 0.31, respectively), there is no power near the 3.6 d period in the \nz~periodogram. \bz~is shown phased with this period in Fig.\ \ref{HD67621_prot} (bottom). We attempted to constrain the period using Hipparcos photometry, but the almost complete absence of variability means that the light curve is dominated by noise. This is consistent with the star's He lines, which also show very low levels of variability.

   \begin{figure}
   \centering
   \includegraphics[width=8.5cm]{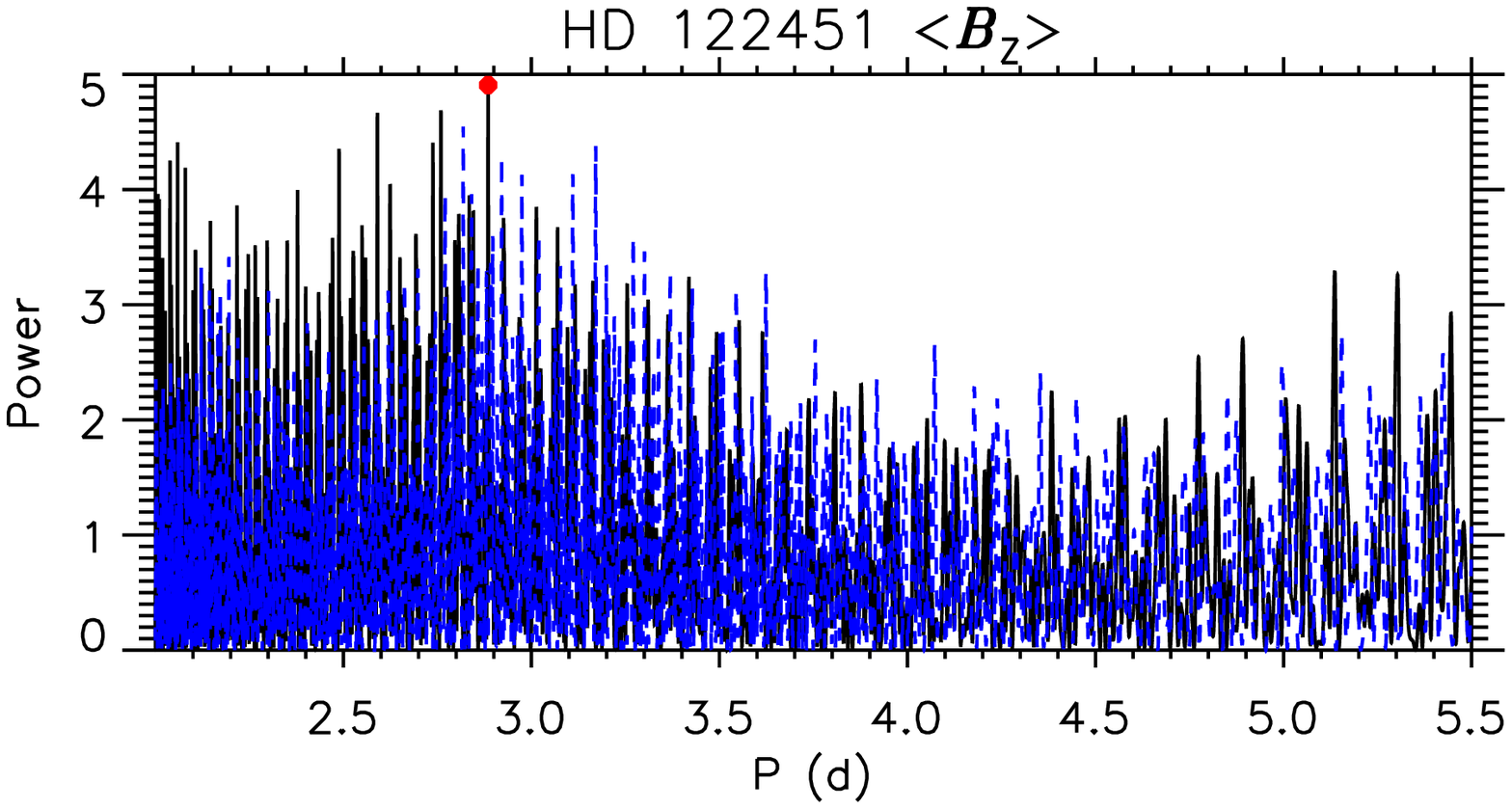} 
   \includegraphics[width=8.5cm]{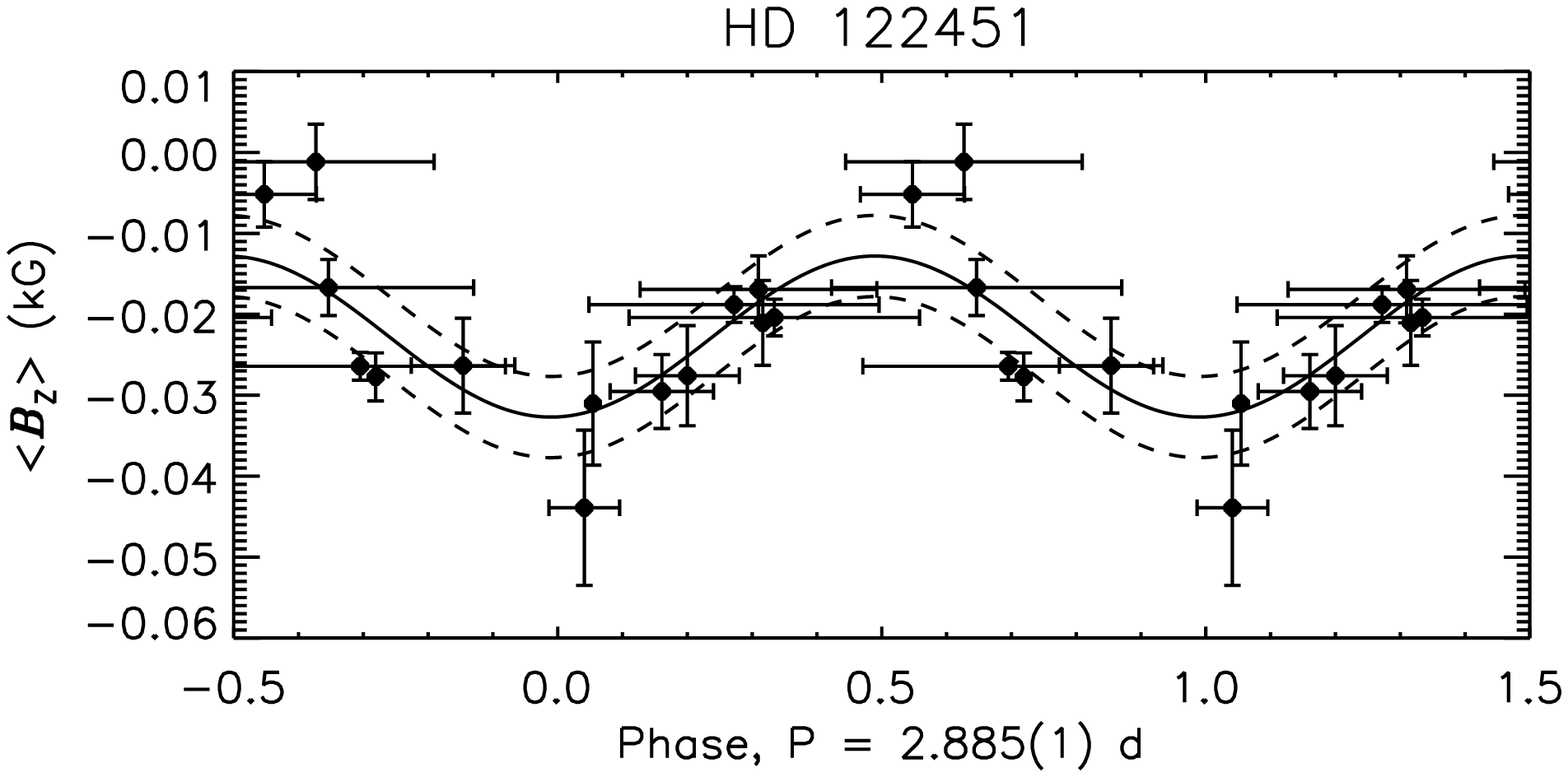}
      \caption[Periodogram and \bz~for HD 122451.]{As Fig.\ \ref{HD25558_prot}, for HD 122451.}
         \label{HD122451_prot}
   \end{figure}

\noindent {\bf HD 122451}: In order to maximize the S/N we used \bz~measurements from LSD profiles in 1~d bins. The lower bound of the period window as defined by the breakup velocity, $\sim 0.5$~d, is at a higher frequency than the Nyquist frequency of $\sim 0.5~{\rm d}^{-1}$, therefore the Nyquist frequency was taken as the lower bound. The highest peak in the \bz~periodogram is at 2.885(1) d (Fig.\ \ref{HD122451_prot}, top). \bz~is shown phased with this period in Fig.\ \ref{HD122451_prot} (bottom). While this period achieves a S/N of 5.4 and is thus formally significant, it must be noted that the FAP is 0.06, which is close to the minimum FAP in the \nz~spectrum, 0.09, so this period should be regarded as tentative. However, it is close to a period identified by \cite{2016A&A...588A..55P} in the BRITE light curve as a possible rotational period, 2.827(2)~d. 

   \begin{figure}
   \centering
\begin{tabular}{c}
   \includegraphics[width=8.5cm]{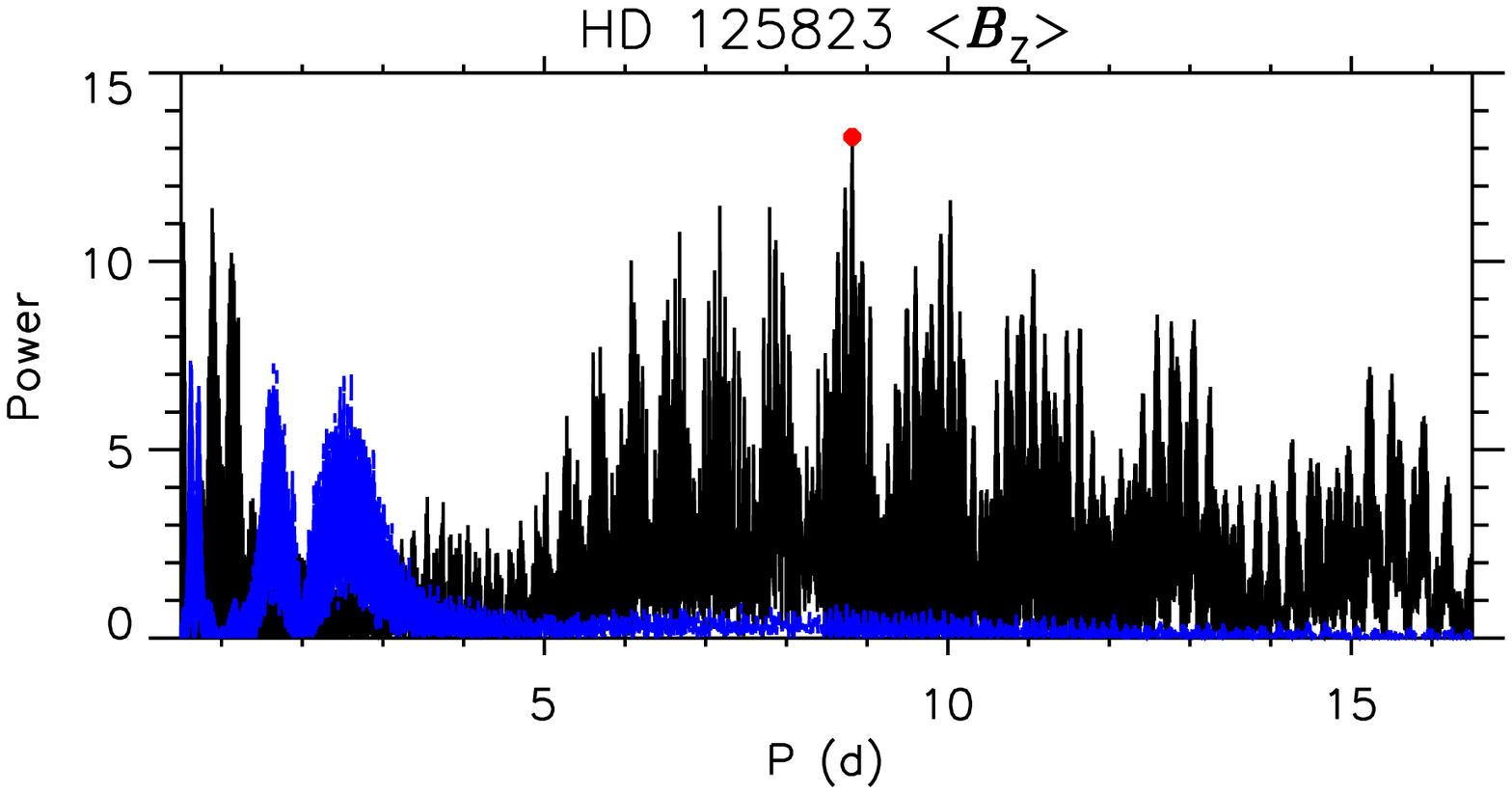} \\
   \includegraphics[width=8.5cm]{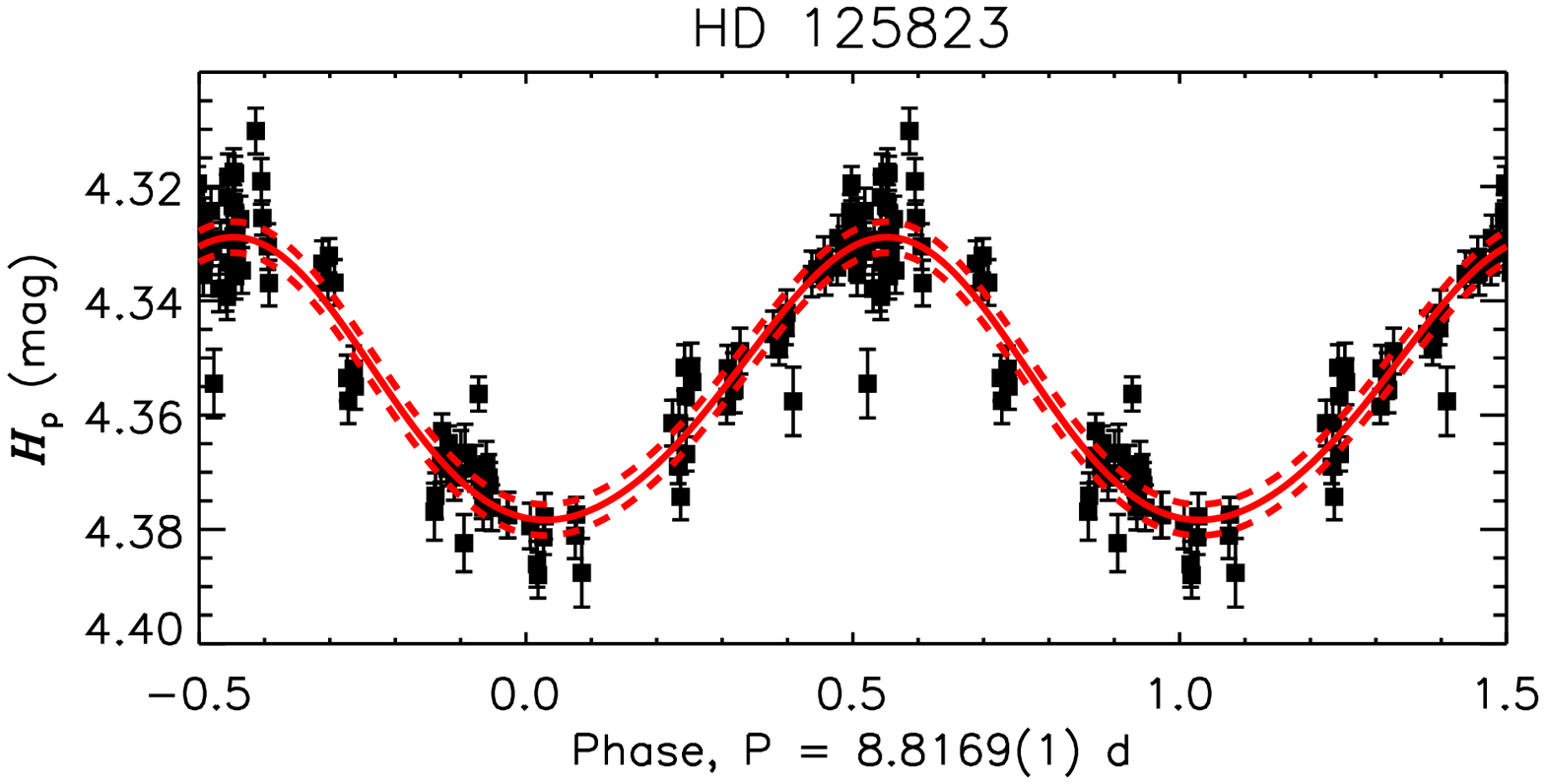} \\
   \includegraphics[width=8.5cm]{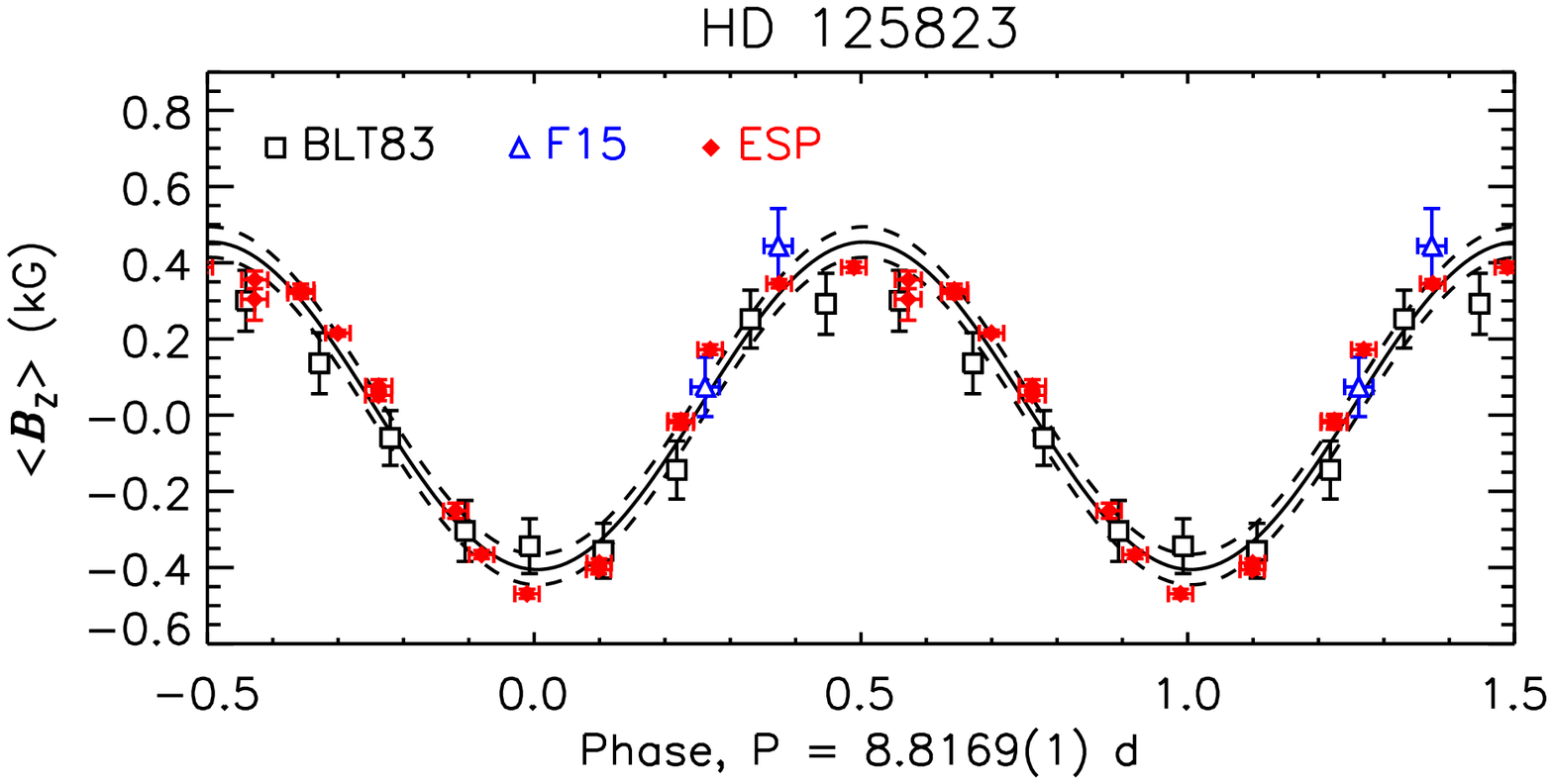}  \\
\end{tabular}
      \caption[Periodogram and \bz~for HD 125823.]{As Fig.\ \ref{HD37017_prot}, for HD 125823.}
         \label{HD125823_prot}
   \end{figure}

\noindent {\bf HD 125823}: The most recent ephemeris for HD 125823 was given by \cite{1996AA...311..230C}, who found a period of 8.8177(1)~d by combining photometric data with an approximately 20 year baseline. This period phases the ESPaDOnS data well if examined in isolation, but there is a systematic phase offset when compared to older data. The highest amplitude in the periodogram for the ESPaDOnS data is at 8.817(4)~d. This period phases the ESPaDOnS \bz~measurements well with the historical data; a periodogram constructed from the combined datasets yields a more precise period, $P_{\rm rot}$=8.8169(1)~d. The FAP of this peak is 2.2$\times10^{-5}$, much lower than the minimum \nz~FAP of 0.33. This period also provides an acceptable phasing of the FORS2 measurements reported by \cite{2015A&A...582A..45F}, who noted they were unable to phase their measurements with the historical data using the \cite{1996AA...311..230C} ephemeris. \bz~and $H_{\rm p}$ are shown phased with this period in Fig.\ \ref{HD125823_prot}. The ephemeris results in the photometric and magnetic extrema coinciding. This is consistent with the strong He-line variability, which changes from He-strong at the positive magnetic pole to He-weak at the negative magnetic pole.

   \begin{figure}
   \centering
   \includegraphics[width=8.5cm]{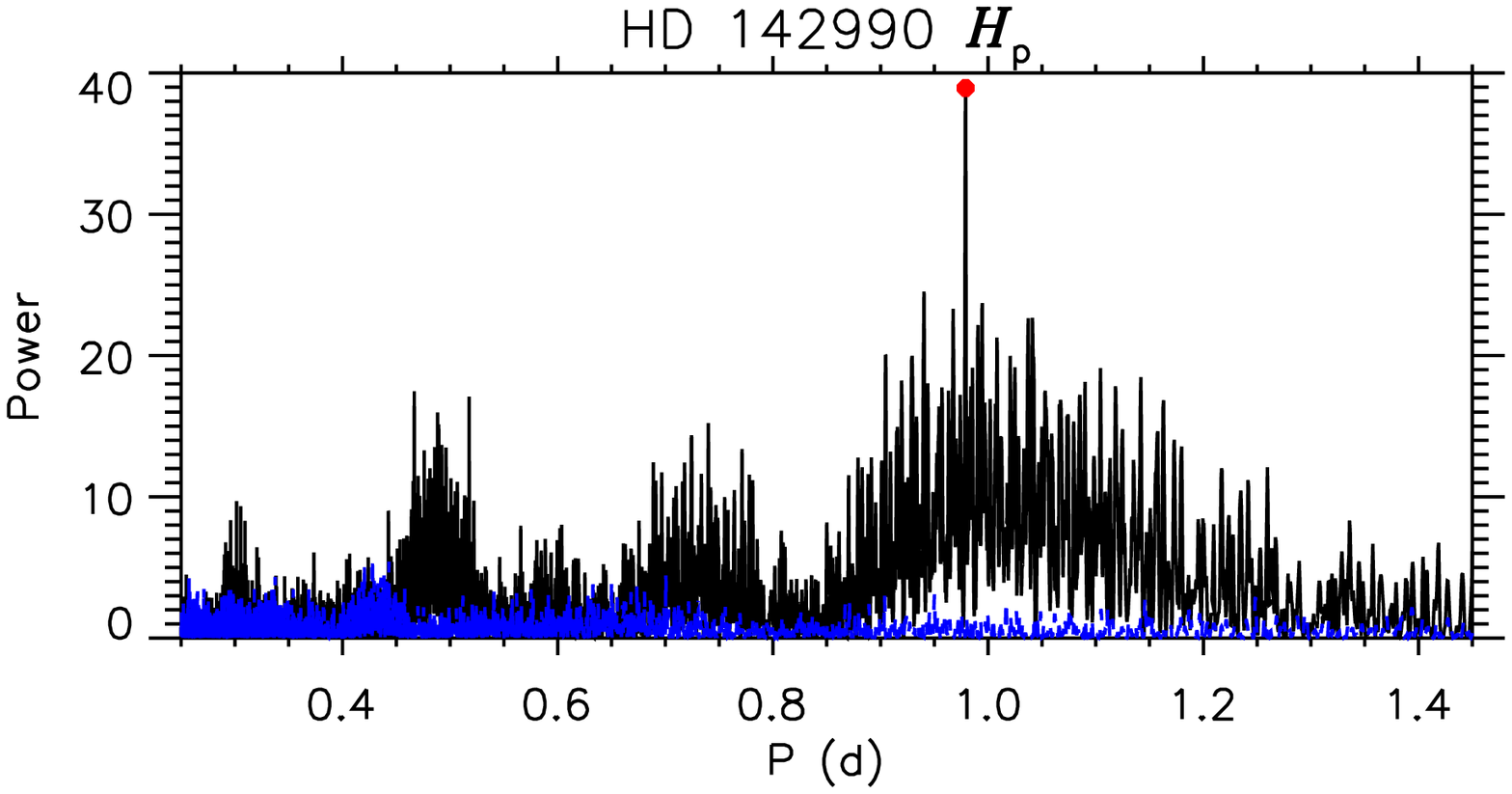} 
   \includegraphics[width=8.5cm]{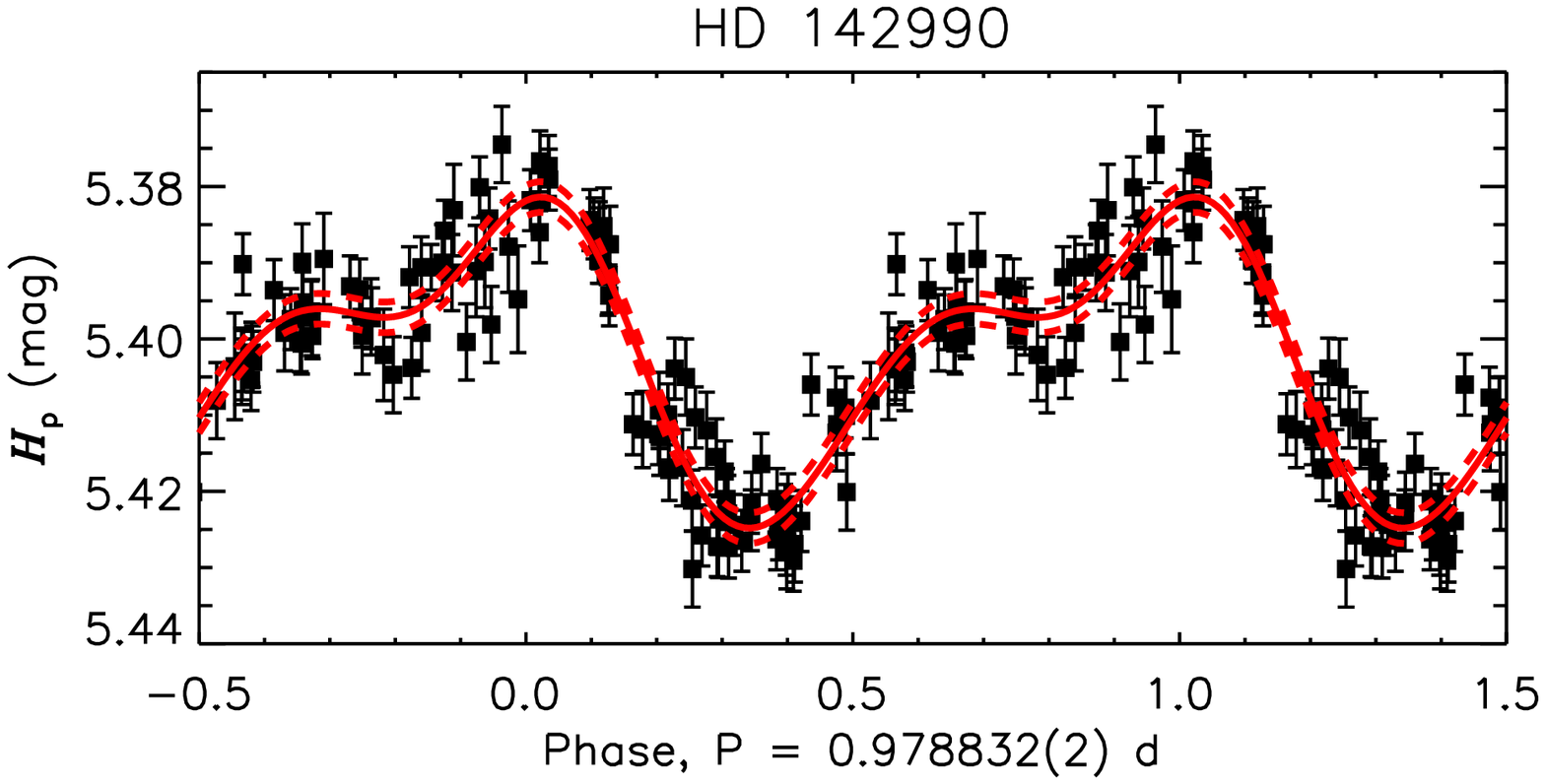} 
   \includegraphics[width=8.5cm]{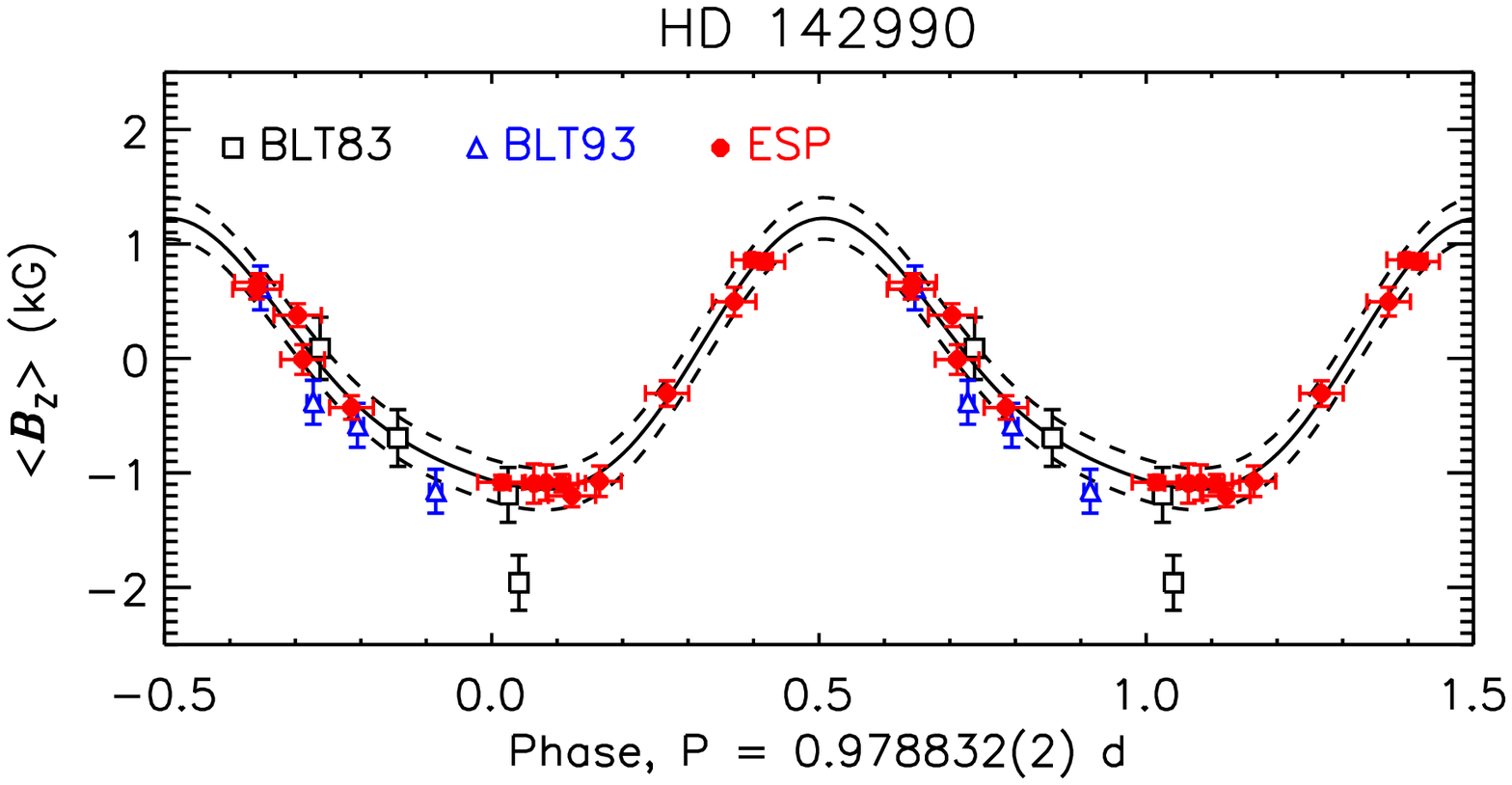} 
      \caption[Periodograms, Hipparocos photometry, and \bz~for HD 142990.]{As Fig.\ \ref{HD35298_prot}, for HD 142990. The sinusoidal fit is of $2^{nd}$ order.}
         \label{HD142990_prot}
   \end{figure}

\noindent {\bf HD 142990}: \cite{2005AA...430.1143B} found a rotation period of 0.97907 d. The magnetic data, even including the new measurements presented here, are rather sparse and cannot be used to identify a unique period within the period window. However, a unique period of 0.97903(5)~d is clearly identified in the periodogram obtained using archival Hipparcos photometry, suggesting the \citeauthor{2005AA...430.1143B} period is indeed correct (Fig.\ \ref{HD142990_prot}, top). The FAP of this peak is $2.3\times10^{-17}$, much lower than the minimum FAP in the null period spectrum of 0.5. This period provides an adequate phasing of the ESPaDOnS \bz~measurements, but is insufficiently precise to phase the new data with the measurements reported by \cite{1983ApJS...53..151B} and \cite{1993AA...269..355B}. A periodogram constructed for all old and new \bz~measurements, limited to a narrow window around 0.979 d, finds maximum amplitude at 0.978832(2)~d, with a FAP of 0.0003 in \bz~and 0.52 in \nz. $H_{\rm p}$~and \bz~are shown phased with this period in the middle and bottom panels of Fig.\ \ref{HD142990_prot}. Maximum light coincides with the negative magnetic pole. A second-order sinusoid has been used to fit \bz, as this provides a better fit than a single-order sinusoid.

   \begin{figure}
   \centering
   \includegraphics[width=8.5cm]{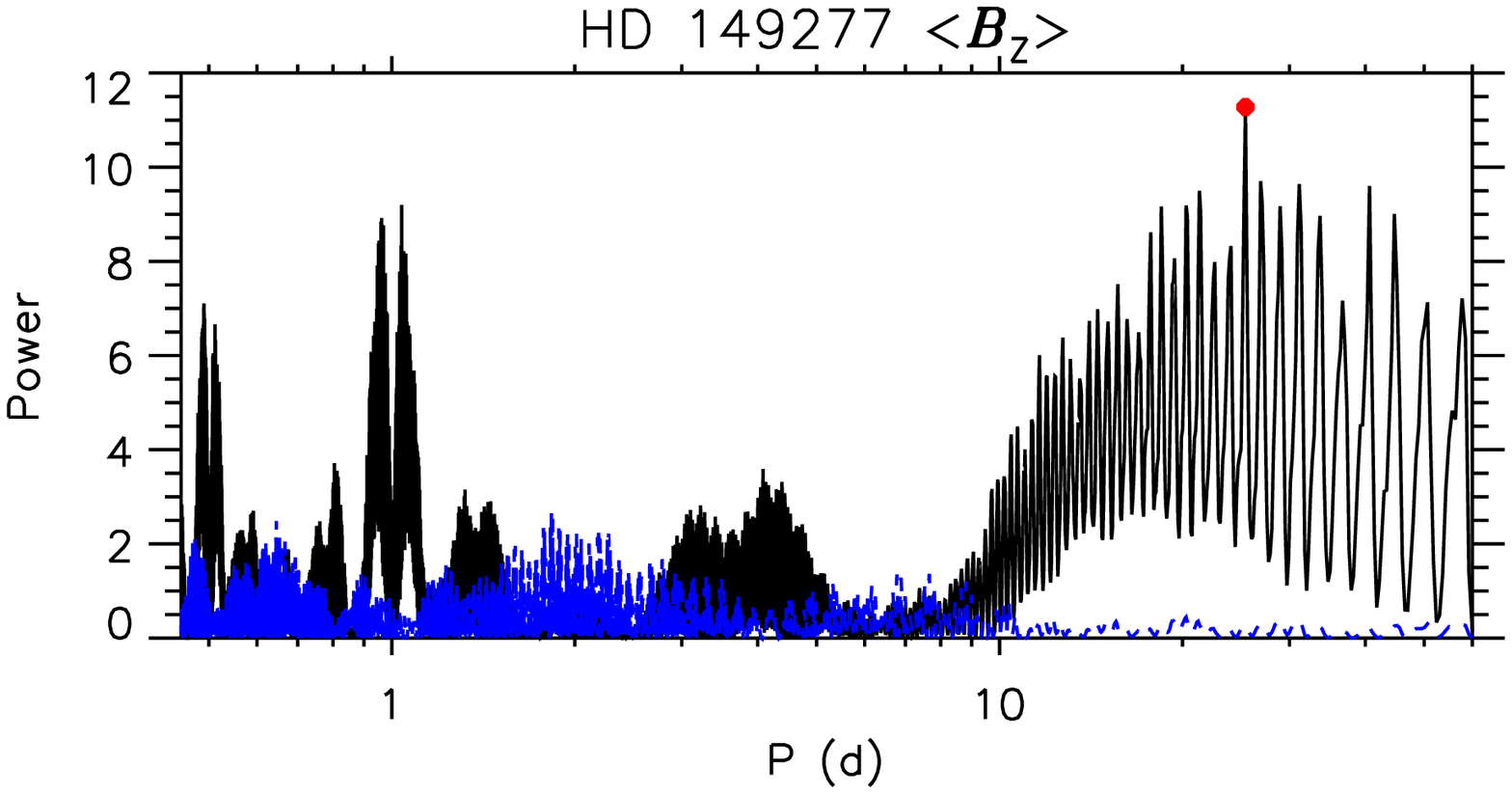} 
   \includegraphics[width=8.5cm]{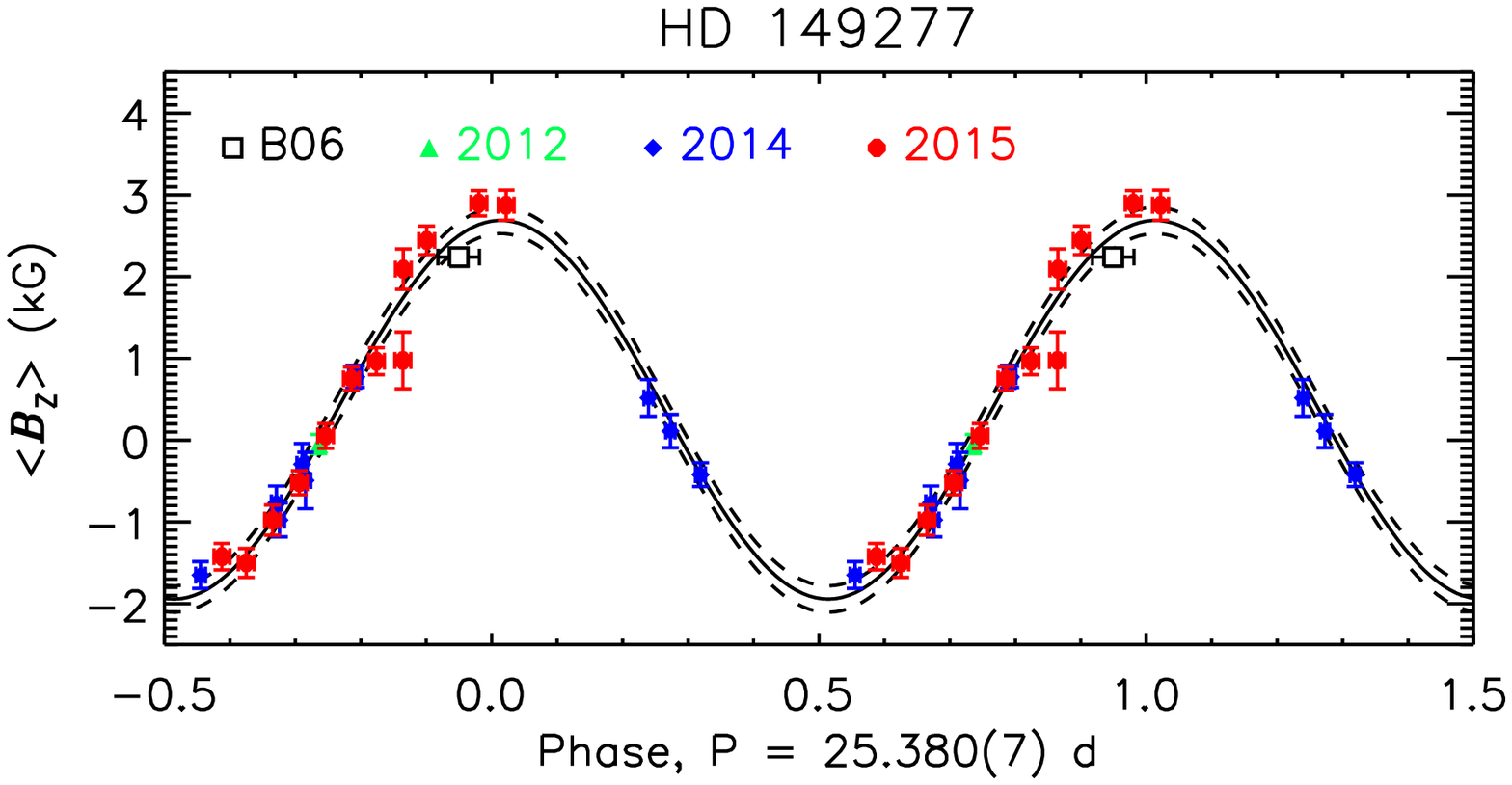} 
      \caption[Periodogram and \bz~for HD 149277.]{As Fig.\ \ref{HD25558_prot}, for HD 149277.}
         \label{HD149277_prot}
   \end{figure}

\noindent {\bf HD 149277}: The strongest peak in the periodogram is at 25.380(7)~d (Fig.\ \ref{HD149277_prot}, top), with a FAP of $9\times10^{-5}$, as compared to a minimum FAP in the \nz~periodogram of 0.02. This period was determined at a lower precision using the ESPaDOnS data obtained before 2015, and with only slight modification successfully phases the newer ESPaDOnS data obtained in 2015, as well as the single FORS1 measurement reported by \cite{bagn2006}. We therefore consider this to be the correct rotation period with high confidence. \bz~is shown phased with this period in Fig.\ \ref{HD149277_prot} (bottom). 

   \begin{figure}
   \centering
   \includegraphics[width=8.5cm]{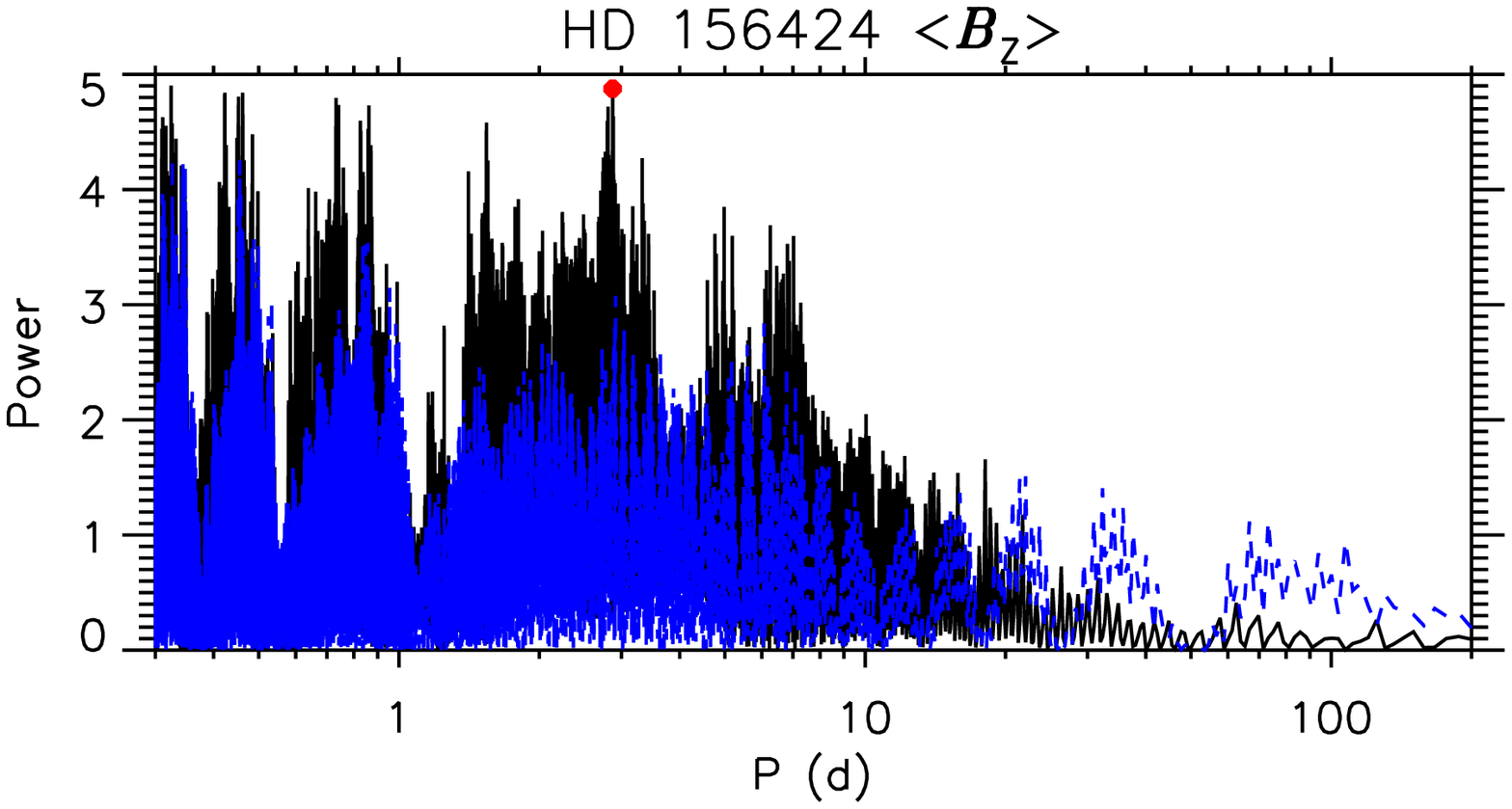} 
   \includegraphics[width=8.5cm]{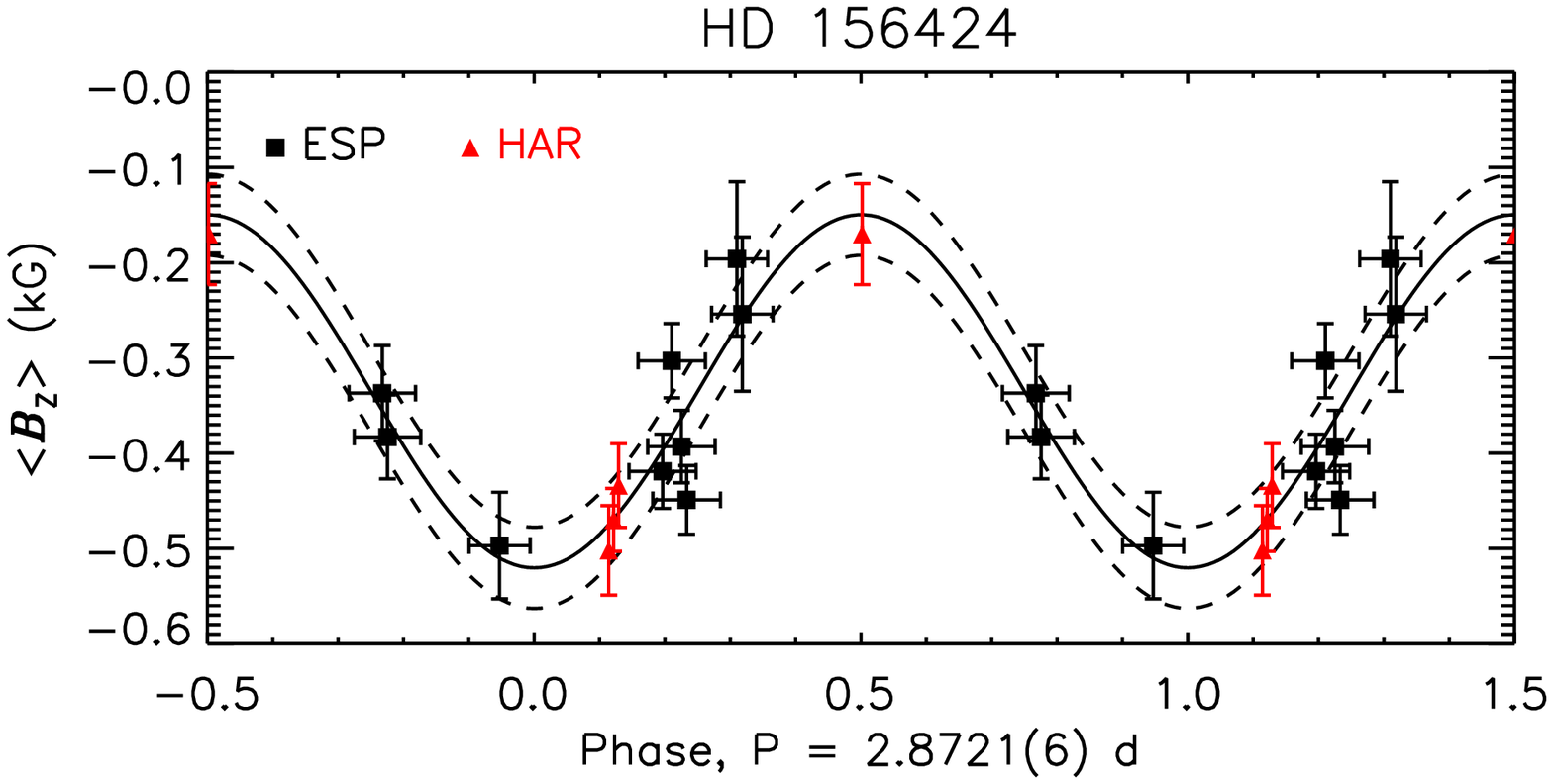}  
      \caption[Periodogram and \bz~for HD 156424.]{As Fig.\ \ref{HD25558_prot}, for HD 156424.}
         \label{HD156424_prot}
   \end{figure}

\noindent {\bf HD 156424}: as this star shows weak emission in the high-velocity wings of its H$\alpha$ line, presumably originating in its magnetosphere, we conducted period analyses using both the magnetic data and H$\alpha$ EWs, with the EWs supplemented by FEROS spectroscopy. The variation in either case is small compared to the error bar. Almost all the H$\alpha$ variation is between datasets, suggesting that it is a consequence of systematic differences between instruments. There is very little power in either periodogram at periods longer than a few days, suggesting that the rotation period must be fairly short despite the star's low \vsini~(see Fig.\ \ref{HD156424_prot}, top). The largest-amplitude peak in the \bz~periodogram without a corresponding peak in the \nz~periodogram is at 2.8721(6)~d. \bz~is shown phased with this period in the bottom panel of Fig.\ \ref{HD156424_prot}. It should be noted that the FAP of the highest peaks in the \bz~period spectrum are around 0.06, not much lower than the minimum in \nz, 0.15.



   \begin{figure}
   \centering
   \includegraphics[width=8.5cm]{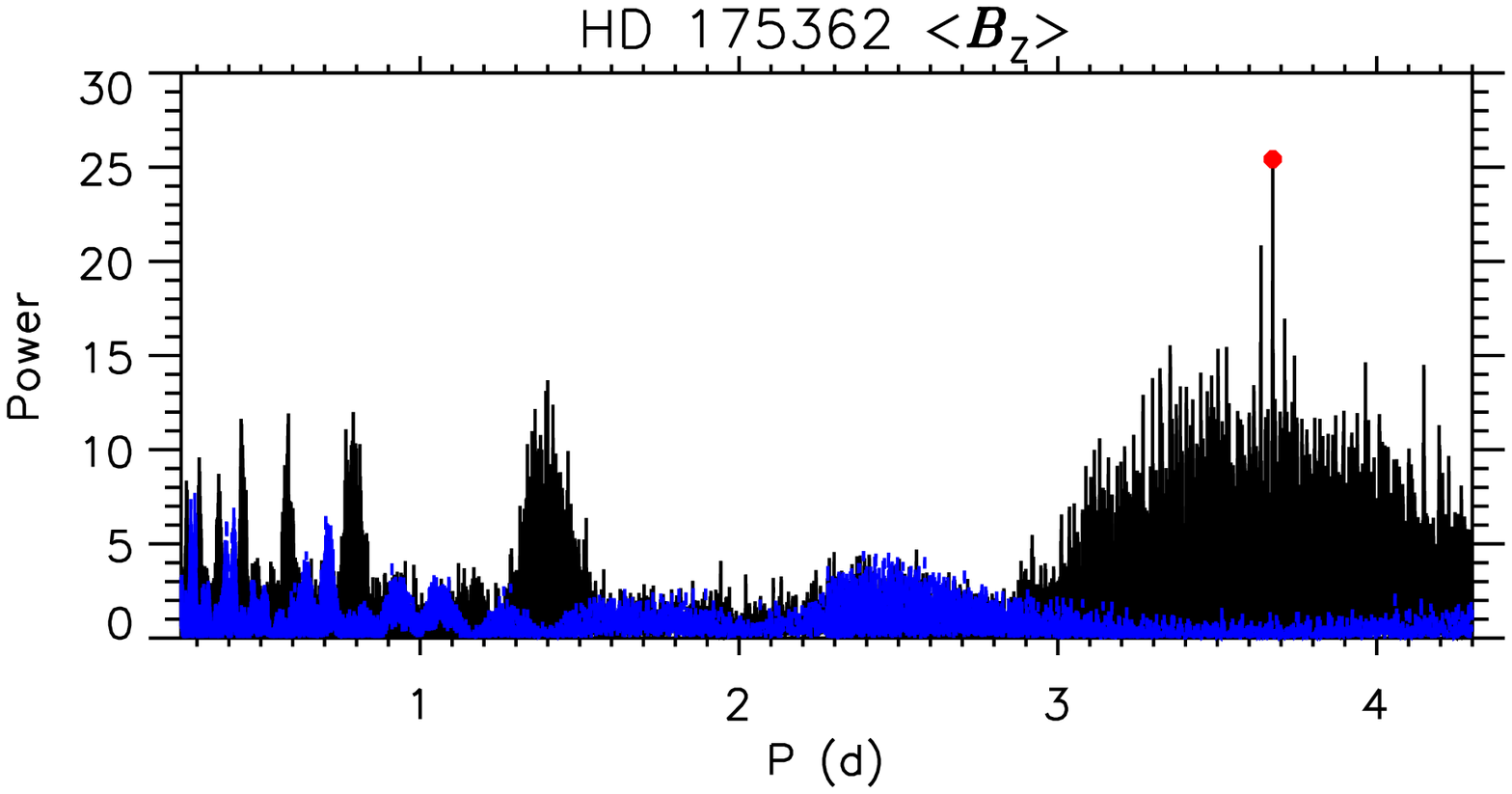} 
   \includegraphics[width=8.5cm]{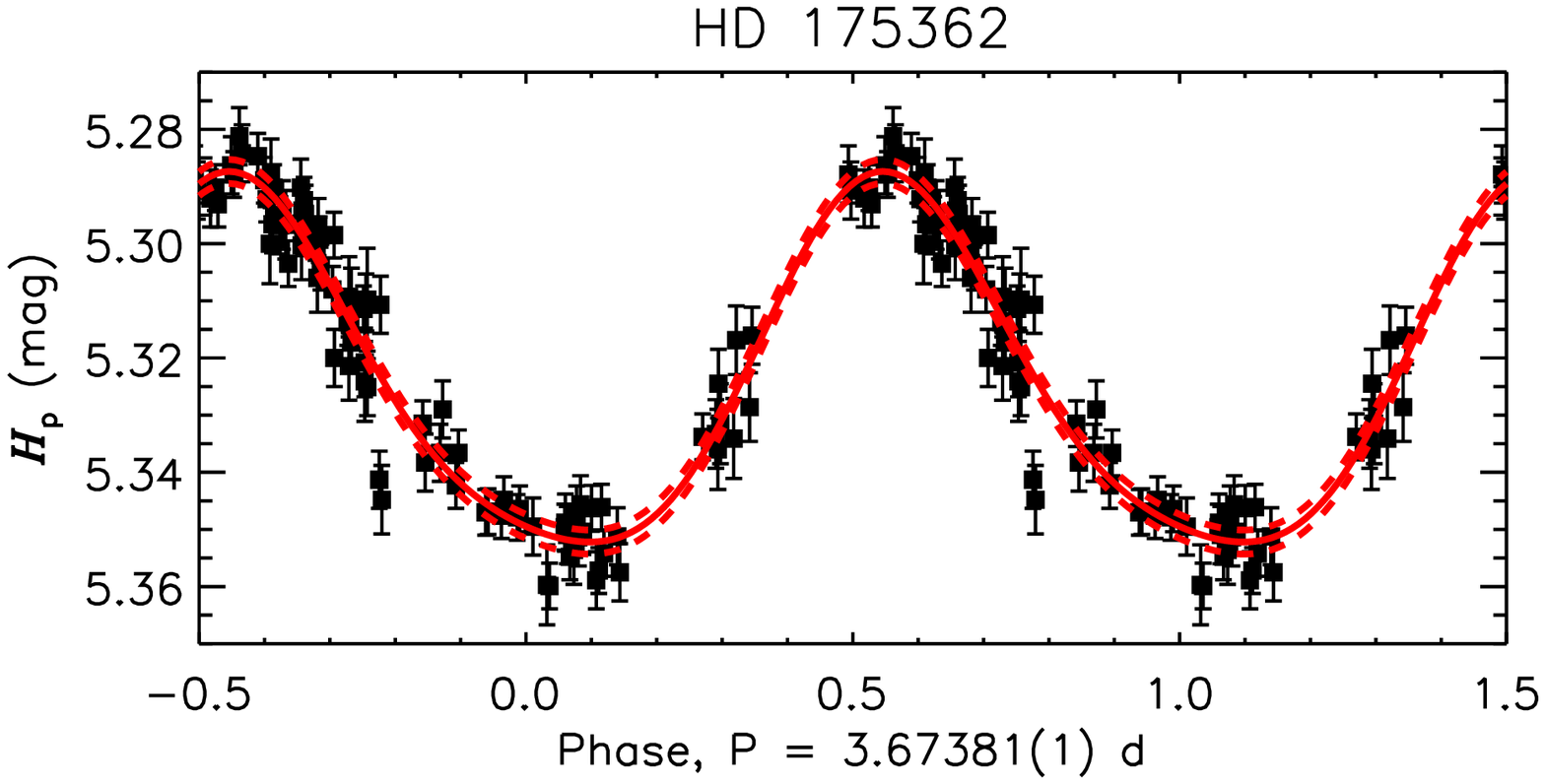}
   \includegraphics[width=8.5cm]{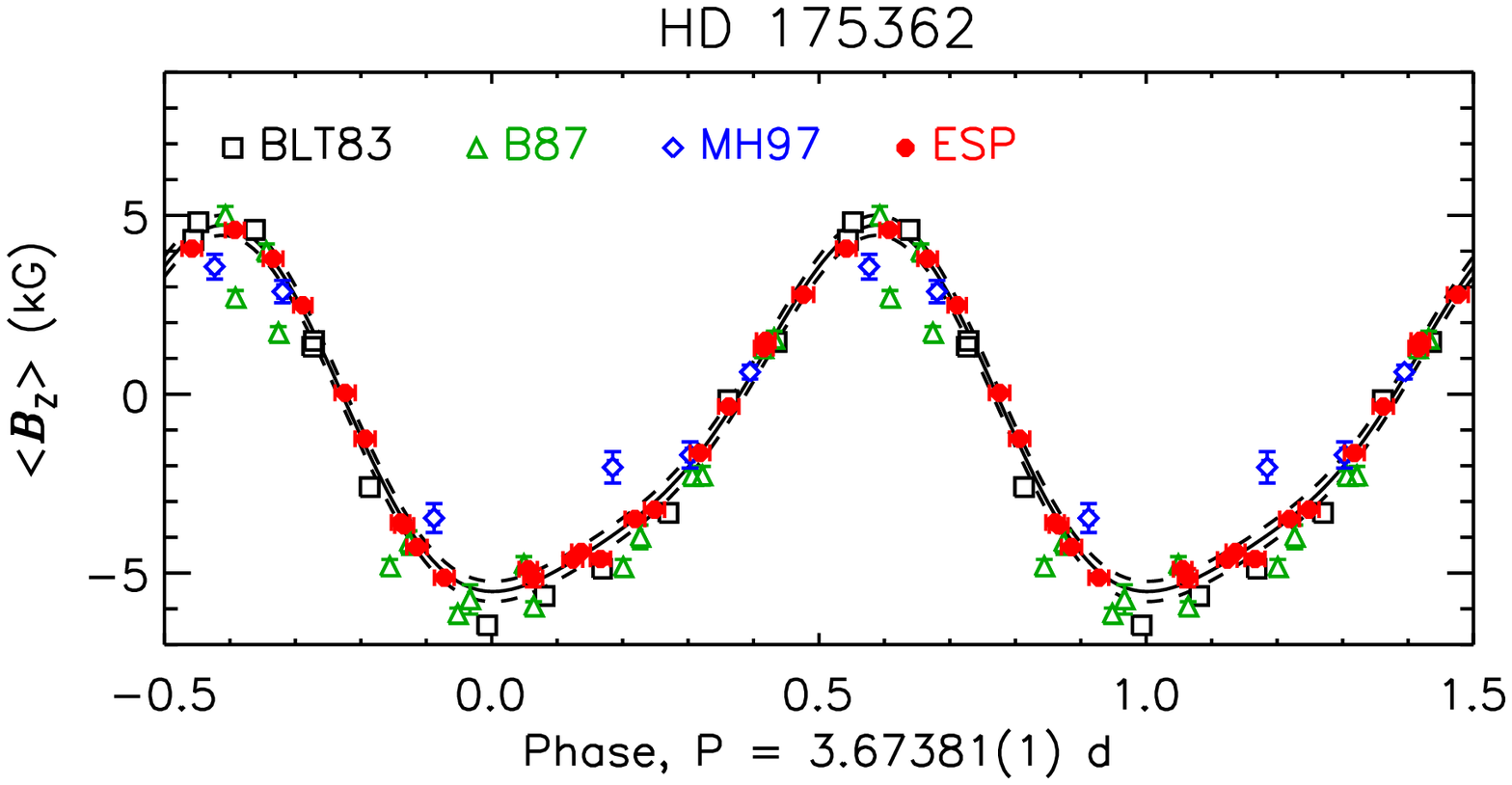} 
      \caption[Periodogram and \bz~for HD 175362.]{As Fig.\ \ref{HD37017_prot} for HD 175362. The sinusoidal fit to \bz~is of $3^{rd}$ order.}
         \label{HD175362_prot}
   \end{figure}

\noindent {\bf HD 175362}: \cite{1987ApJ...323..325B} determined $P_{\rm rot} = 3.6738(4)$~d for this star. Period analysis of the Hipparcos photometry yields a slightly less precise, but compatible period of $3.6732(6)$~d. Combining historical \bz~measurements with ESPaDOnS \bz~measurements enables us to refine the period to $3.67381(1)$~d (Fig.\ \ref{HD175362_prot}, top). \bz~and $H_{\rm p}$ are shown phased with this period in the middle and bottom panels of Fig.\ \ref{HD175362_prot}. Maximum light occurs at the positive magnetic pole. This corresponds well to the He EWs, which are at a minimum at \bz$_{\rm max}$, and a maximum and \bz$_{\rm min}$. 

   \begin{figure}
   \centering
   \includegraphics[width=8.5cm]{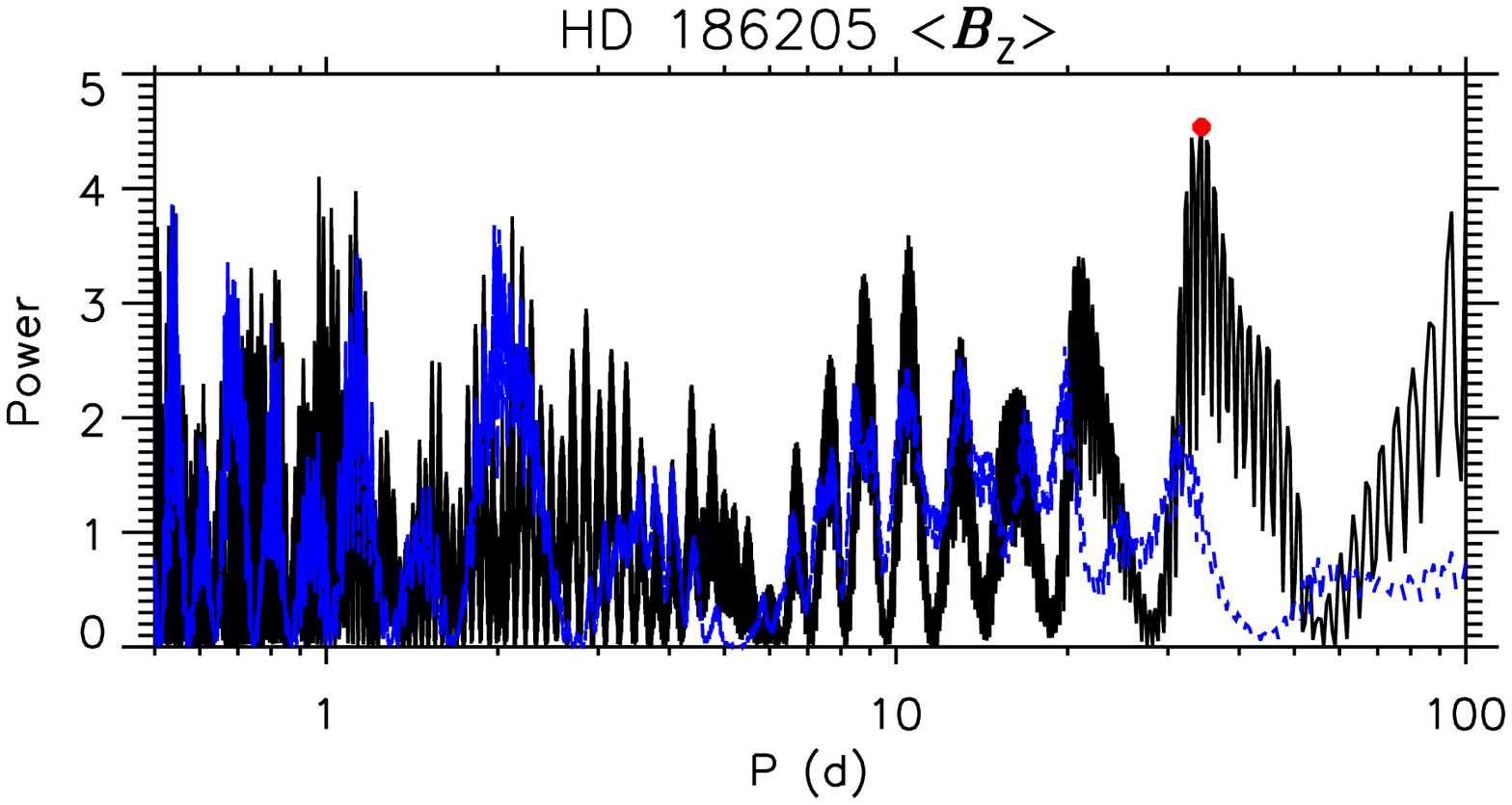} 
   \includegraphics[width=8.5cm]{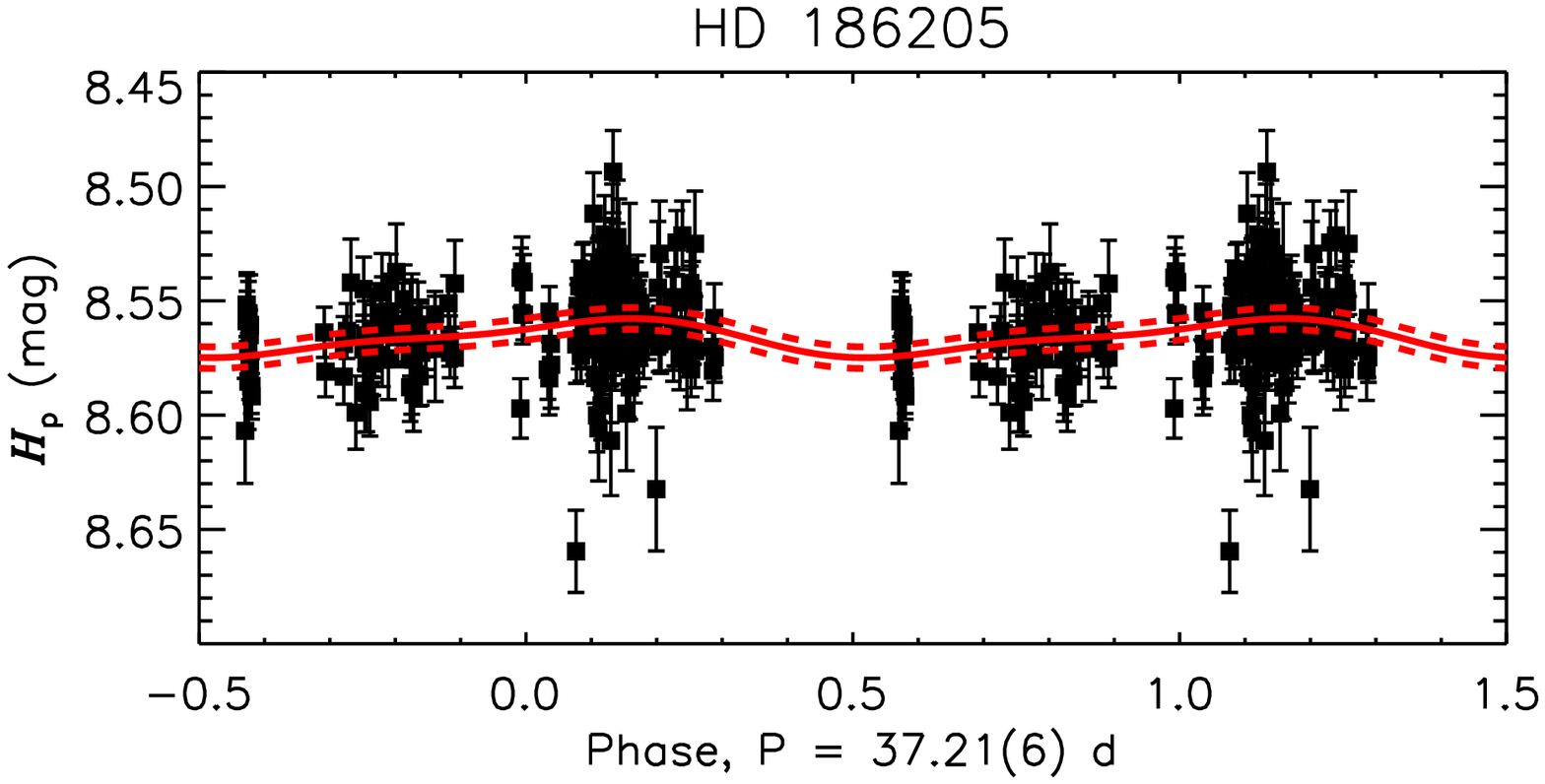}
   \includegraphics[width=8.5cm]{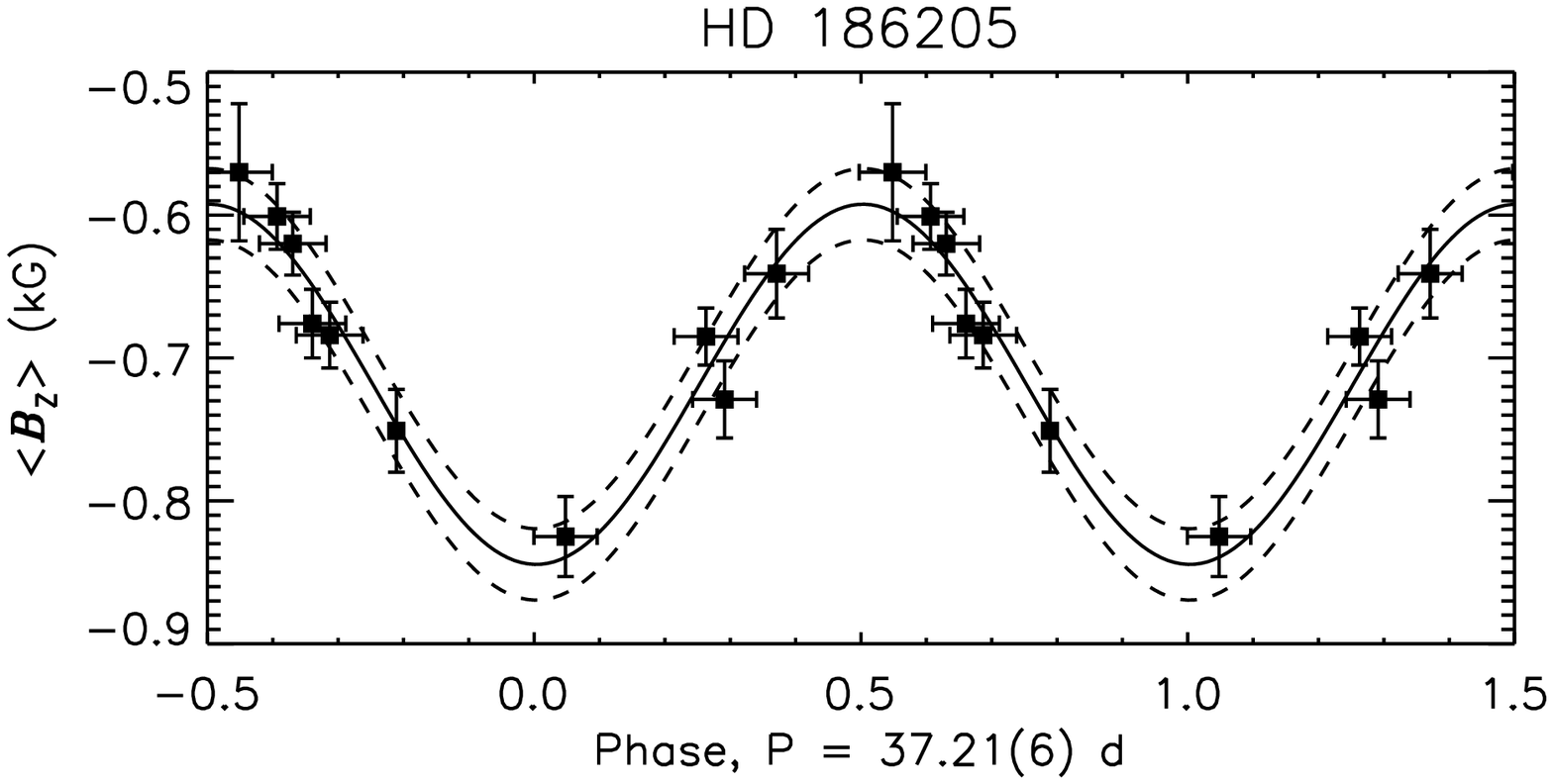} 
      \caption[Periodogram and \bz~for HD 186205.]{As Fig.\ \ref{HD37017_prot} for HD 186205.}
         \label{HD186205_prot}
   \end{figure}

\noindent {\bf HD 186205}: the strongest peak in the \bz~periodogram is at 37.21(6)~d (Fig.\ \ref{HD186205_prot}, top), with a FAP of 0.12 (only slightly lower than the minimum FAP in \nz, 0.20). There is very little power at either much higher or much lower periods, and the long period is consistent with the low \vsini. Furthermore, there is no power in the \nz~periodogram near this period. However, there are numerous nearby peaks which phase the data approximately as well. \bz~is shown phased with this period in Fig.\ \ref{HD186205_prot} (bottom). The semi-amplitude of the star's \bz~variation, $\sim$120~G, is much less than the median error bar in the dimaPol \bz~measurements of 343~G, therefore the dimaPol data could not be used to help constrain $P_{\rm rot}$. The star's He lines show very little variability; consistent with this, the Hipparcos light curve is dominated by noise (middle panel of Fig.\ \ref{HD186205_prot}), and also could not be used to constrain $P_{\rm rot}$. 

   \begin{figure}
   \centering
   \includegraphics[width=8.5cm]{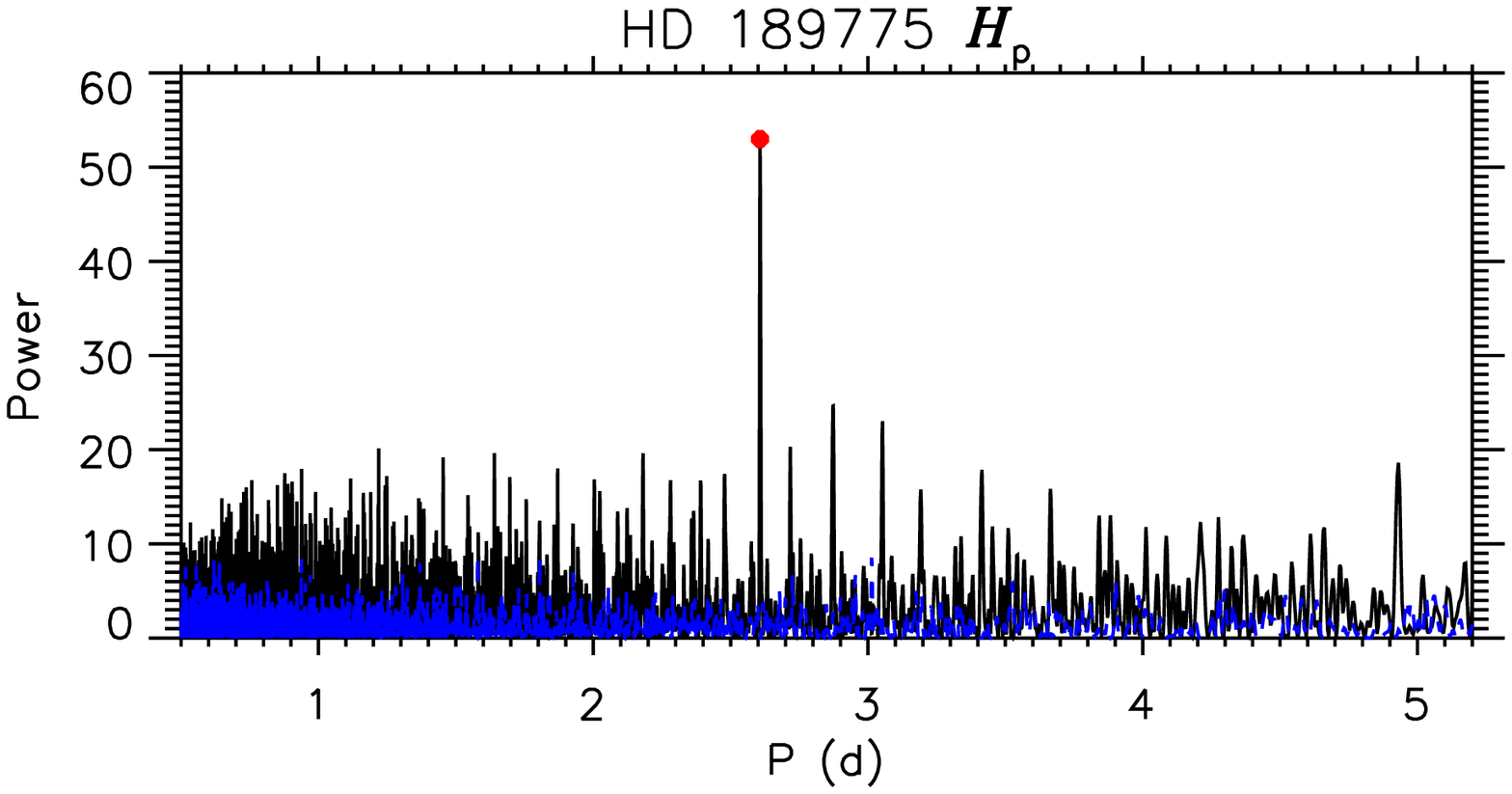}
   \includegraphics[width=8.5cm]{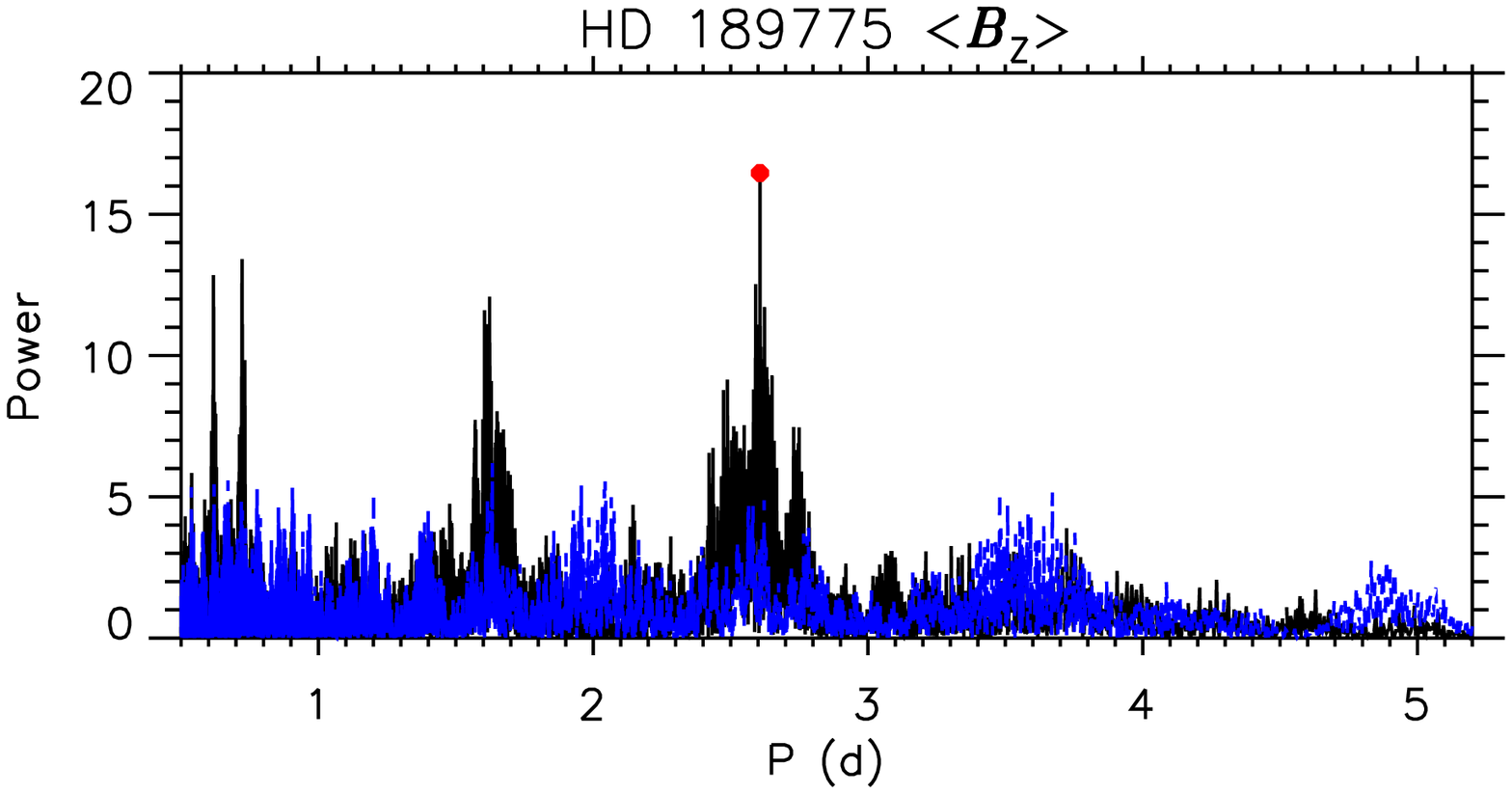} 
   \includegraphics[width=8.5cm]{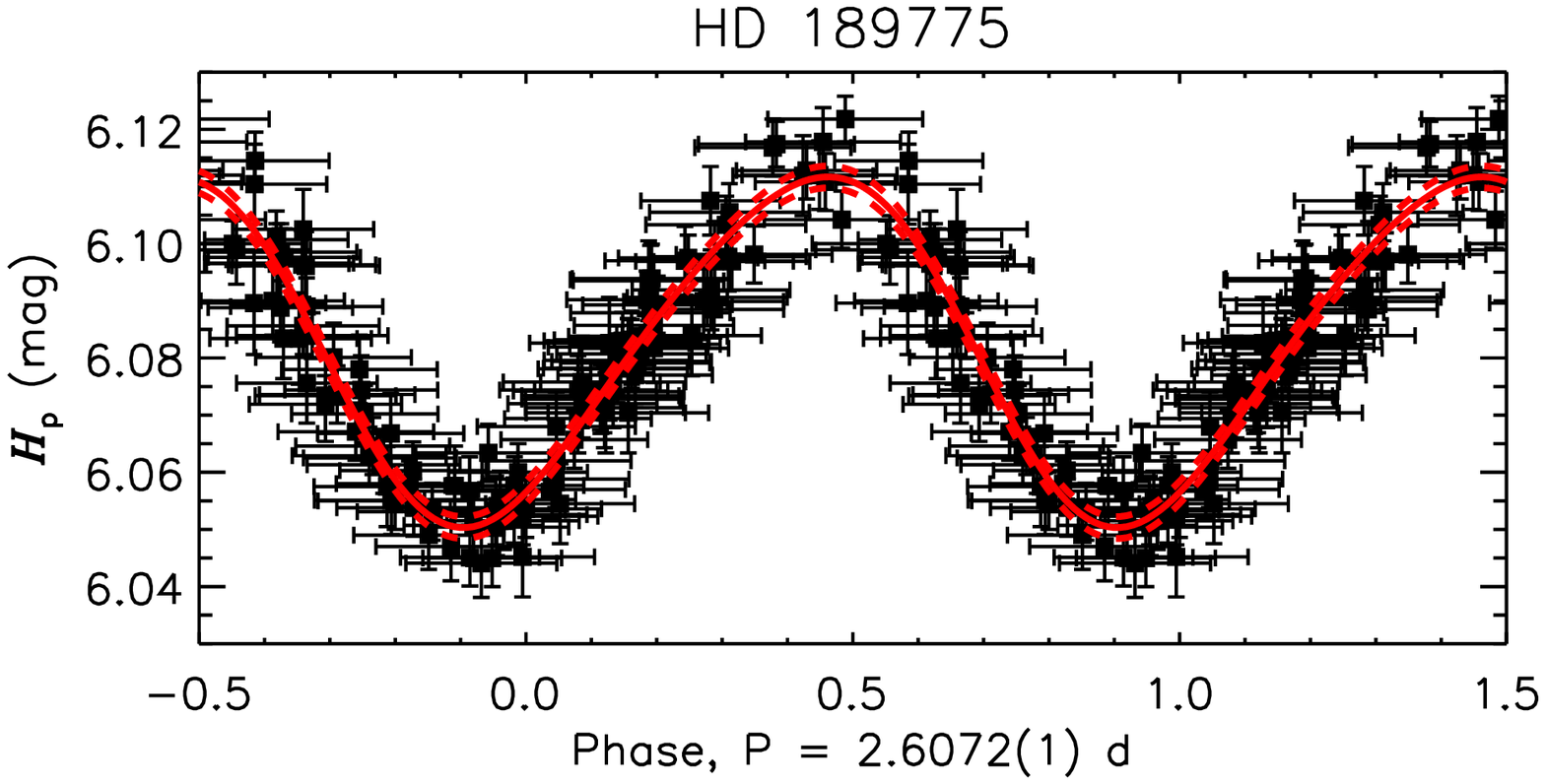}
   \includegraphics[width=8.5cm]{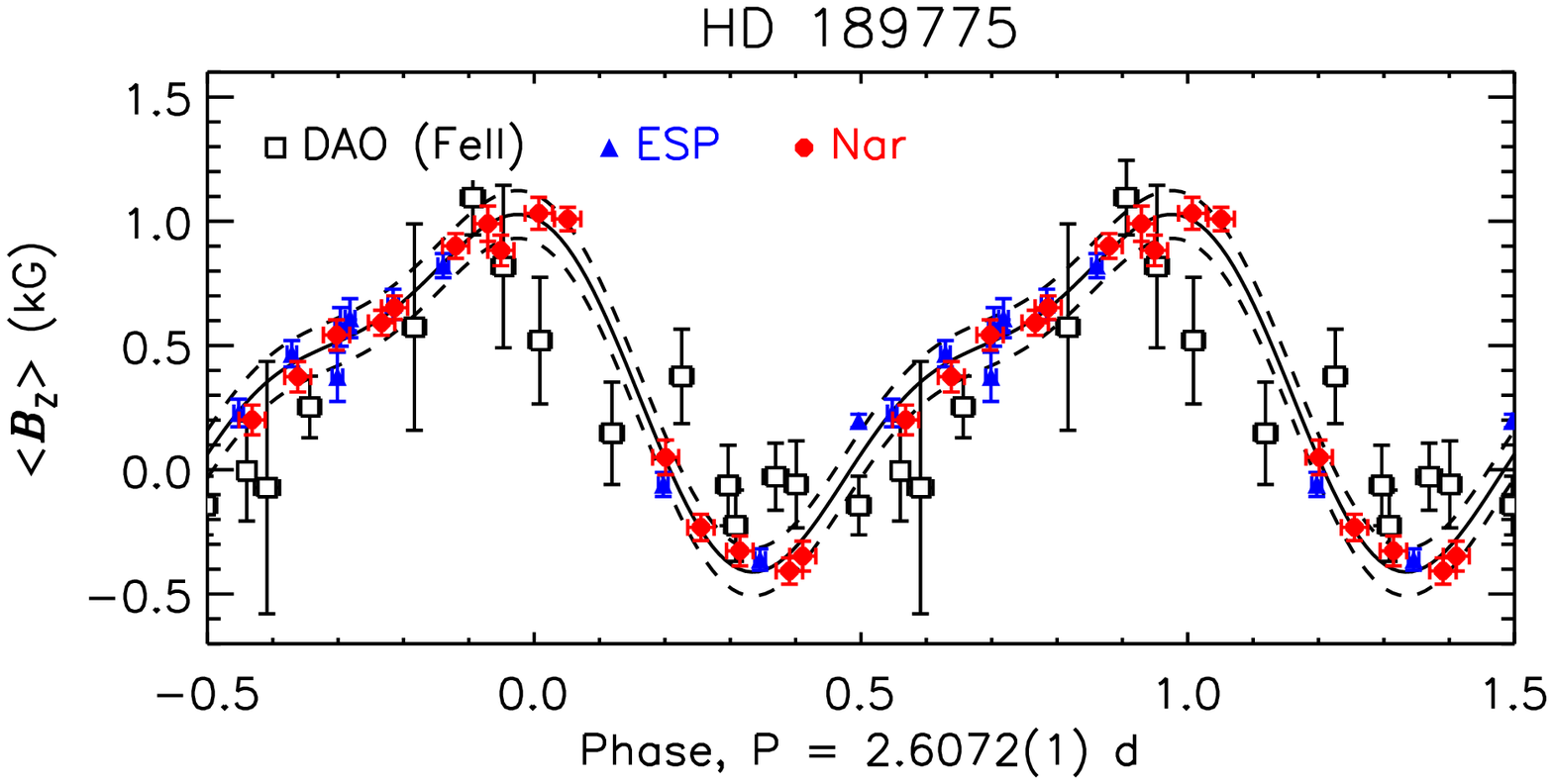}  
      \caption[Periodogram and \bz~for HD 189775.]{As Fig.\ \ref{HD35298_prot}, for HD 189775. The sinusoidal fit to \bz~is of $2^{nd}$ order.}
         \label{HD189775_prot}
   \end{figure}

\noindent {\bf HD 189775}: \cite{petit2013} reported a period of 2.6048~d for this star, based on unpublished magnetic measurements collected with dimaPol. In addition to H$\beta$~\bz~measurements, there are also \bz~measurements conducted using the Fe~{\sc ii} 492.3 nm line. As the H$\beta$~\bz~measurements show significantly more scatter than the Fe~{\sc ii} measurements, we adopt the latter. Hipparcos photometry shows a single peak at 2.6074(3)~d. ESPaDOnS and Narval measurements yield a best-fit period of 2.6071(3)~d. For the combined magnetic data we find 2.6072(1)~d, as shown in the second panel of Fig.\ \ref{HD189775_prot}. The FAP of this peak in the periodogram is $3\times10^{-7}$, lower than the FAP of the highest peak in \nz, 0.1. The Hipparcos light curve and \bz~measurements are shown phased with this period in Fig.\ \ref{HD189775_prot}. A second-order sinusoid was used to fit the magnetic data, as this star's \bz~curve shows evidence of anharmonic behaviour that may be consistent with contributions from non-dipolar magnetic field components. Photometric extrema are somewhat offset from the magnetic extrema, suggesting that the chemical spots do not coincide precisely with the magnetic poles. 

   \begin{figure}
   \centering
\begin{tabular}{c}
   \includegraphics[width=8.5cm]{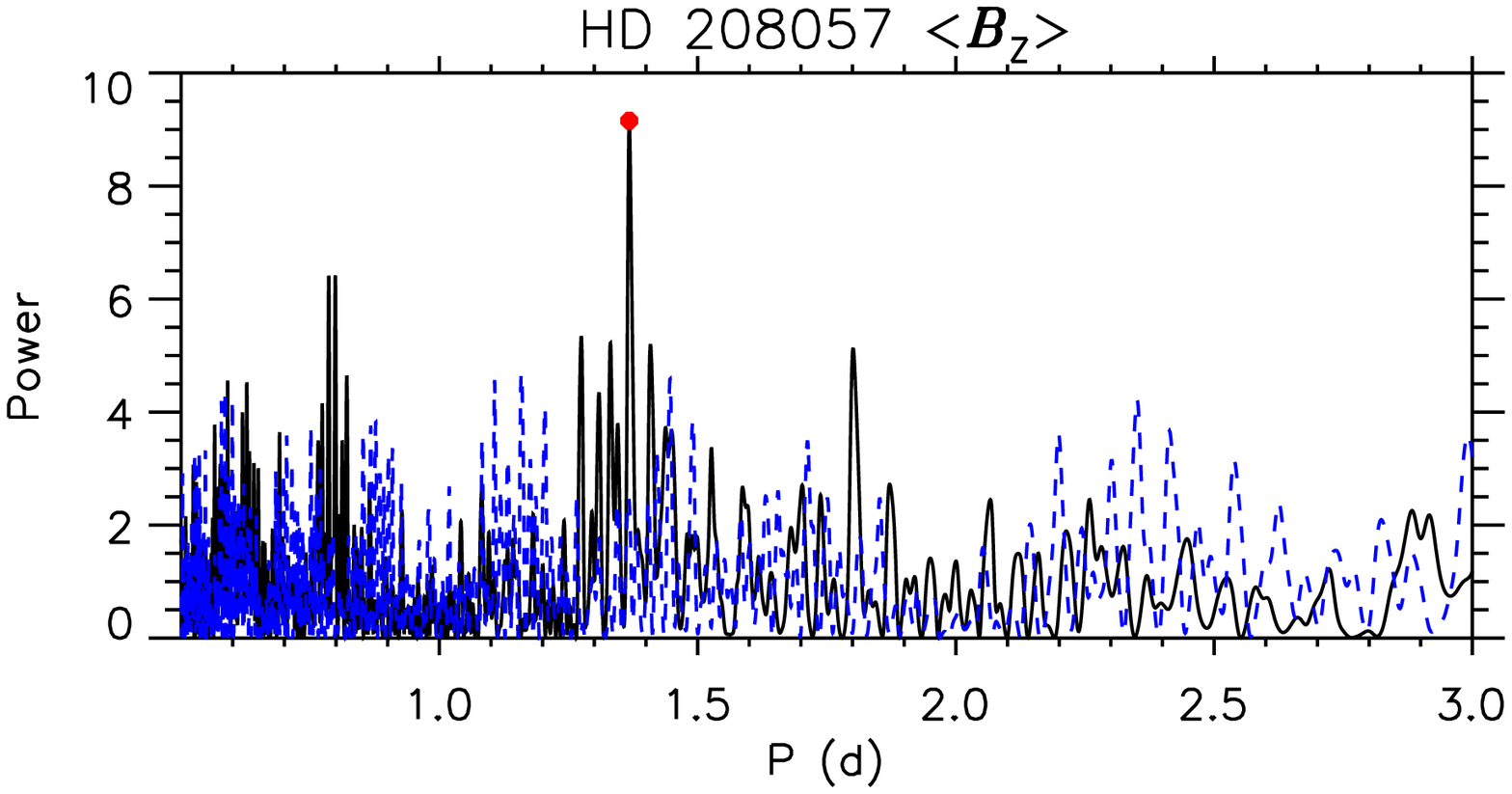} \\
   \includegraphics[width=8.5cm]{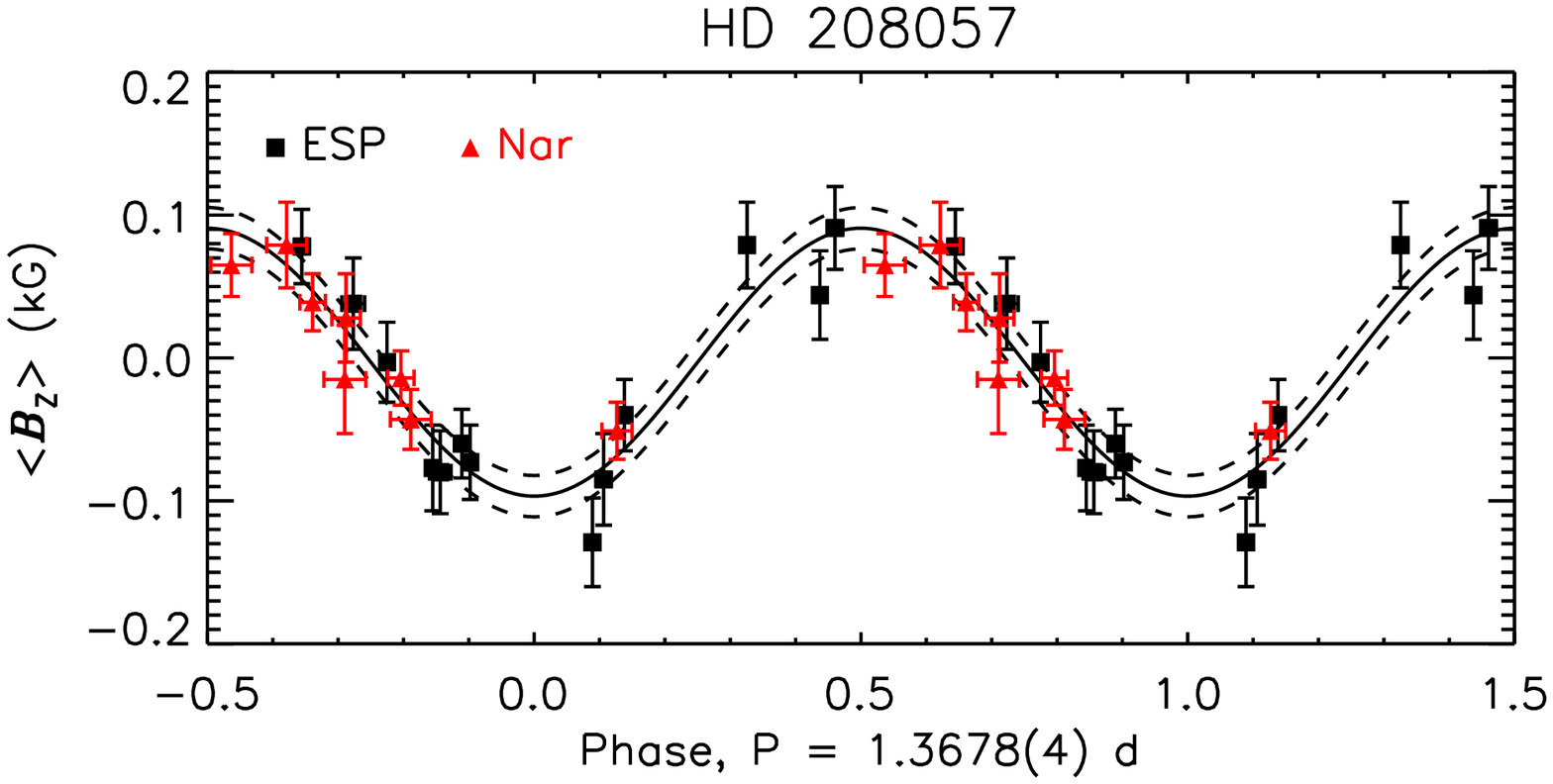} \\
\end{tabular}
      \caption[Periodogram and \bz~for HD 208057.]{As Fig.\ \ref{HD25558_prot} for HD 208057.}
         \label{HD208057_prot}
   \end{figure}

\noindent {\bf HD 208057}: \cite{2009IAUS..259..393H} reported $P_{\rm rot} = 1.441$ d. Using the same data, we find maximum power at 1.3678(4)~d (Fig.\ \ref{HD208057_prot}, top). The light curve is dominated by pulsations, so the Hipparcos photometry could not be used to determine $P_{\rm rot}$. However, it is worth noting that there is no power in the \bz~periodogram at the pulsation periods of 1.23~d and 1.25~d \citep{2015MNRAS.450.1585S}. The FAP of this peak is $7\times10^{-4}$, subtstantially smaller than the minimum FAP in the \nz~period spectrum of 0.12. \bz~is shown phased with this period in Fig.\ \ref{HD208057_prot}. 

\begin{figure}
    \includegraphics[width=8.5cm]{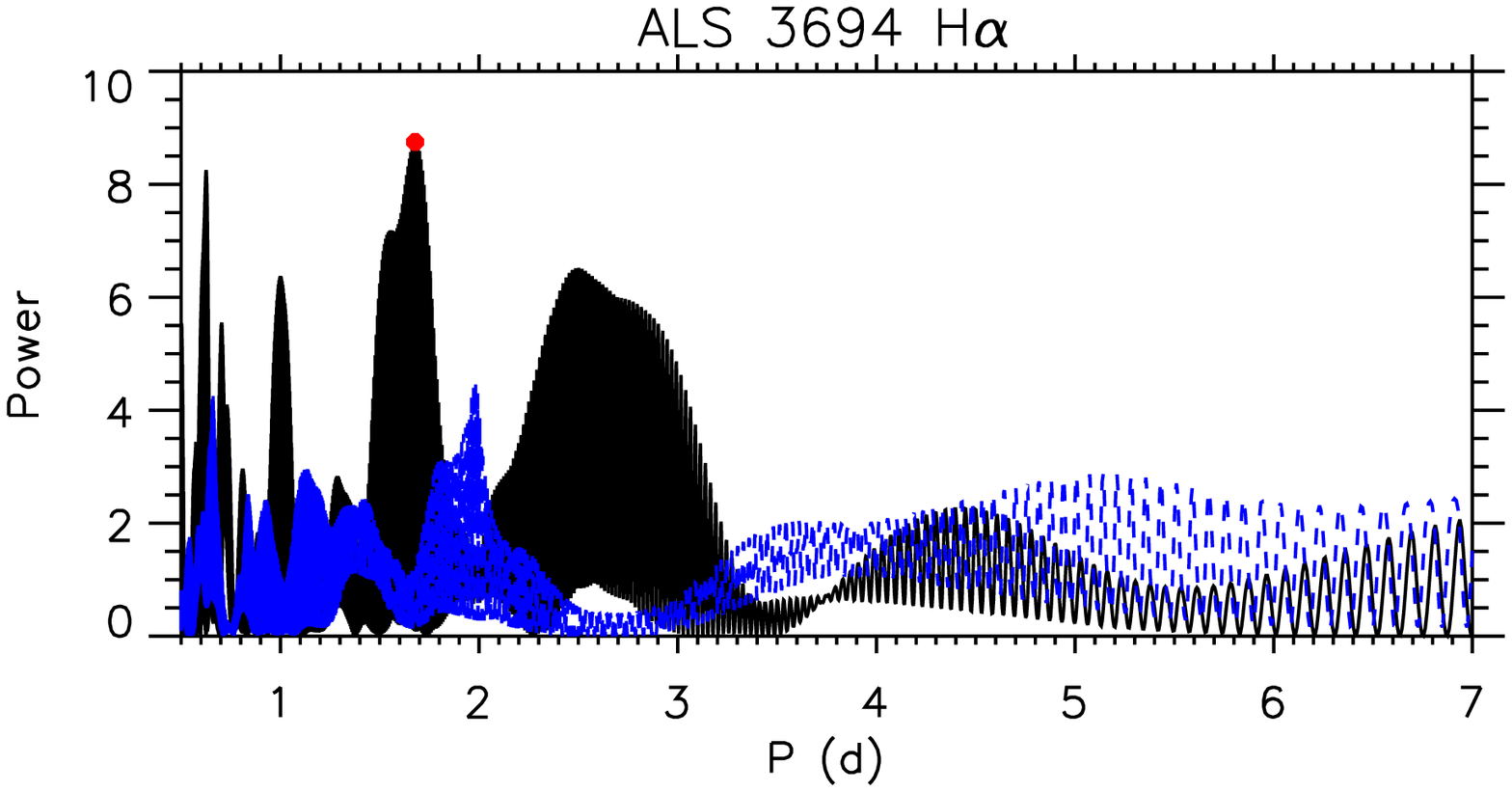}
    \includegraphics[width=8.5cm]{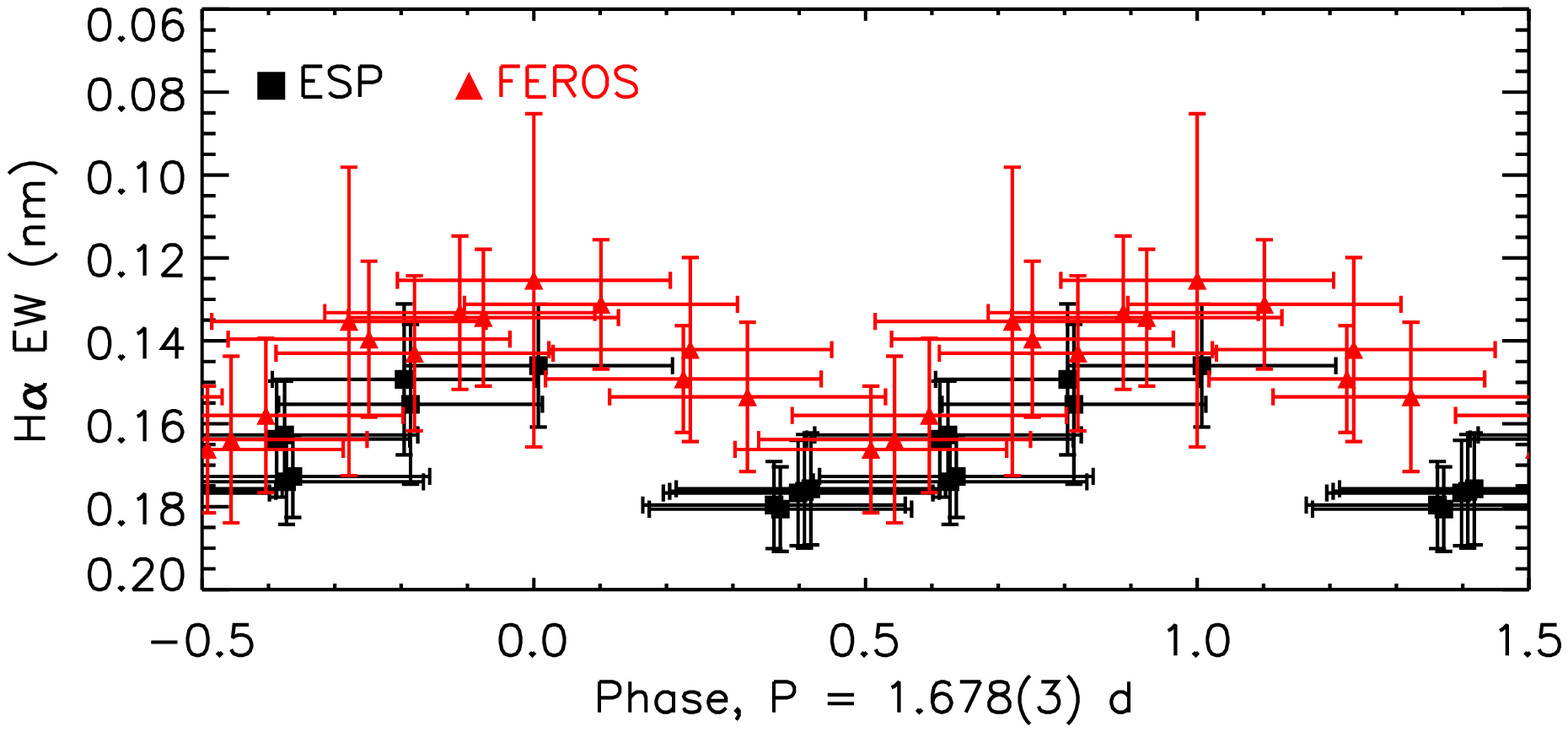}
    \includegraphics[width=8.5cm]{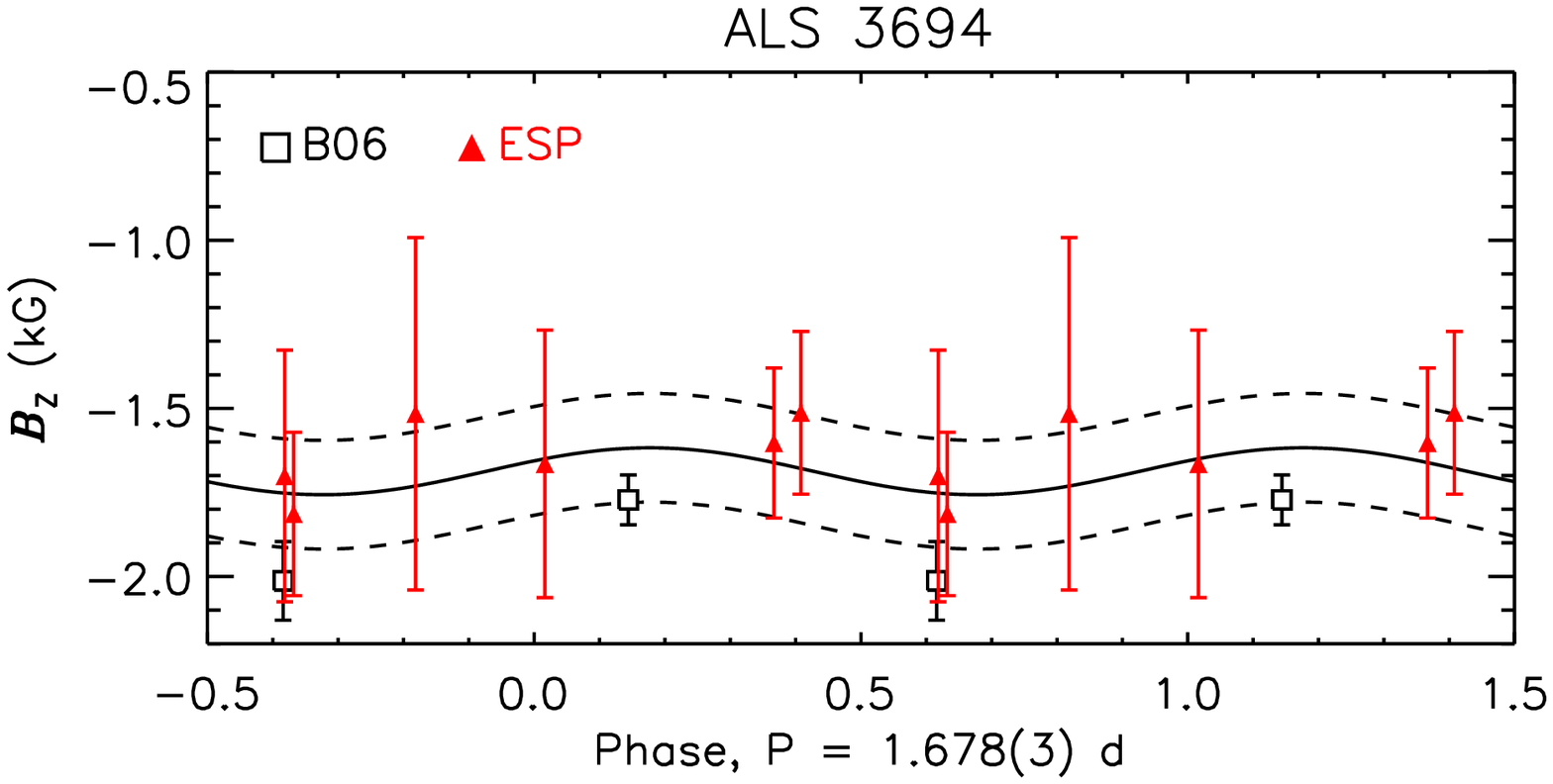}
      \caption[Periodogram, H$\alpha$ EWs, and \bz~for ALS 3694.]{{\em Top}: periodogram for H$\alpha$ EWs of ALS 3694 (solid black line) and synthetic null measurements (dashed blue line). The adopted period is indicated by the red circle. {\em Middle}: H$\alpha$ EWs phased with $P_{\rm rot}$. {\em Bottom}: \bz~phased with $P_{\rm rot}$.}
 \label{ALS3694_prot}
\end{figure}


\noindent {\bf ALS 3694}: no period can be determined from the magnetic data, as $\Sigma_B$ is too low for any statistically significant variation to be detected. However, this star displays H$\alpha$ emission consistent with an origin in a centrifugal magnetosphere \citep{2016ASPC..506..305S}. Combining H$\alpha$ EWs measured from ESPaDOnS and FEROS spectra yields a periodogram with a single strong peak at 1.678(3)~d (Fig.\ \ref{ALS3694_prot}, top). The FAP of this peak is 0.001, much lower than the minimum FAP from the null periodogram, 0.43.  H$\alpha$ EWs and \bz~measurements are shown phased with this period in the middle and bottom panels of Fig.\ \ref{ALS3694_prot}. Minimum H$\alpha$ EW (i.e., maximum emission strength) was used to define JD0, as this should correspond to maximum \bz~(e.g. \citealt{town2005c}). While a sinusoidal variation is not distinguishable in \bz, the magnitude of the ESPaDOnS \bz~measurements is compatible with those published by \cite{bagn2006}.

\end{document}